\pdfoutput=1
\documentclass[pdflatex,epsf,useAMS,usenatbib,usegraphicx]{mn2e}
\usepackage{pdfpages}
\usepackage{epstopdf}
\usepackage{epsfig}  
\usepackage{amssymb, amsmath, graphics}
\usepackage{ulem}
\usepackage{graphicx}
\usepackage{mathrsfs}
\usepackage{nicefrac}
\usepackage{verbatim}
\usepackage{aas_macros}  
\usepackage{longtable}
\usepackage{pdflscape}
\usepackage{appendix}
\usepackage{booktabs}
\usepackage{upgreek}
\usepackage[flushleft]{threeparttable}
\usepackage[percent]{overpic}

\title[SX\,Phe Stars in the Kepler Field ]
{Metal-Rich SX\,Phe stars in the {\it Kepler} Field }

\author[ Nemec, Balona, Murphy {\it et al.} ]
{James\,M.\,Nemec$^{1,2}$\thanks{E-mail: nemec@camosun.ca},
Luis\,A.\,Balona$^{3}$, Simon\,J.\,Murphy$^{4}$, Karen\,Kinemuchi$^{5}$,
\and
and Young-Beom\,Jeon$^{6}$ 
\\
\\
$^{1} $Department of Physics \& Astronomy, Camosun College, Victoria, British Columbia, V8P\,5J2, Canada\\
$^{2} $International Statistics \& Research Corp., Brentwood Bay, British Columbia, V8M\,1R3 Canada \\
$^{3} $South African Astronomical Observatory, PO Box 9, Observatory 7935, Cape Town,  South Africa\\
$^{4} $Sydney Institute for Astronomy (SIfA), School of Physics, The University of Sydney, NSW 2006, Australia\\ 
$^{5} $Apache Point Observatory, Sunspot, New Mexico 88349, USA\\
$^{6} $Korea Astronomy and Space Science Institute, Daejeon, 34055, Korea }

\date{Accepted 2016 November 23. 
      Received 2016 October 29;  
      in original form 2016 August 29}
  
\begin{document}  
  
\maketitle


\begin{abstract} A spectroscopic and photometric analysis has been carried out
for thirty-two {\it candidate} SX\,Phe variable blue straggler stars in the
{\it Kepler}-field (Balona \& Nemec 2012).  Radial velocities (RVs), space
motions ($U,V,W$), projected rotation velocities ($v$\,sin\,$i$), spectral
types, and atmospheric characteristics ($T_{\rm eff}$, $\log g$, [Fe/H],
$\xi_t$, $\zeta_{\rm RT}$, etc.) are presented for 30 of the 32 stars.
Although several stars are metal-weak with extreme halo orbits, the mean [Fe/H]
of the sample is near solar, thus the stars are more metal-rich than expected
for a typical sample of Pop.\,II stars, and more like halo metal-rich A-type
stars (Perry 1969).  Two thirds of the stars are fast rotators with
$v$\,sin\,$i$ $>$ 50~km/s, including four stars with $v$\,sin\,$i$ $>$ 200
km/s.   Three of the stars have (negative) RVs $>$ 250 km/s, five have
retrograde space motions, and 21 have total speeds (relative to the LSR) $>$
400 km/s.  All but one of the 30 stars have positions in a Toomre diagram
consistent with  the kinematics of {\it bona fide} halo stars (the exception
being a thick-disk star).  Observed R{\o}mer time delays,  pulsation frequency
modulations and light curves suggest that at least one third of the stars are
in binary (or triple) systems with orbital periods ranging from 2.3 days to
more than four years.  \end{abstract}


\section{INTRODUCTION}

SX\,Ph\oe necis stars are the Pop.\,II counterparts of  Pop.\,I $\delta$\,Scuti
pulsating variable stars (see Breger 1980; Eggen \& Iben 1989; Nemec 1989;
Nemec \& Mateo 1990a,b).   The best known and nearest examples are the stars
SX~Phe, DY~Peg, BL~Cam and KZ~Hya.  These and other such field SX\,Phe stars
({\it i.e.}, those not in star clusters) generally have the kinematics of halo
(or thick disk) stars, asymmetric and large-amplitude light curves, and low
metallicities (Eggen 1970, 1979; Breger 1975, 1977a,b; McNamara {\it et al.}
1978, 2007).  For many years SX\,Phe stars were suspected of being Pop.\,II
blue stragglers (BSs), but this was established only when short-period variable
stars were found among the BSs in globular clusters (Niss {\it et al.} 1981;
J{\o}rgensen 1982; J{\o}rgensen \& Hansen 1984; Jensen \& J{\o}rgensen 1985;
Nemec \& Harris 1987; DaCosta, Norris \& Villumsen 1986; Nemec 1989; Nemec \&
Cohen 1989;    Mateo {\it et al.} 1990;  Nemec {\it et al.} 1994, 1995;
Sarajedini 1993; Fusi Pecci {\it et al.} 1992;   McNamara 1997; Gilliland {\it
et al.} 1998;  Rodr\'iguez \& L\'opez-Gonz\'alez 2000; Bruntt {\it et al.}
2001).  The leading theory of the formation mechanism for BSs involves mass
transfer in a close binary system (Hoyle 1964, McCrea 1964, Eggen \& Iben 1989,
Leonard 1989).  Since many BSs are now known to be eclipsing binaries (Niss
{\it et al.} 1978; Margon \& Cannon 1989; Mateo {\it et al.} 1990;  Nemec \&
Mateo 1990a,b; Hobbs \& Mathieu 1991; Hodder {\it et al.} 1992;  Kallrath {\it
et al.} 1992; Helt {\it et al.} 1993; Yan \& Mateo 1994;  Nemec {\it et al.}
1995;  Kaluzny {\it et al.} 1996, 2007; Kaluzny 2000; Park \& Nemec 2000;
Preston \& Sneden 2000; Carney {\it et al.} 2001, 2005;  Cohen \& Sarajedini
2012) it follows that SX~Phe stars are also likely to have a binary (or triple
star) nature.

When the first globular cluster SX~Phe stars were discovered they were found to
have pulsation amplitudes, $A_V$, larger than 0.10 mag, and  simple light
curves consistent with radial pulsation and  one or two dominant periods.
Their location in the faint extension of the Cepheid instability strip implies
pulsation due to Eddington's $\kappa$-mechanism operating in the He\,II partial
ionization zone (see Fiorentino {\it et al.} 2014, 2015).  They also were found
to obey a well-defined period-luminosity relationship that can be used to
estimate distances (see Nemec {\it et al.} 1994;  McNamara 1997, 2011; Petersen
\& Hog 1998; Sandage \& Tammann 2006).   Because globular clusters (and Local
Group dwarf galaxies) tend to be distant, the SX~Phe stars in these systems are
quite faint.  Moreover,   the  BSs in globular clusters tend to be centrally
concentrated (Nemec \& Harris 1987; Nemec \& Cohen 1989; Leigh, Sills and
Knigge 2011; Ferraro {\it et al.} 2014) and as a consequence the SX\,Phe stars
are  often  located in the crowded central regions and therefore are much less
amenable to spectroscopic and photometric analysis than those in the field.  

Subsequent discoveries revealed that SX~Phe stars include  lower amplitude
pulsators and stars that exhibit more complex radial and non-radial
oscillations than previously thought to be the case.  Particularly noteworthy
was the observation that the $A_V$ amplitude distribution shown in figure\,2 of
Kaluzny (2000) shows an increase in  the number of SX~Phe stars down to the
detection limit ($\sim$0.02-0.03 mag) of the Warsaw variability surveys, from
which it was  concluded that ``a significant fraction of SX~Phe stars residing
in the survey clusters were most likely missed.'' Over the years many more
faint and pulsationally-complex SX\,Phe stars have been found in 47\,Tuc
(Gilliland {\it et al.} 1998), NGC~3201 (Mazur {\it et al.} 2003), NGC\,5466
(Jeon {\it et al.} 2004), $\omega$~Cen (Olech {\it et al.} 2005) and other
clusters.  The catalogue compiled  by Cohen \& Sarajedini (2012) lists
$\sim$250 SX~Phe stars in 46 galactic globular clusters (see their figure\,5).  

SX~Phe stars are now known to be  ubiquitous, with many having  been identified
in our Galaxy (Ramsey {\it et al.} 2011, Palaversa {\it et al.} 2013, Preston
2015), in the Magellanic Clouds (Soszynski {\it et al.} 2002, 2003) and in most
Local Group dwarf galaxies (Nemec \& Mateo 1990a; Mateo {\it et al.} 1998; Pych
{\it et al.} 2001; Poretti {\it et al.} 2008;   McNamara 2011;  Vivas \& Mateo
2013; Ferraro {\it et al.} 2014; Momany 2014; Fiorentino {\it et al.} 2014,
2015; Coppola {\it et al.} 2015; Mart\'inez-V\'azquez {\it et al.} 2016).
Despite these numerous discoveries discrimination between SX~Phe stars and
$\delta$~Sct stars can be confusing, and is particularly  blurred where
multiple and composite stellar populations have overlapping age, metallicity
and kinematic distributions. 

In 2012 Balona \& Nemec (hereafter BN12)  identified 34 {\it candidate} SX~Phe
stars  in the {\it Kepler} field.  The stars were found by cross-referencing a
list of 1424 {\it Kepler}-field $\delta$ Sct stars (Balona 2014a) with the
UCAC3 proper motion catalog (Zacharias {\it et al.} 2010) and  selecting those
stars with high proper motions, $\mu > 30$ mas/yr, and large tangential
velocities, $V_{\rm t} > 120$ km/s.  Several of the candidates are located more
than 0.5 kpc above the galactic plane thereby strengthening the conclusion that
the sample consists of Pop.\,II stars.  An H-R diagram showing the locations of
the 34 candidate SX~Phe stars (and 1554 $\delta$\,Sct stars) relative to
stellar evolutionary tracks for masses ranging from 0.8 to
2.0\,$\mathscr{M}_\odot$ was  plotted by BN12, where the luminosities and
effective temperatures were taken from the {\it Kepler} Input Catalog (KIC;
Brown {\it et al.} 2011).   Since the physical quantities given in the KIC were
derived mainly from photometric relations established for stars cooler then
7000\,K, the atmospheric parameters given in the KIC were not expected to be
particularly reliable.      BN12 found  that there was little to distinguish
the {\it Kepler} light curves of the candidate SX~Phe stars from those of
$\delta$~Sct stars.  Furthermore, whereas many {\it field} SX~Phe stars have
one or two dominant pulsation modes and rather high amplitudes, almost all of
the {\it Kepler}-field candidates were found to have relatively low amplitudes
and complex Fourier spectra.  Thus, BN12 concluded that previous ground-based
{\it field} SX~Phe star investigations probably suffer from a selection bias
that resulted  in the omission of  stars with the lowest amplitudes and most
complex pulsations.

The goal of the present study was to derive reliable estimates of  the physical
characteristics of the BN12 SX~Phe candidates.  Extensive asteroseismic investigations
using {\it Kepler} photometry have already been reported for two of the stars:
KIC\,11754974, an SX\,Phe pulsator in a 343-day non-eclipsing binary system
(Murphy {\it et al.} 2013b); and KIC\,9244992, an SX~Phe star in which both the
surface and core exhibit slow, nearly-uniform rotation (Saio {\it et al.}
2015).   Since the SX\,Phe stars in the {\it Kepler}-field are unique in having
four years of almost-continuous high-precision photometry, estimation of their
evolutionary status and atmospheric and physical characteristics based on
high-resolution spectra is possible and  obviously desireable.  Moreover, if
the stars are Pop.\,II BSs then determining whether  they are binary or triple
systems (with a range of separations, including coalesced binaries) and
better characterizing their orbital properties becomes important.

In $\S2$ the results of the spectroscopic analyses are presented,  including
atmospheric properties, radial velocities,  space motions  and stellar
population classifications (eg., halo, thick disk).   In $\S$3 frequency
analyses are described for the complete BN12 sample;  the results based on all
the available long- and short-cadence Q0-Q17 {\it Kepler} photometry are
presented, including the finding that one third of the SX~Phe pulsators are  in
binary systems with orbital periods ranging from a few days to more than four
years.  A summary of the paper is given in $\S$4. 


\section{SPECTROSCOPY}

High-resolution  echelle spectra for 32 of the BN12 candidate SX~Phe stars were
acquired in 2013 and 2014 with the ESPaDOnS  spectrograph mounted on the
Canada-France-Hawaii 3.6-m telescope (CFHT).  Two stars fainter than 15th
magnitude, KIC\,5390069 and KIC\,7300184, were too faint to be observed.    A
total of 178 SX~Phe star spectra were taken by the CFHT Queue Service
Observing team, with the number of spectra taken per star ranging from three to
11, over three to seven independent epochs.  Details of the individual spectra
are summarized in {\bf Table~1}.

\begin{table*} \centering \caption{Summary of CFHT and APO spectra.  The stars
are ordered by KIC number (col.1) and the other columns contain: (2) the
spectrum number (CFHT or APO);  (3) the UTC mid-exposure date and time,  (4)
mid-exposure barycentric Julian Day;  (5) exposure time (s);  (6)
signal-to-noise ratio, (7) airmass at the mid-time of the observation;  (8)
barycentric radial velocity (km/s); and (9) projected equatorial velocity, $v
\sin i$ (km/s).  For the SB2 system KIC\,6780873 the RV and $v\sin i$ entries
are for the primary component.  }

\begin{tabular}{lllccccrr}
\hline  
\multicolumn{1}{c}{KIC} & \multicolumn{1}{c}{Spectrum} & \multicolumn{1}{c}{UTC date (mid)}  & \multicolumn{1}{c}{BJD (mid)} & Exp.(s) & S/N  &AM & \multicolumn{1}{c}{RV} & \multicolumn{1}{c}{ $v$\,sin$i$}  \\
\multicolumn{1}{c}{(1)} & \multicolumn{1}{c}{(2)}  &\multicolumn{1}{c}{(3)} & \multicolumn{1}{c}{  (4)   }       &  (5) &  (6) &   (7)  & \multicolumn{1}{c}{ (8) } & \multicolumn{1}{c}{ (9)}   \\
    \hline   
\multicolumn{9}{c}{{\bf (a) {\it Kepler}-field  SX\,Phe stars}} \\
\noalign{\vskip0.1truecm} 
      1162150  &  APO\,(15a)     & 2012/10/25, 04:31:53 & 2456225.6893 &\,600  &  30     & 1.60  &  --11$\pm$9    &   195$\pm$7   \\  
               &  1650050\,(15b) & 2013/08/26, 08:15:58 & 2456530.8490 & 1800  & 148     & 1.06  &  --15$\pm$7    &   226$\pm$2   \\   
               &  1650642\,(15c) & 2013/08/29, 08:49:18 & 2456533.8720 & 1800  & 155     & 1.10  &  --16$\pm$6    &   230$\pm$2   \\   
               &  APO\,(15d)     & 2014/06/13, 05:59:35 & 2456821.7521 & 1200  &  30     & 1.26  &  --13$\pm$6    &   194$\pm$4   \\ 
               &  APO\,(15e)     & 2014/06/13, 06:28:49 & 2456821.7724 & 1200  &  26     & 1.18  &  --13$\pm$9    &   219$\pm$5   \\ 
               &  1733894\,(15f) & 2014/08/18, 11:25:36 & 2456887.9809 & 1200  & 106     & 1.43  &  --17$\pm$5    &   228$\pm$2   \\  
\noalign{\vskip0.1truecm} 
       3456605 &  1650650\,(24a) & 2013/08/29, 11:48:46 & 2456533.9968 & 1350  &  51     & 1.79  &  --7.5$\pm$0.5 & 13.6$\pm$0.8  \\  
               &  1653557\,(24b) & 2013/09/14, 08:02:23 & 2456549.8388 & 1350  &  52     & 1.11  &  --6.4$\pm$0.5 & 10.0$\pm$1.0  \\   
               &  1741394\,(24c) & 2014/09/17, 06:19:24 & 2456917.7672 & 1200  &  52     & 1.06  &  --8.4$\pm$0.5 & 16.2$\pm$0.8  \\  
\noalign{\vskip0.1truecm} 
       4168579 &  1650649\,(23a) & 2013/08/29, 11:24:43 & 2456533.9801 & 1200  &  38     & 1.63  &  33$\pm$6      &   207$\pm$6 \\   
               &  1653555\,(23b) & 2013/09/14, 07:17:38 & 2456549.8077 & 1200  &  37     & 1.07  &  23$\pm$9      &   201$\pm$5 \\
               &  1670335\,(23c) & 2013/11/19, 04:25:45 & 2456615.6832 & 1200  &  45     & 1.21  &  19$\pm$4      &   196$\pm$4      \\
               &  1670336\,(23d) & 2013/11/19, 04:46:26 & 2456615.6976 & 1200  &  39     & 1.26  &  20$\pm$4      &   208$\pm$9   \\
               &  1670337\,(23e) & 2013/11/19, 05:07:07 & 2456615.7119 & 1200  &  39     & 1.34  &  24$\pm$3      &   197$\pm$7      \\
               &  1733877\,(23f) & 2014/08/18, 06:01:35 & 2456887.7560 & 1200  &  40     & 1.24  &  24$\pm$8      &   178$\pm$6  \\ 
\noalign{\vskip0.1truecm} 
       4243461 &  1649174\,(4a)  & 2013/08/22, 07:18:51 & 2456526.8091 & 1200  &  38     & 1.06  &  58.2$\pm$1.3  & 53.0$\pm$2.8   \\  
               &  1649175\,(4b)  & 2013/08/22, 07:39:32 & 2456526.8235 & 1200  &  35     & 1.06  &  57.8$\pm$0.3  & 50.0$\pm$2.2   \\  
               &  1649176\,(4c)  & 2013/08/22, 08:00:13 & 2456526.8378 & 1200  &  35     & 1.07  &  59.0$\pm$1.9  & 50.1$\pm$3.3  \\
               &  1650640\,(4d)  & 2013/08/29, 07:59:07 & 2456533.8368 & 1200  &  37     & 1.09  &  58.8$\pm$1.2  & 56.7$\pm$2.2  \\
               &  1654451\,(4e)  & 2013/09/17, 05:02:14 & 2456552.7129 & 1200  &  46     & 1.08  &  58.6$\pm$0.8  & 52.2$\pm$2.4  \\
               &  1733876\,(4f)  & 2014/08/18, 05:39:33 & 2456887.7403 & 1200  &  36     & 1.20  &  61.8$\pm$0.9  & 55.3$\pm$2.0   \\  
               &  1740694\,(4g)  & 2014/09/14, 05:46:20 & 2456914.7437 & 1200  &  40     & 1.06  &  62.4$\pm$1.4  & 53.3$\pm$3.7   \\  
               &  1753084\,(4h)  & 2014/10/31, 05:09:04 & 2456961.7143 & 1200  &  36     & 1.21  &  62.9$\pm$1.7  & 52.8$\pm$1.6   \\  
\noalign{\vskip0.1truecm} 
       4662336 &  1650048\,(14a) & 2013/08/26, 07:27:05 & 2456530.8151 & 1200  &  48     & 1.07  & --11.4$\pm$0.2 &  81.5$\pm$1.6  \\  
               &  1650049\,(14b) & 2013/08/26, 07:47:46 & 2456530.8295 & 1200  &  52     & 1.06  & --11.1$\pm$0.4 &  84.1$\pm$1.5 \\  
               &  1650647\,(14c) & 2013/08/29, 10:41:23 & 2456533.9499 & 1200  &  60     & 1.39  & --10.1$\pm$0.6 &  82.4$\pm$1.6  \\   
               &  1741395\,(14d) & 2014/09/17, 06:41:53 & 2456917.7827 & 1200  &  52     & 1.07  & --16.9$\pm$1.7 &  84.1$\pm$1.3  \\  
\noalign{\vskip0.1truecm} 
       4756040 &  1650485\,(20a) & 2013/08/28, 09:45:15 & 2456532.9110 & 1200  &  46     & 1.19  & 3.3$\pm$0.7    &  43.0$\pm$1.0 \\  
               &  1650486\,(20b) & 2013/08/28, 10:05:55 & 2456532.9254 & 1200  &  46     & 1.25  & 8.3$\pm$0.5    &  34.0$\pm$1.1 \\
               &  1653556\,(20c) & 2013/09/14, 07:39:01 & 2456549.8226 & 1200  &  46     & 1.10  & 4.7$\pm$1.0    &  43.4$\pm$1.1 \\
               &  1741396\,(20d) & 2014/09/17, 07:04:15 & 2456917.7983 & 1200  &  46     & 1.08  & 3.2$\pm$1.0    &  44.2$\pm$0.6  \\  
\noalign{\vskip0.1truecm} 
       5036493 &  APO\,(26a)     & 2012/10/25, 09:56:58 & 2456225.6652 &\,600  &  17     & 1.28  & --1.8$\pm$1.8  &  24.9$\pm$2.5  \\  
               &  1653322\,(26b) & 2013/09/13, 05:09:41 & 2456548.7192 & 1200  &  80     & 1.16  & --3.4$\pm$1.0  &  20.6$\pm$0.5    \\ 
               &  1655649\,(26c) & 2013/09/22, 07:29:17 & 2456557.8156 & 1200  &  80     & 1.11  & --3.5$\pm$0.7  &  18.4$\pm$0.6   \\  
               &  1741397\,(26d) & 2014/09/17, 07:27:08 & 2456917.8144 & 1200  &  64     & 1.09  & --5.8$\pm$0.4  &  16.1$\pm$0.7   \\  
\noalign{\vskip0.1truecm} 
       5705575 &  1650644\,(22a) & 2013/08/29, 09:37:19 & 2456533.9053 & 1200  &  39     & 1.21  & --31.7$\pm$2.4 &  88.6$\pm$1.7 \\  
               &  1653554\,(22b) & 2013/09/14, 06:56:00 & 2456549.7926 & 1200  &  37     & 1.08  & --33.7$\pm$1.0 &  92.6$\pm$1.8 \\  
               &  1656478\,(22c) & 2013/09/26, 04:57:03 & 2456561.7092 & 1200  &  49     & 1.09  & --35.0$\pm$2.1 &  89.3$\pm$1.7 \\     
               &  1670338\,(22d) & 2013/11/19, 05:28:43 & 2456615.7269 & 1200  &  37     & 1.49  & --30.4$\pm$1.4 &  83.3$\pm$2.2 \\   
               &  1670339\,(22e) & 2013/11/19, 05:49:24 & 2456615.7412 & 1200  &  37     & 1.61  & --31.9$\pm$0.9 &  85.4$\pm$2.1 \\   
               &  1670340\,(22f) & 2013/11/19, 06:10:04 & 2456615.7556 & 1200  &  34     & 1.76  & --31.8$\pm$2.0 &  95.8$\pm$4.4 \\  
               &  1733896\,(22g) & 2014/08/18, 12:09:53 & 2456888.0116 & 1200  &  34     & 1.71  & --42.6$\pm$2.2 &  88.5$\pm$2.6  \\  
               &  1740703\,(22h) & 2014/09/14, 08:52:26 & 2456914.8734 & 1200  &  36     & 1.26  & --47.3$\pm$2.8 &  87.4$\pm$2.0  \\  
               &  1755698\,(22i) & 2014/11/10, 06:54:20 & 2456971.7871 & 1200  &  23     & 1.83  & --40.0$\pm$5.7 &  84.6$\pm$2.8  \\  
\noalign{\vskip0.1truecm} 
       6130500 &  1649381\,(9a)  & 2013/08/23, 07:07:09 & 2456527.8013 & 1200  &  42     & 1.10  & --18.8$\pm$1.0 & 42.9$\pm$1.7 \\  
               &  1649382\,(9b)  & 2013/08/23, 07:27:51 & 2456527.8157 & 1200  &  34     & 1.09  & --19.3$\pm$0.8 & 48.6$\pm$0.9 \\  
               &  1649383\,(9c)  & 2013/08/23, 07:48:33 & 2456527.8301 & 1200  &  35     & 1.08  & --17.1$\pm$1.4 & 52.1$\pm$1.0 \\  
               &  1650648\,(9d)  & 2013/08/29, 11:02:48 & 2456533.9648 & 1200  &  35     & 1.51  & --18.1$\pm$1.3 & 51.2$\pm$1.4  \\  
               &  1654724\,(9e)  & 2013/09/18, 06:31:00 & 2456553.7751 & 1200  &  29     & 1.08  & --19.0$\pm$1.3 & 49.2$\pm$1.1  \\  
               &  1733878\,(9f)  & 2014/08/18, 06:23:21 & 2456887.7710 & 1200  &  36     & 1.20  & --16.0$\pm$1.1 & 47.2$\pm$2.2   \\   
\noalign{\vskip0.1truecm} 
       6227118 &  1653323\,(27a) & 2013/09/13, 05:31:29 & 2456548.7343 & 1200  &  60     & 1.13  &  6.4$\pm$3.2   & 135.9$\pm$1.9 \\   
               &  1655648\,(27b) & 2013/09/22, 07:06:51 & 2456557.8000 & 1200  &  71     & 1.10  &  2.1$\pm$1.8   & 133.0$\pm$2.3  \\  
               &  1740707\,(27c) & 2014/09/14, 10:05:51 & 2456914.9248 & 1200  &  54     & 1.46  &  7.5$\pm$4.4   & 129.4$\pm$2.2  \\  
               &  1755701\,(27d) & 2014/11/10, 07:48:04 & 2456971.8248 & 1200  &  29     & 2.18  &  8.5$\pm$2.0   & 132.5$\pm$3.4  \\  
    \hline      
    \end{tabular}%
  \label{tab:addlabel}%
\end{table*}%

\begin{table*}  
  \centering
  \contcaption{}
\begin{tabular}{lllccccrr}
\hline  
\multicolumn{1}{c}{KIC} & \multicolumn{1}{c}{Spectrum} & \multicolumn{1}{c}{UTC date (mid)}  & \multicolumn{1}{c}{BJD (mid)} & Exp.(s) & S/N  &AM & \multicolumn{1}{c}{RV} & \multicolumn{1}{c}{ $v$\,sin$i$}  \\
\multicolumn{1}{c}{(1)} & \multicolumn{1}{c}{(2)}  &\multicolumn{1}{c}{(3)} & \multicolumn{1}{c}{  (4)   }       &  (5) &  (6) &   (7)  & \multicolumn{1}{c}{ (8) } & \multicolumn{1}{c}{ (9)}  \\
\hline 
       6445601 &  1648978\,(2a)  & 2013/08/21, 11:11:41 & 2456525.9711 & 1200  &  41     & 1.42  & --7.7$\pm$1.6  & 71.3$\pm$2.4 \\    
               &  1648979\,(2b)  & 2013/08/21, 11:32:17 & 2456525.9855 & 1200  &  42     & 1.52  & --8.4$\pm$3.2  & 72.6$\pm$1.8 \\   
               &  1648980\,(2c)  & 2013/08/21, 11:53:04 & 2456525.9999 & 1200  &  41     & 1.65  & --4.0$\pm$0.5  & 70.1$\pm$1.7  \\  
               &  1650645\,(2d)  & 2013/08/29, 09:58:49 & 2456533.9203 & 1200  &  39     & 1.27  & --4.8$\pm$0.9  & 70.4$\pm$3.2  \\  
               &  1654452\,(2e)  & 2013/09/17, 05:23:57 & 2456552.7285 & 1200  &  44     & 1.10  & --4.2$\pm$1.3  & 72.2$\pm$2.1 \\   
               &  1740705\,(2f)  & 2014/09/14, 09:37:09 & 2456914.9045 & 1200  &  38     & 1.42  & --5.7$\pm$1.8  & 70.8$\pm$1.7  \\  
\noalign{\vskip0.1truecm}
       6520969 &  1650643\,(21a) & 2013/08/29, 09:15:49 & 2456533.8904 & 1200  &  49     & 1.18  & --299.8$\pm$1.2  &  $<$$4.2\pm$1.6   \\  
               &  1653553\,(21b) & 2013/09/14, 06:34:45 & 2456549.7778 & 1200  &  41     & 1.08  & --299.7$\pm$0.8  &  $<$$6.9\pm$1.2   \\   
               &  1740704\,(21c) & 2014/09/14, 09:14:52 & 2456914.8890 & 1200  &  38     & 1.35  & --299.4$\pm$0.5  &  $<$$5.4\pm$1.4   \\  
\noalign{\vskip0.1truecm}
       6780873 &  1649178\,(5a)  & 2013/08/22, 08:27:28 & 2456526.8571 & 1200  &  36     & 1.09  &    8.9$\pm$0.3   &  [11]           \\  
               &  1649179\,(5b)  & 2013/08/22, 08:48:09 & 2456526.8714 & 1200  &  34     & 1.10  &    8.4$\pm$0.3   &  [11]           \\  
               &  1649180\,(5c)  & 2013/08/22, 09:08:51 & 2456526.8858 & 1200  &  45     & 1.12  &   10.2$\pm$0.4   &  [11]           \\  
               &  1650646\,(5d)  & 2013/08/29, 10:19:56 & 2456533.9349 & 1200  &  39     & 1.34  &--26.81$\pm$0.13  & 11$\pm$1        \\  
               &  1654453\,(5e)  & 2013/09/17, 05:45:33 & 2456552.7435 & 1200  &  36     & 1.09  &--24.13$\pm$0.15  & 12$\pm$1        \\  
               &  1740695\,(5f)  & 2014/09/14, 06:08:57 & 2456914.7600 & 1200  &  40     & 1.09  &  43.08$\pm$0.14  & 11$\pm$1        \\  
               &  1741393\,(5g)  & 2014/09/17, 05:56:60 & 2456917.7515 & 1200  &  40     & 1.09  &--22.56$\pm$0.14  & 11$\pm$1        \\  
               &  1753085\,(5h)  & 2014/10/31, 05:32:15 & 2456961.7311 & 1200  &  35     & 1.22  &  12.99$\pm$0.20  & 15$\pm$3        \\  
               &  1755692\,(5i)  & 2014/11/10, 04:31:38 & 2456971.6881 & 1200  &  25     & 1.17  & --3.79$\pm$0.18  & 11$\pm$1        \\ 
\noalign{\vskip0.1truecm}
       7020707 &  1650474\,(16a) & 2013/08/28, 05:51:16 & 2456532.7481 & 1200  &  42     & 1.14  & 5.7$\pm$1.4      &  101$\pm$3  \\   
               &  1650475\,(16b) & 2013/08/28, 06:11:58 & 2456532.7625 & 1200  &  43     & 1.12  & 4.4$\pm$3.8      &  109$\pm$3  \\    
               &  1653552\,(16c) & 2013/09/14, 06:12:57 & 2456549.7624 & 1200  &  44     & 1.09  & 0.6$\pm$1.0      &  104$\pm$3  \\  
               &  1733891\,(16d) & 2014/08/18, 10:18:20 & 2456887.9339 & 1200  &  40     & 1.26  & 2.5$\pm$2.9      &  106$\pm$3  \\  
\noalign{\vskip0.1truecm}
       7174372 &  1649377\,(8a)  & 2013/08/23, 05:57:44 & 2456527.7524 & 1200  &  39     & 1.13  & --17.9$\pm$2.3   &  39$\pm$4    \\  
               &  1649378\,(8b)  & 2013/08/23, 06:18:25 & 2456527.7668 & 1200  &  51     & 1.11  & --18.9$\pm$2.5   &  42$\pm$3   \\   
               &  1649379\,(8c)  & 2013/08/23, 06:39:06 & 2456527.7812 & 1200  &  39     & 1.09  & --20.6$\pm$1.8   &  43$\pm$1  \\   
               &  1650637\,(8d)  & 2013/08/29, 06:48:41 & 2456533.7876 & 1200  &  39     & 1.09  & --22.1$\pm$2.4   &  39$\pm$4  \\  
               &  1654454\,(8e)  & 2013/09/17, 06:07:30 & 2456552.7580 & 1200  &  35     & 1.10  & --20.0$\pm$2.9   &  42$\pm$3   \\  
               &  1733884\,(8f)  & 2014/08/18, 07:59:33 & 2456887.8372 & 1200  &  39     & 1.09  & --18.2$\pm$1.5   &  38$\pm$1     \\  
               &  1740698\,(8g)  & 2014/09/14, 07:15:49 & 2456914.8056 & 1200  &  39     & 1.15  & --20.6$\pm$1.6   &  46$\pm$1    \\  
               &  1753086\,(8h)  & 2014/10/31, 05:55:39 & 2456961.7466 & 1200  &  36     & 1.46  & --17.4$\pm$1.7   &  40$\pm$1     \\ 
               &  1755694\,(8i)  & 2014/11/10, 05:18:57 & 2456971.7203 & 1200  &  25     & 1.47  & --17.2$\pm$2.4   &  40$\pm$2    \\ 
\noalign{\vskip0.1truecm}
       7301640 &  1649384\,(10a) & 2013/08/23, 08:10:45 & 2456527.8456 & 1200  &  36     & 1.09  & --10$\pm$6     &   124$\pm$4   \\   
               &  1649385\,(10b) & 2013/08/23, 08:31:27 & 2456527.8600 & 1200  &  37     & 1.09  & --11$\pm$5     &   121$\pm$3   \\  
               &  1649386\,(10c) & 2013/08/23, 08:52:09 & 2456527.8744 & 1200  &  35     & 1.10  &  --8$\pm$5     &   128$\pm$4  \\   
               &  1653559\,(10d) & 2013/09/14, 08:46:24 & 2456549.8695 & 1200  &  36     & 1.21  & --10$\pm$6     &   111$\pm$7  \\   
               &  1733885\,(10e) & 2014/08/18, 08:21:40 & 2456887.8533 & 1200  &  35     & 1.09  & --17$\pm$4     &   122$\pm$4   \\  
\noalign{\vskip0.1truecm}
       7621759 &  1649181\,(6a)  & 2013/08/22, 09:31:44 & 2456526.9018 & 1200  &  33     & 1.14  & 17.8$\pm$3.0    & 77.6$\pm$2.0  \\   
               &  1649182\,(6b)  & 2013/08/22, 09:52:26 & 2456526.9169 & 1200  &  35     & 1.17  & 17.3$\pm$3.5    & 79.9$\pm$2.4 \\    
               &  1649183\,(6c)  & 2013/08/22, 10:13:07 & 2456526.9305 & 1200  &  33     & 1.21  & 18.1$\pm$4.4    & 83.2$\pm$3.4 \\   
               &  1653558\,(6d)  & 2013/09/14, 08:25:14 & 2456549.8548 & 1200  &  33     & 1.18  & 14.5$\pm$3.3    & 75.9$\pm$2.2  \\   
               &  1733881\,(6e)  & 2014/08/18, 06:54:30 & 2456887.7927 & 1200  &  35     & 1.16  & 17.3$\pm$2.4    & 76.9$\pm$1.4  \\  
\noalign{\vskip0.1truecm}
       7765585 &  1653324\,(28a) & 2013/09/13, 05:53:11 & 2456548.7492 & 1200  &  38     & 1.11  &  6.7$\pm$2.7    &  125$\pm$6   \\   
               &  1655645\,(28b) & 2013/09/22, 05:40:49 & 2456557.7401 & 1200  &  45     & 1.10  &  6.9$\pm$1.6    &  125$\pm$6   \\  
               &  1669659\,(28c) & 2013/11/16, 05:27:58 & 2456612.7270 & 1200  &  30     & 1.37  & --14.3$\pm$2.6  &  124$\pm$13  \\
               &  1733882\,(28d) & 2014/08/18, 07:15:60 & 2456887.8076 & 1200  &  34     & 1.14  &  --3.1$\pm$4.1  &  123$\pm$5   \\  
               &  1740696\,(28e) & 2014/09/14, 06:31:20 & 2456914.7757 & 1200  &  36     & 1.09  &    4.1$\pm$4.0  &  122$\pm$3   \\ 
               &  1755699\,(28f) & 2014/11/10, 07:17:38 & 2456971.8036 & 1200  &  20     & 1.91  &  --1.5$\pm$2.0  &  121$\pm$3   \\ 
\noalign{\vskip0.1truecm}
       7819024 &  1650482\,(19a) & 2013/08/28, 08:42:28 & 2456532.8672 & 1200  &  35     & 1.14  & --57.0$\pm$2.1  &  95.1$\pm$5.3 \\   
               &  1650483\,(19b) & 2013/08/28, 09:03:10 & 2456532.8815 & 1200  &  37     & 1.17  & --66.0$\pm$3.5  & 100.0$\pm$2.9 \\   
               &  1650484\,(19c) & 2013/08/28, 09:23:51 & 2456532.8959 & 1200  &  36     & 1.20  & --62.2$\pm$2.2  &  97.8$\pm$1.8 \\
               &  1654189\,(19d) & 2013/09/16, 05:53:05 & 2456551.7487 & 1200  &  56     & 1.10  & --63.5$\pm$2.1  &  90.9$\pm$1.8 \\
               &  1733895\,(19e) & 2014/08/18, 11:47:25 & 2456887.9959 & 1200  &  31     & 1.60  & --58.8$\pm$2.9  &  91.2$\pm$4.5  \\ 
               &  1740702\,(19f) & 2014/09/14, 08:28:38 & 2456914.8568 & 1200  &  34     & 1.23  & --55.7$\pm$2.9  &  97.5$\pm$2.9  \\ 
               &  1753090\,(19g) & 2014/10/31, 06:56:43 & 2456961.7896 & 1200  &  37     & 1.60  & --48.4$\pm$2.4  &  97.9$\pm$2.7  \\ 
               &  1755697\,(19h) & 2014/11/10, 06:29:30 & 2456971.7699 & 1200  &  22     & 1.68  & --62.8$\pm$2.0  &  92.2$\pm$2.3 \\ 
\noalign{\vskip0.1truecm}
       8004558 &  1648976\,(1a)  & 2013/08/21, 10:28:14 & 2456525.9403 & 1200  &  51     & 1.47  &  --260$\pm$4   & 80$\pm$5    \\  
               &  1648977\,(1b)  & 2013/08/21, 10:48:56 & 2456525.9546 & 1200  &  42     & 1.58  &  --261$\pm$4   & 84$\pm$4    \\   
               &  1650636\,(1c)  & 2013/08/29, 06:27:30 & 2456533.7728 & 1200  &  52     & 1.10  &  --254$\pm$5   & 82$\pm$9    \\  
               &  1733883\,(1d)  & 2014/08/18, 07:38:02 & 2456887.8222 & 1200  &  42     & 1.10  &  --254$\pm$5   & 76$\pm$5    \\  
               &  1740697\,(1e)  & 2014/09/14, 06:53:45 & 2456914.7902 & 1200  &  43     & 1.14  &  --251$\pm$4   & 95$\pm$8    \\  
               &  1753087\,(1f)  & 2014/10/31, 06:20:11 & 2456961.7636 & 1200  &  40     & 1.64  &  --251$\pm$4   & 92$\pm$7    \\  
               &  1755693\,(1g)  & 2014/11/10, 04:55:24 & 2456971.7039 & 1200  &  26     & 1.40  &  --242$\pm$4   & 94$\pm$7     \\  
    \hline      
    \end{tabular}%
  \label{tab:addlabel}%
\end{table*}%

\begin{table*}  
  \centering
  \contcaption{}
\begin{tabular}{lllccccrr}
\hline  
\multicolumn{1}{c}{KIC} & \multicolumn{1}{c}{Spectrum} & \multicolumn{1}{c}{UTC date (mid)}  & \multicolumn{1}{c}{BJD (mid)} & Exp.(s) & S/N  &AM & \multicolumn{1}{c}{RV} & \multicolumn{1}{c}{ $v$\,sin$i$}  \\
\multicolumn{1}{c}{(1)} & \multicolumn{1}{c}{(2)}  &\multicolumn{1}{c}{(3)} & \multicolumn{1}{c}{  (4)   }       &  (5) &  (6) &   (7)  & \multicolumn{1}{c}{ (8) } & \multicolumn{1}{c}{(9)} \\
\hline
       8110941 &  1653325\,(29a) & 2013/09/13, 06:15:24 & 2456548.7647 & 1200  &  44     & 1.10  &   4.3$\pm$0.2   & $<$$7.8\pm$0.6 \\ 
               &  1655646\,(29b) & 2013/09/22, 06:03:09 & 2456557.7557 & 1200  &  48     & 1.10  &   4.4$\pm$0.4   & $<$$7.2\pm$0.5 \\  
               &  1669658\,(29c) & 2013/11/16, 05:05:27 & 2456612.7114 & 1200  &  34     & 1.29  &   4.4$\pm$0.5   & $<$$7.6\pm$0.5 \\   
               &  1741400\,(29d) & 2014/09/17, 08:34:35 & 2456917.8611 & 1200  &  39     & 1.22  &   4.3$\pm$0.2   & $<$$7.1\pm$2.1  \\ 
\noalign{\vskip0.1truecm}
       8196006 &  1653326\,(30a) & 2013/09/13, 06:37:34 & 2456548.7803 & 1200  &  41     & 1.10  & --4.6$\pm$1.6   & 90.5$\pm$1.6 \\  
               &  1655848\,(30b) & 2013/09/23, 09:54:21 & 2456558.9164 & 1200  &  46     & 1.54  & --2.4$\pm$1.9   & 96.4$\pm$2.3  \\ 
               &  1670136\,(30c) & 2013/11/18, 04:51:24 & 2456614.7017 & 1200  &  36     & 1.23  & --2.2$\pm$1.8   & 93.3$\pm$2.6  \\  
               &  1741401\,(30d) & 2014/09/17, 08:57:55 & 2456917.8776 & 1200  &  38     & 1.24  & --6.1$\pm$3.0   & 92.8$\pm$2.3  \\  
\noalign{\vskip0.1truecm} 
       8330910 &  1648981\,(3a)  & 2013/08/21, 12:15:07 & 2456526.0154 & 1200  &  44     & 1.59  &    20$\pm$7     & 220$\pm$6  \\   
               &  1648982\,(3b)  & 2013/08/21, 12:35:50 & 2456526.0298 & 1200  &  43     & 1.73  &    22$\pm$15    & 222$\pm$6   \\  
               &  1654192\,(3c)  & 2013/09/16, 06:58:02 & 2456551.7944 & 1200  &  51     & 1.10  &    24$\pm$9     & 231$\pm$6  \\   
               &  1741402\,(3d)  & 2014/09/17, 09:20:29 & 2456917.8933 & 1200  &  43     & 1.29  &    19$\pm$9     & 224$\pm$5  \\  
\noalign{\vskip0.1truecm} 
       9244992 &  1649185\,(7a)  & 2013/08/22, 10:45:22 & 2456526.9530 & 1200  &  32     & 1.27  & --15.8$\pm$0.4  & $<$7.4$\pm$0.6 \\  
               &  1649186\,(7b)  & 2013/08/22, 11:06:03 & 2456526.9674 & 1200  &  33     & 1.32  & --15.9$\pm$0.5  & $<$7.2$\pm$0.4 \\  
               &  1649187\,(7c)  & 2013/08/22, 11:26:44 & 2456526.9818 & 1200  &  32     & 1.40  & --15.9$\pm$0.3  & $<$5.8$\pm$0.5\\   
               &  1654191\,(7d)  & 2013/09/16, 06:36:30 & 2456551.7794 & 1200  &  36     & 1.11  & --16.8$\pm$0.6  & $<$5.8$\pm$0.5\\   
               &  1733887\,(7e)  & 2014/08/18, 08:50:09 & 2456887.8731 & 1200  &  34     & 1.11  & --15.4$\pm$0.4  & $<$7.1$\pm$0.5  \\   
\noalign{\vskip0.1truecm} 
       9267042 &  1649575\,(12a) & 2013/08/24, 08:39:11 & 2456528.8647 & 1200  &  42     & 1.16  &    --8$\pm$4    & 109.8$\pm$3.4  \\  
               &  1649576\,(12b) & 2013/08/24, 08:59:53 & 2456528.8790 & 1200  &  37     & 1.19  &   --11$\pm$5    & 115.4$\pm$9.8 \\   
               &  1649577\,(12c) & 2013/08/24, 09:20:35 & 2456528.8934 & 1200  &  37     & 1.23  &    --9$\pm$3    & 115.2$\pm$2.9 \\   
               &  1650639\,(12d) & 2013/08/29, 07:37:38 & 2456533.8217 & 1200  &  47     & 1.12  &   --16$\pm$4    & 120.7$\pm$3.0 \\   
               &  1654455\,(12e) & 2013/09/17, 06:30:10 & 2456552.7740 & 1200  &  37     & 1.13  &    --8$\pm$3    & 124.2$\pm$4.1 \\   
               &  1733889\,(12f) & 2014/08/18, 09:34:01 & 2456887.9029 & 1200  &  39     & 1.21  &   --14$\pm$5    & 118.9$\pm$4.2 \\ 
               &  1740699\,(12g) & 2014/09/14, 07:38:17 & 2456914.8215 & 1200  &  39     & 1.19  &   --16$\pm$7    & 120.4$\pm$4.4 \\  
               &  1754691\,(12h) & 2014/11/07, 04:36:12 & 2456968.6912 & 1200  &  22     & 1.26  & --10.5$\pm$0.9  &  94.8$\pm$3.9 \\  
               &  1754692\,(12i) & 2014/11/07, 04:57:36 & 2456968.7061 & 1200  &  21     & 1.31  & --12.2$\pm$0.7  &  97.3$\pm$2.1 \\  
               &  1754693\,(12j) & 2014/11/07, 05:19:03 & 2456968.7210 & 1200  &  19     & 1.39  & --11.3$\pm$0.3  & 103.4$\pm$0.9 \\  
               &  1755695\,(12k) & 2014/11/10, 05:42:19 & 2456971.7369 & 1200  &  24     & 1.55  & --11.8$\pm$1.3  & 113.0$\pm$3.9  \\  
\noalign{\vskip0.1truecm} 
       9535881 &  1650651\,[25a] & 2013/08/29, 12:11:48 & 2456534.0127 & 1200  &  41     & 2.01  &  7.1$\pm$0.7    & 59.4$\pm$1.2 \\   
               &  1654190\,[25b] & 2013/09/16, 06:15:04 & 2456551.7642 & 1200  &  45     & 1.12  &  6.7$\pm$0.5    & 51.7$\pm$1.2  \\
               &  1741398\,[25c] & 2014/09/17, 07:49:37 & 2456917.8298 & 1200  &  45     & 1.17  &  6.9$\pm$1.3    & 53.1$\pm$0.7   \\  
               &  1753092\,[25d] & 2014/10/31, 07:27:07 & 2456961.8111 & 1200  &  27     & 1.68  &  2.9$\pm$1.4    & 59.1$\pm$1.6   \\  
\noalign{\vskip0.1truecm} 
       9966976 &  1653327\,(31a) & 2013/09/13, 06:59:20 & 2456548.7952 & 1200  &  47     & 1.12  & --2$\pm$5       & 122.6$\pm$1.5  \\  
               &  1655845\,(31b) & 2013/09/23, 08:43:48 & 2456558.8673 & 1200  &  53     & 1.32  & --1$\pm$5       & 125.6$\pm$2.4  \\  
               &  1670137\,(31c) & 2013/11/18, 05:12:58 & 2456614.7167 & 1200  &  38     & 1.35  & --1$\pm$6       & 122.3$\pm$2.2 \\   
               &  1741399\,(31d) & 2014/09/17, 08:12:28 & 2456917.8458 & 1200  &  42     & 1.19  &   3$\pm$6       & 123.7$\pm$2.2 \\  
\noalign{\vskip0.1truecm} 
      10989032 &  1653560\,(32a) & 2013/09/14, 09:08:06 & 2456549.8845 & 1200  &  34     & 1.31  &  --44$\pm$2     & 44.4$\pm$1.6 \\ 
               &  1655844\,(32b) & 2013/09/23, 08:21:29 & 2456558.8517 & 1200  &  43     & 1.28  &  --43$\pm$3     & 47.4$\pm$1.6 \\  
               &  1670135\,(32c) & 2013/11/18, 04:29:40 & 2456614.6866 & 1200  &  34     & 1.25  &  --32$\pm$2     & 47.6$\pm$1.9   \\   
               &  1740691\,(32d) & 2014/09/14, 05:11:51 & 2456914.7205 & 1200  &  36     & 1.22  &  --14$\pm$2     & 41.8$\pm$0.9   \\ 
               &  1740708\,(32e) & 2014/09/14, 10:28:24 & 2456914.9403 & 1200  &  31     & 1.63  &   --8$\pm$2     & 46.0$\pm$1.4   \\ 
               &  1741390\,(32f) & 2014/09/17, 05:24:18 & 2456917.7290 & 1200  &  35     & 1.18  &  --15$\pm$2     & 45.2$\pm$1.4    \\  
               &  1741405\,(32g) & 2014/09/17, 09:51:03 & 2456917.9142 & 1200  &  33     & 1.50  &  --22$\pm$2     & 47.4$\pm$1.4   \\ 
               &  1755702\,(32h) & 2014/11/10, 08:10:06 & 2456971.8404 & 1200  &  18     & 2.50  &  --37$\pm$3     & 47.9$\pm$2.0  \\ 
\noalign{\vskip0.1truecm} 
      11649497 &  1649572\,(11a) & 2013/08/24, 07:34:50 & 2456528.8198 & 1200  &  39     & 1.16  & --21.7$\pm$0.3  & $<$5.6$\pm$0.3 \\  
               &  1649573\,(11b) & 2013/08/24, 07:55:31 & 2456528.8342 & 1200  &  39     & 1.17  & --21.1$\pm$0.3  & $<$5.9$\pm$0.4 \\  
               &  1649574\,(11c) & 2013/08/24, 08:16:13 & 2456528.8486 & 1200  &  42     & 1.18  & --21.4$\pm$0.3  & $<$5.3$\pm$0.6  \\ 
               &  1650638\,(11d) & 2013/08/29, 07:16:26 & 2456533.8069 & 1200  &  39     & 1.16  & --21.2$\pm$0.3  & $<$4.7$\pm$0.4  \\  
               &  1654456\,(11e) & 2013/09/17, 06:52:03 & 2456552.7891 & 1200  &  38     & 1.19  & --21.3$\pm$0.3  & $<$5.8$\pm$0.3 \\   
               &  1733888\,(11f) & 2014/08/18, 09:12:12 & 2456887.8876 & 1200  &  40     & 1.22  & --21.9$\pm$0.2  & $<$6.1$\pm$0.3  \\   
\noalign{\vskip0.1truecm} 
      11754974 &  APO\,(13a)     & 2012/10/28, 05:53:24 & 2456228.7455 &\,900  &  19     & 2.44  & --301$\pm$4     &  27.4$\pm$2.2  \\  
               &  APO\,(13b)     & 2012/10/28, 06:10:57 & 2456228.7576 &\,900  &  22     & 2.71  & --299$\pm$4     &  25.0$\pm$2.5  \\  
               &  1649578\,(13c) & 2013/08/24, 09:43:25 & 2456528.9093 & 1200  &  60     & 1.31  & --300.2$\pm$1.0 &  25.7$\pm$2.1  \\  
               &  1649579\,(13d) & 2013/08/24, 10:04:06 & 2456528.9236 & 1200  &  66     & 1.37  & --306.8$\pm$2.0 &  36.6$\pm$1.4 \\  
               &  1650641\,(13e) & 2013/08/29, 08:21:22 & 2456533.8521 & 1200  &  62     & 1.20  & --313.2$\pm$2.2 &  25.7$\pm$2.5  \\  
               &  1733890\,(13f) & 2014/08/18, 09:55:49 & 2456887.9180 & 1200  &  57     & 1.28  & --302.1$\pm$0.8 &  25.5$\pm$1.5 \\  
               &  1740700\,(13g) & 2014/09/14, 08:00:31 & 2456914.8370 & 1200  &  59     & 1.26  & --301.3$\pm$0.9 &  25.6$\pm$1.6 \\  
               &  1755696\,(13h) & 2014/11/10, 06:05:49 & 2456971.7536 & 1200  &  35     & 1.66  & --302.3$\pm$1.7 &  31.3$\pm$1.7  \\  
\hline
    \end{tabular}%
  \label{tab:addlabel}%
\end{table*}%

\begin{table*}  
  \centering
  \contcaption{}
\begin{tabular}{lllccccrr}
\hline  
\multicolumn{1}{c}{KIC/HD} & \multicolumn{1}{c}{Spectrum} & \multicolumn{1}{c}{UTC date (mid)}  & \multicolumn{1}{c}{BJD (mid)} & Exp.(s) & S/N  &AM & \multicolumn{1}{c}{RV} & \multicolumn{1}{c}{ $v$\,sin$i$}  \\
\multicolumn{1}{c}{(1)} & \multicolumn{1}{c}{(2)}  &\multicolumn{1}{c}{(3)} & \multicolumn{1}{c}{  (4)   }       &  (5) &  (6) &   (7)  & \multicolumn{1}{c}{ (8) } & \multicolumn{1}{c}{ (9)}  \\
\hline
      12643589 &  1650476\,[17a] & 2013/08/28, 06:33:57 & 2456532.7776 & 1200  &  48     & 1.19  & --61.3$\pm$0.3  & 29.2$\pm$0.5 \\   
               &  1650477\,[17b] & 2013/08/28, 06:54:38 & 2456532.7920 & 1200  &  38     & 1.18  & --61.6$\pm$0.5  & 28.8$\pm$0.4 \\   
               &  1650478\,[17c] & 2013/08/28, 07:15:20 & 2456532.8063 & 1200  &  38     & 1.18  & --61.7$\pm$0.4  & 28.5$\pm$0.5 \\   
               &  1654187\,[17d] & 2013/09/16, 05:10:20 & 2456551.7188 & 1200  &  37     & 1.20  & --61.2$\pm$0.4  & 28.5$\pm$0.5 \\    
               &  1733892\,[17e] & 2014/08/18, 10:40:16 & 2456887.9489 & 1200  &  35     & 1.41  & --59.5$\pm$0.7  & 28.5$\pm$0.7 \\   
\noalign{\vskip0.1truecm} 
      12688835 &  1650479\,(18a) & 2013/08/28, 07:38:01 & 2456532.8221 & 1200  &  35     & 1.18  & --26$\pm$10     & 217$\pm$15 \\   
               &  1650480\,(18b) & 2013/08/28, 07:58:43 & 2456532.8365 & 1200  &  35     & 1.19  & --20$\pm$10     & 239$\pm$15 \\   
               &  1650481\,(18c) & 2013/08/28, 08:19:25 & 2456532.8508 & 1200  &  35     & 1.21  & --28$\pm$12     & 238$\pm$15 \\
               &  1654188\,(18d) & 2013/09/16, 05:31:20 & 2456551.7334 & 1200  &  37     & 1.19  & --25$\pm$10     & 234$\pm$15 \\
               &  1733893\,(18e) & 2014/08/18, 11:02:28 & 2456887.9643 & 1200  &  31     & 1.48  & --51$\pm$10     & 221$\pm$5  \\  
\noalign{\vskip0.2truecm} 
\multicolumn{9}{c}{  {\bf (b) IAU Radial Velocity Standard Stars }} \\  
\noalign{\vskip0.1truecm} 
    HD\,144579   &  1733875\,(a)   &  2014/08/18, 05:26:16 & 2456887.2267  & 30 & 182 & 1.08 & --59.17$\pm$0.41        & 3.65$\pm$0.13     \\  
                 &  1733879\,(b)   &  2014/08/18, 06:36:44 & 2456887.2757  & 30 & 165 & 1.17 & --58.98$\pm$0.41        & 3.66$\pm$0.11   \\   
                 &  1740690\,(c)   &  2014/09/14, 04:56:45 & 2456914.2062  & 30 & 186 & 1.19 & --58.97$\pm$0.41        & 3.66$\pm$0.34   \\
                 &  1741391\,(d)   &  2014/09/17, 05:38:18 & 2456917.2351  & 30 & 177 & 1.35 & --59.05$\pm$0.42        & 3.76$\pm$0.24   \\
\noalign{\vskip0.1truecm}
    HD\,154417   &  1733874\,(a)   &  2014/08/18, 05:19:15 & 2456887.2218  & 30 & 213 & 1.06 & --16.61$\pm$0.13        & 5.14$\pm$0.17   \\   
                 &  1733880\,(b)   &  2014/08/18, 06:39:14 & 2456887.2775  & 30 & 217 & 1.09 & --16.50$\pm$0.13        & 4.85$\pm$0.14    \\   
                 &  1740692\,(c)   &  2014/09/14, 05:27:38 & 2456914.2277  & 30 & 230 & 1.15 & --16.69$\pm$0.13        & 4.33$\pm$0.16   \\
                &  1741392\,(d)   &  2014/09/17, 05:42:09 & 2456917.2378  & 30 & 224 & 1.22 & --16.44$\pm$0.18        & 4.37$\pm$0.14   \\
\noalign{\vskip0.1truecm} 
    HD\,171391 &  1733886\,(a)   &  2014/08/18, 08:36:14 & 2456887.3586  & 30 & 349 & 1.26 &    7.69$\pm$0.20     & 3.67$\pm$0.28 \\   
               &  1740693\,(b)   &  2014/09/14, 05:31:31 & 2456914.2304  & 30 & 391 & 1.16 &    7.75$\pm$0.19     & 3.63$\pm$0.19 \\   
               &  1741389\,(c)   &  2014/09/17, 05:09:09 & 2456917.2148  & 30 & 384 & 1.16 &    7.72$\pm$0.19     & 3.97$\pm$0.38 \\   
               &  1753083\,(d)   &  2014/10/31, 04:54:05 & 2456961.2044  & 30 & 380 & 1.52 &    7.78$\pm$0.19     & 3.58$\pm$0.16 \\   
               &  1754690\,(e)   &  2014/11/07, 04:19:19 & 2456968.1803  & 30 & 111 & 1.48 &    7.96$\pm$0.19     & 4.04$\pm$0.50 \\   
\noalign{\vskip0.1truecm} 
    HD\,182572 &  1740701\,(a)   &  2014/09/14, 08:14:35 & 2456914.3436  & 30 & 282 & 1.15 &--100.13$\pm$0.34  & 2.96$\pm$0.12 \\  
               &  1741403\,(b)   &  2014/09/17, 09:34:00 & 2456917.3988  & 30 & 349 & 1.59 &--100.07$\pm$0.33  & 2.81$\pm$0.15 \\  
               &  1753088\,(c)   &  2014/10/31, 06:36:40 & 2456961.2756  & 30 & 327 & 1.56 & --99.66$\pm$0.32  & 2.87$\pm$0.12 \\
               &  1755691\,(d)   &  2014/11/10, 04:16:12 & 2456971.1781  & 30 & 180 & 1.12 & --99.53$\pm$0.33  & 2.95$\pm$0.12 \\  
\noalign{\vskip0.1truecm} 
    HD\,187691 &  1740706\,(a)   &  2014/09/14, 09:51:18 & 2456914.4108  & 30 & 288 & 1.46 &    0.22$\pm$0.12       & 3.45$\pm$0.14 \\   
               &  1741404\,(b)   &  2014/09/17, 09:36:41 & 2456917.4006  & 30 & 324 & 1.44 &    0.22$\pm$0.12        & 4.03$\pm$0.15 \\   
               &  1753091\,(c)   &  2014/10/31, 07:11:56 & 2456961.3001  & 30 & 323 & 1.65 &    0.21$\pm$0.12        & 3.94$\pm$0.14 \\
               &  1754694\,(d)   &  2014/11/07, 05:33:51 & 2456968.2320  & 30 & 143 & 1.24 &    0.36$\pm$0.12        & 2.76$\pm$0.14 \\
               &  1754695\,(e)   &  2014/11/07, 05:35:02 & 2456968.2328  & 30 & 126 & 1.24 &    0.34$\pm$0.12        & 3.47$\pm$0.17 \\
               &  1754696\,(f)   &  2014/11/07, 05:36:12 & 2456968.2337  & 30 & 143 & 1.25 &    0.38$\pm$0.12        & 3.31$\pm$0.19 \\
               &  1755700\,(g)   &  2014/11/10, 07:32:58 & 2456971.3147  & 30 & 129 & 2.55 &    0.37$\pm$0.12        & 4.08$\pm$0.18 \\
    \hline     
    \end{tabular}%
  \label{tab:addlabel}%
\end{table*}%

 
In 2013 ESPaDOnS was used in `star+sky' mode (spectral resolving power
$\lambda/\Delta\lambda$$\sim$66000) with spectral resolution (gaussian FWHM of
instrumental profile) $\sim$0.0083\,nm at 550\,nm, 2.5 pixels per resolution
element and velocity resolution 4.5\,km/s.  In 2014  `star-only' mode was used,
which widened the stellar spectra, increased the resolving power to
$\sim$80000, reduced the FWHM of the instrumental profile to $\sim$0.0070\,nm,
and improved the velocity resolution to 3.8\,km/s.    Each spectrum covers the
wavelength range 370-1030\,nm, spread over 40 orders.  Almost all the
observations were made under photometric conditions.  Starting in 2014 several
IAU faint radial velocity standard stars, selected from the list given by Udry
{\it et al.} (1999), were observed each night in addition to the program stars
(Table\,1).   All raw spectra were preprocessed using the
`Libre-Esprit'/`Upena' CFHT pipeline (see Donati {\it et al.} 1997).  The
reductions included  barycentric velocity corrections,  small radial velocity
corrections based on the telluric lines, and rectification of  the spectra.
The Versatile Wavelength Analysis (VWA) `rainbow' program (Bruntt {\it et al.}
2010a,b) was used to merge the overlapping echelle orders and to improve, where
necessary, the normalization of the continuum level.   Sample  spectra for
the 32 observed stars are shown in {\bf Figure\,1}. 

High-resolution spectra also were taken at the Apache Point Observatory (APO)
with the ARCES echelle spectrograph on the Astrophysical Research Consortium
(ARC) 3.5-m telescope.  Some of the
spectra were of well-known field SX\,Phe stars taken for comparison purposes,
including three RV standard stars selected again from the Udry {\it et al.}
list.  The resolution of the APO spectra, $R \sim 31500$ (or 2.5
pixels), is lower than that for the CFHT spectra, and the wavelength coverage
was from 320 to 1000\,nm, dispersed over 107 orders.  The default slit (size
$1.6\arcsec\times3.2\arcsec$) was used for all the observations, and the
readout noise typically was less than 7 e$^-$/pixel with a gain of 3.8
e$^-$/ADU.  All the  spectra were preprocessed with IRAF.  Because of their
lower resolution and lower signal-to-noise ratios the APO  spectra were used
mainly for additional radial velocity and rotation velocity information.
Details of the APO observations of the {\it Kepler}-field SX~Phe stars also are
given in Table~1.

While acquiring the APO spectra,  KIC\,1162150 (*15) and KIC\,6227118 (*27)
were seen in the 3.5-m telescope finder fields to be optical doubles with very
close, relatively-bright neighbours.  Finding charts made from the DSS digital
sky survey are shown  in {\bf Figure\,2} for these and two other stars.
KIC\,1162150 is seen to be located among a clump of relatively bright stars,
and the image of KIC\,6227118 is quite asymmetric.  It is  possible that
photometric observations (and possibly our spectroscopy) for these two stars
are contaminated by light from the neighbour stars.



\begin{figure*} \begin{center}
\begin{overpic}[width=18.0cm]{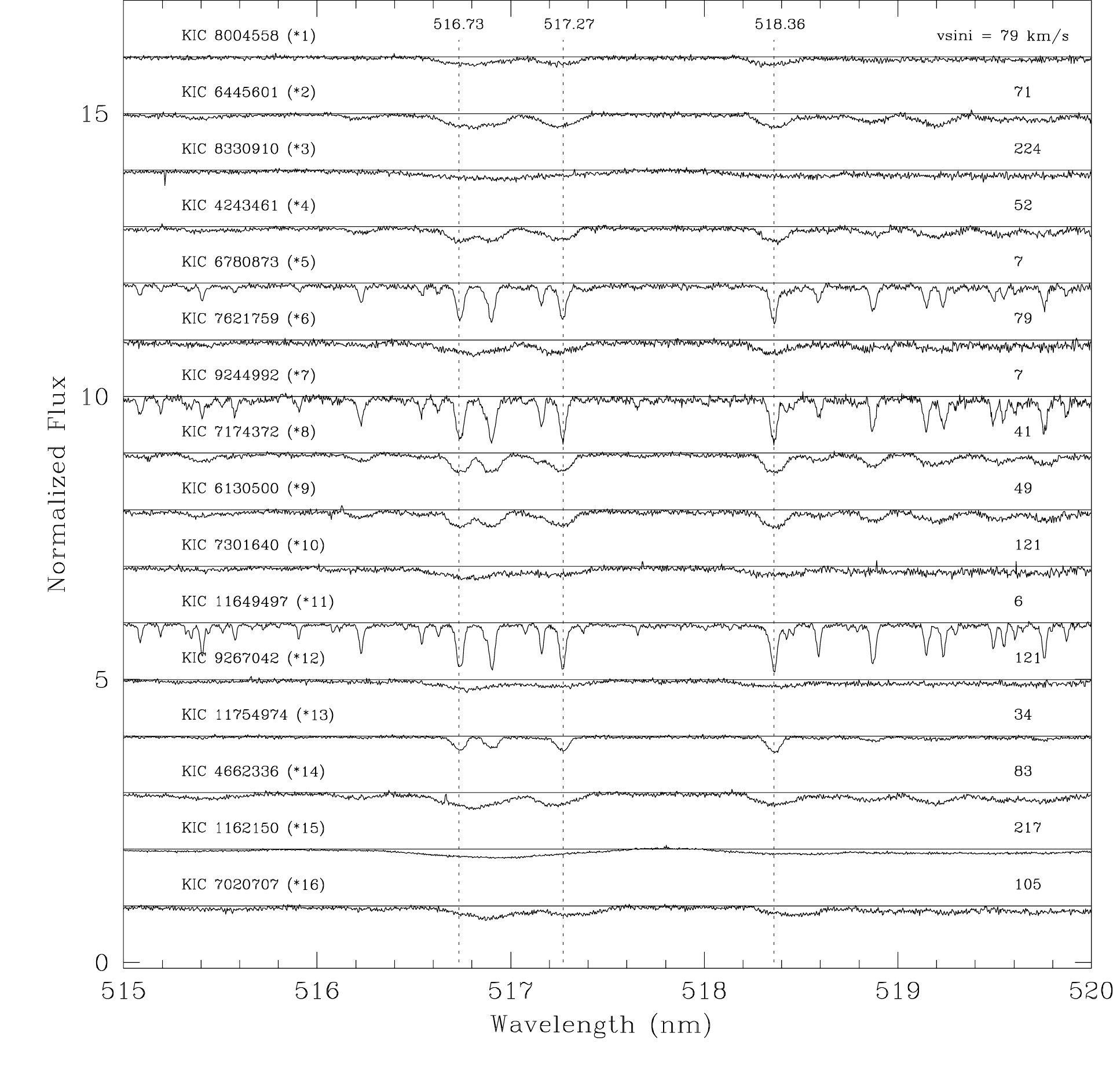} \put(1,1){} \end{overpic}
\end{center} \caption{ESPaDOnS (CFHT) spectra for the wavelength region 515 to
520 nm, illustrating the wide range of line broadenings.  Each spectrum is
identified by the KIC number (and in parentheses the CFHT star number). The dashed
vertical lines identify the Mg triplet lines, and on the
right hand side of the diagram the average $v \sin i$ (km/s) values are given.
The ordering is by CFHT star number.  } \label{Mgtriplet1}
\end{figure*}

\begin{figure*} \begin{center} \contcaption{}
\begin{overpic}[width=18.0cm]{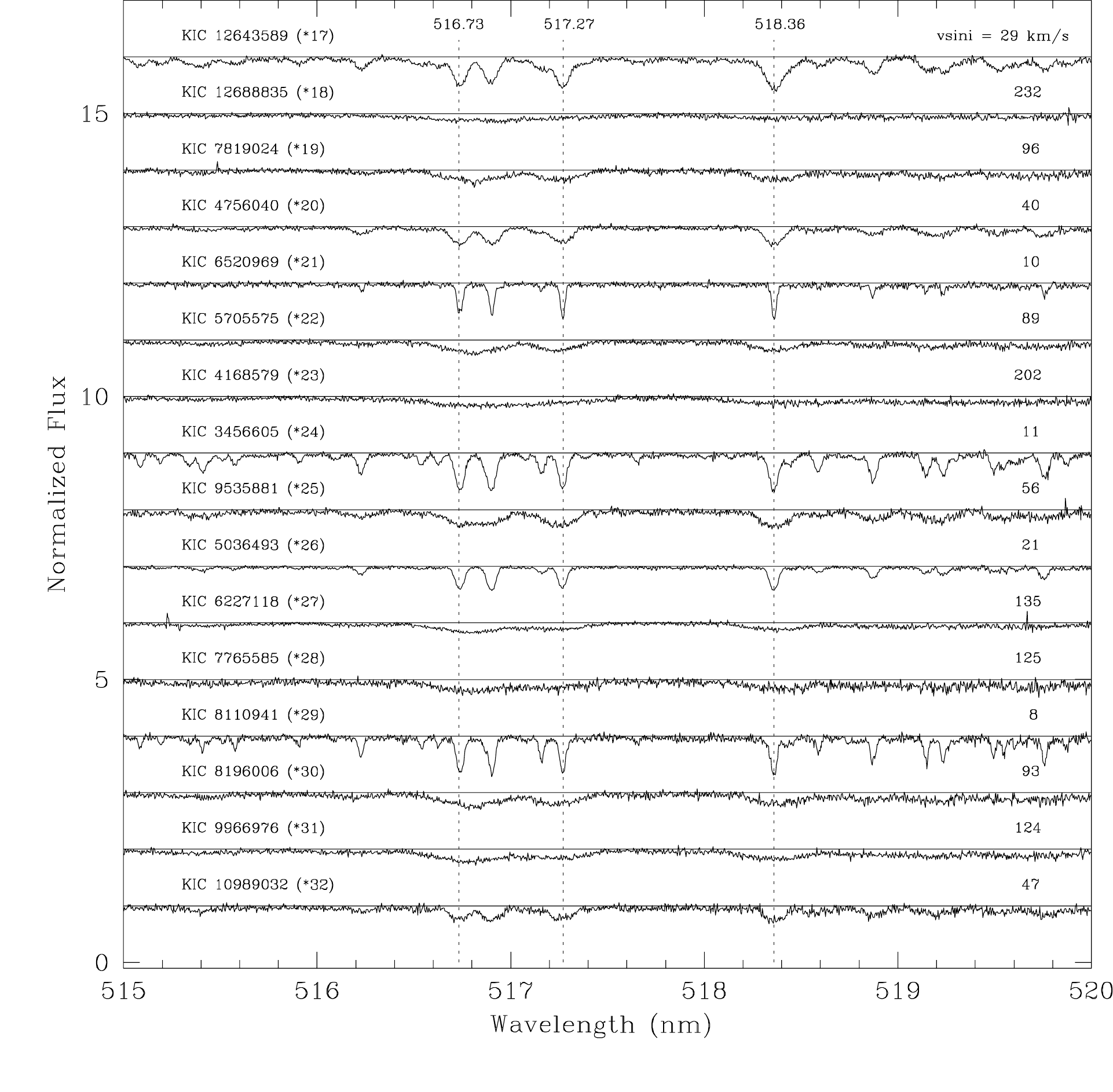}\put(1,1){} \end{overpic}
\end{center} \label{Mgtriplet2} \end{figure*}

\begin{figure} \centering 
\begin{overpic}[width=3.9cm]{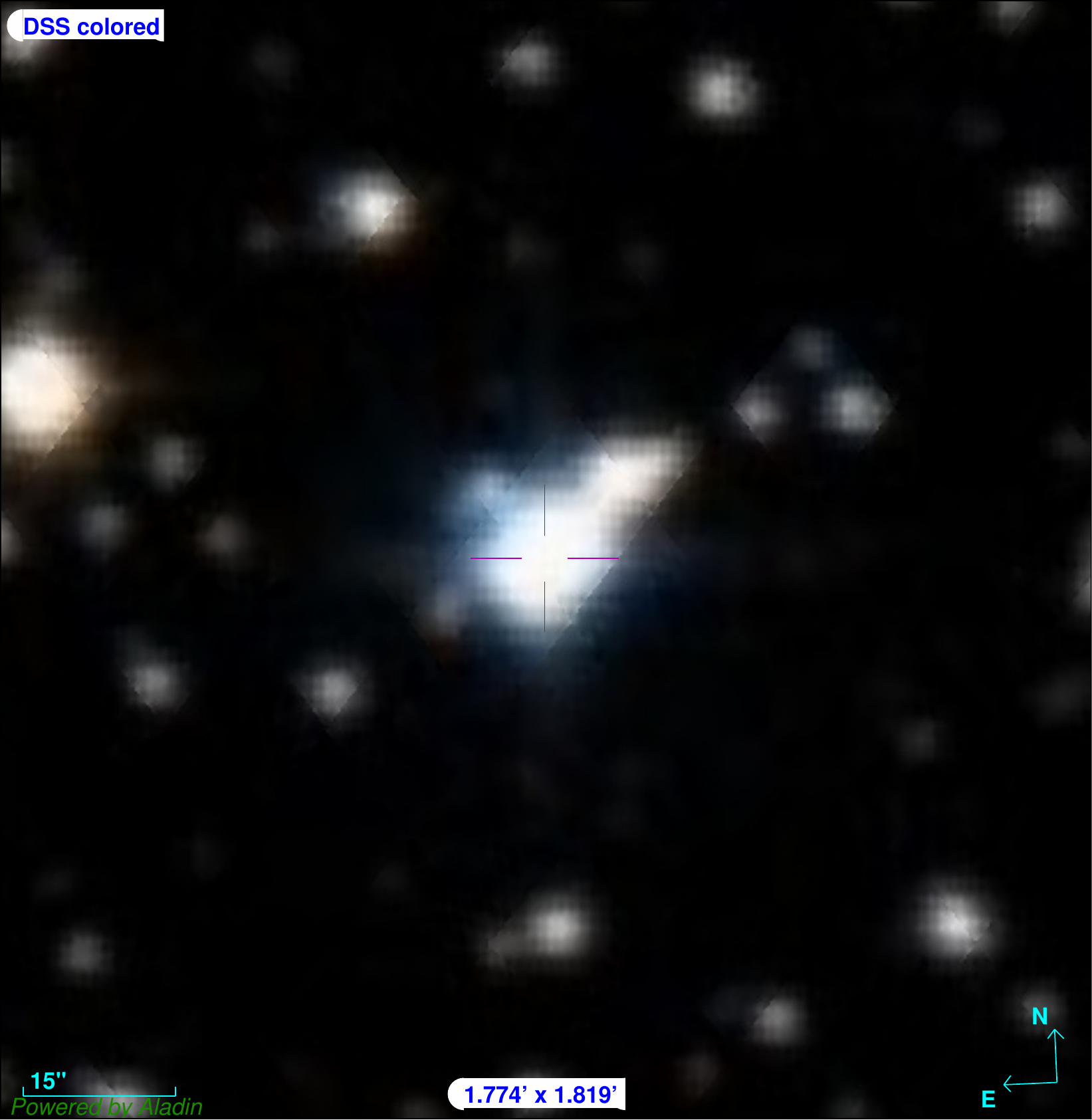}\end{overpic}
\begin{overpic}[width=3.9cm]{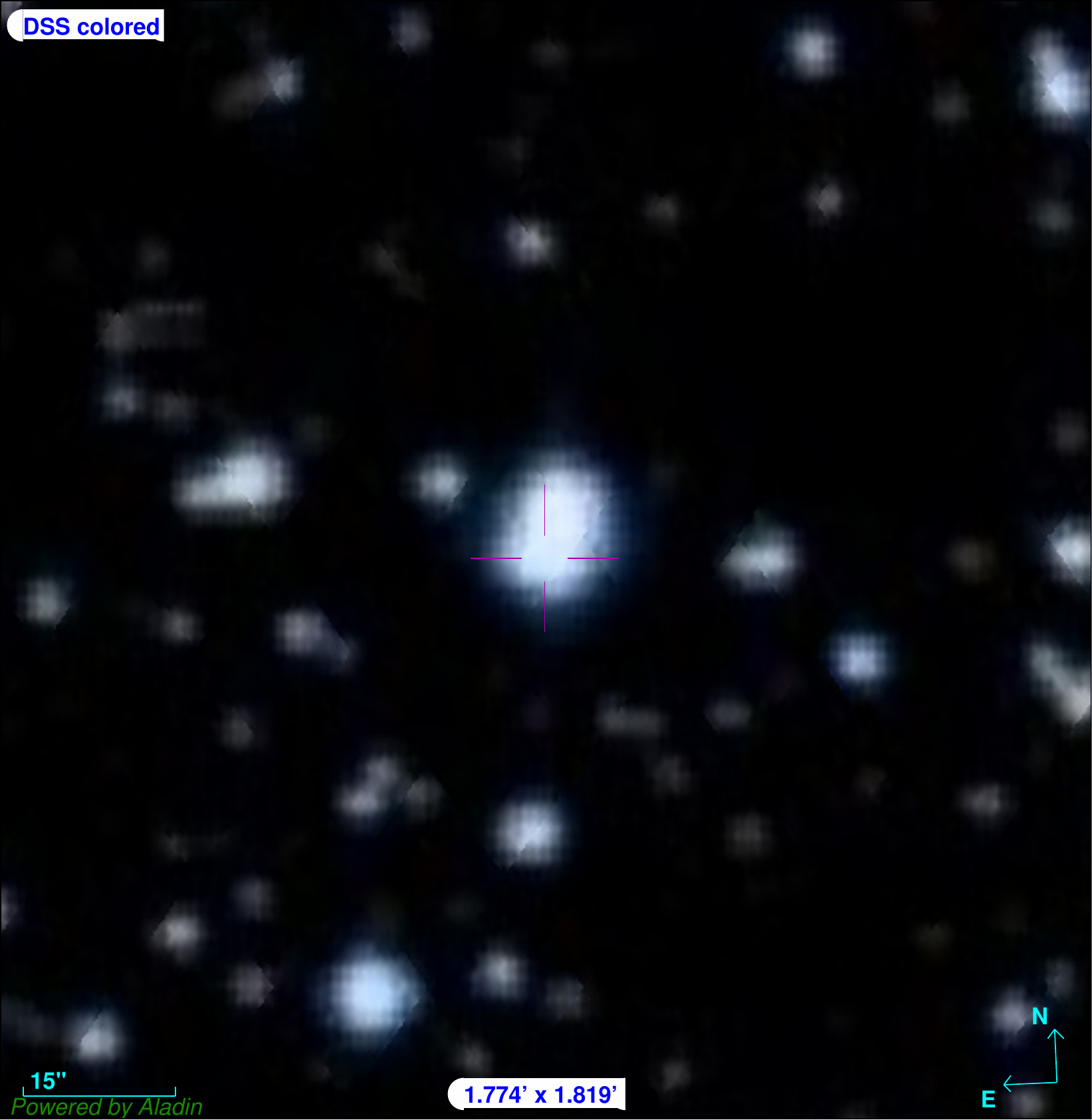}\end{overpic}
\vskip0.1truecm
\begin{overpic}[width=3.9cm]{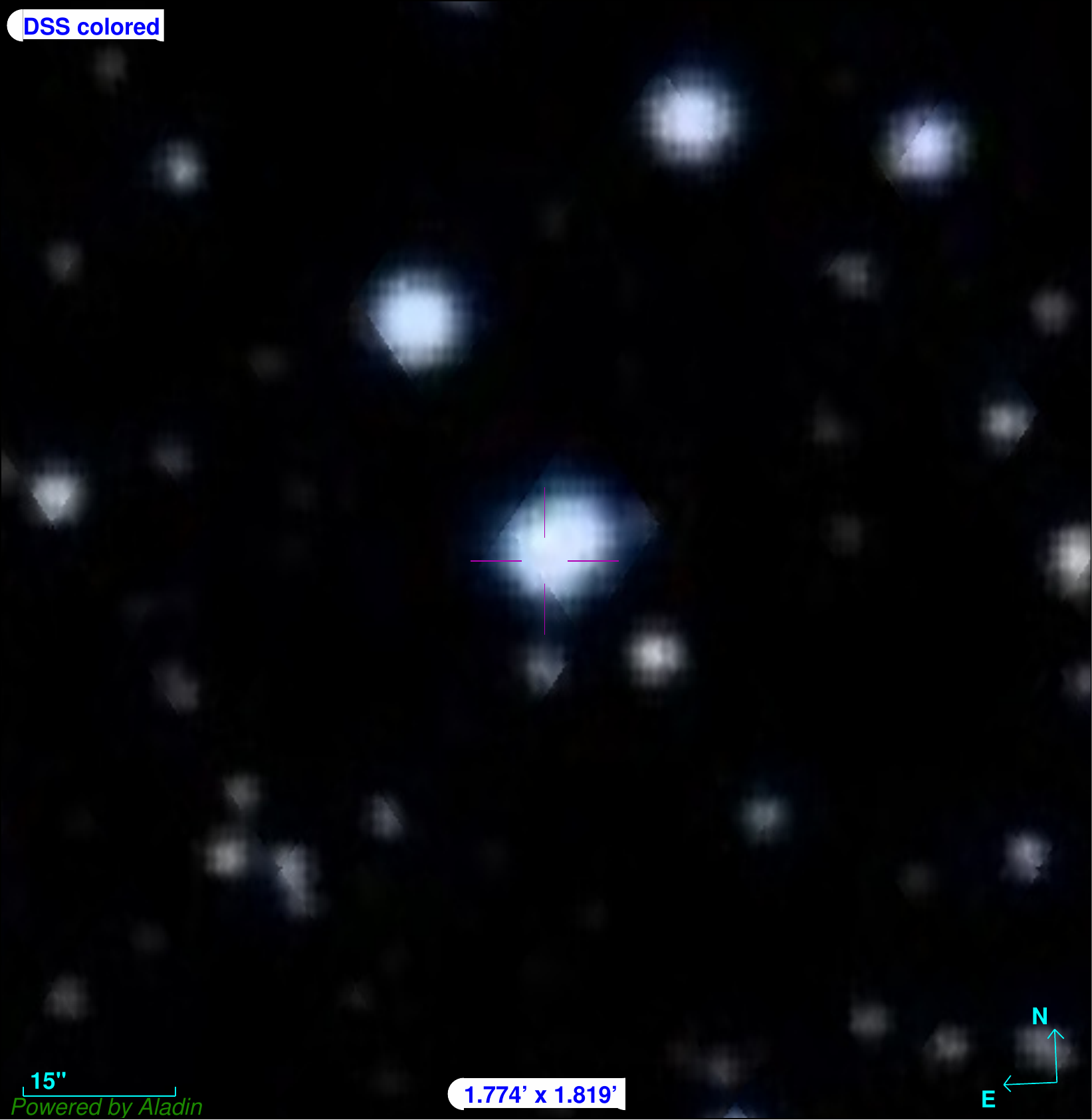}\end{overpic}
\begin{overpic}[width=3.9cm]{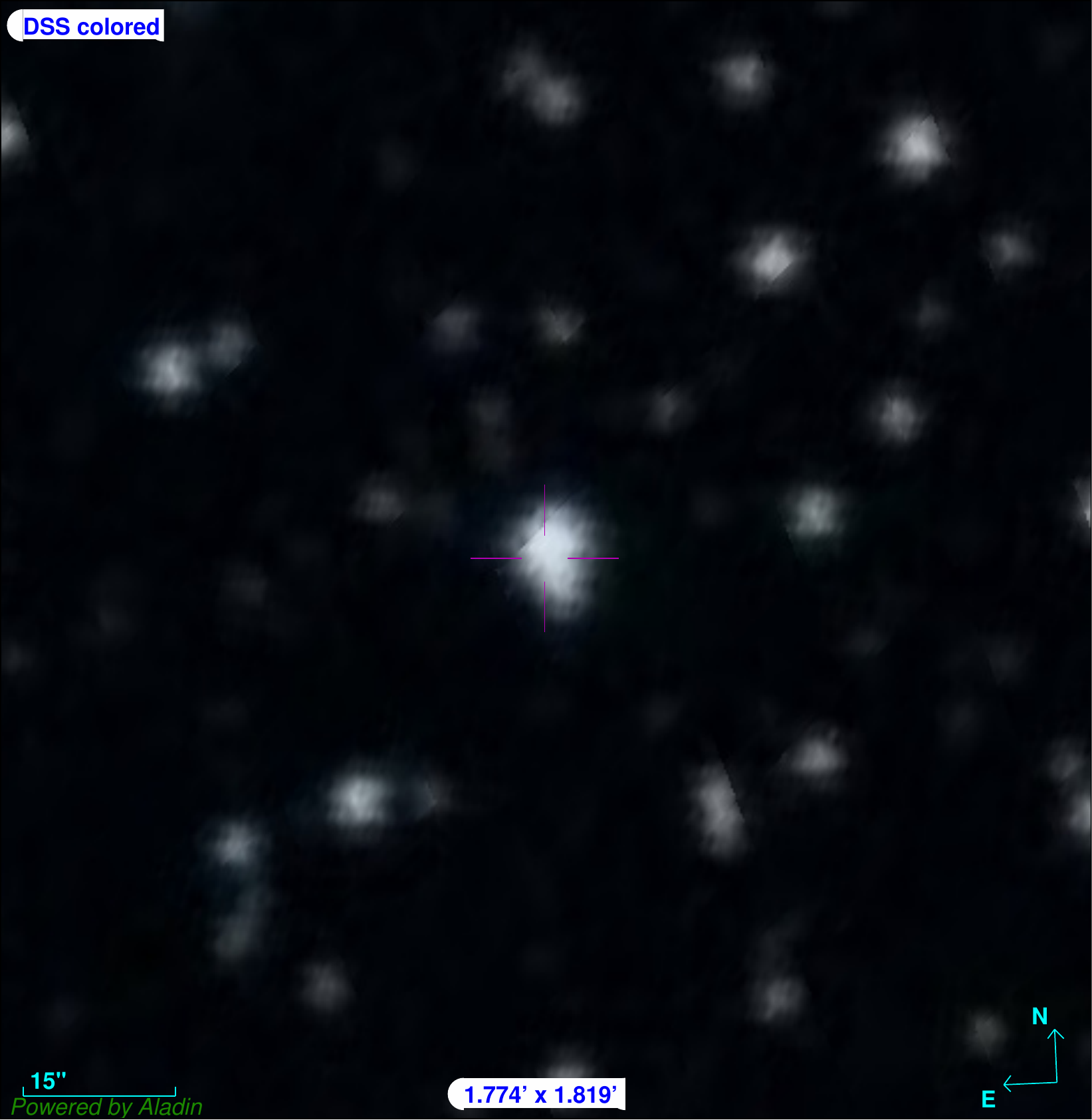}\end{overpic} 
\caption{Finding charts for four stars that appear non-circular due to close
neighbours: KIC\,1162150 (*15, upper left), KIC\,5036493 (*26, upper right),
KIC\,6227118 (*27, lower left) and KIC\,7621759 (*6, lower right).  All are
colour DSS images from the Aladin Sky Atlas, 1.774$\times$1.819 arcmin across,
with North to the top and East to the left.    For KIC\,5036493 the brighter
bluer star to the South is the SX~Phe candidate. } 
\label{findingcharts}
\end{figure}

\subsection{Radial Velocities}

Barycentric radial velocities (RVs) were measured for each  spectrum using the
cross correlation methods implemented in the routines `fxcor' and `xcsao'  in
IRAF (Tonry \& Davis 1979; Kurtz \& Mink 1998) and the `rcros' program of
D\'iaz {\it et al.} (2011), and by fitting synthetic spectra.  Since only minor
differences were found using the different methods we chose to adopt the
`fxcor' values, where the cross correlation function (CCF) was based on
appropriate template synthetic spectra.   The resultant RVs for  the individual
spectra are presented in Table\,1,  and mean RVs are given in Table\,2.

The RV zero-points were checked using the IAU standard star observations.  When
the mean RVs for the five standards observed  with ESPaDOnS were compared with
those given by Udry {\it et al.} (1999) the mean difference (ours minus the
Udry {\it et al.} value) was 0.38$\pm$0.06 km/s.  This mean is comparable with
the $\pm$0.3 km/s uncertainty in the Udry {\it et al.} RV values.  Our
velocities also show good agreement with those from other studies:  for HD\,171391
Massarotti {\it et al.} (2008) found RV=7.59$\pm$0.05 km/s,  compared with our
mean value (five spectra) of 7.78$\pm$0.05 km/s; and for HD\,182572 Nidever
{\it et al.} (2002) measured RV$=-100.29$\,km/s, compared with our
$-99.84\pm0.15$ km/s (four spectra).  Comparisons  for the three IAU RV
standard stars observed at APO  likewise showed excellent agreement between our
RVs and the standard values: mean difference between our estimates  and the
Udry {\it et al.} values amounted to $-0.4\pm0.4$\,km/s. 

\subsubsection{Stars with High Radial Velocities}

Three of the  stars have large negative RVs: KIC\,6520969 (*21) with
RV$=-299.5\pm0.1$ km/s;  KIC\,8004558 (*1) with RV$=-254.1\pm1.5$ km/s;  and
KIC\,11754974 (*13) with RV$=-307\pm4$ km/s.  Two of the stars (KIC\,8004558
and 11754974) were already of interest because of their resemblance to field
SX~Phe stars (see Table\,3 of BN12).  Since all three stars have retrograde
motions and low metallicities (see below, Figs.\,6 and 13) there is little
doubt that they are {\it bona fide} SX~Phe stars.  Their metal abundances are
comparable to those of the SX~Phe  BSs found in metal-rich globular clusters,
for example, 47~Tuc (Gilliland {\it et al.} 1998; Bruntt {\it et al.} 2001) and
M71 (Hodder {\it et al.} 1992;  McCormac {\it et al.} 2013).  The 47~Tuc stars
observed with HST also resemble the {\it Kepler} stars in exhibiting
multiperiodic variations and of having similarly low amplitudes ($\sim$5-50
mmag).   

KIC\,11754974  is known  to reside in a 343\,d binary system (Murphy {\it et
al.} 2013b), and, based on an analysis of the complete set of {\it Kepler}
Q0-Q17 photometry, we show below that KIC\,8004558 also is a photometric
time-delay binary (orbital period $\sim$262\,d).  In both cases the variable
star is thought to be the more massive primary ($\mathscr{M} \sim 1.5
\mathscr{M}_{\odot}$) and the secondary has a mass
$\sim$0.5$\mathscr{M}_{\odot}$.   No photometric or spectroscopic evidence
could be found to suggest that KIC\,6520969 (*21) is a binary system.

\subsubsection{Radial Velocity Variations}

There are several reasons why one might expect some of the program stars to
show RV variations:  (1) all of the stars exhibit light variations due to
radial and non-radial pulsations (usually multiperiodic); (2) at least ten of
the pulsators are in binary systems (see $\S3.3$ below); and (3) the two
misclassified stars, KIC\,9535881  and KIC\,12643589, are known to be close
eclipsing binaries\footnote{Squared parentheses around the CFHT star numbers
for KIC\,9535881 [*25] and KIC\,12643589 [*17] have been used throughout this
paper to distinguish these two stars from the 32 {\it bona fide} candidate
SX~Phe pulsators.}.

Since the spectroscopic part of this study was intended only as a first-look
survey the sampling design is far from optimal for analyzing temporal
variations in the spectra, and only a relatively small number of spectra were
taken for each star (typically 5 and at most 11).  Moreover, owing to the wide
range of line widths of the sample stars the uncertainties of the individual RV
measurements range from $\sim$$\pm10$\,km/s for the four rapidly rotating stars
to only a few tenths of a km/sec for the narrow-lined metal-rich (and standard)
stars that have  high signal-to-noise ratio (S/N) spectra.  Furthermore, the
spacing of the observations is not ideal for identifying RV variations, in
particular: (1) the $\sim$300 day gaps between the 2012, 2013 and 2014
observations are a potential cause of aliasing difficulties; (2) many of the
stars were observed only once in 2014; and (3) the spacing may be inappropriate
for detecting certain orbital and pulsation periods ({\it i.e.} cycle-count and
Nyquist aliasing).  Finally, the IAU RV standards observed at CFHT were
observed only in 2014, and thus year-to-year offsets in the RVs are possible
(although this seems unlikely given the constancy of the RVs for several
program stars).     

Despite these potential problems, time-series graphs show some evidence of RV
variability for approximately half of the 30 candidate stars with spectra.
Stars were ranked by RV range, where $\Delta$RV = RV(max) -- RV(min).  In the
absence of variability, and assuming a normal distribution of RVs,
$\Delta$RV/$\sqrt n \sim \sigma_{\rm RV}$, where $\sigma_{\rm RV}$ is the
standard deviation of the RVs and $n$ is the number of measured RVs for a given
star.   For the non-variable standard stars a typical value for
$\Delta$RV/$\sqrt n$ is $\sim$0.1, which is comparable to the uncertainties
given in Table\,1.

KIC\,6780873\,(*5) has narrow spectral lines and the largest RV range,
$\Delta$RV = 69.9$\pm$0.3 km/s.   {\bf Figure\,3} shows that five of its nine
spectra exhibit line doubling, identified by two distinct peaks  in the
cross-correlation function; thus it is clearly a double-lined spectroscopic
binary (SB2).  In Table\,1  only the RVs for the primary star (assumed to be
the SX~Phe star) are given;  these were calculated using the stronger and wider
of the two CCF peaks ({\it i.e.}, the peak with the larger area and larger
height above the baseline).  The observed RVs of the SX~Phe star vary from
--26.81$\pm$0.13 km/s to +43.08$\pm$0.14 km/s.  The RVs for both components,
noted in Fig.3 and plotted in {\bf Figure\,4}, are consistent with KIC\,6780873
being a close binary with a 9.161$\pm$0.001\,d orbital period (see $\S3.3$).
The widths of the CCFs suggest that $v\sin i$ for the primary is larger than
that of  the secondary, 11$\pm$2 km/s versus 5$\pm$2 km/s.   The observation of
spectral lines due to the secondary suggests that the less luminous companion
star is a low-mass main-sequence star and not a giant or white dwarf.  

The other SX~Phe candidates with large $\Delta$RV values are:  KIC 12688835
(*18),   10989032 (*32),  7765585 (*28),  7819024 (*19),  8004558 (*1), 5705575
(*22),  11754974 (*13) and 9267042 (*12).    All except KIC\,12688835 and
KIC\,7765585  are binary systems (see $\S3.3$).  The large $\Delta$RV of
KIC\,12688835 (*18) is based on four 2013 spectra with RV$\sim$\,--25 km/s and
a single 2014 spectrum with RV=$-51\pm10$\,km/s;  owing to the large
uncertainties which result from  broad spectral lines this evidence for its RV
variability must be considered tentative.  Likewise, the evidence for the RV
variation of  KIC\,7765585 (*28) hinges primarily on one of the RVs being
significantly different from the rest.  The measured $\Delta$RV =
$14\pm6$\,km/s for KIC\,11754974\,(*13), the 344-d non-eclipsing binary studied
by Murphy {\it et al.} (2013b), is consistent with the photometric $K_1$ value of
$8.2\pm0.2$\,km/s (see $\S$3.3);  the significant RV differences seen for the
three August 2013 spectra, if real, hint at additional variability. 

Two other time-delay binaries exhibit small RV ranges  consistent with their
photometrically predicted small $K_1$ values (see Table\,9): KIC\,4243461\,(*4)
was found to have $\Delta$RV = 5.1$\pm$2.0 km/s, which is to be compared with
the predicted $K_1$ = 5.3$\pm$0.2 km/s;  and KIC\,9966976 (*31) has $\Delta$RV
= $5\pm5$ km/s (the large uncertainty being due to its large $v \sin i$)
compared with the predicted $K_1$=0.64$\pm$ 0.04 km/s.   A third time-delay
binary, KIC\,7300184, which was not observed spectroscopically, also is
predicted (based on $K_1$=0.02$\pm$0.02 km/s) to have a small RV range.

Finally, several of the above-mentioned stars show broadened CCFs and structure
that is (or may be) due to RV variation in a binary system (see
$\S3.3$), but might also be caused by rotation, large amounts of
macrotubulence, noise, etc.  The CCFs for four such stars are shown in {\bf
Figure\,5}.  As noted earlier KIC\,8004558 (*1) is a high-velocity 262\,d
binary system; if the CCFs for its eight spectra are fitted with a Gaussian the
RVs appear to shift from $-260$$\pm$4 to $-242$$\pm$4\,km/s, consistent with binary
motion.  KIC\,10989032 (*32) exhibits the second largest $\Delta$RV
(37.5$\pm$1.2 km/s), and  the {\it Kepler} Q0-Q17 light curve shows it to be a
close semi-detached binary system with an orbital period of 2.3\,d.  If it is a
single-line spectroscopic binary and its broadening is due to stellar rotation
then $v\sin i \sim 45$$\pm$2 km/s (see $\S2.3$).  The star is listed in the
Villanova Eclipsing Binary (EB) Catalog\footnote{The Villanova Eclipsing Binary
catalog (Pr\u sa {\it et al.} 2011; Slawson {\it et al.} 2011; Matijevic {\it
et al.} 2012) is available on-line at {\tt http://keplerebs.villanova.edu}.}
where its $P_{\rm orb}$ is given as 2.305097~days.  The RV curve derived from
the eight available spectra appears to be sinusoidal and thus suggests that the
orbits have been circularized (see $\S$3 below).    

Higher SNR spectroscopic observations and more sophisticated analysis methods
(such as the `broadening functions'  advocated by Rucinski 1999) might help to
better assess the RV variations of the program stars.

\begin{figure} \centering 
\begin{overpic}[width=8.0cm]{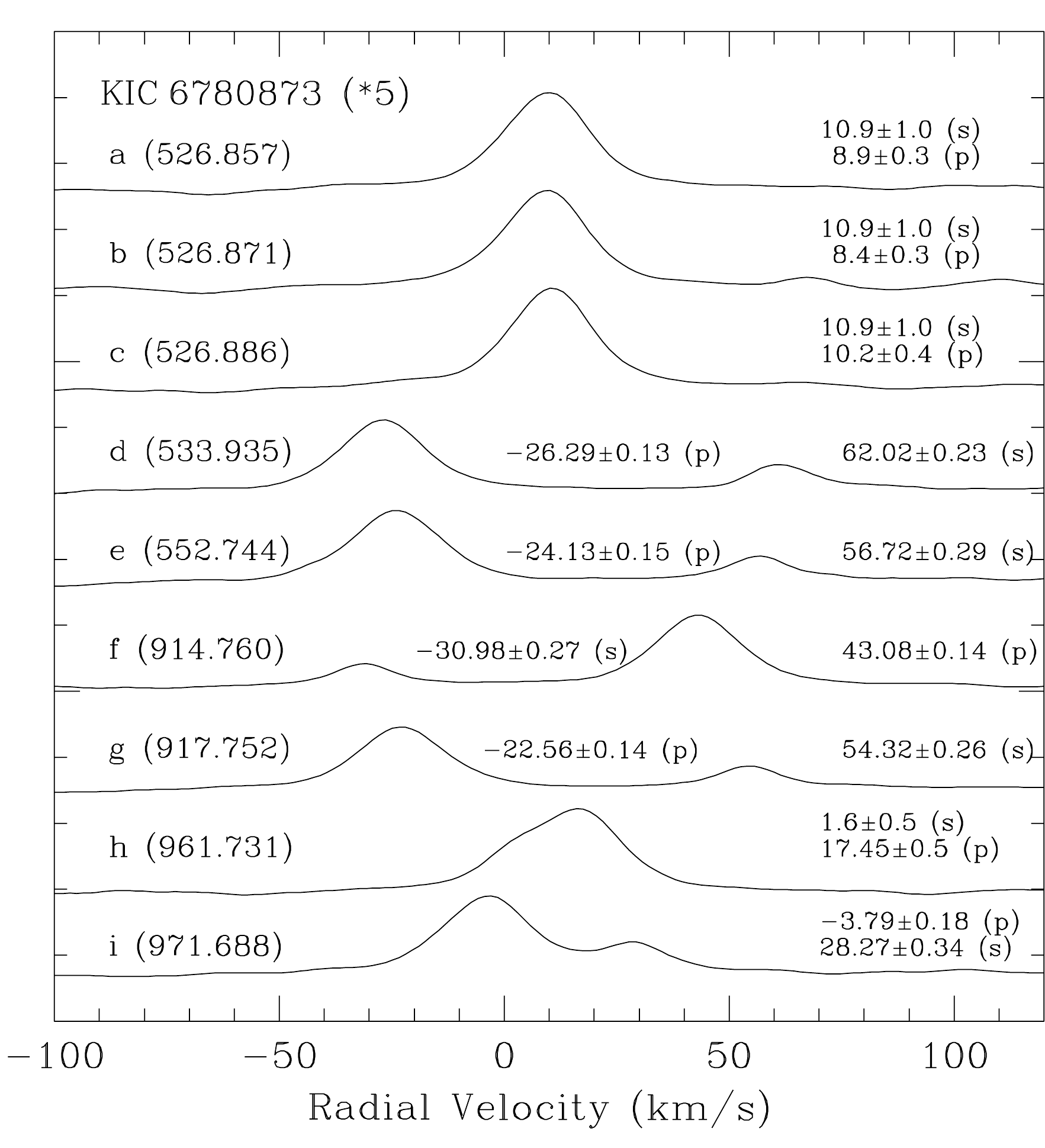} \end{overpic}
\caption{CCFs for the nine spectra of KIC\,6780873, a 9.3-day close binary
system shown here to be double-lined.  Each CCF has been labelled with a letter
(a-i) identifying the measured spectrum, and the time of the observations (BJD
minus 2456000).  Using the peak strengths, each CCF `bump' has been identified
as being due to either the primary (p) or secondary (s) star.  Also given
(right side) are the RVs (km/s) for each component.   } \label{Star5RVcurves1}
\end{figure}

\begin{figure} \centering
\begin{overpic}[width=8.0cm]{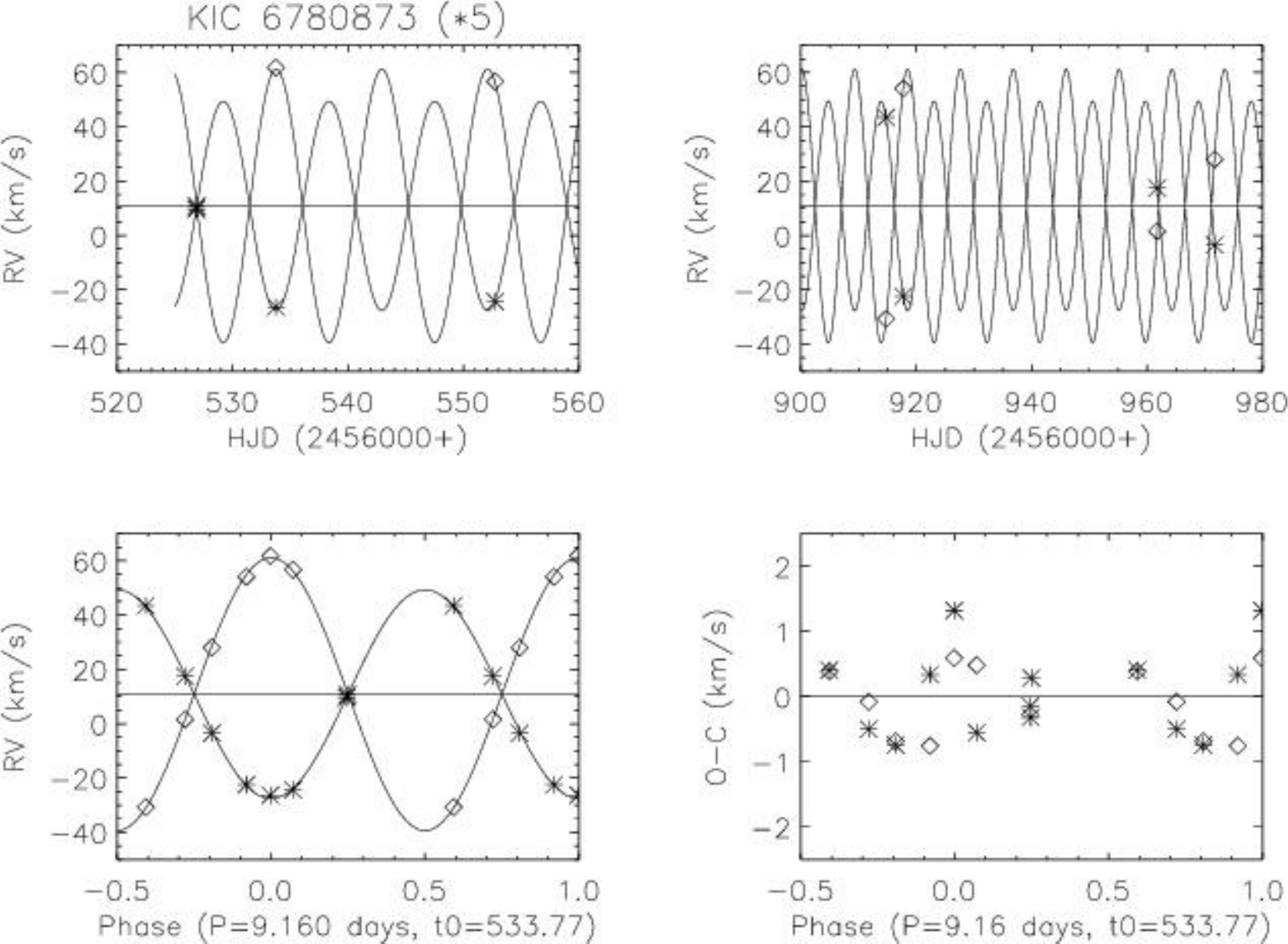} \end{overpic}
\caption{The top two panels show the individual RVs for KIC\,6780873 (*5), the 2013
data on the left and the 2014 data on the right,  overplotted with model curves
assuming eccentricity $e$=0, orbital period $P_{\rm orb}$=9.16\,d, HJD time of
maximum RV for the secondary = 2456533.77,  RV semi-amplitudes for the primary
and secondary $K_1$ = 38.5 km/s and $K_2$ = 50.3 km/s,  and systemic radial
velocity $\gamma$ = 10.885 km/s.  The bottom left panel shows the observed RVs
(asterisks for the primary and open diamonds for the secondary) and the
corresponding primary and secondary phased RV curves.  The bottom right panel
shows the O-C residuals, where the averages for the primary and secondary are
0.009 km/s and $-0.011$ km/s, respectively, and corresponding standard
deviations of the residuals about the means, 0.65 and 0.60 km/s.  }
\label{Star5RVcurves2} \end{figure}

\begin{figure} \begin{center} 
\begin{overpic}[width=4.1cm]{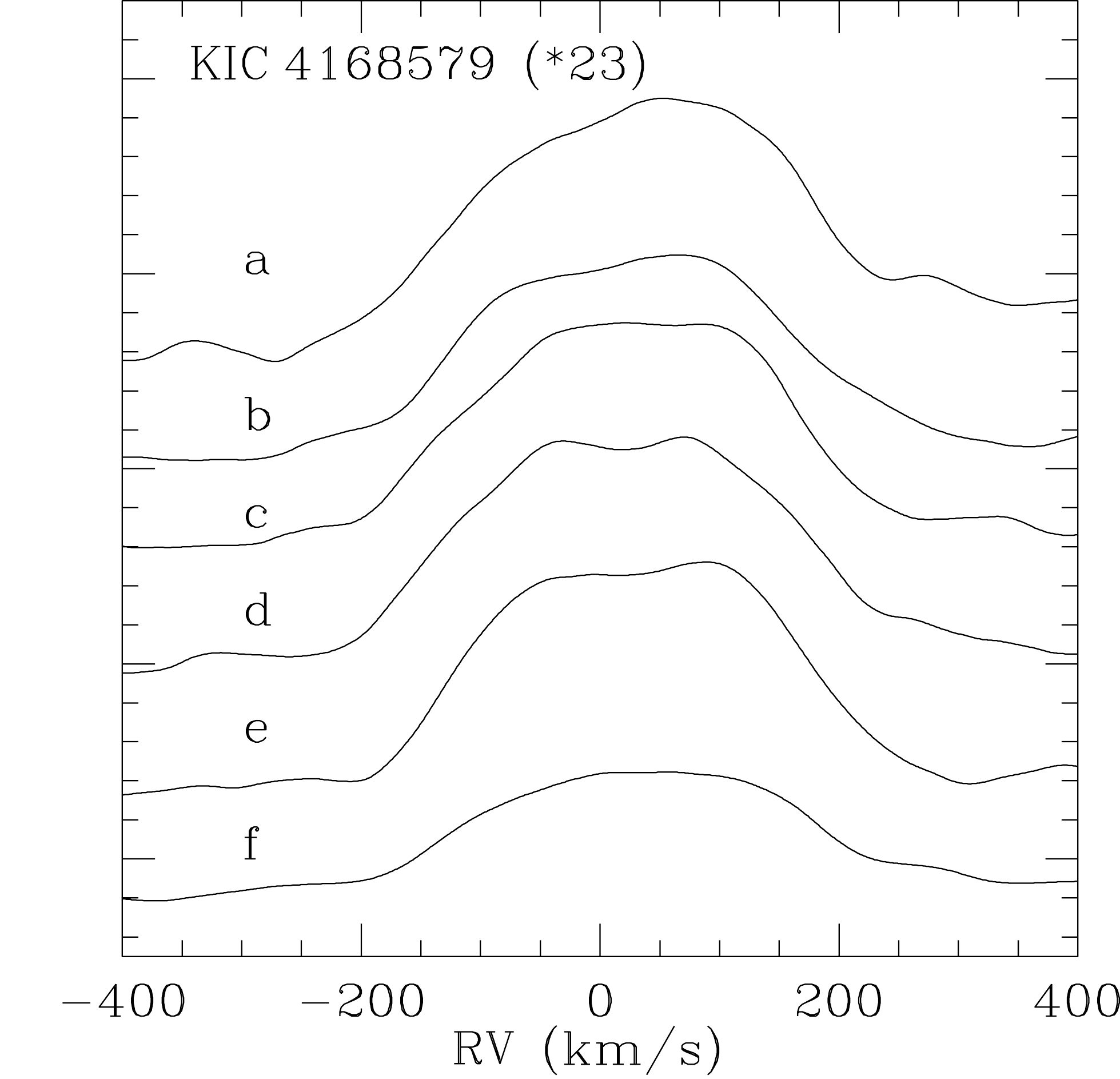} \end{overpic}
\begin{overpic}[width=4.1cm]{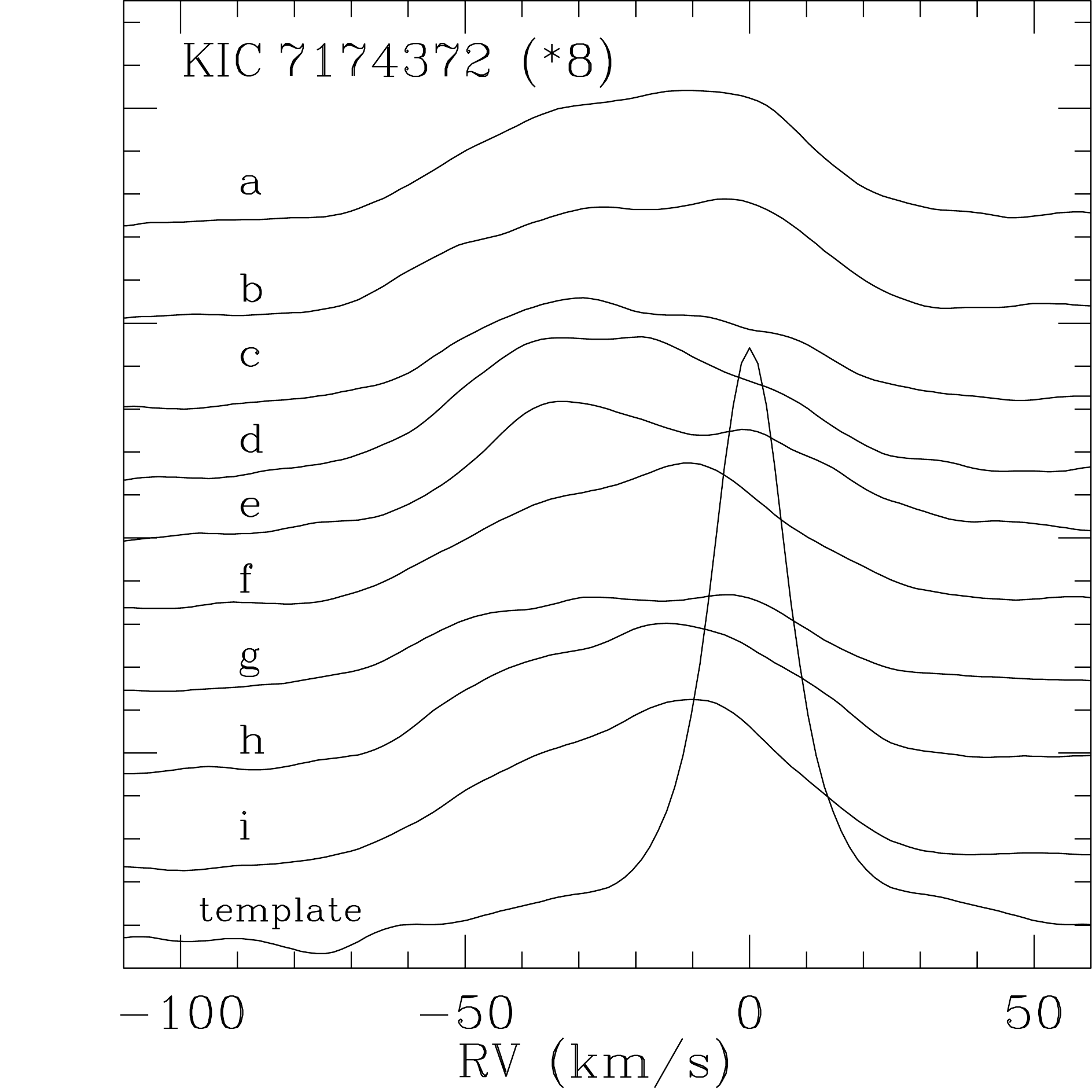} \end{overpic} 
\end{center} \begin{center}
\begin{overpic}[width=4.1cm]{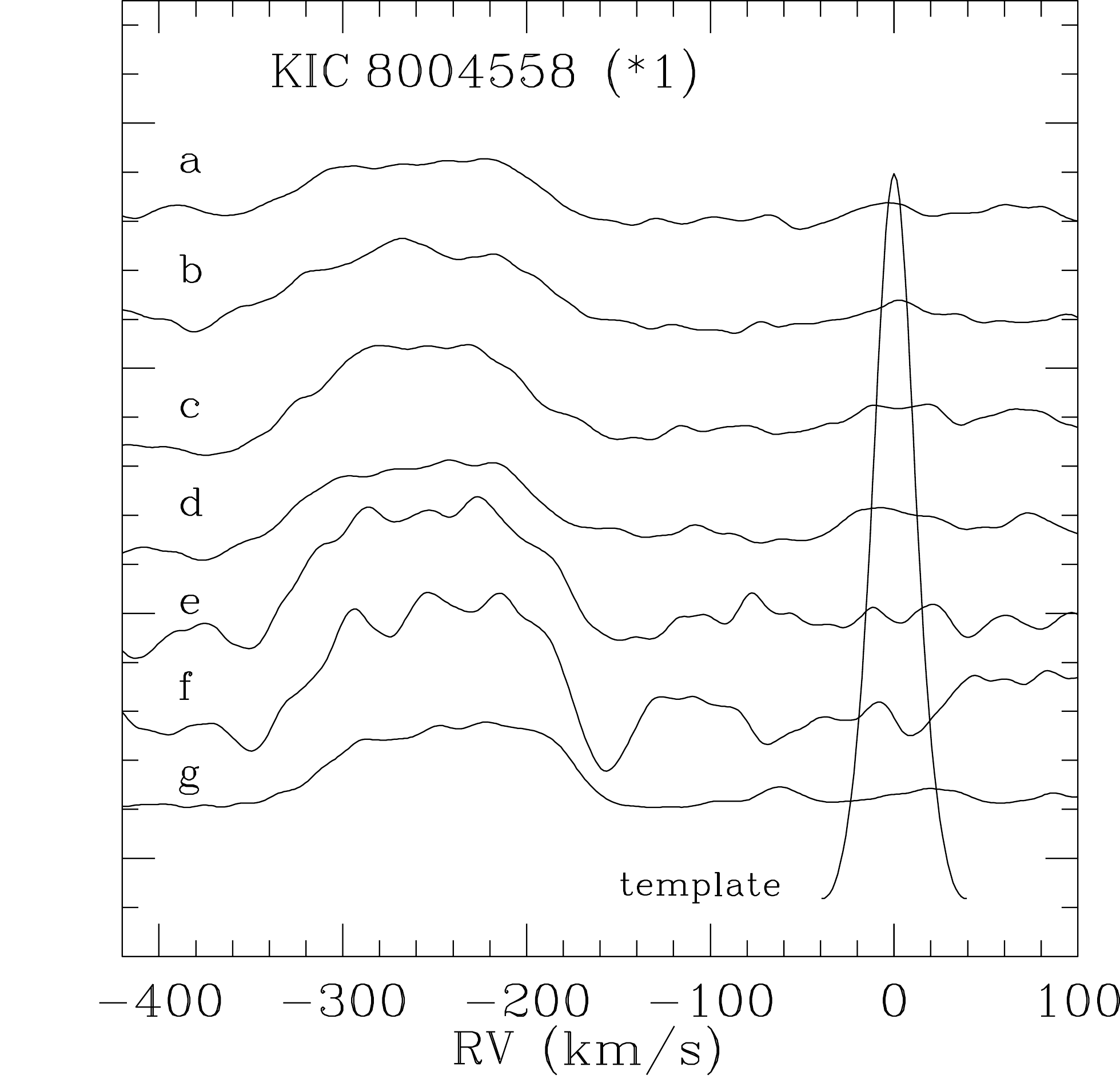} \end{overpic}
\begin{overpic}[width=4.1cm]{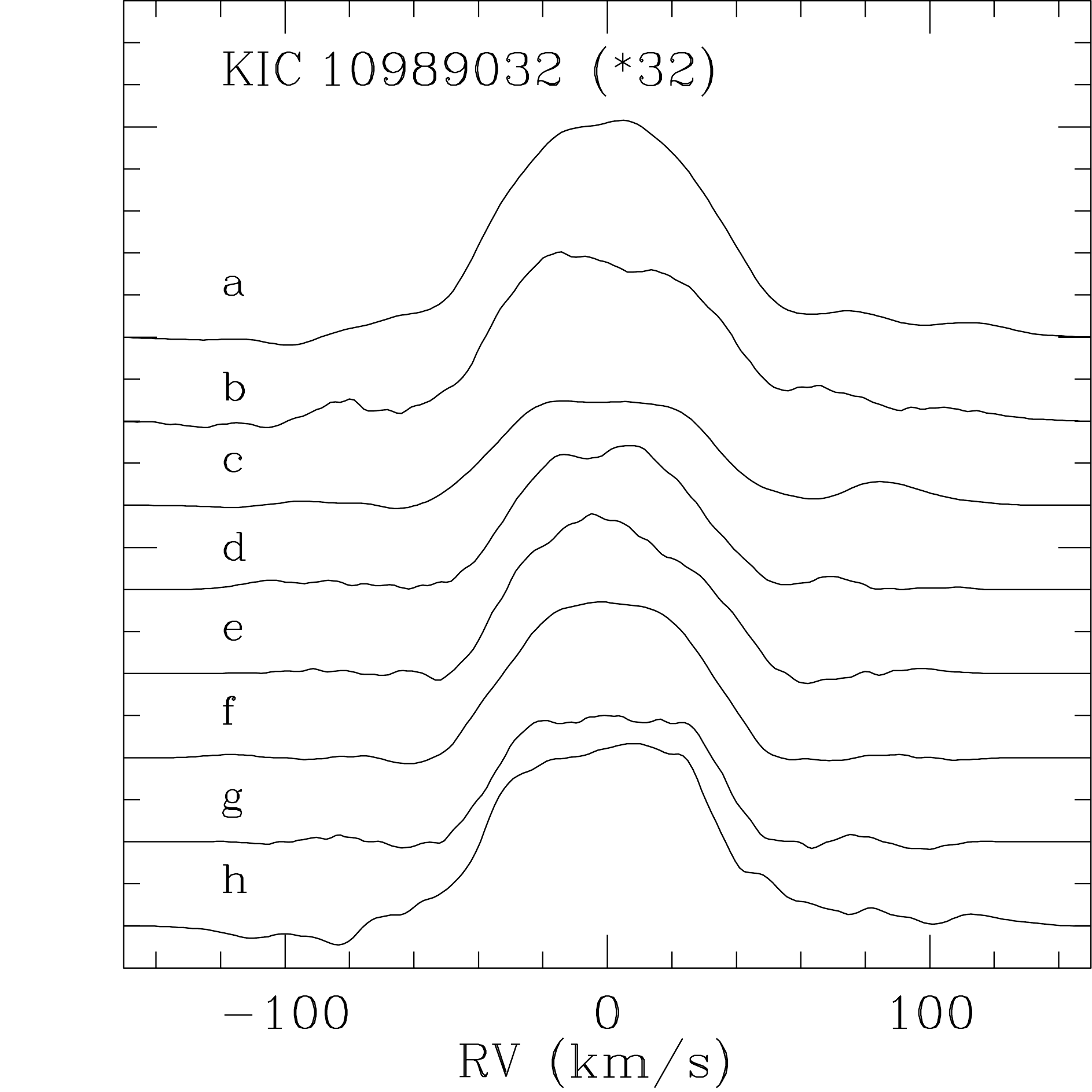} \end{overpic} 
\end{center} \caption{CCFs for
four of the BN12 stars, where the `letter' labels on the left identify the
individual spectra (see Table~1).  For KIC\,10989032 (*32, KOI\,7397), which is
a close binary system with an orbital period of 2.3\,d, the CCFs have been
Doppler shifted so that the mean RV is zero.   KIC\,7174372 (*8) and
KIC\,8004558 (*1) also are probable binary systems (see $\S$3.3).   Also shown
for KIC\,7174372 and KIC\,8004558 are the much narrower autocorrelation CCFs
for the template synthetic spectra.  } \label{Stars23_8_1_32} \end{figure}


\subsection{ Distances and Space Motions }
 
Membership in the galactic halo stellar population is one of the defining
features of an SX~Phe star.  Height  above the galactic plane and space motions
are often used to discriminate between Pop.\,II and Pop.\,I stars, where stars
with vertical height  greater than $\sim$500\,pc and total speed greater than
$\sim$200 km/s almost certainly are Pop.\,II  stars.

Galactic coords ($l$, $b$), $E(B$--$V)$ reddenings, distances, $d$\,(pc), and
vertical heights above the galactic plane, $z$\,(pc), for the program stars are
given in {\bf Table 2}.  The KIC distances are derived from the Sloan
photometry which gives estimates of $T_{\rm eff}$ and $\log g$.  The radii
in the KIC follow from the gravities and model masses, and the $T_{\rm
eff}$ and radius information gives the luminosity, $L$.  From $L$ and the
observed apparent magnitude Brown {\it et al.} (2011) calculate the distance,
from which the $E(B$--$V)$ reddening and the $A_V$ extinctions are estimated
using the simple exponential fall-off model given by their equations 8-13 (see
BN12).  The only program star for which the KIC does not give $E(B$--$V)$, and
consequently distance information, is KIC\,11754974, the star studied by Murphy
{\it et al.} (2013b). 

Brown {\it et al.} (2011) warned that where there is a clumpy distribution of
interstellar matter  ``the result is systematic misclassification'' in the KIC
due to  ``a scattered and confused relation between $T_{\rm eff}$ and color,
and other failings''.  For this reason new distance and reddening estimates
were derived.  These were made using the on-line `cumulative reddening {\it
vs.} distance' tool created for the `3D Dust Mapping' project (see Green {\it
et al.} 2014, 2015) available at {\tt http://argonaut.skymaps.info/query}.   This tool
requires as input  a direction indicator ($l,b$ or RA,DEC) and the distance
modulus,  $\mu_V$ (=$m_V$--$M_V$), and calculates $E(B$--$V)$ and $d$\,(pc),
where the interstellar extinction is based on dust maps (see Schlegel {\it et
al.} 1998, Schlafly \& Finkbeiner 2011) rather than an exponential model.  

For each star $\mu_V$ was derived using the apparent visual magnitude $m_V$
given in column\,4 of Table\,5, and by estimating the absolute visual
magnitude, $M_V$, by assuming for the mass the value
1.5\,$\mathscr{M}_{\odot}$ ({\it i.e.}, estimated mean for
SX~Phe stars) and by substituting the spectroscopic estimates
of $\log g$ and $T_{\rm eff}$ (given in Table\,4, columns 4 and 6) into the
following equations: 

$$\log (R/R_{\odot}) = 0.5 \, [ \log \mathscr{M}/\mathscr{M}_{\odot} - \log(g/g_{\odot})], 
$$ $$\log (L/L_{\odot}) = 2 \log(R/R_{\odot}) + 4 \log(T_{\rm eff}/T_{\rm eff,\odot}),$$  
$$ M_{V} =   4.79 - 2.5 \log(L / L_{\odot} ).    $$

\noindent For the Sun we adopted $T_{\rm eff,\odot}=5772$\,K and  $\log
g_{\odot} = 4.438$, and for the program stars the bolometric corrections, which
are small, were ignored.  The derived  reddenings, distances and
vertical heights are given next to the KIC values in Table 2. 

The new reddenings are smaller than the KIC values for all but six of the
stars, and in general the distances are greater\footnote{Increasing
(decreasing) the assumed mass by 0.25\,$\mathscr{M}_{\odot}$ has little impact
on $\mu_V$ and $d$(pc):  the $\mu_V$ is larger (smaller) by $\sim$0.17 mag, and
the distance is larger (smaller) by $\sim$150\,pc;  the effect of such a change
on   $E(B$--$V)$ depends on the distance and  gradient in the   $E(B$--$V)$
{\it vs.}  $\mu_V$ graph, but is usually small.}. 
The average reddening is 0.12\,mag, the average distance is
2.1\,kpc, and the average height above the galactic plane is 400\,pc. 
The new reddenings and
distances improve upon the KIC values in two ways:  (1) $T_{\rm eff}$ and $\log
g$ now are based on measurements of high-dispersion spectra, whereas the KIC
estimates are based on photometry and have been shown, at least for cooler
stars, to be systematically too low by over 200\,K (Pinsonneault {\it et al.}
2012;  see below); and (2) the patchy distribution of the interstellar gas and
dust, which is evident in the appearance of multiple discrete interstellar
lines in several of the CFHT spectra\footnote{The distinction between  stellar
and interstellar Na\,I\,D lines  is clear for almost all the stars.  This is
especially true for the three stars with large RVs where the interstellar
Na\,I\,D lines are shifted $\sim$0.5\,nm to the right of the stellar lines.
For several of the stars two or three individual sets of interstellar lines are
seen, each presumably arising from a separate interstellar cloud.  For several
of the stars the interstellar Na\,I lines are saturated, an extreme example
being KIC\,9244992 (*7), which has one of the largest $E_{B-V}$ reddenings.  A
large range in line strengths was  observed for the interstellar neutral
potassium (K\,I) line at 769.8\,nm.  Surprisingly, KIC\,3456605 with
$E_{B-V}=0.14$ has a particularly strong and broad interstellar K\,I line.
Detailed quantitative analysis of the interstellar lines, such as that
performed by Poznanski {\it et al.} (2012), might be useful for investigating
these and other spectral features but is beyond the scope of the present paper.
}, is now taken into account.

\begin{table*} \caption{Locations and kinematics for the 34 SX~Phe candidates.
The columns contain:  (1-2) KIC and CFHT star numbers; (3) galactic latitude
and longitude (degrees); (4-5) $E(B$--$V)$ reddening (mag) and distance
$d$\,(pc), from the KIC and based on dust maps;  (6) height above the galactic
plane $z$\,(pc), KIC values and assuming the new distances; (7) mean radial
velocity $\pm$ standard error (with number of measured spectra  given in
parentheses) - for binary systems the systemic RVs are given and 
\underline{underlined}; (8) space motions, $U,\,V,\,W$, and total speed, $T$,
all relative to the Local Standard of Rest. }

\begin{tabular}{rccrrrrl}
\hline
\multicolumn{1}{c}{KIC} &
\multicolumn{1}{c}{CFHT}&
\multicolumn{1}{c}{Galactic Coords. } &  
\multicolumn{1}{c}{ $E(B$--$V)$ } &
\multicolumn{1}{c}{$d$\,(pc)} &
\multicolumn{1}{c}{$z$\,(pc)} &
\multicolumn{1}{c}{ $<$${\rm RV}$$>$ } &
\multicolumn{1}{c}{($U,V,W,T$)  } \\
\multicolumn{1}{c}{no.}& 
\multicolumn{1}{c}{no.}&
\multicolumn{1}{c}{$l$ (J2000) $b$ }&
\multicolumn{1}{c}{KIC, new} & 
\multicolumn{1}{c}{KIC, new} &
\multicolumn{1}{c}{KIC, new} &
\multicolumn{1}{c}{ (km/s) } &
\multicolumn{1}{c}{(km/s)} \\
\multicolumn{1}{c}{(1)} & 
\multicolumn{1}{c}{(2)} &
\multicolumn{1}{c}{(3)} &
\multicolumn{1}{c}{(4)} &
\multicolumn{1}{c}{(5)}&
\multicolumn{1}{c}{(6)}&
\multicolumn{1}{c}{(7)}&
\multicolumn{1}{c}{(8)} \\
\hline
   1162150 & 15 & 69.5124, ~9.7968 & 0.160, 0.07   & ~729, 1060   &  124, 180   & $-14.8\pm0.9$\,(6)    & ~~99, --5, 234, 254      \\ 
   3456605 & 24 & 72.2362, ~8.1478 & 0.203, 0.14   & ~956, 1690   &  135, 240   &  $-7.4\pm0.6$\,(3)    & --335, --70, --186, 389    \\  
   4168579 & 23 & 72.6974, ~8.8075 & 0.257, 0.17   & 1644, 3620   &  252, 550   &  $22.9\pm1.7$\,(6)    & --345, 70, --841, 912       \\  
   4243461 & 4  & 69.8106, 14.9474 & 0.153, 0.11   & ~983, 1580   &  254, 410   & \underline{$57.2\pm1.8$}\,(8)    & --331, --169, 456, 588       \\
   4662336 & 14 & 73.0933, ~9.2758 & 0.204, 0.12   & 1059, 1760   &  171, 280   & $-11.3\pm0.4$\,(4)    & --455, --93, --233, 520    \\  
   4756040 & 20 & 73.2980, ~8.9596 & 0.206, 0.11   & 1048, 1780   &  163, 280   &   $5.9\pm1.4$\,(4)    & --149, 35, --343, 376        \\  
   5036493 & 26 & 74.8907, ~6.7575 & 0.223, 0.27   & ~996, 1890   &  117, 220   &  $-4.9\pm0.7$\,(4)    & --355, --62, --187, 406      \\
   5390069 & -- & 75.5152, ~6.5051 & 0.174, 0.19   & ~693, 4200   &  ~78, 480   &\multicolumn{1}{c}{--} & \multicolumn{1}{c}{ -- } \\ 
   5705575 & 22 & 73.4695, 11.2094 & 0.209, 0.09   & 1412, 2220   &  275, 430   &\underline{$-38.1\pm1.0$}\,(9)    & --364, 125, --1214, 1274      \\  
   6130500 & 9  & 74.6166, ~9.9930 & 0.212, 0.12   & 1235, 2170   &  214, 380   & $-18.3\pm0.5$\,(6)    & ~147, 145, --586, 622       \\
   6227118 & 27 & 76.0714, ~7.6323 & 0.364, 0.12   & 4172, ~850   &  539, 110   &   $5.3\pm1.7$\,(4)    & --68, 22, --125, 144    \\  
   6445601 & 2  & 74.5439, 10.9821 & 0.185, 0.08   & 1026, 2260   &  195, 430   &  $-4.5\pm0.5$\,(6)    & --416, --191, 439, 634        \\
   6520969 & 21 & 74.1754, 11.9396 & 0.212, 0.08   & 1718, 2850   &  355, 590   &$-299.5\pm0.1$\,(3)    & ~112,  --395, 571, 703       \\
   6780873 & 5  & 74.9284, 11.0447 & 0.153, 0.06   & ~729, 2090   &  140, 400   & \underline{$10.1\pm0.3$}\,(9)    & --603,--128, --27, 617       \\   
   7020707 & 16 & 73.5340, 14.8428 & 0.166, 0.04   & 1236, 1970   &  317, 510   &   $2.4\pm1.3$\,(4)    & --297, --122, 196, 376       \\
   7174372 & 8  & 71.9764, 18.9932 & 0.136, 0.05   & 1135, 2970   &  369, 970   & \underline{$-19.2\pm0.6$}\,(9)     & --1120, --404, 102, 1195       \\  
   7300184 & -- & 76.8302, ~8.7823 & 0.181, 0.27   & ~832, 4240   &  127, 650   &\multicolumn{1}{c}{ -- } & \multicolumn{1}{c}{ -- }  \\ 
   7301640 & 10 & 77.0021, ~8.5969 & 0.212, 0.26   & 1068, 2110   &  160, 320   & $-12.0\pm1.8$\,(5) & ~~95, 93, --431, 451      \\
   7621759 & 6  & 76.8876, ~9.5983 & 0.207, 0.19   & 1120, 2550   &  187, 430   &  $17.0\pm0.6$\,(5) & ~543, 128, 189, 588           \\
   7765585 & 28 & 77.0330, ~9.6755 & 0.115, 0.18   & ~455, 1820   &  ~76, 310   &   $1.1\pm3.3$\,(6) & ~707, 158, 128, 736            \\
   7819024 & 19 & 75.4361, 13.1002 & 0.184, 0.05   & 1310, 1950   &  297, 440   & \underline{$-63.7\pm4.4$}\,(8) & --298, --61, --259, 399     \\  
   8004558 & 1  & 72.7607, 20.3549 & 0.135, 0.05   & 1519, 1430   &  528, 500   & \underline{$-254.1\pm1.5$}\,(7) & --198, --290, --92, 363     \\  
   8110941 & 29 & 77.6587, ~9.6836 & 0.184, 0.15   & ~910, 1760   &  153, 300   &   $4.4\pm0.3$\,(4) & --52, --62, 414, 422        \\
   8196006 & 30 & 79.3018, ~7.0045 & 0.198, 0.32   & ~850, 1820   &  104, 220   &  $-3.5\pm0.8$\,(4) & --133, 26, --311, 339         \\
   8330910 & 3  & 79.6884, ~6.8192 & 0.228, 0.31   & 1036, 1770   &  123, 210   &  $20.9\pm1.1$\,(4) & ~572, 142, --9, 590        \\
   9244992 & 7  & 80.6380, ~7.8965 & 0.288, 0.22   & 1858, 1470   &  255, 200   & $-15.9\pm0.2$\,(5) & --37, 40, --334, 338       \\
   9267042 & 12 & 75.8193, 17.8478 & 0.156, 0.05   & 2170, 1580   &  665, 480   & \underline{$-11.5\pm1.0$}\,(11) & ~506, --124,  774, 933   \\  
   9535881 &[25]& 79.4210, 11.1577 & 0.168, 0.07   & ~864, 1940   &  167, 380   &  \underline{$5.0\pm0.1$}\,(4) & ~247, 55, 71, 263          \\  
   9966976 & 31 & 80.7107, 10.1734 & 0.221, 0.09   & 1396, 1400   &  247, 250   & \underline{$-0.1\pm1.1$}\,(4) & ~362, 62, 68, 374           \\  
  10989032 & 32 & 82.0254, 11.1154 & 0.230, 0.10   & 1931, 2710   &  372, 520   & \underline{$-25.0\pm0.3$}\,(8) & --997, --236, 436, 1113   \\ 
  11649497 & 11 & 79.6609, 19.5242 & 0.138, 0.04   & 1374, 1950   &  459, 650   & $-21.5\pm0.2$\,(6) & --246, 48, --273, 371        \\
  11754974 & 13 & 80.5370, 17.8823 &  --~~, 0.04   & ~~--~~, 1250 & ~~--~~, 380 &\underline{$-307\pm4$}\,(8) & --379, --393, 76, 552     \\   
  12643589 &[17]& 82.5739, 17.7885 & 0.111, 0.05   & ~592, 1200   &  181, 370   & \underline{$-60.9\pm0.1$}\,(5) & ~568, 144, --365, 691       \\  
  12688835 & 18 & 82.7967, 17.5845 & 0.159, 0.03   & 2445, 1890   &  739, 570   & $-31.2\pm7.6$\,(5) & ~318, 14, 34, 321          \\
\\ 
\hline
\end{tabular}
\medskip
\end{table*}

\begin{figure} 
\centering 
\begin{overpic}[width=8cm]{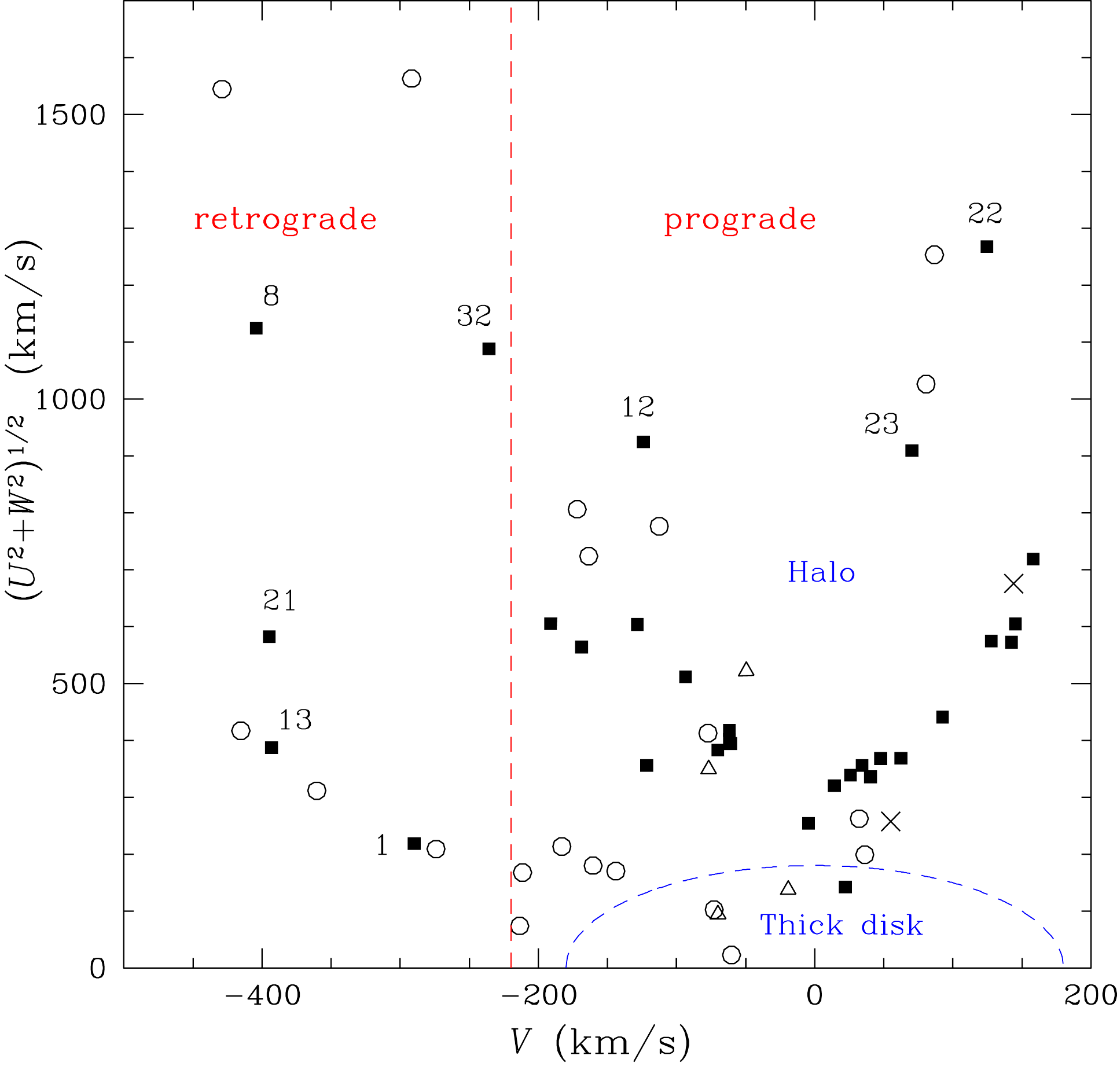} \end{overpic}
\caption{Toomre diagram for the candidate SX~Phe stars in the {\it Kepler}
field (solid squares).  For comparison purposes 24 of the {\it Kepler}-field
RR~Lyrae stars are also plotted (open circles for metal-poor, open triangles
for metal-rich).  The vertical line at $V = -220$ km/s separates stars
undergoing retrograde and prograde motion (asymmetric drift), and the curved
line at total space velocity $T$=180 km/s separates the thick disk and halo
stars.  The labels for the most extreme SX~Phe stars are the CFHT star numbers.
The two close binaries that were misclassified as candidate SX~Phe stars,
KIC\,9535881 [*25] and KIC\,12643589 [*17], are plotted as crosses.}
\label{Toomre} \end{figure}


Galactic  $U,V,W$ space velocities  were estimated  by combining the mean RVs
given in Table\,2 (col.\,7), the new distances (col.\,5), and the proper motion
information summarized in BN12.  The velocities are with respect to the Local
Standard of Rest (LSR), assuming for the solar motion the values
($U,V,W$)$_{\odot}$ = ($-8.5\pm0.29, 13.38\pm0.43, 6.49\pm0.26$) km/s
(Co\c{s}kuno\u{g}lu {\it et al.} 2011).  The calculations were made using the
general method given by Johnson \& Soderblom (1987), as implemented in the IDL
`Astrolib' routine ``gal\_uvw'', which assumes  $U$  positive in the
anti-center direction, $V$  positive in the direction of galactic rotation, and
$W$  positive in the direction of the North Galactic Pole ({\it i.e.},
left-handed Galactic system).  Also calculated were total space motions, $T$,
equal to $(U^2+V^2+W^2)^{1/2}$.  The resulting velocities are given in
column\,8 of Table~2.


The `Toomre diagram' is a graphical summary of $U,V,W$ space motions that is
useful for discriminating  halo, thick disk and thin disk stars (Sandage \&
Fouts 1987, Venn {\it et al.} 2004).  Of course, population discrimination
based solely  on kinematic information is subject to  error because the
underlying  velocity distributions for the different  stellar populations
overlap by various amounts (Nemec \& Linnell Nemec 1991, 1993), a problem that
is compounded when the complexity of the distributions is increased by possible
galaxy mergers and local streaming events (see, for example, Bensby {\it et
al.} 2007; Carollo {\it et al.} 2007, 2010).  The Toomre diagram  also ignores
chemical composition and age differences (see Carollo {\it et al.} 2016).
Despite these limitations the diagram has proven to be useful for identifying
Pop.\,II stars (see, for example, Schuster {\it et al.} 2012, and Ram\'irez
{\it et al.} 2013). 

A Toomre diagram for the candidate SX~Phe stars is plotted in {\bf Figure\,6}.
The (red) dashed vertical line at $-220$ km/s separates the stars with prograde
motions from those with retrograde motions, and the (blue) dashed curve divides
the halo population stars from the thick disk stars which  have total space
velocities $80 < T < 180$ km/s.   Also plotted in
Fig.\,6 are points for 24 of the {\it Kepler}-field RR~Lyrae stars, where the
$U,V,W$ velocities were calculated using the proper motions and distances given
in  BN12 and the mean velocities from Nemec {\it et al.} (2013).  The
metal-poor RR~Lyrae stars ({\it i.e.}, those with [Fe/H] $< -1.0$ dex) are
shown as open circles, and the four `metal-rich' RR~Lyrae stars (V782~Cyg,
V784~Cyg, V2470~Cyg and KIC~11125706) are plotted as open triangles.

Based on the new distances and velocities we conclude that:  (1) half of the
stars have $z$-heights greater than 400\,pc;  (2) in the Toomre diagram all the
stars (except possibly KIC\,6227118 which is located just inside the thick-disk
boundary) have total space motions $T$$>$180 km/s, including 16 stars with $T$
greater than 500 km/s;  (3) five of the  stars orbit the Galaxy in a retrograde
direction\footnote{Other well-known field SX~Phe stars on retrograde orbits
include SX~Phe itself, BL~Cam (=GD~428), and KZ~Hya (see Table~1 of Nemec \&
Mateo 1990b).  Given the extreme retrograde motion of the SDSS halo blue
straggler J1300+0422 (Tillich {\it et al.} 2010) it would be of interest to see
if it too is pulsationally unstable.};   and (4) the space motions of the
SX~Phe stars, in particular the fraction on retrograde orbits, are not
substantially different from those of the {\it Kepler}-field RR~Lyrae stars.
Each of these findings supports the argument that most, if not all, of the
SX~Phe candidate stars belong kinematically to the galactic halo stellar
population, a conclusion that remains unchanged regardless of whether the
$z$-heights and space motions are based on the new spectroscopic values or on
the information in the KIC.  Finally, if the galactic halo consists of ``two
broadly everlapping structural components'' as advocated by Carollo {\it et al.}
(2007, 2016) then the stars with retrograde orbits may be members of the
outer-halo component.

\begin{table*} 
\caption{Spectral types, luminosity classes,  mean projected rotational
velocities, $<$$v \sin i$$>$ (km/s),  and mean radial-tangential macroturbulent
velocities, $<$$\zeta_{\rm RT}$$>$ (km/s), for the {\it Kepler}-field candidate
SX~Phe stars.  The columns contain: (1) KIC number; (2) CFHT number; (3-4)
Right Ascension and Declination (J2000);  (5-7) spectral type based on the
K-line, the Hydrogen lines, and the metal lines; (8) luminosity class;  (9)
weighted average of the $v \sin i$ values given in Table~1, with the number of
spectra measured given in parentheses; (10) the mean $v \sin i$ values derived
using the goodness-of-fit method; and (11) the average $\zeta_{\rm RT}$ derived
using the goodness-of-fit method.  In the last two columns the uncertainty is the standard
deviation of the mean, and in (11) the number of lines that were measured is given in
parentheses. }
\label{Table3}
\begin{tabular}{rcccrccccrcr}
\hline
\multicolumn{1}{c}{KIC} &
\multicolumn{1}{c}{CFHT} &
\multicolumn{2}{c}{RA     (J2000)      DEC} & 
\multicolumn{1}{c}{ } &
\multicolumn{3}{c}{Spectral Type} &
\multicolumn{1}{c}{Lum.} &
\multicolumn{2}{c}{$<v \sin i>$ } &
\multicolumn{1}{c}{ $<\zeta_{\rm RT}>$ } \\ 
\cline{3-4} \cline{6-8} \cline{10-11}
\multicolumn{1}{c}{no.}&
\multicolumn{1}{c}{no.}& 
\multicolumn{1}{c}{h : m : s} &
\multicolumn{1}{c}{$^{\circ}$ : $'$ : $''$} &
\multicolumn{1}{c}{ } & K & H & M & 
\multicolumn{1}{c}{Class} &
\multicolumn{1}{c}{FT} &
\multicolumn{1}{c}{GOF}  &
\multicolumn{1}{c}{GOF} \\
\multicolumn{1}{c}{(1)} & 
\multicolumn{1}{c}{(2)} &
\multicolumn{1}{c}{(3)} &
\multicolumn{1}{c}{(4)} &
\multicolumn{1}{c}{   } &
\multicolumn{1}{c}{(5)} &
\multicolumn{1}{c}{(6)} &
\multicolumn{1}{c}{(7)} &
\multicolumn{1}{c}{(8)} &
\multicolumn{1}{c}{(9)} &
\multicolumn{1}{c}{(10)} &
\multicolumn{1}{c}{(11)} \\
\hline
   1162150 & 15 & 19:25:16.66 & +37:17:22.84 &&   -      &  {\bf F1} &  {\bf A7}   & V       &  $225\pm5$\,(6)    & \multicolumn{1}{c}{$\dots$} & \multicolumn{1}{c}{$\dots$}        \\ 
   3456605 & 24 & 19:38:34.22 & +38:30:45.46 &&   F0     &   F0      &   F0        & II/IIIa &  $13.7\pm1.7$\,(3) & $11\pm3$   & $16\pm3$\,(20)  \\ 
   4168579 & 23 & 19:36:41.30 & +39:13:34.36 && {\bf A8} &  {\bf F1} &  {\bf A9}   & V       &  $197\pm4$\,(6)    &\multicolumn{1}{c}{$\dots$}  & \multicolumn{1}{c}{$\dots$}        \\ 
   4243461 &  4 & 19:01:45.58 & +39:19:09.88 &&   A9     &   A9      &   A9        & V       &  $53.2\pm0.8$\,(8) & $53\pm3$   & $29\pm6$\,(08)  \\ 
   4662336 & 14 & 19:35:29.34 & +39:47:26.20 &&   A9     &   A9      &   A9        & III     &  $83.2\pm0.6$\,(4) &  \multicolumn{1}{c}{$\dots$}   &  \multicolumn{1}{c}{$\dots$}   \\ 
   4756040 & 20 & 19:37:26.19 & +39:49:17.59 && {\bf A7} &  {\bf F1} &  {\bf F0}   & V       &  $42.3\pm2.1$\,(4) & $48\pm2$   & $18\pm4$\,(05)  \\ 
   5036493 & 26 & 19:51:29.03 & +40:07:58.53 && {\bf A3} &  {\bf A7} &  {\bf A3}   & V       &  $19.0\pm1.1$\,(4) & $18\pm1$   & $15\pm1$\,(27)  \\ 
   5705575 & 22 & 19:27:19.60 & +40:59:57.71 && {\bf A5} &  {\bf A9} &  {\bf A6}   & V       &  $88.2\pm1.1$\,(9) &\multicolumn{1}{c}{$\dots$}& \multicolumn{1}{c}{$\dots$}                  \\ 
   6130500 &  9 & 19:35:48.64 & +41:27:13.79 &&  -       &   -       &   F0        & IV:     &  $49.3\pm1.1$\,(6) & $48\pm1$   & $25\pm2$\,(20)  \\ 
   6227118 & 27 & 19:50:33.67 & +41:35:04.07 && {\bf A6} &  {\bf A8} &  {\bf A5}   & V       &  $133\pm2$\,(4)    &\multicolumn{1}{c}{$\dots$}&\multicolumn{1}{c}{$\dots$}  \\ 
   6445601 &  2 & 19:30:55.42 & +41:50:28.11 &&   F2     &   F2      &   F2        & V       &  $71.3\pm0.4$\,(6) & $70\pm2$   & $25\pm4$\,(20)         \\ 
   6520969 & 21 & 19:25:27.08 & +41:56:30.11 && {\bf A3} &  {\bf A7} &  {\bf A3}   & V       &$<$$5.8\pm0.8$\,(3) & $<$$8\pm1$     & $15\pm2$\,(35)  \\ 
   6780873 &  5 & 19:31:32.23 & +42:12:22.83 &&   F1     &   F1      &   F1        & IV      &  $<$$8\pm1$\,(3)   & $<$$8\pm1$     & $16\pm1$\,(13)  \\ 
   7020707 & 16 & 19:09:46.88 & +42:35:07.20 &&   -      &  {\bf F1} &  {\bf F0}   & V       & $105\pm2$\,(4)     &\multicolumn{1}{c}{$\dots$}& \multicolumn{1}{c}{$\dots$}   \\ 
   7174372 &  8 & 18:45:48.25 & +42:43:36.90 &&   A9     &   A9      &   A9        & III     &  $41.6\pm1.0$\,(9) & $41\pm1$   & $26\pm2$\,(21)  \\ 
   7301640 & 10 & 19:48:28.67 & +42:51:41.00 &&   A9     &   A9      &   A9        & V       & $123\pm2$\,(5)     &\multicolumn{1}{c}{$\dots$}& \multicolumn{1}{c}{$\dots$}   \\ 
   7621759 & 6  & 19:43:22.63 & +43:14:43.47 &&   F1     &   F1      &   F1        & IV/V    &  $77.7\pm1.0$\,(5) &  \multicolumn{1}{c}{$\dots$}    &  \multicolumn{1}{c}{$\dots$}   \\ 
   7765585 & 28 & 19:43:22.71 & +43:24:29.74 &&   A9     &   A9      &   A9        & V       & $122\pm1$\,(6)     &\multicolumn{1}{c}{$\dots$}& \multicolumn{1}{c}{$\dots$}  \\ 
   7819024 & 19 & 19:22:38.01 & +43:33:08.38 && {\bf A9} &  {\bf F1} &  {\bf A9}   & V       &  $95.1\pm1.3$\,(8) &\multicolumn{1}{c}{$\dots$}& \multicolumn{1}{c}{$\dots$}   \\ 
   8004558 &  1 & 18:40:04.07 & +43:52:18.06 && {\bf A2.5}& {\bf F0} &  {\bf A2}   & V       &  $84.2\pm2.6$\,(7) &  \multicolumn{1}{c}{$\dots$}   &   \multicolumn{1}{c}{$\dots$}   \\ 
   8110941 & 29 & 19:44:57.95 & +43:57:15.65 &&   F0     &   F0      &   F0        & III     &$<$$7.5\pm0.2$\,(4) & $<$$6\pm1$ & $11\pm1$\,(19)  \\ 
   8196006 & 30 & 20:02:23.14 & +44:01:32.08 &&  -       &  {\bf F0} &  {\bf F2}   & V       &  $92.6\pm1.3$\,(4) &\multicolumn{1}{c}{$\dots$}& \multicolumn{1}{c}{$\dots$}\\
   8330910 &  3 & 20:04:23.03 & +44:15:22.01 &&  -       &   -       &   -         & -       &  $224\pm3$\,(4)    &\multicolumn{1}{c}{$\dots$}&  \multicolumn{1}{c}{$\dots$}    \\ 
   9244992 &  7 & 20:01:57.43 & +45:37:15.59 &&  F0      &   F0      &   F0        & II      &$<$$6.7\pm0.3$\,(5) & $<$$6\pm1$ & $12\pm3$\,(13)  \\ 
   9267042 & 12 & 18:58:52.07 & +45:44:57.85 && {\bf A7} &  {\bf A7} &  {\bf A3}   & V       & $106\pm3$\,(11)    &\multicolumn{1}{c}{$\dots$}& \multicolumn{1}{c}{$\dots$} \\ 
   9535881 &[25]& 19:42:14.31 & +46:10:40.94 &&  F0.5    &  F0.5     &   F0.5      & V       &  $54.6\pm1.7$\,(4) &  $62\pm2$  &   $18\pm5$\,(09)  \\ 
   9966976 & 31 & 19:50:46.42 & +46:49:28.71 &&   F0     &    F0     &    F0       & V       & $123\pm1$\,(4)     &\multicolumn{1}{c}{$\dots$}& \multicolumn{1}{c}{$\dots$}  \\ 
  10989032 & 32 & 19:49:34.46 & +48:24:38.88 &&   A5     &    A5     &    A5       & V       &  $45.0\pm0.9$\,(8) & $44\pm1$   &  $15\pm2$\,(26)  \\ 
  11649497 & 11 & 18:56:55.73 & +49:44:33.42 &&  -       &  {\bf A6} &  {\bf F0}   & III     &$<$$5.6\pm0.2$\,(6) &$<$$5\pm1$  &  $12\pm2$\,(07)  \\ 
  11754974 & 13 & 19:08:15.95 & +49:57:15.56 && {\bf A3} &  {\bf F1} &  {\bf A2}   & V       &  $28.8\pm1.7$\,(8) & $31\pm1$   & $13\pm2$\,(21)  \\ 
  12643589 &[17]& 19:13:17.13 & +51:43:35.59 &&   -      &  {\bf F5} &  {\bf F5.5} &  V      &  $28.7\pm0.1$\,(5) & $32\pm2$   & $18\pm2$\,(06)    \\ 
  12688835 & 18 & 19:15:01.07 & +51:50:55.34 && {\bf A3} &  {\bf A7} &  {\bf A3}   & V       & $230\pm5$\,(5)     &\multicolumn{1}{c}{$\dots$}& \multicolumn{1}{c}{$\dots$}  \\ 
\\
\hline
\end{tabular}
\medskip
\end{table*}

\subsection{Spectral Types}

SX~Phe (and $\delta$\,Sct) stars usually have spectral types in the range A3 to
F2, corresponding to surface temperatures ranging from $\sim$8600 to
$\sim$6900\,K.  They also  define the blue and red edges of the
instability strip at absolute magnitudes  $<$$M_V$$>$ $\sim$ 1.5 to 3.5 ({\it
i.e.}, $L/L_{\odot}$ from $\sim$15 to 3).  The SX~Phe stars found among  BSs
in globular clusters tend to have, on average, lower luminosities and masses
than  Pop.\,I $\delta$~Sct stars.  

Spectral types and luminosity classes for the program stars were determined
using the rectified CFHT spectra and are given in  {\bf Table\,3} (cols.\,5-8).
Separate classifications were made  based on the appearance of the hydrogen
lines, the Ca\,II K-line,  and the overall metallic-line spectrum, using the
following criteria: {\bf (H)} the hydrogen Balmer lines have their maximum
strength at A2 with the cores and the wings of the lines decreasing in strength
as the temperatures decrease; {\bf (K)} the Ca\,II K line at 393.3\,nm
increases in strength towards later spectral types, starting from a line depth
comparable to that of the Mg\,II 448.1\,nm line near A0, to a depth that
surpasses that of H$\epsilon$ at F0; and {\bf (M)} the line strengths of
neutral metals steadily increase towards later spectral types\footnote{The
G-band at 430\,nm, which is a useful temperature indicator for mid-to-late
F-type stars, was also used to assess the spectral type of KIC\,12643589 [*17],
one of the two misclassified SX~Phe stars and the star with the latest spectral
type.   Its G-band spectral type is F6\,V, which is to be compared with the
F5\,V type from H\,$\gamma$,  and  F5/F6 from the metal lines.  All the
spectral indicators are consistent with a relatively cool temperature,
$\sim$6500\,K.}.

For the early-A stars, the luminosity class was established using the hydrogen
lines,  which are luminosity sensitive and permit discrimination within the
main-sequence band.    For the early-F stars, the metal lines, which are
sensitive to both temperature and luminosity, were used --  the primary
luminosity indicator being the Fe/Ti\,$\lambda$4172--9 blend.  The most
uncertain luminosity classes are those for the late-A stars where the hydrogen
and metal-line criteria do not quite overlap (see Gray \& Garrison 1989).
Inspection of Table\,3 reveals that most of the stars are of luminosity class V
and thus are on or near the upper main sequence. Of particular interest is the
main sequence character of KIC\,6227118 (*27), whose distance, $\log g$ and
$\log L/L_{\odot}$ in the KIC, were found to be outliers by BN12 (see their
Table\,1), and the high-luminosity character of KIC\,9244992 (*7), which is
consistent with its apparently advanced evolutionary state (see fig.\,2 of
BN12, and Saio {\it et al.} 2015).

For  single main sequence stars application of any one of the above indicators
usually gives  the same spectral  type;  however,  such agreement is not
necessarily expected  for A-F stars, especially those having chemical
peculiarities, such as the Am stars (at least half of which are spectroscopic
binaries),  or for  SX\,Phe stars, whose  blue straggler nature and
possibility of coalescence or mass transfer in binary systems complicates
matters.  The greatest disparities in spectral type  were found for
KIC\,8004558 and KIC\,11754974, both of which are binary systems (see $\S$3.3). 

Assignment of spectral type can also be problematic for spectra that have broad
metal lines (see Fig.\,2).  Since rotation  broadens spectral lines, rapid
rotators (at least those with $i$ significantly greater than 0$^{\circ}$) have
lines that are wider and correspondingly shallower than those for slow
rotators.  Placing too much emphasis on line depth rather than width  may lead
to an erroneous classification.  To mitigate this problem, and improve upon the
metal-line spectral types, the spectra of the rapid rotators were compared with
the spectra of rapidly rotating standard stars selected from the high $v\sin i$
standard star lists provided by Gray \& Garrison (1989) and Gray \& Corbally
(1994, 2009).

Owing to instrumental artefacts, arising perhaps as  remnants of the merging
of the orders in the echelle spectra,  certain lines were less useful than
others  for classification purposes.  For instance, the H\,$\delta$ line was
often unnatural in shape.  For this reason,  the assigned hydrogen line type
was always determined from H\,$\gamma$, and occasionally checked against
H\,$\delta$.  Often it was  difficult to find the continuum near the Ca\,II
K-line, in which case  a Ca\,II K-line type is not provided in Table\,3.  In
general the least weight was placed on the K-line when it was used in the
classifications.

\begin{figure*} \centering 
\begin{overpic}[width=8cm]{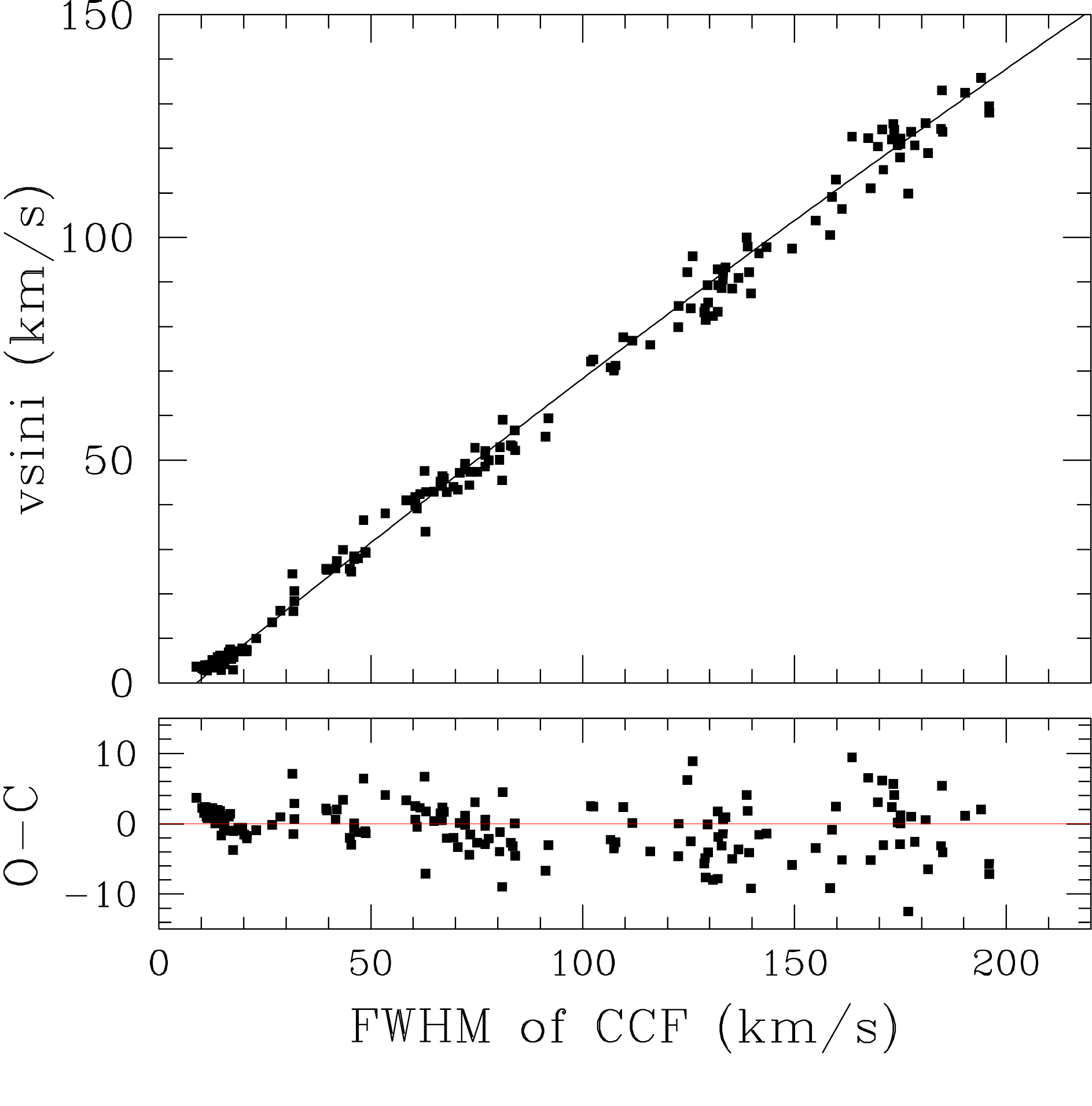} \end{overpic}
\begin{overpic}[width=8cm]{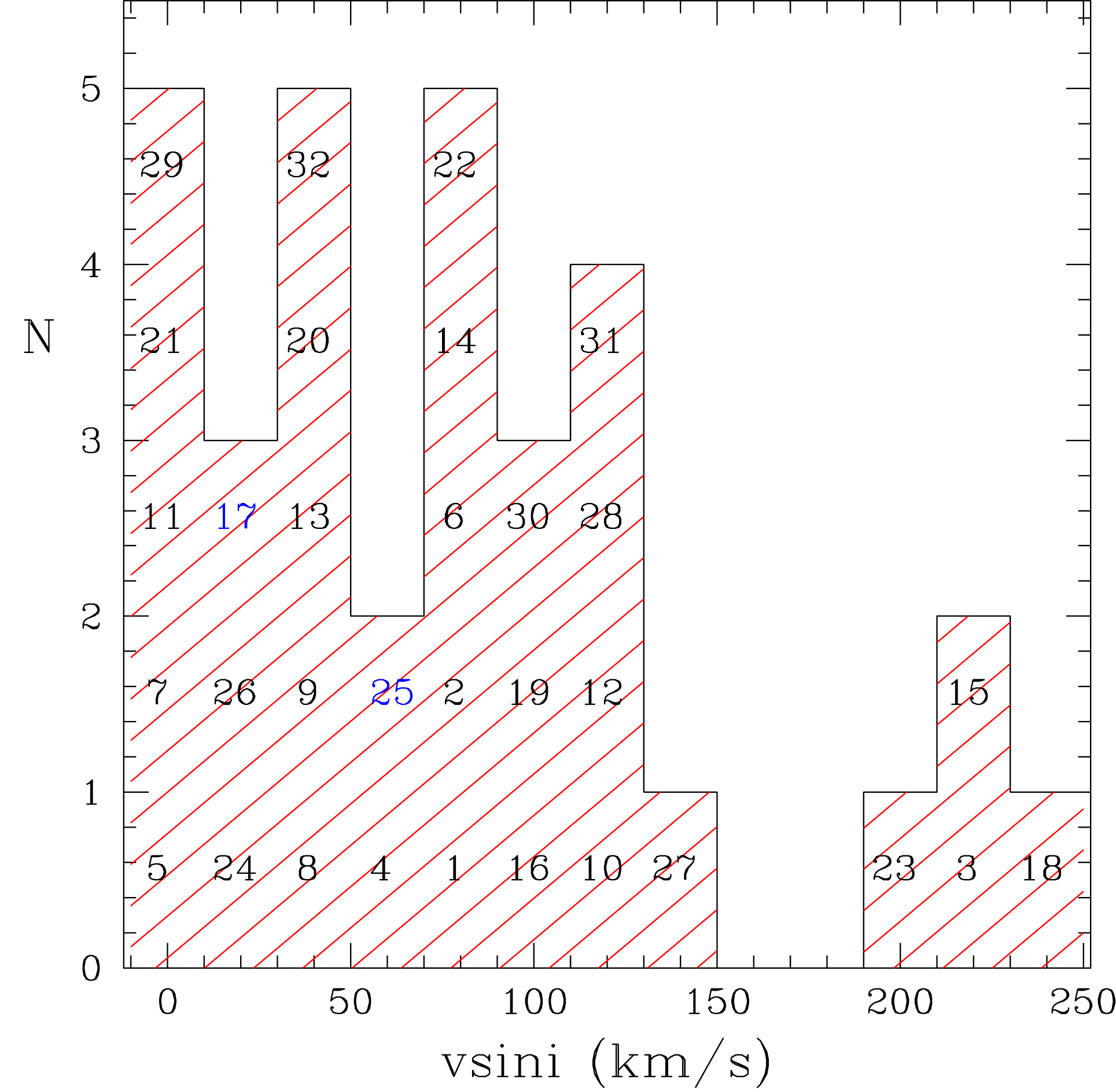} \end{overpic} 
\caption{Left: The upper panel
shows the correlation between the $v\sin i$ values derived using the D\'iaz {\it
et al.} (2011) Fourier transform method, and the gaussian FWHM of the
CCF\,(object-template).  The lower panel shows the residuals from the quadratic
fit, the rms scatter about zero amounting to $\sim$6 km/s.   Right:
Histogram of the derived mean $v \sin i$ values for the {\it Kepler}-field
candidate SX~Phe stars.  The labels are the CFHT star numbers (see Table~1) and are
coloured blue for the two close binaries misclassified as candidate SX~Phe
stars: KIC\,9535881 [*25] and KIC\,12643589 [*17].  } \label{vsini}
\end{figure*}

%
\subsection {Rotational and Macroturbulent Velocities}

The spectra  exhibit a wide range of line widths, with two thirds of the stars
having broad lines (see Fig.\,2).  Stellar rotation is usually the dominant
broadening mechanism;   however macroturbulence and  pulsations may also be
contributing factors.  For stars with narrow lines (which tend to be  the
slowest rotators but may, like Procyon, be fast rotators seen pole-on) the line
profile shapes are determined mainly by instrumental and natural broadening,
photospheric thermal motions, Coulomb interactions of neighbouring particles,
and microturbulence.

\subsubsection{Projected equatorial rotation velocities}

The projected equatorial rotation velocity, $v \sin i$, was measured for each
spectrum  using the Fourier transform (FT) method pioneered by Carroll (1928,
1933) and further developed by Gray (1973, 1975, 1978), Smith \& Gray (1976)
and others (Dravins 1982;   Reiners \& Schmitt 2002; Reiners \& Royer 2004;
Sim\'on-D\'iaz \& Herrero 2007, 2014)\footnote{The $v$ in $v \sin i$ is the
equatorial velocity, and the inclination angle, $i$, is the angle between the
observer's line-of-sight and the direction of the rotation axis.  This $i$ is
not to be confused with the orbital inclination angle used in $\S3$ to describe
the binary systems, where $i$ is the angle between the line-of-sight and the
line perpendicular to the orbital plane.}.   Since most of the available
spectra have signal-to-noise ratios $<$100 no attempt was made to measure
differential rotations or to investigate line bisectors.  The $v \sin i$ values
given in Table\,1 (col.\,9) were derived using  the `rcros' program of D\'iaz
{\it et al.} (2011).  This algorithm calculates the cross-correlation function
(CCF) between an object spectrum and a template spectrum, where  user-specified
wavelength intervals (e.g., 200-400\AA) replace  individual lines.  The
$v\sin\,i$ values follow from the location of the first zero in the FT of the
CCF central maximum (assuming a linear limb darkening law with $\epsilon=0.6$,
and taking account of wavelength dependence).  For our spectra, the method has
several advantages over measuring individual line profiles, the greatest being
that in the FTs the first zeros were usually  well defined with sidelobe
signatures significantly higher than the background noise level.  Using
simulations D\'iaz {\it et al.} concluded that ``for the usual values of S/N
and instrumental broadening, the variation in the first zero position caused by
additional broadening and noise is below 1\%."   Selection of the template
spectra depended on the amount of line broadening.  For the spectra with broad
lines both narrow- and broad-lined templates appropriate for A- and early-F
spectral types were used, including synthetic spectra with $v\sin i$ values
equal to 10, 50 and 100 km/s.  For the stars with narrow-lined spectra
(including the RV standards) the adopted template spectrum consisted of either
a synthetic solar spectrum, an A-star spectrum, or the  spectrum of one of the
very narrow-lined program or standard stars.  

The derived  $v \sin i$ values correlate well with the measured Gaussian
full-width at half-maximum (FWHM) values of the CCFs (see left panel of {\bf
Figure\,7}).  Because the four fastest rotators have CCFs that deviate from  a
Gaussian distribution they have been excluded from the graph; no corrections
were made for the $\sim$4 km/s instrumental broadening.  The quadratic
least-squares fit is given by $y = -2.859\times 10^{-4} x^2 + 0.781 x - 6.888,$
where $x$ is the FWHM of the CCF, and $y$ is the value of $v\sin i$.  The
residuals have an rms scatter of approximately 6 km/s.  The uncertainties in
the $v\sin i$ values depend on the width of the CCF central maximum, the height
of the CCF peak, and the noise in  the CCF.  The largest random errors occur
for the fastest rotators with the broadest lines.    For the narrow-lined
spectra the rotational broadening is comparable in magnitude to the
instrumental and other broadenings mentioned above and hence the $v \sin i$
values are upper limits.

Weighted-average $v \sin i$ values based on the FT method applied to all the
spectra are given in column\,9 of {\bf Table~3}, and  a histogram of the
$<$$v\sin i$$>$ values is plotted in the right panel of Figure\,7.   Although
the number of stars is relatively small, the distribution appears approximately
uniform  between 0 and 150 km/s, with the four fastest rotators having $<$$v
\sin i$$>$ values greater than 195 km/s.  The four slowest rotators have
$<$$v\sin i$$>$ values smaller than 8 km/s, and about two-thirds  of the stars
have $v\sin i$ values larger than 50 km/s.  Since  main-sequence A-type stars
generally have $v \sin i$ values ranging from the resolution limit of the
instrumentation to nearly as high as the rotational break-up limit $\sim$350
km/s, and because F-type stars tend to rotate much more slowly (see Fig.\,18.21
of Gray 2005, Fig.\,1 of Royer {\it et al.} 2007, and Fig.\,11 of Bruntt {\it
et al.} 2010b), the observed range of line broadenings meets expectations for a
sample of A- and early-F type stars.  

\begin{figure} \begin{center}
\begin{overpic}[width=8.7cm]{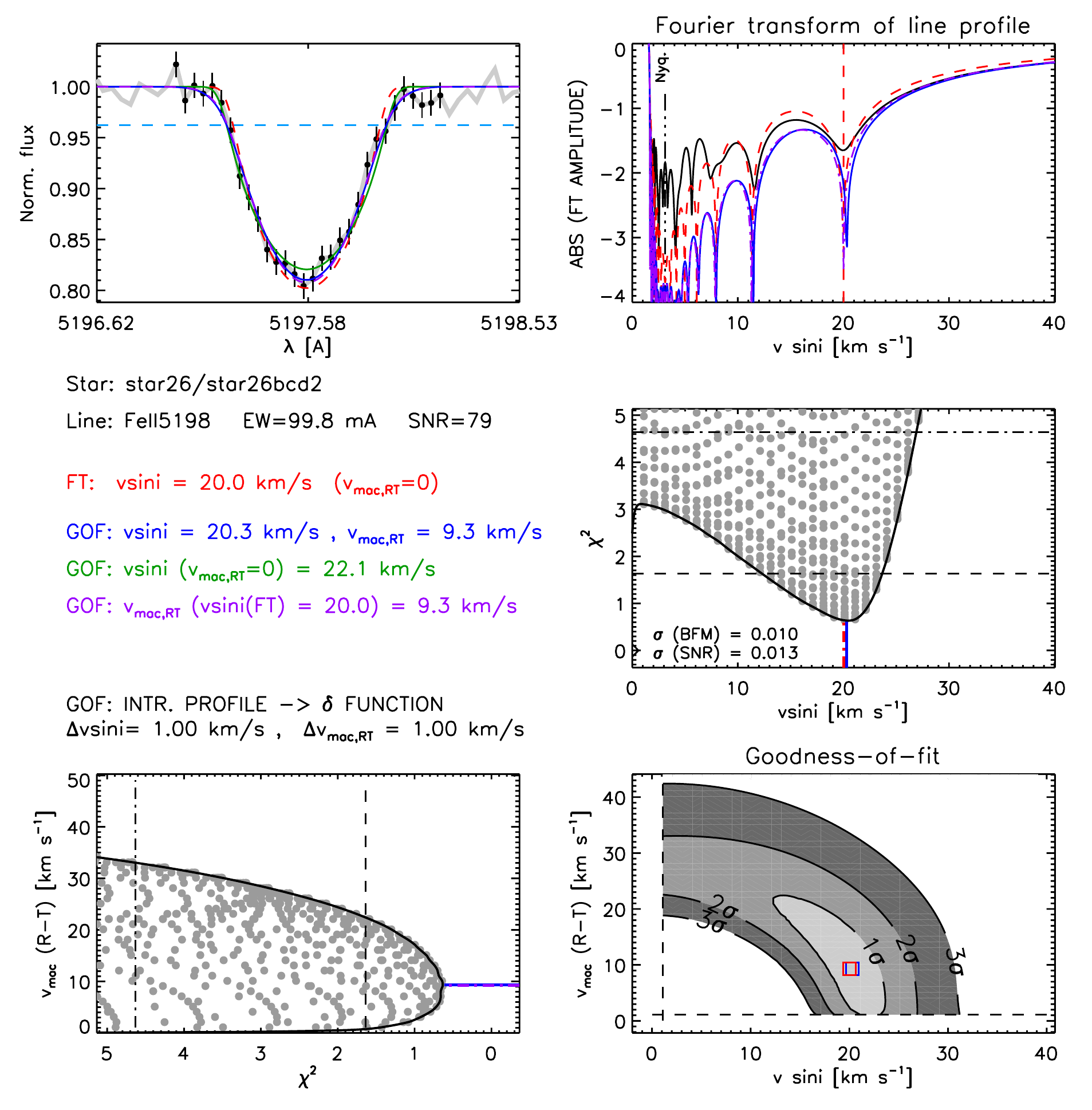} \end{overpic}
\end{center}
\caption{Example output from the `IACOB-broad tool', which was used
to estimate  radial-tangential macroturbulent velocity,
$\zeta_{\rm RT}$, and the projected rotation velocity, $v\sin i$.    
Here, the strong Fe\,II line at 5197.58\AA \, was measured
in the coadded spectrum of KIC\,5036493\,(*26).   The resulting FT and GOF
$v\sin i$ values for this line are in excellent agreement with the  $<$$v\sin
i$$>$ value derived using `rcros', 19.0$\pm$1.1 km/s.  The resulting
$\zeta_{\rm RT}$ of 9$\pm$8 km/s (the uncertainty follows from the contour plot
in the bottom right panel) is smaller than, but consistent with, the mean value
of $<$$\zeta_{\rm RT}$$>$ = 15$\pm$1 km/s derived from measurements of this
line and 26 other lines in the same spectrum.  } \label{iacob} \end{figure}

\subsubsection{Macroturbulent velocities}


The existence of granulation cells  larger than the mean free path of a photon
({\it i.e.,} macroturbulence; see Gray 2005, Chapter 17) is often invoked to
explain  line broadening in excess of that attributed to  rotation.  It has
been suggested that gravity waves, possibly originating with non-radial
oscillations (Lucy 1976; de Jager 1990;  Cantiello {\it et al.} 2009;
Sim\'on-D\'iaz {\it et al.} 2010; Aerts {\it et al.} 2009, 2014; Balona 2011;
Schiode {\it et al.} 2013; Grassitelli {\it et al.} 2015), may cause or
contribute to macroturbulent broadening.  In many cases the amount  of  line
broadening  due to   macroturbulence may be as much or more than that caused  by
rotation (e.g., Markova {\it et al.} 2014). 

The radial-tangential macroturbulent velocity dispersion, $\zeta_{\rm RT}$
(Gray 1973, 1975, 1978), was estimated for  sixteen BN12 stars using the
`IACOB-broad' tool of Sim\'on-D\'iaz \& Herrero (2014).  Artificial line
profiles were fitted to  unblended high-SNR lines in the observed spectra by
varying  $v \sin i$ and $\zeta_{\rm RT}$, where the combined  Fourier transform
(FT) and goodness-of-fit (GOF) methodology was used to infer the optimum
(minimum $\chi^2$) values of $v \sin i$ and $\zeta_{\rm RT}$ for each line.
These values were then averaged over lines to give the final estimates  for
each star, which are given in the last two columns of Table\,3. 

An application of the IACOB-broad tool is illustrated in {\bf Figure\,8}.  The
upper-left panel shows observed and fitted profiles for the 5197.58\,{\AA} FeII
line in the spectrum of KIC\,5036493 (*26).   The  upper-right panel shows the
Fourier transform of the observed line profile, where the first dip corresponds
to the $v \sin i$ value if $\zeta_{\rm RT} = 0$;    the middle-right and
lower-left panels show  $\chi^2$ plots;  and the middle-left panel summarizes
the FT and GOF estimates.   Here, and for most of the other stars, the FT and
GOF values are consistent.    Plots similar to the bottom-right contour graph,
where the `banana-shaped' contours reflect  the relation  $v_{\rm tot}^2 =
[(v\sin\,i)^2 + \zeta_{\rm RT}^2)]^{1/2}$ introduced by Saar \& Osten (1997),
have  been presented by Ryans {\it et al.} (2002), Dall {\it et al.} (2010),
and Bruntt {\it et al.} (2010a).  Since the GOF $v \sin i$ values agree
extremely well with those derived using the `rcros' FT method   (see $\S2.4.1$)
the latter were adopted, and only the  $\zeta_{\rm RT}$  retained.

{\bf Figure\,9} shows that the derived macroturbulent velocity dispersions fall
in the range 10-30 km/s, with  an apparent trend of increasing  $\zeta_{\rm
RT}$ with increasing $v \sin i$.  Without knowing the inclination angle it is
hard to know whether there is correlation between the  equatorial velocity and
$\zeta_{\rm RT}$.

For main sequence stars the `granulation boundary' that marks the onset of
convection (B\"ohm-Vitense 1958; B\"ohm-Vitense \& Canterna 1974; Gray \& Nagel
1989; Paxton {\it et al.} 2011, 2013, 2015) occurs near spectral type F0,
corresponding to $B$--$V$$\sim$0.3 and $T_{\rm eff}\sim7000$\,K.  This boundary
appears to be coincident with the red edge of the Cepheid instability strip
(B\"ohm-Vitense \& Nelson 1976).  Since the stars  considered here are mainly
mid-to-late A-type  pulsators with (B-V)$_0 < 0.30$ mag (see Table\,4) they
tend  to lie on the hot side of the granulation boundary in the instability
strip.  Such stars might be expected to have reversed-C shaped line bisectors
(Gray \& Toner 1986; Gray 2009) and possibly significant atmospheric velocity
fields (Landstreet 1999; Landstreet {\it et al.} 2009).  In addition, many  may
have chemical abundance anomalies (Preston 1974; Adelman 2004) that are
undoubtedly related to  magnetic fields (Donati \& Landstreet 2009) and
diffusion (Michaud {\it et al.} 1976).

\begin{figure} \begin{center}
\begin{overpic}[width=8.7cm]{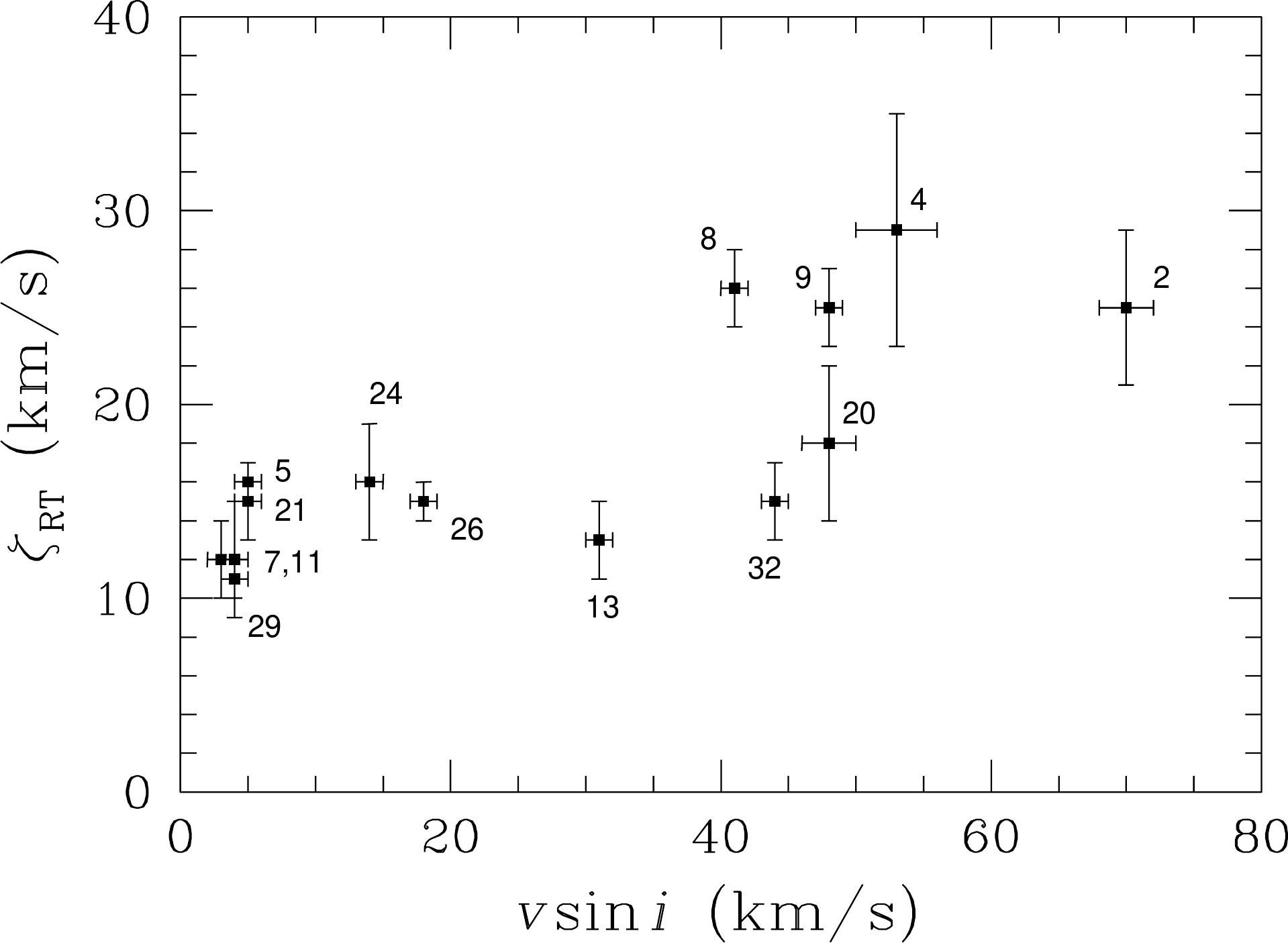} \end{overpic}
\end{center}
\caption{ Macroturbulent velocity dispersion, $\zeta_{\rm RT}$, versus projected equatorial velocity, $v \sin i$,
for 14 SX~Phe stars. The labels are the CFHT star numbers.  } 
\label{vsini_vmac} \end{figure}

The hotter and more luminous OB-stars (for which microturbulence is assumed to
be negligible) tend to have very large $\zeta_{\rm RT}$ values, possibly as
high as $\sim$150 km/s, which appear to increase  with increasing $T_{\rm eff}$
and  $v \sin i$ (see Ryans {\it et al.} 2002; Simon-Diaz \& Herrero 2014;
Markova {\it et al.} 2014).  On the cool side of the granulation boundary the
F, G and K-type stars (see Gray 1988) tend to show an increase in  $\zeta_{\rm
RT}$  with increasing temperature (and luminosity).  For the Sun $\zeta_{\rm
RT}\sim$3.5 km/s, and for F5\,V stars values reach $\sim$6 km/s  (see
Fig.\,17.10 of Gray 2005; Bruntt {\it et al.} 2010b; Doyle {\it et al.} 2014).
Fig.\,3 of Valenti \& Fischer (2005) shows that for stars with  $T_{\rm
eff}=6200$\,K the upper limit of $\zeta_{\rm RT}$ is $\sim$9-10 km/s.  Such
cool stars  tend to be slow rotators (see above), have C-shaped line bisectors
(Gray \& Nagel 1989; Gray 1989, 2009), and show chromospheric emission in UV
spectra (B\"ohm-Vitense \& Dettmann 1980; Gray \& Toner 1986).   Simple
interpolation  suggests  that our program stars might be expected to  have
intermediate $\zeta_{\rm RT}$ values, which is,  in fact, borne out by the
observation that the  average  $<$$\zeta_{\rm RT} > = 18\pm2$ km/s for the
measured stars.   

Recent precise $v \sin i$ and $\zeta_{\rm RT}$ measurements made by Gray (2014)
for five narrow-lined A0-A2 main sequence stars show them to have much  lower
$\zeta_{\rm RT}$ values than the hot but slightly cooler stars studied here.
It  is perhaps significant that our sample stars are all pulsating while the
stars studied by Gray are outside the instability strip and do not appear to
pulsate.  This observation lends support to the notion that the relatively high
macroturbulent velocity dispersions of the SX\,Phe stars are due to (or at
least related to) non-radial pulsations.



\begin{table*} \caption{Atmospheric characteristics  for 32 of the 34 candidate
SX~Phe stars in the {\it Kepler} field.  For all the stars the first row gives
the SME estimates, and when VWA/MOOG estimates (based on the EW/COG method) were
also made they are given on the 2nd row.  Assumed and uncertain values are given in
parentheses.  The KIC and CFHT star numbers are
given in the first two columns.  The third to ninth columns give, respectively,
the derived surface gravity, effective temperature, microturbulent velocity and
metal abundance.  For comparison purposes the table also contains the KIC and
Huber {\it et al.} (2014) photometrically-based values.  }

\begin{tabular}{rcrlcrlrcc}
\hline
\multicolumn{1}{c}{KIC} & \multicolumn{1}{c}{CFHT} &   \multicolumn{2}{c}{$\log g$  (cm/s$^2$)} &&
\multicolumn{2}{c}{$T_{\rm eff}$ (K)}  & \multicolumn{1}{c}{$\xi_t$\,(km/s) } &  \multicolumn{2}{c}{ [Fe/H]  (dex) } \\ \cline{3-4}\cline{6-7}\cline{9-10}
\multicolumn{1}{c}{no.} & no. &  KIC, H14 &  This paper  && \multicolumn{1}{c}{KIC, H14} & \multicolumn{1}{c}{This paper} & \multicolumn{1}{c}{This paper}  &\multicolumn{1}{c}{KIC, H14} & This paper    \\                              
\multicolumn{1}{c}{ (1) }  & \multicolumn{1}{c}{ (2) } & \multicolumn{1}{c}{(3)}  &\multicolumn{1}{c}{(4)}     && 
\multicolumn{1}{c}{(5)}  & \multicolumn{1}{c}{ (6) } &\multicolumn{1}{c}{(7)} & \multicolumn{1}{c}{ (8)} & (9)   \\
\hline
1162150 &  15  & 3.46, 3.49&$3.61\pm0.10$  && 6871, 7090 & $7390\pm50$   & ($7\pm2$)   &  $-0.01, +0.02$ & $-0.20\pm0.30$ \\
3456605 &  24  & 3.94, 3.94&$3.99\pm0.05$  && 7112, 7353 & $7460\pm90$   & $3.8\pm0.1$   &  $-0.18, -0.18$ & $+0.50\pm0.10$ \\
        &      &           &$4.2\pm0.2$    &&            & $7630\pm50$   & $2.8\pm0.2$   &                 & $+0.58\pm0.11$ \\ 
4168579 &  23  & 3.79, 3.84&$3.57\pm0.08$  && 7534, 7757 & $7750\pm90$   & $5.3\pm2.0$   &  $-0.09, -0.1 $ & $+0.08\pm0.20$ \\
4243461 &  4   & 4.15, 4.20&$4.33\pm0.05$  && 6918, 7159 & $7740\pm50$   & $6.8\pm0.2$   &  $-0.24, -0.22$ & $+0.11\pm0.15$ \\
4662336 &  14  & 3.89, 3.88&$3.95\pm0.05$  && 7211, 7452 & $7450\pm50$   & $5.9\pm0.1$   &  $-0.13, -0.14$ & $-0.05\pm0.15$ \\
4756040 &  20  & 4.09, 4.09&$4.08\pm0.05$  && 7603, 7837 & $7720\pm50$   & $4.9\pm0.3$   &  $-0.01, -0.02$ & $-0.15\pm0.20$ \\
        &      &           &$4.1\pm0.2$    &&            & $7650\pm100$  & $2.8\pm0.2$   &                 & $-0.12\pm0.15$  \\ 
5036493 &  26  & 3.95, 3.97&($4.7\pm0.2$)  && 7962, 8148 & $8430\pm100$  & $2.7\pm0.2$   &  $+0.00, +0.07$ & $-0.02\pm0.15$   \\
        &      &           & $4.5\pm0.2$   &&            & $8250\pm100$  & $2.4\pm0.3$   &                 & $-0.22\pm0.20$   \\
5705575 &  22  & 3.99, 4.03&$4.00\pm0.05$  && 7583, 7785 & $7860\pm50$   & ($10\pm2$)    &  $-0.16, -0.18$ & $-0.1\pm0.2$ \\
6130500 &  9   & 4.14, 4.11&$4.1\pm0.2$    && 7581, 7801 & $7820\pm100$  & $5.3\pm1.0$   &  $-0.01, -0.02$ & $-0.07\pm0.20$ \\
6227118 &  27  & 1.03, 2.84&($4.5\pm0.3$)  && 7217, 7197 & $7500\pm200$  & $4.2\pm1.0$   &  $-0.15, -0.5 $ & ($+0.04\pm0.20$) \\
6445601 &  2   & 4.09, 4.11&$3.90\pm0.15$  && 7186, 7419 & $7450\pm60$   & $6.0\pm0.5$   &  $-0.17, -0.16$ & $+0.3\pm0.2$   \\
6520969 &  21  & 3.96, 3.78&  (3.80)       && 8355, 8572 & $8250\pm50$   & $2.7\pm0.3$   &  $-0.13, -0.2 $ & $-0.7\pm0.2$ \\
        &      &           &$4.3\pm0.2$    &&            & $7730\pm100$  & $2.4\pm0.2$   &                 & $-0.84\pm0.16$  \\ 
6780873 &  5   & 4.35, 4.36&$4.03\pm0.08$  && 6874, 7158 & $7530\pm50$   & $3.7\pm0.4$   &  $-1.09, -1.1 $ & ($+0.0\pm0.3$)   \\
        &      &           &$4.3\pm0.2$    &&            & $7200\pm100$  & $1.3\pm0.3$   &                 & ($+0.16\pm0.20$)   \\ 
7020707 &  16  & 4.00, 3.95&$3.93\pm0.05$  && 7447, 7689 & $7560\pm50$   & ($10\pm2$)    &  $-0.21, -0.2 $ & $-0.2\pm0.2$ \\
7174372 &  8   & 4.10, 4.10&$3.4\pm0.2$    && 7228, 7457 & $7740\pm100$  & $3.4\pm0.5$   &  $-0.30, -0.3 $ & $-0.2\pm0.2$ \\
        &      &           &$3.7\pm0.2$    &&            & $7380\pm100$  & $2.1\pm0.2$   &                 & $+0.11\pm0.20$  \\
7301640 &  10  & 4.07, 4.13&$4.2\pm0.2$    && 7014, 7239 & $8000\pm100$  & ($14\pm2$)    &  $-0.01, +0.02$ & $+0.45\pm0.15$ \\
7621759 &  6   & 4.05, 4.02&  (4.02)       && 6966, 7316 & $7800\pm80$   & ($11\pm2$)    &  $-0.26, -0.24$ &\multicolumn{1}{c}{($-0.24$)}      \\
7765585 &  28  & 4.80, 4.31&  (4.31)       && 6714, 6853 & $7800\pm150$  & ($9\pm2$)     &  $+0.15, +0.07$ & $+0.15\pm0.10$ \\
7819024 &  19  & 4.10, 4.10&$4.13\pm0.10$  && 7534, 7758 & $7690\pm50$   & $7.0\pm1.5$   &  $-0.27, -0.32$ & $+0.00\pm0.20$ \\
8004558 &  1   & 3.90, 3.89&$4.3\pm0.2$    && 7674, 7899 & $7950\pm100$  & $5.2\pm0.4$   &  $-0.45, -0.44$ & $-0.3\pm0.2$   \\
8110941 &  29  & 4.13, 4.13&$4.13\pm0.05$  && 6839, 7177 & $7350\pm150$  & $2.9\pm0.5$   &  $-0.09, -0.06$ & $+0.1\pm0.2$ \\
        &      &           &$4.0\pm0.2$    &&            & $7300\pm150$  & $2.5\pm0.5$   &                 & $+0.06\pm0.20$  \\ 
8196006 &  30  & 4.25, 4.28&$4.27\pm0.05$  && 7015, 7232 & $7740\pm65$   & ($10\pm2$)    &  $-0.44, -0.44$ & $+0.16\pm0.20$ \\
8330910 &  3   & 4.06, 4.07&  (4.13)       && 7379, 7599 & $8000\pm60$   & ($7\pm2$)     &  $-0.20, -0.24$ & $-0.1\pm0.3$   \\
9244992 &  7   & 3.51, 3.52&$3.52\pm0.15$  && 6592, 6902 & $7550\pm100$  & $4.5\pm0.7$   &  $-0.14, -0.16$ & $+0.1\pm0.3$ \\
9267042 &  12  & 3.80, 3.76&$4.28\pm0.10$  && 8128, 8354 & $8200\pm80$   & ($8\pm2$)     &  $-0.11, -0.14$ & $-0.02\pm0.20$ \\
9535881 & [25] & 4.18, 4.07&$3.97\pm0.05$  && 7161, 7397 & $7520\pm70$   & $6.1\pm0.8$   &  $-0.22, -0.2 $ & $-0.03\pm0.15$ \\
9966976 &  31  & 3.93, 3.92&$4.32\pm0.06$  && 7638, 7872 & $7740\pm50$   & ($7\pm2$)   &  $+0.05, +0.07$ & $+0.02\pm0.10$ \\
10989032&  32  & 4.07, 4.08& (4.10)        && 8610, 8851 & $8630\pm100$  & ($5\pm2$)     &  $-0.01, +0.07$ & $+0.3\pm0.3$ \\
11649497&  11  & 4.02, 4.03&$3.98\pm0.05$  && 7656, 7904 & $7650\pm80$   & $4.4\pm0.1$   &  $-0.10, -0.04$ & $+0.2\pm0.2$ \\
11754974&  13  &  -- , 3.98&$3.98\pm0.05$  &&  -- , 7231 & $7500\pm50$   & $2.3\pm0.2$   &  -- , $+0.01$   & $-1.1\pm0.2$   \\
        &      &           &$4.0\pm0.2$    &&            & $7300\pm150$  & $2.1\pm0.3$   &                 & $-1.4\pm0.2$   \\
12643589& [17] & 4.44, 4.39&$4.40\pm0.15$  && 6501, 6657 & $7000\pm50$   & $4.3\pm0.4$   &  $-0.26, -0.34$ & $+0.14\pm0.20$ \\
12688835&  18  & 3.86, 3.81&$4.3\pm0.4$    && 8531, 8759 & $8330\pm80$   & ($7\pm3$)   &  $-0.28, -0.3 $ & $-0.15\pm0.10$ \\
\hline
\end{tabular}
\end{table*}

\begin{figure} \begin{center}
\begin{overpic}[width=8.5cm]{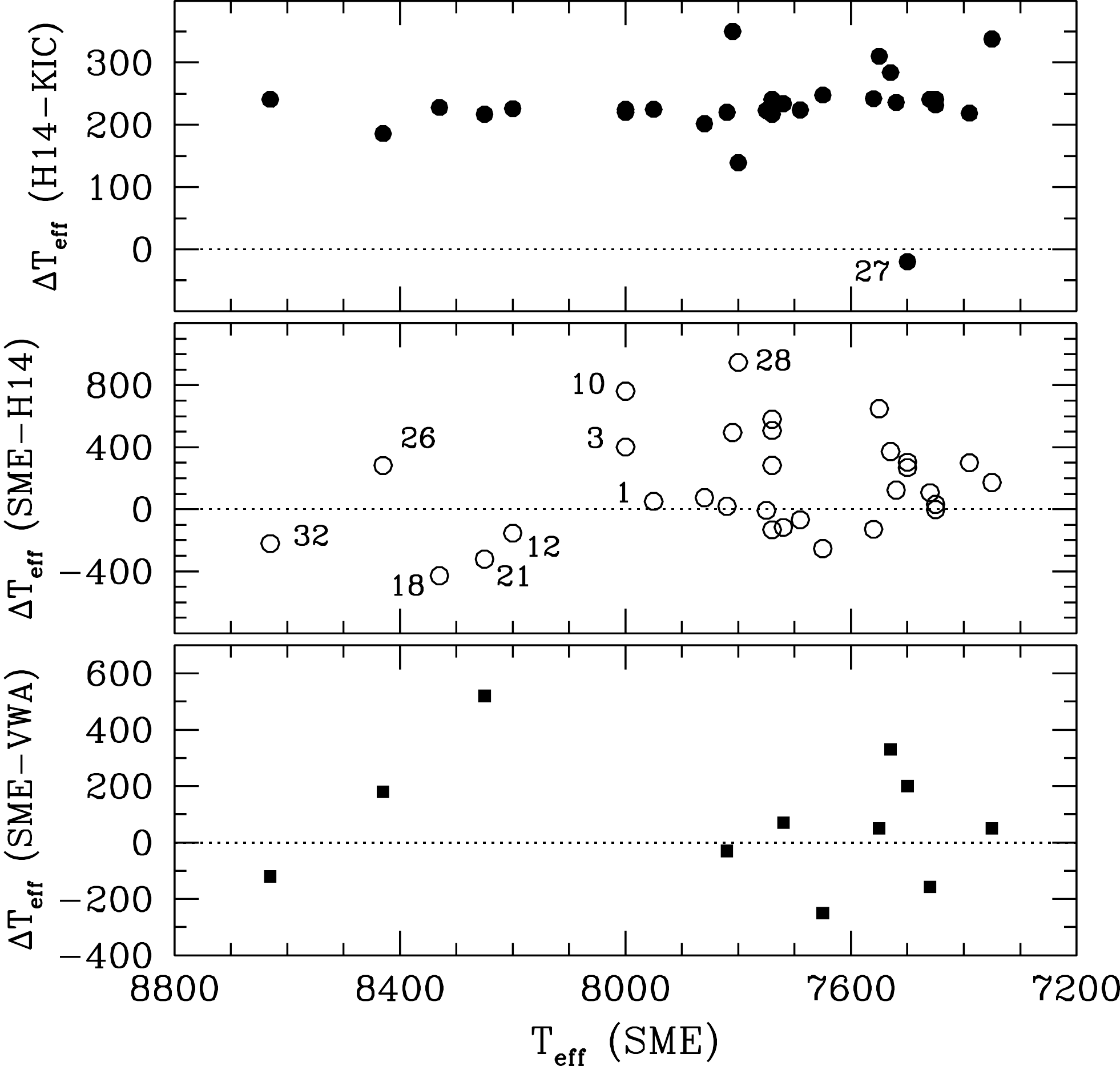} \end{overpic}
\end{center}
\caption{{\bf (Top)} Comparing the KIC and H14 photometric effective
temperatures for the SX\,Phe candidates reveals a systematic difference of
$\sim$240\,K, in agreement with, and extending to hotter temperatures, the
conclusion reached earlier by Pinsonneault {\it et al.} (2012).  KIC\,6227118
(*27) is an obvious outlier.  {\bf (Middle)} Comparison of the SME
spectroscopic T$_{\rm eff}$ and the  H14 photometric T$_{\rm eff}$ values --
nine stars have been labelled with the CFHT star number; {\bf (Bottom)}
Comparing the spectroscopically derived effective temperatures.  }
\label{Teff_SME_KIC_H14} \end{figure}

\subsection{Other Atmospheric Characteristics}

In addition to  $v \sin i$ and $\zeta_{\rm RT}$ velocities, several other
atmospheric characteristics were derived from the spectra.    These include:
effective temperature, $T_{\rm eff}$;  surface gravity, $\log g$;
microturbulent velocity, $\xi_t$, and metal abundance, [Fe/H].   
Since the {\it Kepler} SX~Phe sample consists of stars exhibiting a wide range
of line widths and metallicities,  stars with radial speeds as high as 300
km\,s$^{-1}$, close and wide binary systems, etc., and  spectra with a wide range of
SNRs, it was necessary to employ several different  methods for
the spectral data analyses.   In general
synthetic spectra were fitted to  observed spectra, where  $\chi^2$ minimization
techniques were used  to optimize atmospheric parameters.  In all cases 1D plane-parallel
atmospheres in LTE were assumed. Specifically, the ATLAS9 model atmospheres of
Kurucz (see Castelli \& Kurucz 2004) were used for the radiative transfer
calculations.  Basic spectral line information (excitation potentials,
log$gf$ values) was taken from the VALD3 website (Piskunov {\it et al.} 1995; Kupka
{\it et al.} 1999), and for  the assumed composition mix the Asplund {\it et al.}
(2009) solar abundances  were adopted.   

All the spectra (both individual and coadded) were measured using the
`Spectroscopy Made Easy' (SME) program of Valenti \& Piskunov (1996).
Initially,  wavelength ranges were limited to  the same `windows' used by
Valenti \& Fischer (2005), supplemented by regions around the first three
Balmer lines.  The results were then refined using  simultaneous fitting of
the 25 wavelength windows as advocated by Brewer {\it et al.}  (2015 - see
their Table\,2), where the initial estimates served  as starting values.  A
sample fit for a portion of the KIC\,7020707 spectrum can be found in Nemec {\it
et al.} (2015). 

For those narrow-lined stars for which equivalent widths (EWs) of unblended
lines could be reliably measured, the curve-of-growth (COG) method, also known
as the equivalent width (EW) method,  was used to derive a second set of
estimated atmospheric characteristics.  The analyses were performed using both
MOOG (Sneden 1973) and VWA (Bruntt {\it et al.} 2002).    For the MOOG analyses
the EWs were measured using ARES (Sousa {\it et al.} 2007, 2015), while for the
VWA analyses they were measured as part of the reductions.   To help ensure
comparability of the results from VWA and MOOG  every attempt was made to
measure the same iron lines; usually there was good agreement.  VWA diagnostic
diagrams were plotted, including derived iron abundances, A(Fe) (= log\,{\rm
N}$_{\rm Fe}/{\rm N}_{\rm tot}$), versus measured EW, and  A(Fe) versus lower
excitation potential (EP).  The $T_{\rm eff}$ values were derived by requiring
independence of A(Fe) and EP.  The microturbulent velocities $\xi_t$ were
derived by requiring independence of A(Fe) and EW.  And the $\log g$ were
derived by requiring similar mean A(Fe) values for the Fe\,I and Fe\,II lines
({\it i.e.}, ionization equilibrium). 

The results of the spectral analyzes are summarized in {\bf Table\,4}, where
the first row for each star summarizes the parameter estimates derived using
SME, and the second row gives, if available, the results of the most reliable
VWA/MOOG analyses;  also given are the photometric estimates from the KIC and
from Huber {\it et al.} (2014).    When individual parameters were assumed,
such as for the difficult-to-measure surface gravity, the values that were
assumed are given in parentheses; otherwise derived values (with uncertainties)
are recorded.  Since the $T_{\rm eff}$ and $\log g$ estimates ignore variation
over the pulsation cycle the true uncertainties are difficult to estimate and
may be larger than the reported values.

\begin{table*}    \centering \caption{Mean magnitudes and colours for the {\it
Kepler}-field SX~Phe candidates.  The columns contain:  (1-2) KIC and CFHT star
numbers;  (3) mean {\it Kepler} $Kp$ magnitude from the KIC; (4-6) mean Johnson
$V$ magnitudes, $B-V$ colours and $U-B$ colours -- the top row is from Everett
{\it et al.} (2012), and the lower rows are new photometric observations;  (7-8) dereddened
$B-V$ and $U-B$ colours, assuming both the KIC and new reddenings (given in
Table~2);  (9) Gunn-Thuan (SDSS) $g$--$r$ colour from
the KIC catalog and from the Kepler-INT Survey (KIS);  (10-12) $r$--$i$, $H$--$K$ and
$J$--$K$ colours given in the KIC. }
\begin{tabular}{lccccccccrrr}
\hline
\multicolumn{1}{c}{ KIC }  & \multicolumn{1}{c}{ CFHT}     &\multicolumn{1}{c}{$Kp$}   &\multicolumn{1}{c}{$V$}  & \multicolumn{1}{c}{ $B-V$}  & \multicolumn{1}{c}{$U-B$}  &
\multicolumn{1}{c}{($B$--$V$)$_0$ }     & \multicolumn{1}{c}{($U$--$B$)$_0$ }     &\multicolumn{1}{c}{ $g-r$ } &  $r-i$ &  $H-K$  & $J-K$  \\
\multicolumn{1}{c}{no.}  & no.  & (KIC)  & \multicolumn{3}{c}{ Everett {\it et al.} (2012),}    & KIC, new     &   KIC, new    &   KIC, KIS  & (KIC) & (KIC) &  (KIC)  \\
 &  &        & \multicolumn{3}{c}{   new (this paper)             }    &               &             &       &       &         \\
\multicolumn{1}{c}{ (1) }  & \multicolumn{1}{c}{ (2) }  & \multicolumn{1}{c}{(3)} &\multicolumn{1}{c}{(4)}  & \multicolumn{1}{c}{ (5) }  &
\multicolumn{1}{c}{(6)}    & \multicolumn{1}{c}{ (7) }  &\multicolumn{1}{c}{(8)}  & \multicolumn{1}{c}{(9)} &  \multicolumn{1}{c}{(10)} &  \multicolumn{1}{c}{(11)}  &  \multicolumn{1}{c}{(12)} \\
    \hline 
    1162150 & 15 & 11.240 &  --    &~~--~~ & ~~--~~ &~~--~~,~~--~~  &  ~~--~~,~~--~~   &  0.233, 0.352 &  0.012 &  0.032 & 0.292 \\  
    3456605 & 24 & 13.108 & 13.181 & 0.452 & ~0.154 & ~0.25, 0.31   &  ~~0.01,  ~0.05  &  0.235, 0.422 &  0.015 &  0.058 & 0.208 \\  
    4168579 & 23 & 13.612 & 13.720 & 0.373 & ~0.021 & ~0.12, 0.20   &  --0.16, --0.10  &  0.205, 0.347 &  0.032 &  0.028 & 0.172 \\  
    4243461 & 4  & 13.786 & 13.856 & 0.360 & ~0.070 & ~0.21, 0.25   &  --0.04, --0.01  &  0.185, 0.361 &--0.009 &  0.019 & 0.170 \\  
            &    &        & 13.765 & 0.335 & ~0.043 & ~0.18, 0.23   &  --0.07, --0.04  &               &        &        &       \\  
    4662336 & 14 & 13.105 & 13.258 & 0.389 & ~0.070 & ~0.19, 0.27   &  --0.08, --0.02  &  0.225, 0.394 &  0.030 &  0.032 & 0.198 \\ 
    4756040 & 20 & 13.315 & 13.396 & 0.345 & ~0.019 & ~0.14, 0.24   &  --0.13, --0.06  &  0.160, 0.371 &--0.012 &  0.005 & 0.181 \\ 
    5036493 & 26 & 12.553 & 12.626 & 0.315 & ~0.110 & ~0.09, 0.05   &  --0.05, --0.08  &  0.103, 0.242 &--0.049 &  0.167 & 0.138 \\ 
    5390069 & -- & 15.110 & 15.241 & 0.475 & ~0.132 & ~0.30, 0.29   &  ~~0.01,  ~0.00  &  0.310, 0.411 &  0.075 &  0.059 & 0.275 \\  
    5705575 & 22 & 13.692 & 13.718 & 0.341 & ~0.011 & ~0.13, 0.25   &  --0.14, --0.05  &  0.162, 0.313 &--0.019 &--0.058 & 0.075 \\  
            &    &        & 13.579 & 0.338 & ~0.041 & ~0.13, 0.25   &  --0.11, --0.04  &               &        &        &       \\  
    6130500 & 9  & 13.869 & 13.880 & 0.331 & ~0.054 & ~0.12, 0.21   &  --0.10, --0.03  &  0.174, 0.321 &--0.010 &  0.017 & 0.128 \\  
    6227118 & 27 & 12.932 & 13.013 & 0.366 & ~0.066 & ~0.00, 0.25   &  --0.20, --0.02  &--0.015, 0.306 &--0.016 &  0.090 & 0.177 \\  
    6445601 & 2  & 13.595 & 13.679 & 0.383 & ~0.050 & ~0.20, 0.30   &  --0.08, --0.01  &  0.216, 0.411 &  0.028 &  0.036 & 0.215 \\  
    6520969 & 21 & 13.422 & 13.486 & 0.187 &--0.061 &--0.03, 0.11   &  --0.21, --0.12  &  0.023, 0.222 &--0.066 &  0.026 & 0.126 \\  
    6780873 & 5  & 13.746 & 13.827 & 0.386 & ~0.049 & ~0.23, 0.33   &  --0.06, ~~0.01  &  0.231, 0.434 &  0.069 &  0.041 & 0.214 \\  
            &    &        & 13.773 & 0.447 &--0.052 & ~0.29, 0.39   &  --0.16, --0.13  &               &        &        &        \\  
    7020707 & 16 & 13.433 &  --    &~~--~~ & ~~--~~ & ~~--~, ~--~   &  ~~--~~, ~~--~~  &  0.141, 0.310 &--0.016 &  0.007 & 0.144 \\  
    7174372 & 8  & 13.621 & 13.633 & 0.374 & ~0.109 & ~0.24, 0.32   &  ~~0.01, ~~0.07  &  0.162, 0.345 &  0.002 &  0.053 & 0.165 \\  
            &    &        & 13.640 & 0.331 & ~0.053 & ~0.20, 0.28   &  --0.04, --0.03  &               &        &        &       \\  
    7300184 & -- & 15.430 & 15.563 & 0.504 & ~0.163 & ~0.32, 0.23   &  ~~0.03, --0.03  &  0.377, 0.485 &  0.100 &  0.061 & 0.288 \\  
    7301640 & 10 & 13.862 & 13.971 & 0.367 & ~0.125 & ~0.16, 0.11   &  --0.03, --0.06  &  0.285, 0.430 &  0.039 &  0.046 & 0.180 \\  
    7621759 & 6  & 13.912 & 14.033 & 0.467 & ~0.002 & ~0.26, 0.28   &  --0.15, --0.13  &  0.267, 0.446 &  0.070 &--0.044 & 0.137 \\  
    7765585 & 28 & 13.980 & 14.085 & 0.451 & ~0.165 & ~0.34, 0.27   &  ~~0.08, ~0.04   &  0.288, 0.395 &  0.039 &  0.096 & 0.263 \\  
            &    &        & 14.010 & 0.424 & ~0.095 & ~0.31, 0.24   &  ~~0.01, ~0.02   &               &        &        &       \\  
    7819024 & 19 & 13.799 & 13.886 & 0.317 &--0.078 & ~0.13, 0.27   &  --0.21, --0.11  &  0.140, 0.327 &--0.002 &  0.024 & 0.179 \\  
            &    &        & 13.781 & 0.325 &--0.055 & ~0.14, 0.28   &  --0.19, --0.13  &               &        &        &       \\  
            &    &        & 13.793 & 0.309 &--0.051 & ~0.13, 0.26   &  --0.18, --0.13  &               &        &        &         \\  
    8004558 & 1  & 13.350 & 13.296 & 0.310 &--0.121 & ~0.18, 0.26   &  --0.22, --0.16  &  0.053, 0.251 &--0.040 &  0.071 & 0.163 \\  
            &    &        & 13.390 & 0.255 &--0.123 & ~0.12, 0.21   &  --0.22, --0.20  &               &        &        &       \\  
    8110941 & 29 & 13.749 & 13.798 & 0.449 & ~0.082 & ~0.27, 0.30   &  --0.05, --0.03  &  0.286, 0.472 &  0.049 &  0.068 & 0.253 \\  
    8196006 & 30 & 13.810 & 13.903 & 0.435 & ~0.079 & ~0.24, 0.12   &  --0.06, --0.15  & \multicolumn{1}{l}{ 0.265, -- }   &  0.069 &  0.095 & 0.278 \\  
    8330910 & 3  & 13.457 & 13.520 & 0.389 & ~0.072 & ~0.16, 0.08   &  --0.09, --0.15  &  0.216, 0.407 &  0.033 &  0.050 & 0.229 \\  
    9244992 & 7  & 13.998 & 14.206 & 0.571 & ~0.191 & ~0.28, 0.35   &  --0.02, ~~0.03  &  0.429, 0.561 &  0.156 &  0.110 & 0.286 \\  
    9267042 & 12 & 13.424 & 13.447 & 0.233 &--0.004 & ~0.08, 0.18   &  --0.12, --0.04  &  0.001, 0.175 &--0.117 &--0.023 & 0.086 \\  
            &    &        & 13.409 & 0.202 & ~0.006 & ~0.05, 0.15   &  --0.11, --0.07  &               &        &        &       \\  
    9535881 &[25]& 13.402 & 13.466 & 0.349 & ~0.010 & ~0.18, 0.28   &  --0.11, --0.04  &  0.204, 0.395 &  0.030 &  0.006 & 0.163 \\  
    9966976 & 31 & 13.491 & 13.577 & 0.356 & ~0.039 & ~0.14, 0.27   &  --0.12, --0.03  &  0.173, 0.278 &--0.025 &  0.027 & 0.142 \\  
            &    &        & 13.517 & 0.285 & ~0.018 & ~0.06, 0.20   &  --0.14, --0.06  &               &        &        &       \\  
   10989032 & 32 & 13.866 & 13.933 & 0.240 & ~0.056 & ~0.01, 0.14   &  --0.11, --0.02  &  0.013, 0.199 &--0.106 &--0.034 &--0.002\\  
   11649497 & 11 & 13.432 & 13.454 & 0.252 & ~0.089 & ~0.11, 0.21   &  --0.01, ~~0.06  &  0.083, 0.270 &--0.059 &--0.035 & 0.105 \\  
   11754974 & 13 & 12.678 & 12.605 & 0.255 &--0.158 & ~~--~, ~0.22  & ~~~--~~,~--0.19  &  0.148, 0.234 &--0.046 &--0.010 & 0.128 \\  
            &    &        & 12.570 & 0.259 &--0.174 & ~~--~, ~0.22  & ~~~--~~,~--0.25  &               &        &        &       \\  
   12643589 &[17]& 13.754 & 13.858 & 0.457 & ~0.028 & ~0.35, 0.41   &  --0.05, --0.01  &  0.305, 0.459 &  0.071 &  0.023 & 0.267 \\  
            &    &        & 13.831 & 0.493 &--0.028 & ~0.38, 0.44   &  --0.11, --0.11  &               &        &        &       \\ 
            &    &        & 13.815 & 0.489 &--0.004 & ~0.38, 0.44   &  --0.08, --0.08  &               &        &        &       \\ 
            &    &        & 13.833 & 0.481 &--0.016 & ~0.37, 0.43   &  --0.10, --0.10  &               &        &        &       \\ 
   12688835 & 18 & 13.801 & 13.806 & 0.169 & ~0.025 & ~0.01, 0.14   &  --0.09, ~~0.00 &--0.078, 0.178 &--0.072 &--0.118 &--0.018\\  
 \hline
    \end{tabular}%
  \label{tab:addlabel}%
\end{table*}

\begin{figure*} 
\begin{overpic}[width=7.9cm]{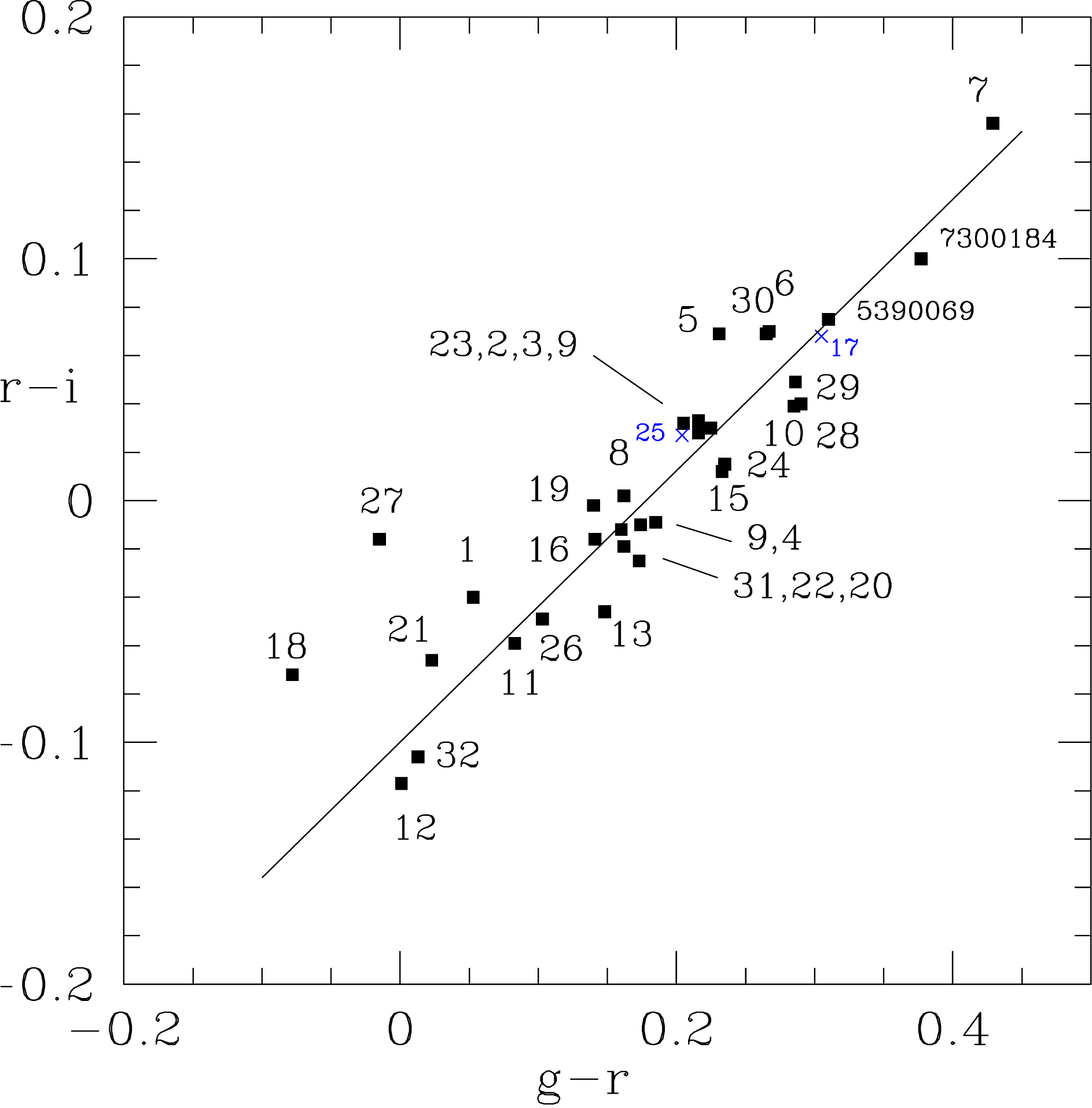} \end{overpic}
\begin{overpic}[width=7.9cm]{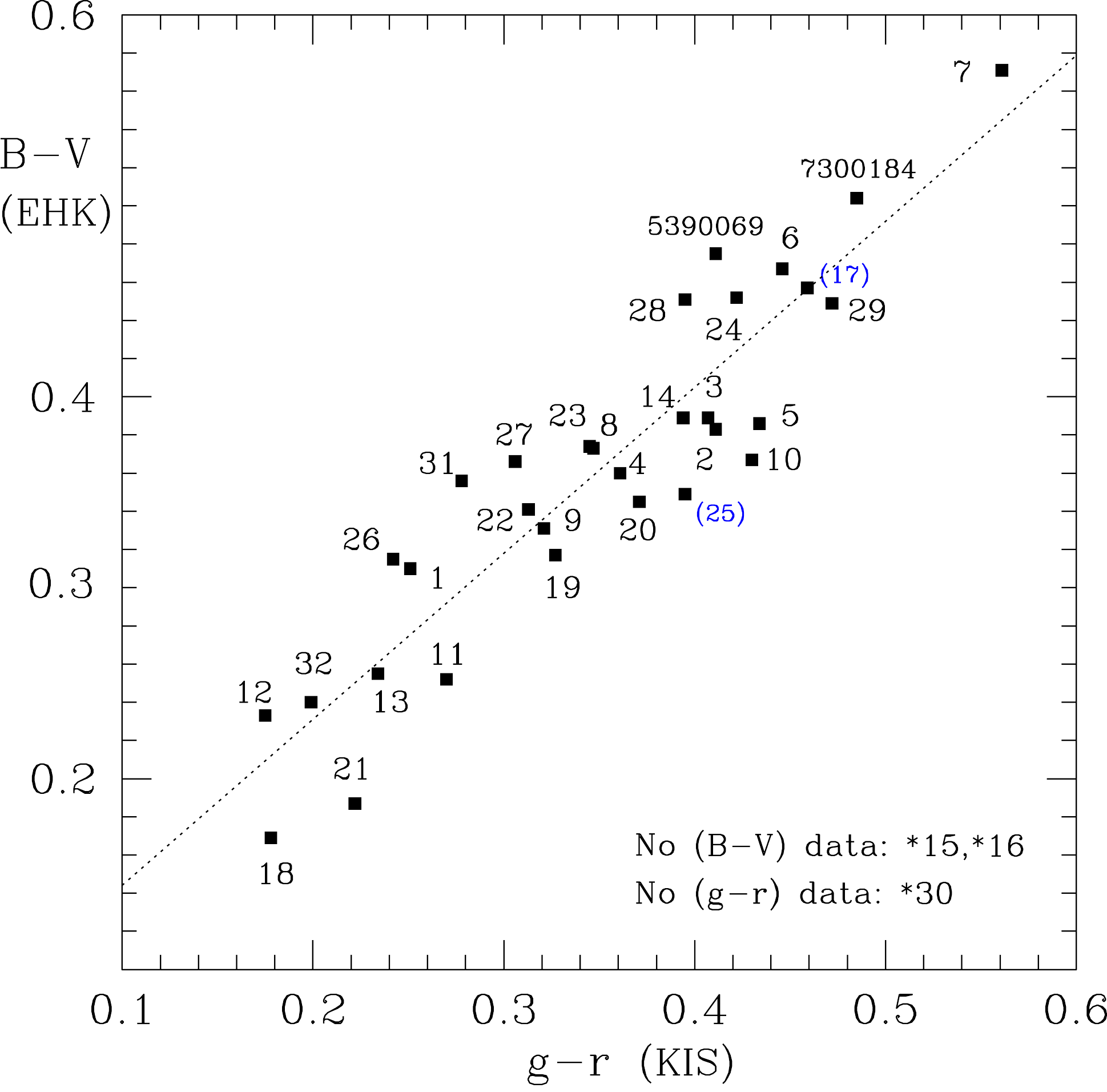} \end{overpic} 
\begin{overpic}[width=7.9cm]{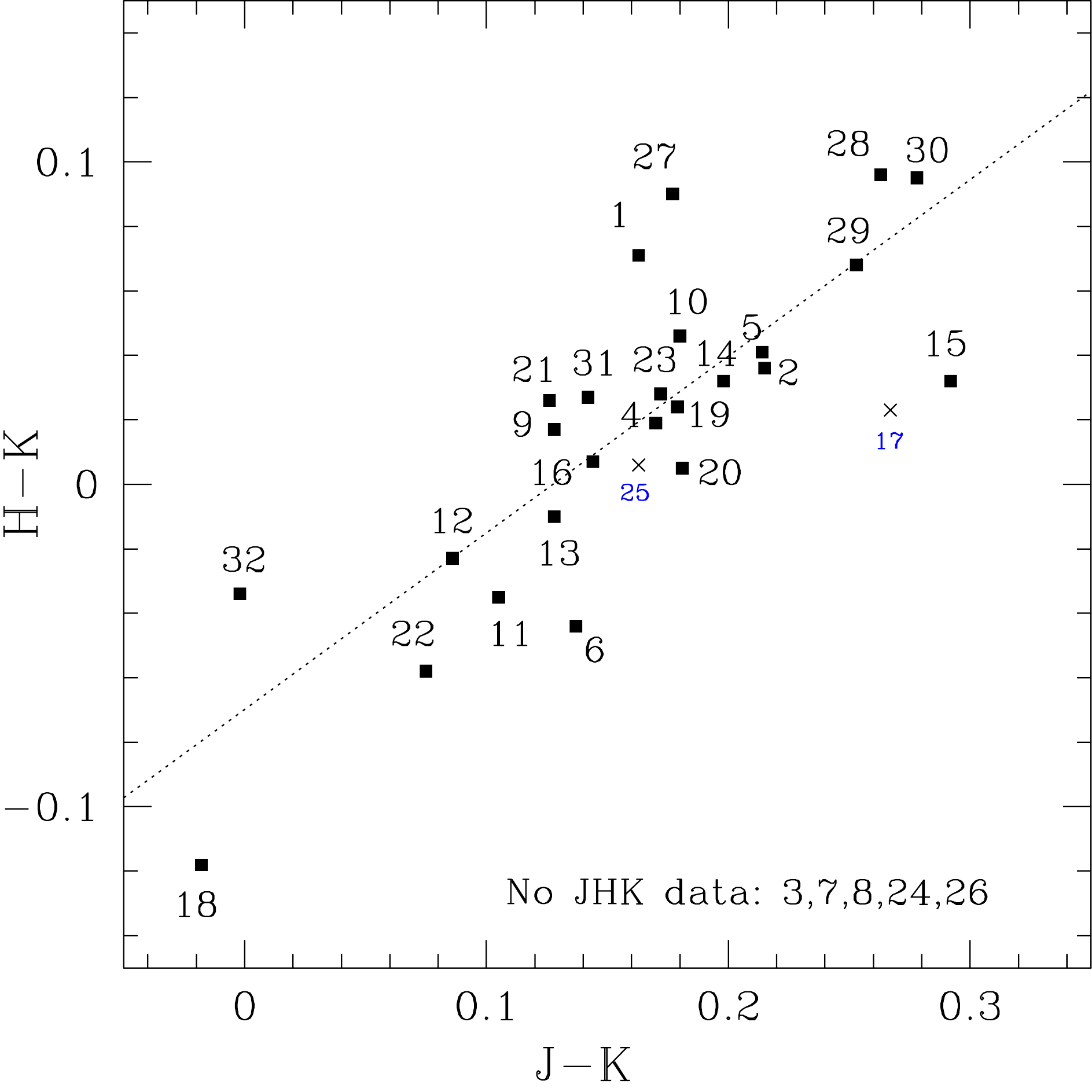} \end{overpic}
\begin{overpic}[width=8.0cm]{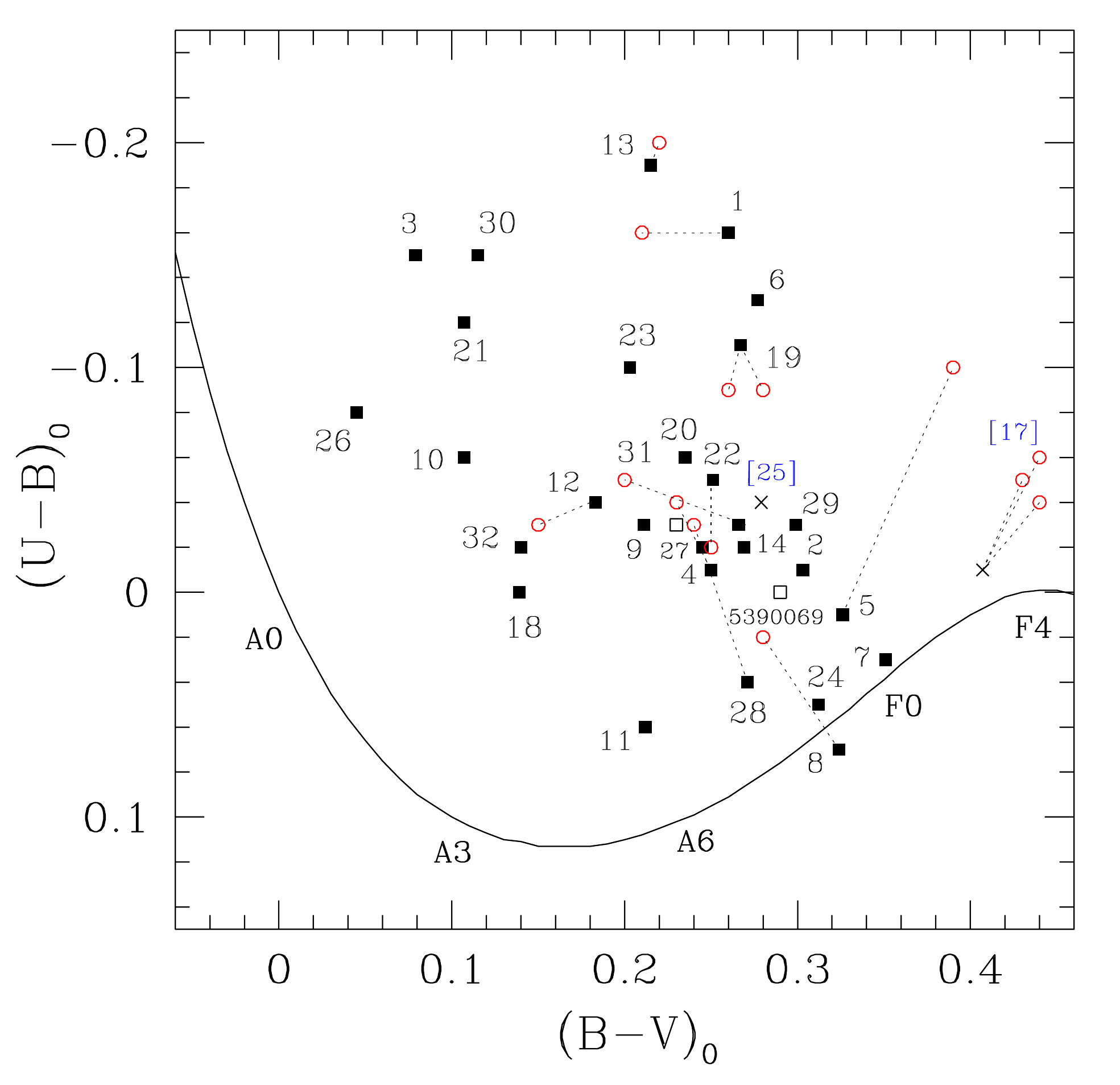} \end{overpic}
\caption{Two-colour diagrams for the
candidate SX\,Phe stars.  In each panel the stars have been labelled with the
CFHT star numbers, except the two stars too faint to have been observed
spectroscopically which have been labelled with their KIC numbers.   {\bf (Top
left)} Gunn-Thuan griz (SDSS) system two-colour diagram, r$-$i versus g$-$r
(both from the KIC), for the candidate SX\,Phe stars.   The equation of the
line, r$-$i = 0.5611\,(g$-$r)$-$0.0998, was derived excluding the two outliers,
KIC\,6227118 (*27) and KIC\,12688835 (*18), and excluding the two misclassified
non-SX\,Phe stars, KIC\,12643589 (*17) and KIC\,9535881 (*25).  {\bf (Top
right)} Two-colour diagram for the candidate SX~Phe stars, comparing the B$-$V
colours from Everett {\it et al.} (2012) with the g$-$r colours given in the
Greiss {\it et al.} (2012) {\it Kepler}-INT survey (KIS).  The equation of the
line, B$-$V = 0.870\,(g$-$r)$+$0.057, with $R^2=0.85$, was derived after
excluding the two misclassified stars, KIC\,12643589 (*17) and KIC\,9535881
(*25).  For KIC~4662336 (*14) the KIC g$-$r colour, 0.225, transformed to the
KIS system using the relation g$-$r(KIS) $=$0.817 g$-$r(KIC)$+0.199$, {\it
i.e.}, 0.383, rather than the too-red KIS colour, 0.584, was used.  {\bf
(Bottom left)} Near-infrared (2MASS) two-colour diagram for the candidate
SX~Phe stars.     The equation of the line is: H$-$K =
0.5478\,(J$-$K)$-$0.0698, with $R^2=0.68$.  {\bf (Bottom right)}
Reddening-corrected (U-B)$_0$ vs (B-V)$_0$  diagram.  The adopted   $E_{B-V}$
reddenings are the new values given in Table\,2, and the  $E_{U-B}$ reddenings
were calculated using the relation (U--B)$_0$ = (U--B) -- 0.72 E$_{B-V}$.  The
curve is the standard relation for unreddened main sequence stars; the solid
black squares represent the  Everett {\it et al.} (2012) photometry
reddening-corrected with the new reddenings; and the red open circles
correspond to the new Bohyunsan Observatory $UBV$ photometric observations. }
\label{two-colour four-panel} \end{figure*}

\subsubsection{Effective Temperatures}

In BN12 the primary source of temperature information was the KIC, where the
$T_{\rm eff}$ values were derived from Gunn-Thuan (SDSS) $griz$ and 2MASS $JHK$
photometric magnitudes and colours.   Since then several  photometric
investigations have provided  additional information about  the temperatures
(and other atmospheric characteristics) of the  {\it Kepler}-field program
stars: Pinsonneault {\it et al.} (2012) found that, for stars with $T_{\rm
eff}$ in the range 4000-6500\,K, the KIC temperature scale needed a correction
of about $+200$\,K; Greiss {\it et al.} (2012) presented the {\it Kepler} Isaac
Newton 2.5-m Telescope Survey (KIS) consisting of stellar photometry through
$U,g,r,i$ and H$\alpha$ filters for over $\sim$50\% of the {\it Kepler} field;
Everett {\it et al.} (2012) presented  Johnson $U,B,V$ photometry for over 4
million sources in the {\it Kepler} field;  and Huber {\it et al.} (2014,
herafter H14) constructed  a catalogue of revised stellar properties for over
196,000 stars in the {\it Kepler} field.  In {\bf Table~5} much of this new
photometric data,  in particular colour indices relevant for estimating
effective temperatures, have been summarized for our program stars.

The spectroscopic (SME, VWA/MOOG) and photometric (KIC, H14) effective
temperatures for the program stars are compared in {\bf Figure\,10}.   The
stars are seen to have (SME) temperatures ranging  from  7300\,K to 8600\,K
({\it i.e.}, 3.863 $< \log T_{\rm eff} <$ 3.934\,K). The upper temperature
matches well the blue edge of the theoretical $\delta$~Sct instability strip
(e.g., Dupret {\it et al.} 2004);  hwoever,  the lower temperature is a few hundred
Kelvins hotter than the theoretical red edge.   The top panel reveals a
systematic offset of $\sim$240\,K between the KIC and H14 photometric
temperatures, consistent with the Pinsonneault {\it et al.} upward  revision of
the KIC temperatures, and extending its range to temperatures
$\sim$8600\,K.  The most extreme outlier  is the crowded optical double
KIC\,6227118 (*27).    The middle panel shows that for most of the cooler stars,
the SME  temperatures are hotter than the H14 temperatures, the average
difference being $\sim$300\,K; the largest discrepancies  occur for KIC\,
7765585 (*28) and KIC\,7301640 (*10), both of which are rapid rotators.  For
the five hottest stars, the H14 temperatures appear to be $\sim$200\,K hotter
than the SME temperatures, {\it i.e.}, more in line with the original KIC
temperatures.   In the bottom panel the SME and VWA/MOOG spectroscopic
temperatures are compared.  Apart from a scatter $\sim$200\,K there is
reasonable agreement.  The largest discrepancy, $\sim$500\,K, is for
KIC\,6520969\,(*21).   We suspect that a large part of the observed difference
can be attributed to the different wavelength intervals that were measured. 

The $T_{\rm eff}$'s from asteroseismology, 7100$\pm$150\,K  for KIC\,11754974
(Murphy {\it et al.} 2013b) and 6622\,K for KIC\,9244992 (Saio {\it et al.}
2015), are both somewhat cooler than the spectroscopic values:  the
KIC\,11754974 value is within the measuring errors, but the asteroseismology
value for KIC\,9244992 is $\sim$1000~K cooler than that derived here!  

In order of decreasing temperature the five hottest stars ($T_{\rm
eff}$\,$>$\,8100\,K) appear to be:   KIC\,10989032 (*32), KIC\,5036493 (*26),
KIC\,12688835 (*18), KIC\,6520969 (*21) and KIC\,9267042 (*12).  The hottest of
these is  a 2.3\,d semidetached binary (see $\S3$).  All  five stars have
early-A spectral types (see Table\,3) and are correspondingly hot in the H14
study.  The five coolest stars (excluding the misclassified binary
KIC\,12643589) appear to be KIC\,6780873 (*5), which is the newly discovered
SB2 system discussed above, the two faint stars not observed spectroscopically
(KIC\,5390069 and KIC\,7300184),  KIC\,3456605 (*24), and KIC\,7174372 (*8).

To establish a relative $T_{\rm eff}$ ranking four photometric colour-colour
graphs have been plotted in {\bf Figure\,11}.  The top-left panel compares the
$r$--$i$ and $g$--$r$ colours, two of the main  indices used by the KIC and by
H14.  The top-right panel is a plot of the Everett {\it et al.} (2012) $B$--$V$
colours versus the {\it Kepler}-INT $g$--$r$ colours, and the bottom-left panel
shows the near-infrared  $H$--$K$ {\it vs.} $J$--$K$ two-colour diagram.  In
all three graphs significant linear correlations are seen, which appear to
confirm that  KIC\,10989032 (*32), KIC\,12688835 (*18) and KIC\,9267042 (*12)
are among the hottest stars in the sample  (KIC\,5036493 was not measured in
the near-IR). 

The bottom-right panel of Figure\,11 is a two-colour plot of  $U$--$B$ {\it
vs.} $B$--$V$, where, unlike the other three panels, the broad-band colours
have been dereddened using the new reddenings in Table\,2 (col.\,4).  The
($B$--$V$)$_0$ ordering should, therefore, be closer to a temperature ranking.
The graph includes the well-known $U$--$B$, $B$--$V$ two-colour main sequence
curve (in this case based on the colours given in Table 1.1 of B\"ohm-Vitense
1989) with the corresponding spectral types given along the curve.  Also
plotted  for 11 of the program stars are new $UBV$ colours based on photometric
observations made in October 2014 with the 1.8-m Bohyunsan Observatory
telescope (the magnitudes and colours are given in Table 5). The new colours
agree well with the Everett {\it et al.} (2012) colours, the largest
differences occurring, as one might expect, for the SB2 system KIC\,6780873
(*5).  Note that at a given ($B$--$V$)$_0$ colour the program stars have
($U$--$B$)$_0$ colours that lie well above the main sequence curve (but well
below the blackbody curve which passes through the point [($B$--$V$)$_0$,
($U$--$B$)$_0$] = [+0.20, --0.68].  As seen clearly in Fig.16 of Preston \&
Sneden (2000), there is, unfortunately, considerable overlap of curves of
constant metal abundance and apparent age; as a consequence, the UV-excess is
rendered ``an ambiguous indicator of abundance'' in this situation.

\subsubsection{Surface Gravities}
 
The surface gravities derived from the CFHT spectra are listed in column\,4 of
Table\,4. When there are many overlapping echelle spectral orders $\log g$ is
not an easy parameter to measure  (Smalley 2004; Catanzarro {\it et al.} 2011);
therefore,  and the derived values are not in many cases well constrained.  In
Table\,4 the values measured using SME are in the top row, and VWA (or MOOG)
values are  in the row below;  assumed or uncertain values are enclosed
parentheses.  Also listed are the KIC and H14 gravities that were derived by
matching observed photometric colours to stellar atmosphere models.  All
stars, except KIC\,6227118 (*27) and KIC\,7765585 (*28), have  KIC and H14
values that are practically identical (note: the KIC does not give $\log g$ for
KIC\,11754974, but all other surface gravity derivations, including that by
Murphy {\it et al.} 2013b, suggest a value close to 4.0).  It is important to
recall that Brown {\it et al.} (2011) warned against relying on the KIC
estimates of $\log g$ for hot main sequence stars, such as those being
investigated here.   

A comparison of the photometric and spectroscopic $\log g$ values shows that
for all but the hottest stars there is  reasonably good agreement, the
mean difference between the H14 and SME values being 0.05, with a standard
deviation of the differences amounting to 0.29.  For the hottest stars the
spectroscopic $\log g$ values  tend to be significantly larger than the
photometric values (which has the effect of increasing the derived temperatures
and metal abundances).  No attempt was made to constrain the spectroscopic
$\log g$ values using the photometric values (see Torres {\it et al.} 2012).

The surface gravity of KIC\,6227118 (*27) clearly is a problem.  The KIC value,
$\log g$=1.03, is exceptionally small, resulting in an unrealistically high
luminosity, log\,$L/L_\odot$$\sim$4.0.   The revised gravity given by H14,
$\log g = 2.84$, is higher but is still  much smaller than  expected for an
SX~Phe (or $\delta$~Sct) star.  KIC\,6227118 is  an outlier in the top panel of
Fig.\,10 -- presumably because  it is an optical double (see footnote 1) and
highly reddened (see Table\,2).  The gravity derived using SME, $\log g = 4.5
\pm 0.3$, is quite uncertain but is more consistent with the star being close
to the main sequence.  

Two stars for which the photometric and spectroscopic gravities differ
significantly  are  KIC\,5036493 (*26)  and KIC\,7174372 (*8).  For
KIC\,5036493 the SME and VWA estimates of $\log g$ are both greater than $\log
g = 4.3$, the expected value for an A5 zero-age main sequence star, and larger
than the H14 photometric value of 3.97.   The spectroscopic values  for
KIC\,7174372 suggest that $\log g$ is closer to 3.5 (which is consistent with
the A9\,III spectral type) than  the H14 value of 4.1. 

In Fig.\,2 of BN12 two other stars appear to have low gravities (and therefore
high luminosities since  for stars of a given temperature $g \propto L^{-1}$):
KIC\,1162150 (*15),  and KIC\,9244992 (*7).  Both stars are relatively cool and
may represent stars that have evolved away from the  main sequence.  Although
KIC\,1162150 has a relatively bright neighbour to its northwest  and   an even
closer faint blue star to its southeast, the spectroscopically and
photometrically derived  $\log g$ values are in close agreement, and  are  near
$\log g=3.5$.    In contrast, there was a lack of agreement between the
spectrosopic and photometric estimates  for KIC\,9244992, the star studied in
detail by Saio {\it et al.} (2015).  The  $\log g \sim 4.5$ derived assuming
ionization balance  differs by 1.0 from the SME and photometric values, which
are all  close to 3.5.   For now, little can be said except that the  value,
$\log g = 4.0$, derived by Saio {\it et al.}, lies midway between the SME and
VWA spectroscopic estimates.

\begin{figure*} \begin{center}
\begin{overpic}[width=7.5cm]{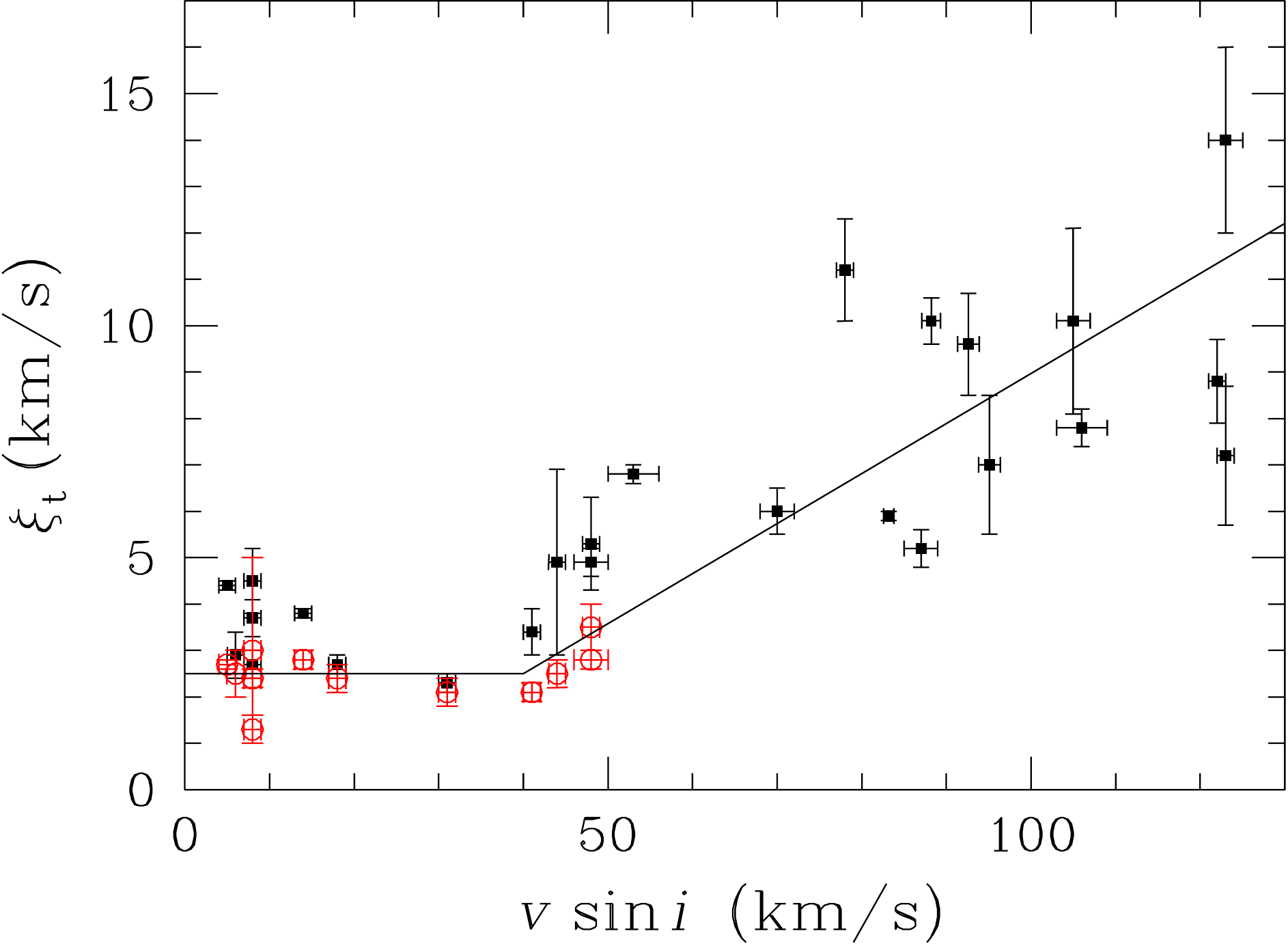} \end{overpic}
\begin{overpic}[width=7.7cm]{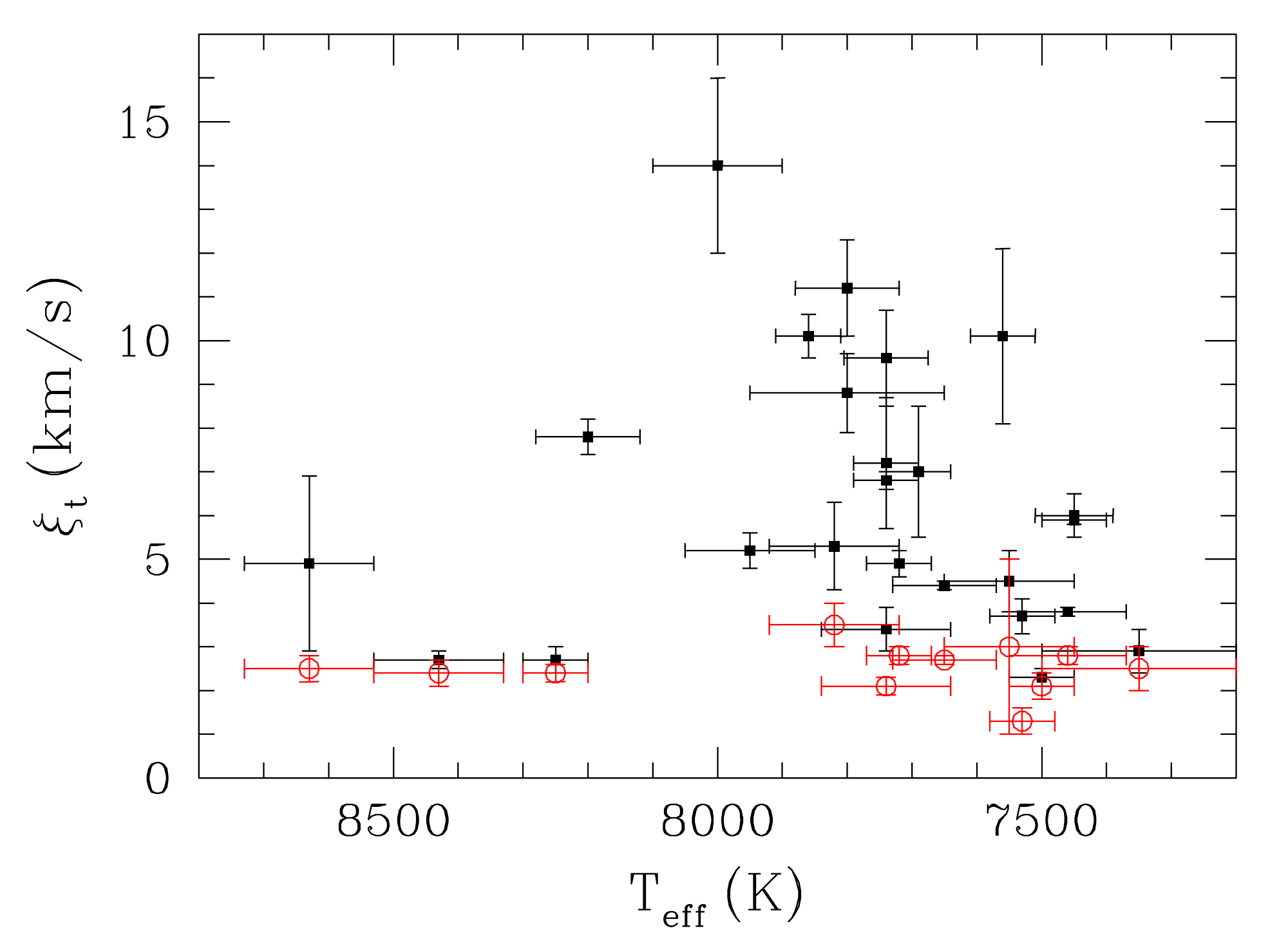} \end{overpic}
\end{center} 
\caption{ Microturbulent velocity $\xi_t$ {\it vs.} projected equatorial
velocity $v\sin i$ (left) and {\it vs.} SME effective temperature (right). The
black filled squares are the SME values, and the red open circles are VWA
values.  Both graphs exclude the four very rapidly rotating stars ({\it i.e.},
those with $v \sin i>$190 km/s),  the problematic star KIC\,6227118 (*27), and
the two misclassified stars.    } \label{vsini_vmic_teff} \end{figure*}

\subsubsection{Microturbulent Velocities}

The concept of microturbulence was introduced by Struve \& Elvey (1934) as a
means of deriving, using the curve-of-growth method, consistent chemical
abundances from weak and strong spectral lines (see Gray 1988, 2005;
B\"ohm-Vitense 1989, Landstreet {\it et al.} 2009,  Aerts {\it et al.} 2014).
Physically, the microturbulent velocity, $\xi_t$, is related to the mean free
path of a photon through small convection cells.  In practice  $\xi_t$
accounts for excess (Gaussian) line broadening over and above that of thermal
broadening.

For each of the program stars  $\xi_t$ was estimated using SME.  For the 12
stars with the narrowest lines, estimates were also derived using VWA and MOOG  (by
adjusting $\xi_t$ until the derived abundances were independent of EW).  The
resulting $\xi_t$ values are given in column\,7 of Table\,4, where, for each
star, the SME value is given in the first row and the VWA/MOOG value is in the
second row. 

{\bf Figure\,12} shows  that there is a  systematic difference between the SME
and VWA microturbulent velocities, with the SME values (black filled squares)
being $\sim$1-2 km/s larger than the VWA values (red open circles), which are
more in line with expectation for A-type stars.  The left panel shows that the
broad line (more rapidly rotating) stars, for which only SME measurements were
made, exhibit a significant trend of increasing $\xi_t$ with increasing $v \sin
i$.  Whether this trend is real, or a result of the increasing difficulty of
measuring $\xi_t$ as  line blending increases,  is uncertain.  The right panel
shows that for the slow rotators, {\it i.e.}, the stars with narrow lines and
$v \sin i < 50$ km/s (red open circles), there is no apparent dependence of
$\xi_t$ on surface temperature.  

How do these results compare with previous $\xi_t$ measurements for similar
stars?  Gray (1988) discusses variations of microturbulence across the HR
diagram and concludes  that ``there is no strong change in $\xi$ with effective
temperature within a given luminosity class";  for main sequence stars the
adopted value for $\xi_t$ was 1~km/s.  More recently, Gray (2014) concludes
from his analysis of five slowly rotating early-A stars that ``an upper limit
of $\lesssim2$ km\,s$^{-1}$ is placed on the microturbulence dispersion".  On
the other hand, Landstreet (1998) found $\xi_t$ values as large as 4.5 km/s in
his analysis of later A-type stars.  More recently, Landstreet {\it et al.}
(2009) concluded (see their Fig.\,2) that $\xi_t$ reaches a maximum of $\sim$4
km/s near 8000\,K, falling as temperatures increase or decrease,  with the
highest values occuring where chemical peculiarities (such as are seen in Am
stars) are exhibited.  Smalley's (2004, Fig.4) graph shows a variation of
$\xi_t$ with $T_{\rm eff}$, suggesting $\xi_t$ values $\sim$3 km/s for stars
with temperatures in the range 7500-8300\,K.  Thus, while $\xi_t$ values
$\sim$4-5 km/s are known for A-type stars from high-dispersion high SNR
spectra,  values as large as 10-15 km/s (as derived here using SME) are
unusually high, suggesting that they may  be the result of difficulties in
measuring $\xi_t$ for fast rotators with broad and blended lines.

\begin{figure} \begin{center}
\begin{overpic}[width=8.5cm]{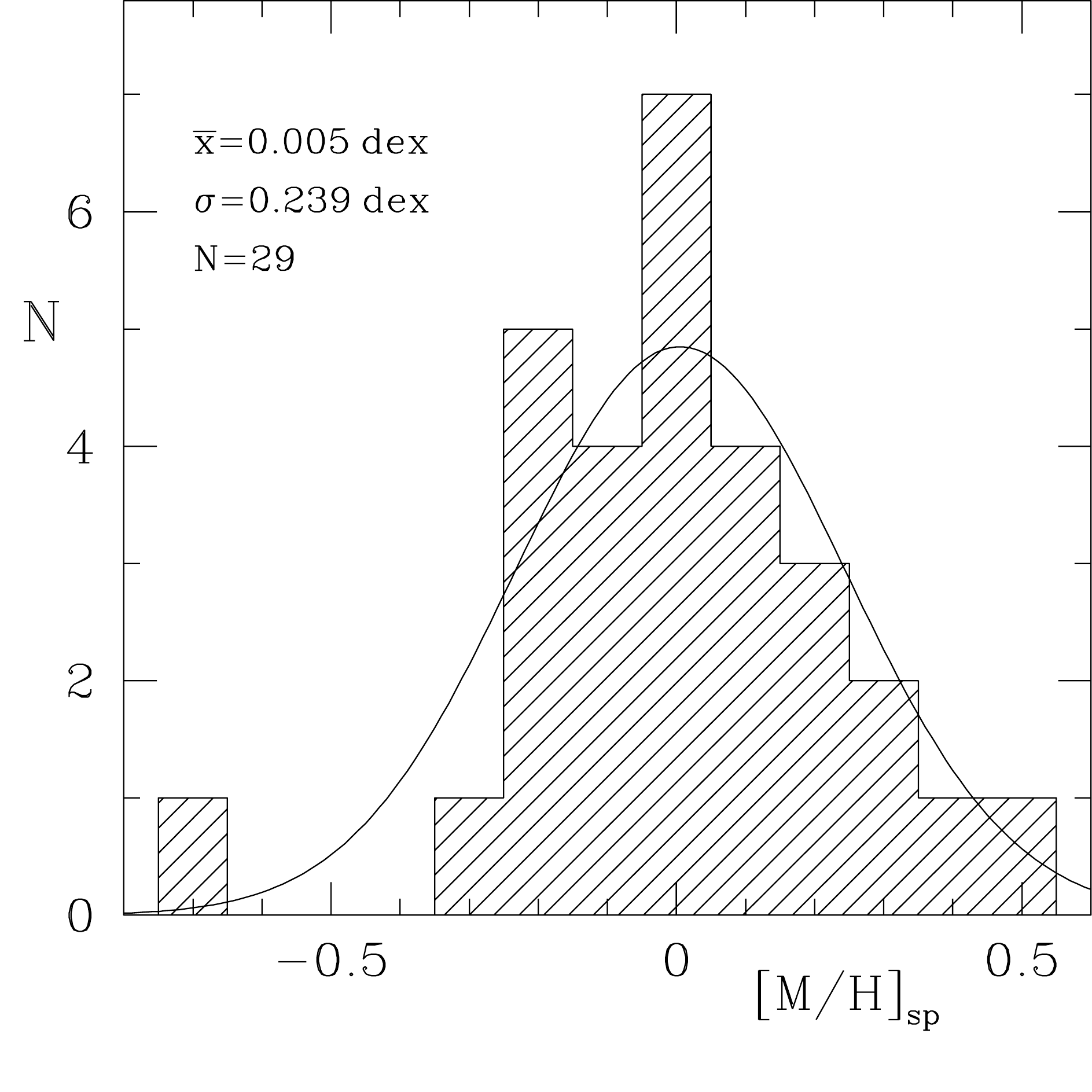} \end{overpic}
\begin{overpic}[width=8.5cm]{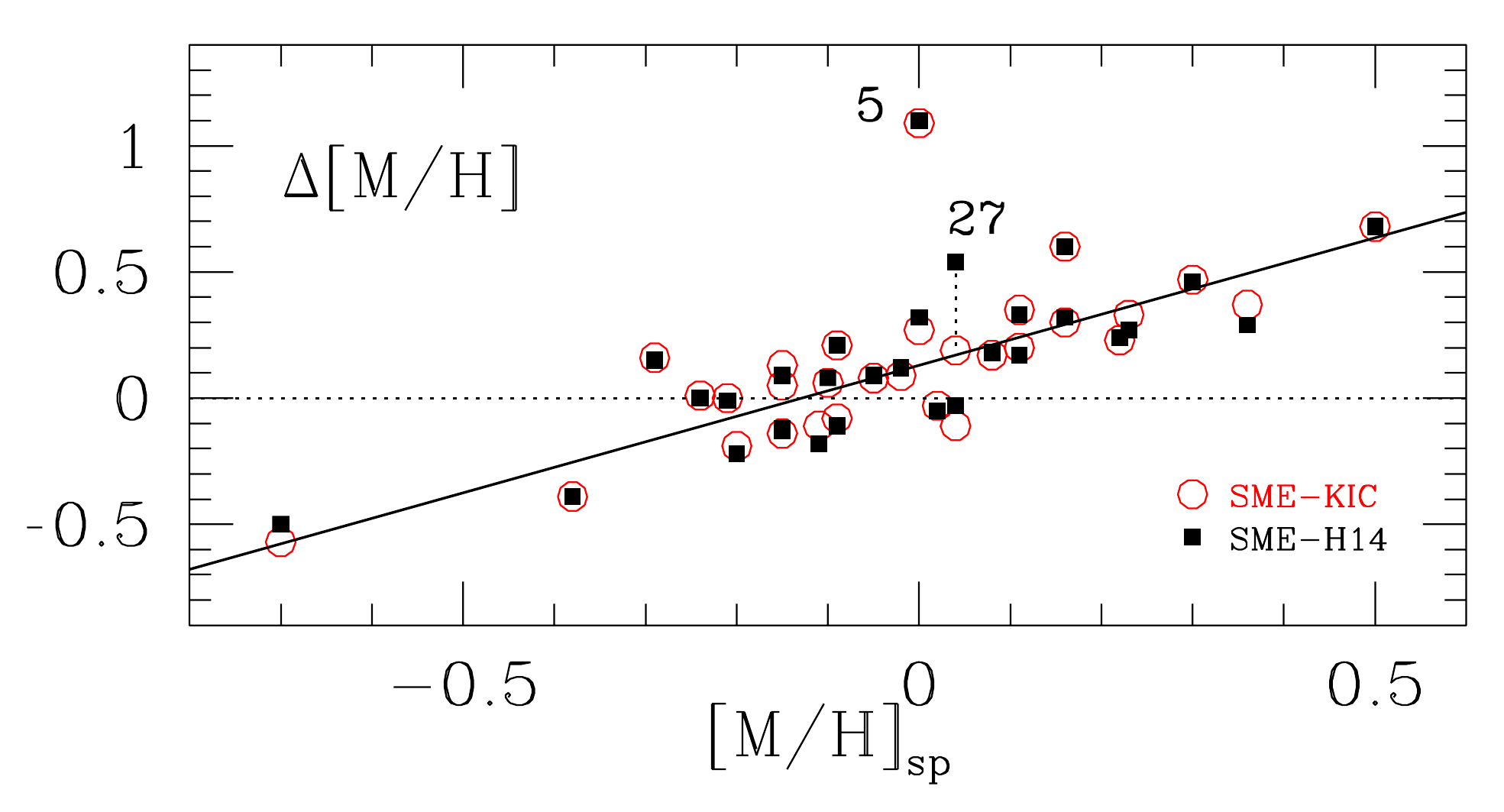} \end{overpic}
\end{center} 
\caption{ (Top panel) Histogram of spectroscopic metal abundances, [M/H]$_{\rm
sp}$, for 29 of the BN12 SX~Phe candidates,  fitted with a Gaussian of mean
0.005\,dex and standard deviation 0.239\,dex;  note that KIC\,11754974 is
off-scale at [M/H]$_{\rm sp} = -1.2\pm0.3$\,dex.    (Bottom panel) Comparing
the spectroscopic metallicities from SME and the photometry-based KIC (red
circles) and H14 (black squares) metallicities:  a well-defined linear trend is
seen.  The equation of the fitted line, after excluding the double-lined
spectroscopic binary KIC\,6780873 (*5), is $\Delta {\rm [M/H]} = 0.949\,{\rm
[M/H]}_{\rm sp} + 0.127,$ with $R^2 = 0.73$.  } \label{FeH_histogram}
\end{figure}

\subsubsection{Metal abundances}

The spectral synthesis method (as implemented in SME) was used to measure metal
abundances for  the program stars.  For those narrow-lined stars with unblended
lines and reliable equivalent widths (EWs), the curve-of-growth method (as
implemented in VWA and MOOG) was also  used to measure [Fe/H].  Most of the
measurements were made of co-added spectra, but when high SNR spectra were
available the abundance calculations were repeated  using individual spectra,
and  the results were folded into those from the coadded spectra.
Unfortunately our spectra were too noisy at near-UV wavelengths for reliable
CaII K-line EWs to be measured and used for [Fe/H] determinations (see Preston
\& Sneden 2000).  The final derived abundances are given in the last column of
Table~4.     

The most metal-poor star in the BN12 sample is KIC\,11754974 (*13), the 343\,d
time-delay binary (Murphy {\it et al.} 2013b).  The mean abundance estimated
from our spectra,  [Fe/H] = --1.2$\pm$0.3\,dex, is lower than (but within the
measurement error of) the abundance  $-0.5\pm0.5$\,dex reported by Murphy {\it
et al.} that was based on lower-dispersion spectra.  Other low metal abundance
stars include the narrow-lined star KIC\,6520960 (*21) with [Fe/H] =
--0.8$\pm$0.2 dex, and KIC\,8004558 (*1) with  [Fe/H] = --0.3$\pm$0.2 dex.  All
three stars are halo objects with retrograde galactic orbits and  exhibit
ultraviolet (UV) excesses in the bottom-right panel of Fig.\,11 (see also
Fig.\,1a of Preston {\it et al.} 1994).    The other two  stars on retrograde
orbits, KIC\,7174372 (*8) and KIC\,10989032 (*32), are not metal poor and do
not show UV excesses. 

Discrepancies between the SME and VWA/MOOG abundances are most likely due to
difficulties associated with line blending,  the fact that our spectra have
only moderate SNRs, and  the differences between  the methods ({\it e.g.}, SME
includes several `metals' in addition to iron in the various wavelength
windows, whereas the VWA and MOOG analyses were restricted to iron lines).  For
those stars with broad lines, that is, with $v \sin i$ between $\sim$50 and 230
km/s, accurate measurement of  [Fe/H] became increasingly difficult  as $v \sin
i$ increased.  Given the size of the overall uncertainties the small
differences between [M/H] and [Fe/H] have been ignored.   

The KIC and H14 photometric  metal abundances tend to be within $\pm$0.10\,dex
of each other.   The largest difference (0.35 dex) occurs for KIC\,6227118
(*27), the star identified earlier as  an outlier with respect to other
quantities.  Excluding this star from the calculations, the mean difference
between the KIC and H14 [Fe/H] values is $0.00$\,dex, with a standard deviation
for the 31 stars of only $0.04$\,dex. 

A histogram of the spectroscopic metallicities is shown in the upper panel of
{\bf Figure\,13},  and the spectroscopic and photometric metal abundances are
compared in the lower panel.   The differences, $\Delta$[M/H], are in the sense
`SME minus H14' (solid black squares) and `SME minus KIC' (open red circles).
The observed  trend implies  that for metal-poor stars the
KIC/H14 abundances are more metal-rich than the spectroscopic abundances, and
for the metal-rich stars, the KIC/H14 abundances are too metal-poor.    For
example, the KIC and H14 metallicities for KIC\,3456605 (*24) are both
$-0.18$\,dex,  compared with the SME value of $+0.50\pm0.10$ dex; and for
KIC\,11754974 (*13) the KIC metallicity is +0.01\,dex compared with the SME
value of $-1.1\pm0.2$ dex.  The largest difference is for KIC\,6780873 (*5),
the newly-discovered SB2 system discussed above.  For it, both the KIC and H14
give $-1.1$\,dex,  compared with the SME metallicity $0.0\pm0.3$\,dex and the
VWA abundance $+0.16\pm0.20$ dex; no attempt was made to disentangle the
spectral lines and so the uncertainties should be considered to be minimum
values.

\section{{\it KEPLER} PHOTOMETRY }

The {\it Kepler} photometry  available at the time of the BN12 analysis
comprised long cadence (LC) data from quarters Q0-Q5, and limited short cadence
(SC) data for three stars:  KIC\,1162150 (Q4.3),  KIC\,9267042 (Q3.3) and
KIC\,11754974 (Q3.1).  Since then  three additional years of LC photometry have
become available , including two additional full quarters (Q6, Q7) of SC
photometry for KIC\,11754974;  the latter  have been analyzed in detail by
Murphy {\it et al.} (2013b) and Balona (2014b).  BN12 give pulsation
frequencies  for the four stars that most closely resemble large-amplitude
field SX\,Phe stars, as well as periodograms for the ten stars located well
above the galactic plane. 

Periodograms (also referred to as `Fourier transforms') based on the  four
years of available {\it Kepler} photometry (Q0-Q17) are presented below for all
34 candidate stars.  Included are the two stars too faint to have been observed
spectroscopically, and the two close eclipsing binaries misclassified as
candidate SX~Phe stars.  Pulsation frequencies, amplitudes and rotation periods
have been derived from the periodograms and subsequently analyzed.   A
significant result is the discovery of numerous binary systems and the
derivation of their orbits.   


\begin{figure}
\begin{center}
\begin{overpic}[width=7.5cm]{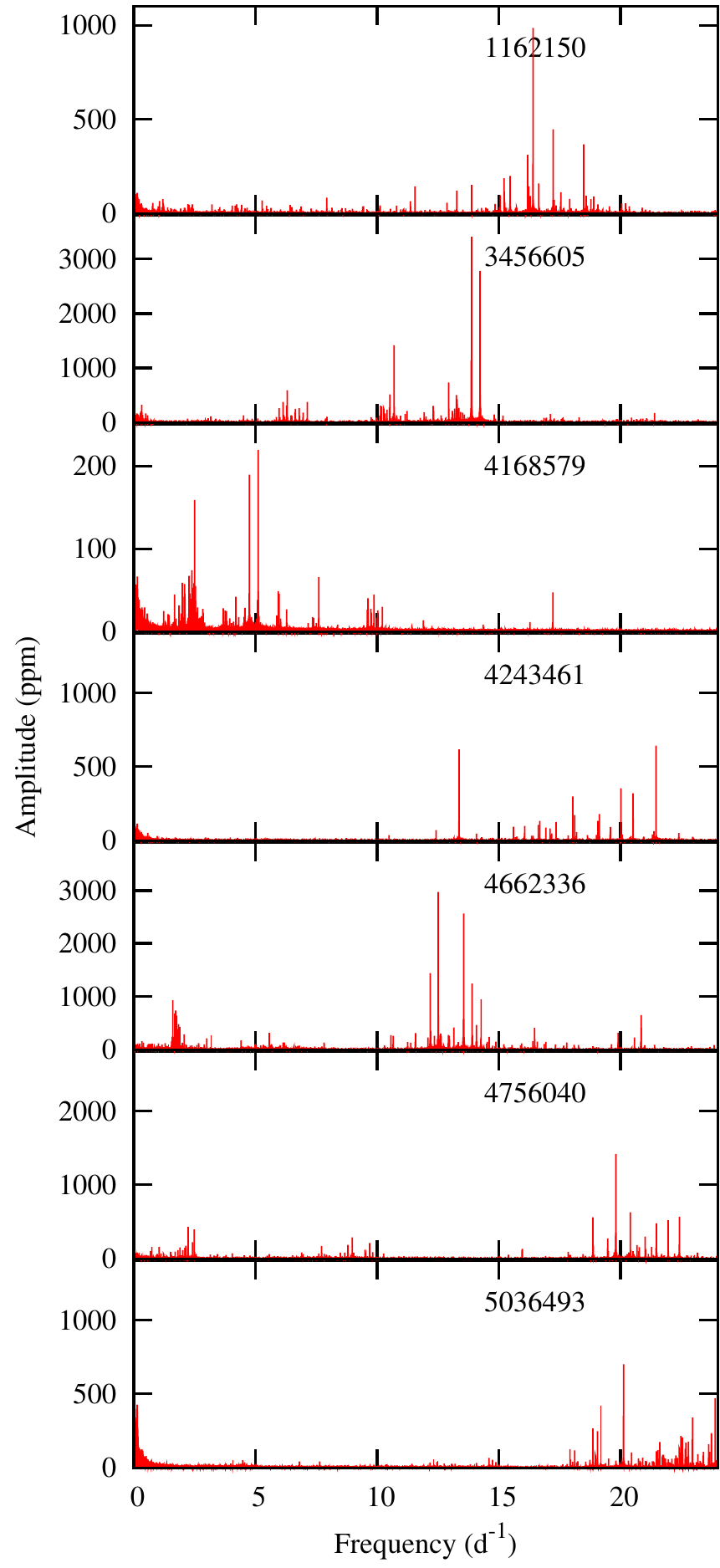} \end{overpic} 
\end{center}
\caption{Periodograms for the 34 candidate SX~Phe stars.  The graphs are ordered by KIC number,  
and the frequency maximum has been set to 24.46 d$^{-1}$, the Nyquist frequency 
of the {\it Kepler} LC data.   The number of  brightness
measurements that were analyzed usually amounted to $\sim$52000 long-cadence
points per star, sampled over $\sim$1470 days.    {\it Erratum}:  in Fig.\,3 of BN12 there are two panels with the label
`11649497';  the lower label is correct, and the upper label should have been
`10989032'.}
\label{pspec1}
\end{figure}

\begin{figure}
\begin{center}
\contcaption{} 
\begin{overpic}[width=7.5cm]{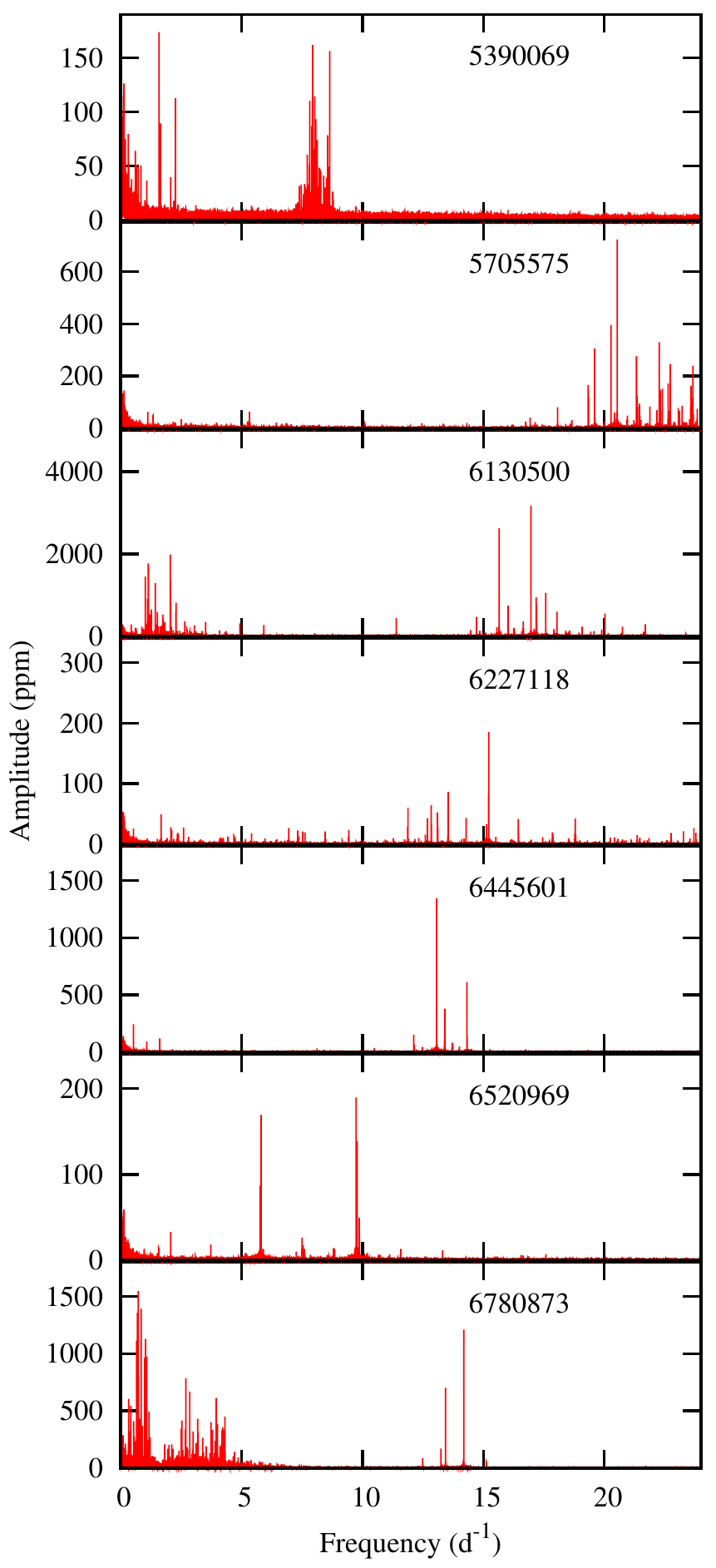} \end{overpic}  
\end{center}
\label{pspec2}
\end{figure}

\begin{figure}
\begin{center}
\contcaption{}
\begin{overpic}[width=7.5cm]{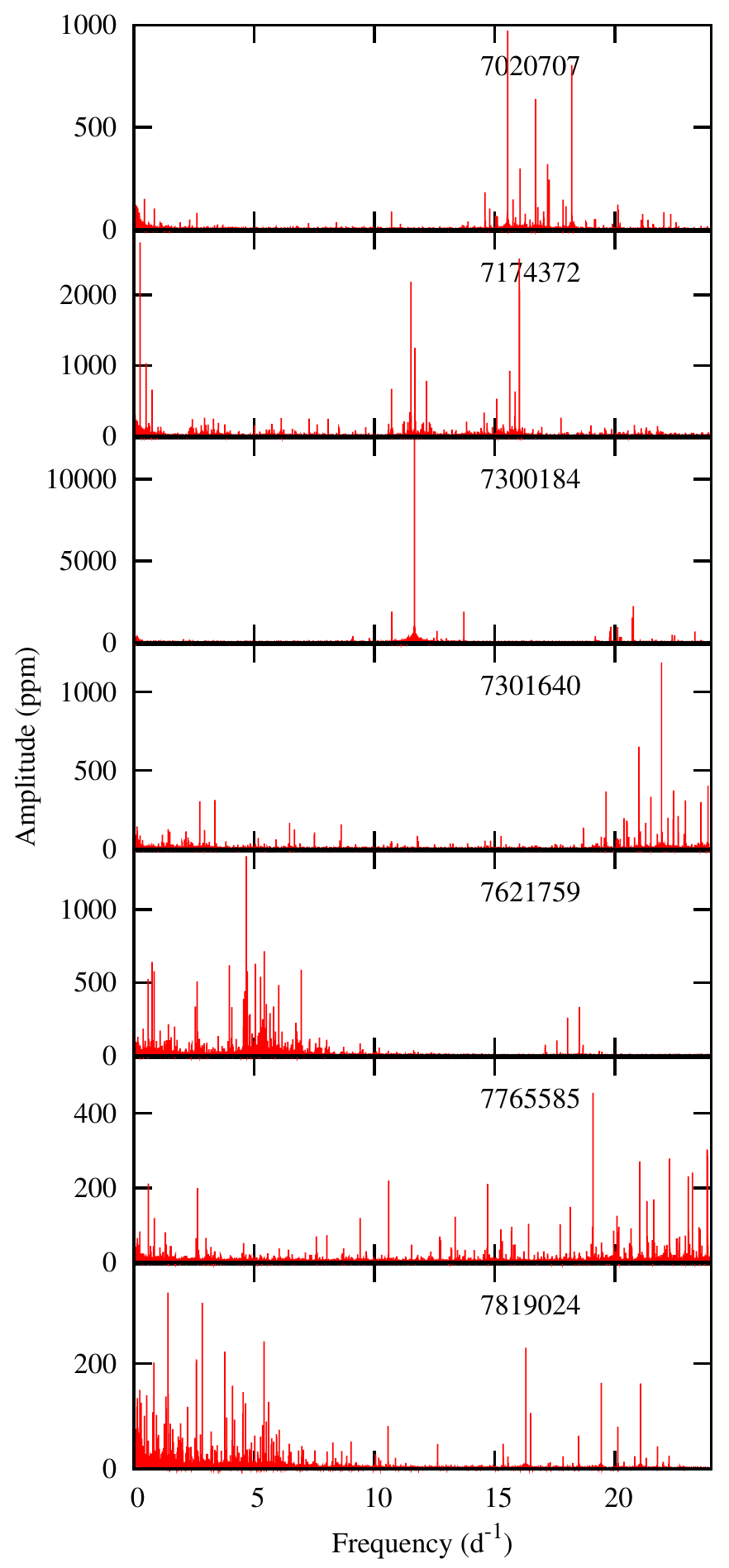}  \end{overpic}
\end{center}
\label{pspec3}
\end{figure}

\begin{figure}
\begin{center}
\contcaption{}
\begin{overpic}[width=7.5cm]{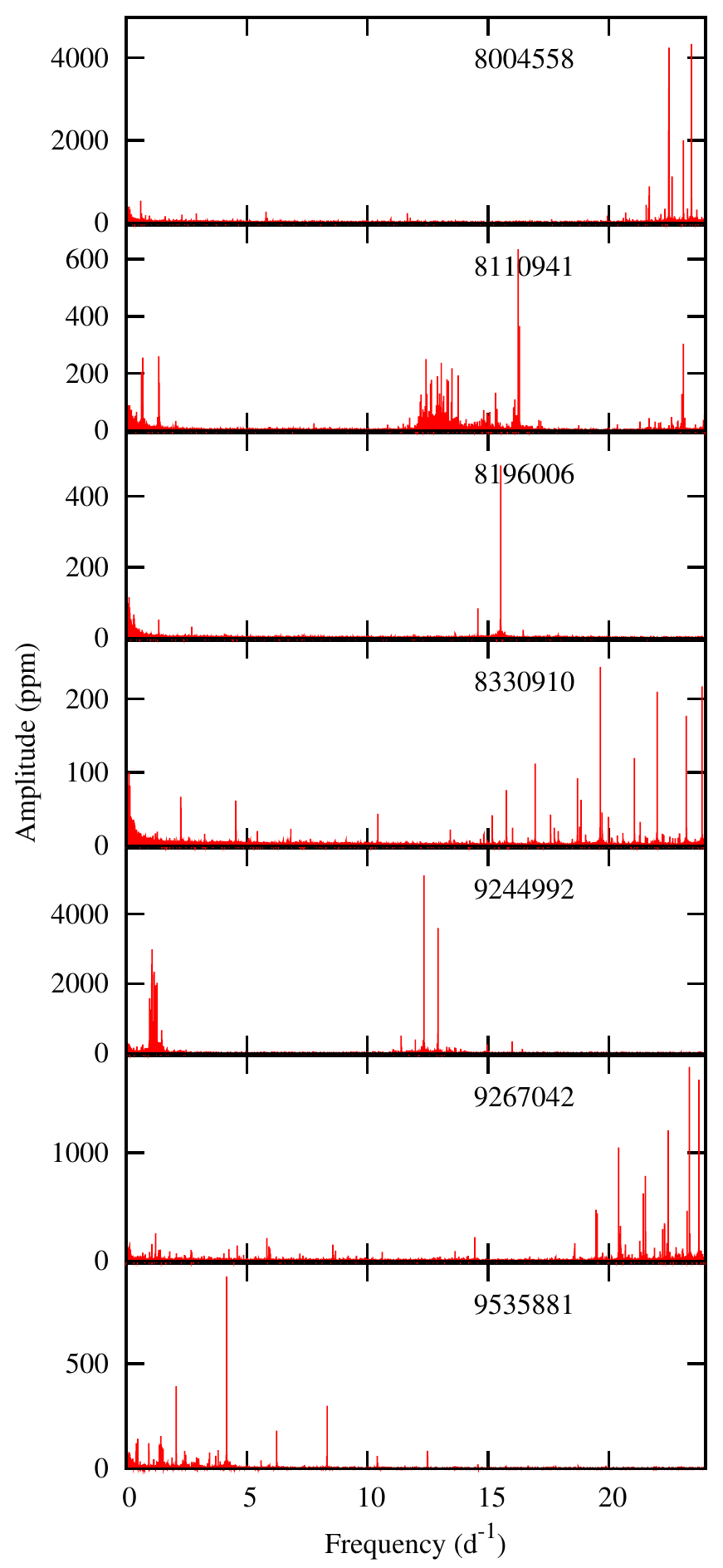} \end{overpic} 
\end{center}
\label{pspec4}
\end{figure}

\begin{figure}
\begin{center}
\contcaption{}
\begin{overpic}[width=7.5cm]{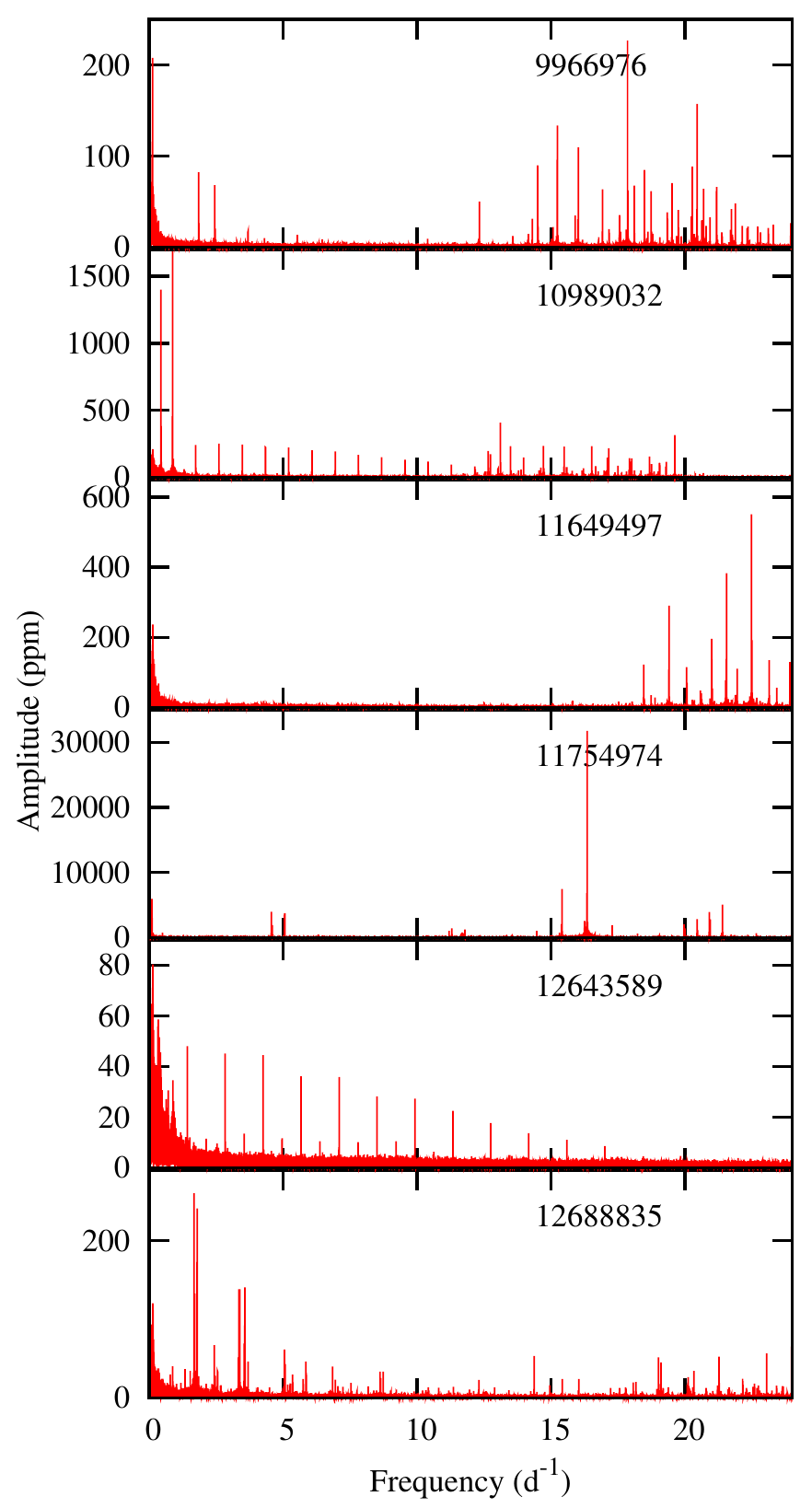} \end{overpic} 
\end{center}
\label{pspec5}
\end{figure}

\begin{figure}
\begin{center}
\begin{overpic}[width=7.5cm]{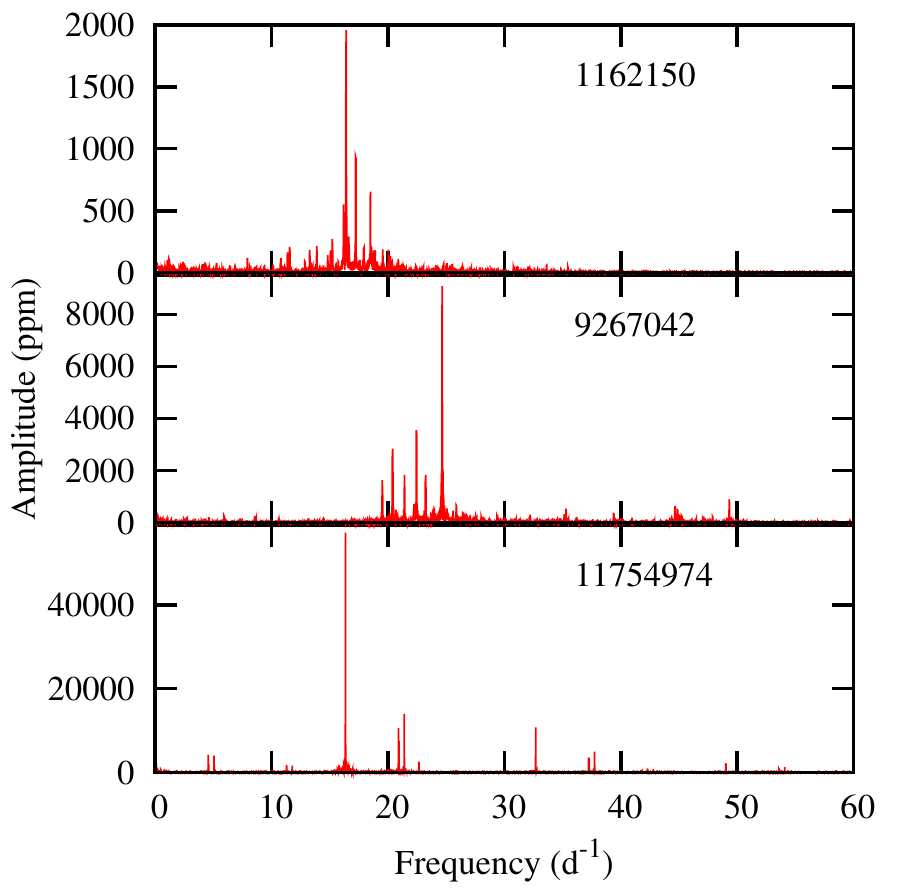} \end{overpic}  
\end{center}
\caption{Periodograms derived from the short-cadence  {\it Kepler} photometry
for KIC\,1162150 (*15, Q4.3, $\sim$42000 points), KIC\,9267042 (*12, Q3.3, $\sim$42000 points) and KIC\,11754974 (*13;
Q3.1,Q6,Q7; $\sim$300000 points).  The maximum frequency of the graphs, 60 d$^{-1}$, is well below 
the 734 d$^{-1}$ Nyquist frequency of the SC data.   }
\label{pspecs}
\end{figure}

\subsection{Pulsations }

Periodograms derived using the  {\it Kepler} LC and SC photometry are shown in
{\bf Figures 14} and {\bf 15}, respectively.    The frequency searches were
made using the Lomb-Scargle algorithm (Press \& Rybicki 1989).  Six of the
stars (KIC\,4168579, 4662336, 4756040, 9966976, 11649497 and 11754974) were
located on  {\it Kepler}'s failed Module\,3 and thus every fourth quarter is
missing from the LC photometry.   The gaps slightly altered the spectral
windows but otherwise were not found to cause serious problems.

Another surmountable problem was the discrimination between `real' and `alias'
peaks in the LC periodograms  due to  aliasing  of pulsation frequencies higher
than the LC Nyquist frequency at 24.462\,d$^{-1}$ ({\it i.e.}, at half the
sampling rate of $\sim$49 photometric observations per day).  The problem of
identifying `super-Nyquist' frequencies  has been discussed  by Murphy (2012),
Murphy {\it et al.} (2013a), Chaplin {\it et al.} (2014) and others.   For the
three stars with SC photometry such aliasing was not a problem owing to the
high Nyquist frequency of the SC data, 734\,d$^{-1}$ (= half the sampling rate
of 1468 points per day), and the fact that  SX~Phe and $\delta$~Sct stars have
in the past been found to have maximum frequencies under 100\,d$^{-1}$.  The SC
periodogram for KIC\,9267042 clearly reveals a dominant {\it real}
super-Nyquist frequency at 24.66\,d$^{-1}$ (responsible for the strong
24.26\,d$^{-1}$ {\it alias} peak seen in the LC periodogram).

For the majority of the stars only  LC photometry is available.  To identify
the super-Nyquist frequencies, searches from 0-50\,d$^{-1}$  were  conducted
using both Lomb-Scargle periodograms and the Fourier methods implemented in the
{\small PERIOD04} program of Lenz \& Breger (2005).  The criterion used to
discriminate between  real and alias frequencies  was based on the assumption
that the alias (which, for real super-Nyquist frequencies occurs at twice the
LC Nyquist frequency minus the real frequency) has a smaller  peak amplitude
than the corresponding real frequency.  This classification rule  appears to be
borne out by the detailed analyses discussed in the references given above.
By applying this rule  half of the program stars were found to have
super-Nyquist frequencies, with the  highest dominant real frequency occurring
at 54.7\,d$^{-1}$ (for KIC\,6520969).

In {\bf Table 6} the pulsation frequencies, $\nu_n$ (d$^{-1}$), and amplitudes,
$A_n$ ($\mu$mag), for the periodogram peaks with the largest amplitude and
greatest significance are given for  the program stars.  The numbers in the
table are based on  {\small PERIOD04}  analyses of the Q0-Q17 LC (and any
available SC) photometry.  In general, the uncertainties in the frequencies are
smaller than 10$^{-5}$\,d$^{-1}$,  while the amplitude uncertainties depend on
the method used to construct the periodogram and may amount to as much as
10$\%$.

CoRoT and {\it Kepler} observations have shown that most, if not all,
$\delta$\,Sct stars exhibit both low and high frequency pulsations (Balona
2014a).  All 32 of the  {\it Kepler}-field SX~Phe candidates  show frequencies
$>$5\,d$^{-1}$ (by definition), and $\sim$25 of the stars also exhibit
significant pulsation at low frequencies.  Since most $\delta$\,Sct stars are
too hot to possess a significant outer convection zone expectation was that the
low frequencies are unlikely to be due to the convective blocking mechanism
that is thought to drive the $\gamma$\,Dor pulsations seen at frequencies 0.3
$<$ $\nu_{\rm puls}$ $<$ 3.3 d$^{-1}$ (Guzik {\it et al.} 2000).  However, the
asteroseismic analysis of KIC\,9244992 by Saio {\it et al.} (2015) showed that
at least one of the stars exhibits rotationally split core $g$-modes
characteristic of $\gamma$\,Dor pulsators, as well as surface $p$-mode
multiplets.

Detailed examination of the periodograms provides fundamental information about
pulsation modes, a means for identifying binary systems (from time delays and
phase modulations), and a framework for making inferences (which involves
identification of  rotationally-split multiplets or light variations due to
rotational modulation and the presence of possible starspots).   A brief
summary of the main findings for each star follows;  many of the stars deserve
more  detailed study.

\vskip0.3truecm

{\noindent \bf KIC\,1162150} (*15) - The strongest frequencies are between 15
and 20 \,d$^{-1}$, all of which are below the LC Nyquist frequency.  The three
highest peaks occur at 16.407, 17.232 and 18.490\,d$^{-1}$ and give rise to a beat pattern of 
(variable) period $\sim1.0-1.3$\,d (best seen in the SC light curve).
\vskip0.3truecm

{\noindent \bf KIC\,3456605} (*24) - The two strongest peaks occur at 13.8771
and 14.2277\,d$^{-1}$, with a third strong frequency at 10.6943\,d$^{-1}$.  Two
of these, 10.6943 and 13.8771 d$^{-1}$, have a ratio 0.77 which might be the
ratio of the fundamental to the first overtone radial mode (and therefore
indicative of the star's mass).  All the frequencies appear to be sub-Nyquist.
\vskip0.3truecm

{\noindent \bf KIC\,4168579} (*23) - The periodogram is dominated by low
frequencies near 2.5, 5.0, 7.5 and 10\,d$^{-1}$, and a distinct, but not very
strong, $p$-mode frequency at 17.22\,d$^{-1}$.  Quite possibly the former are
harmonics (retrograde modes?) caused by the pulsator's rapid rotation ($<$$v
\sin i$$>$$\sim$200 km/s), in which case the rotation frequency is
2.49\,d$^{-1}$.  However other modes also are possible (for example, see Breger
{\it et al.} 2013).  Also present at low frequencies is an interesting group of
peaks symmetric in spacing (and amplitude) at $\nu_1 = 5.87206$, $\nu_2 =
5.93402$, $\nu_3 = 5.96988$ and $\nu_4 = 6.03183$\,d$^{-1}$.  While $\nu_4 -
\nu_3 = \nu_2 - \nu_1 = 0.06195$\,d$^{-1}$, the meaning of this pattern is not
clear.  All the significant frequencies appear to be sub-Nyquist.  Visual
inspection of the periodogram gives the impression of equal spacing among some
peaks, but closer study reveals that the spacings are not equal.
\vskip0.3truecm

{\noindent \bf KIC\,4243461} (*4) - Many peaks are present with frequencies
higher than 13\,d$^{-1}$ and amplitudes in excess of 200\,ppm.  Frequency
modulations also are seen.  The peak of highest amplitude, at 21.458\,d$^{-1}$,
may be detectable from the ground.  Also seen are a large number of frequencies
below 4\,d$^{-1}$, all with amplitudes less than 30\,ppm; since the star is
cool these are probably $g$ modes.  No evidence for spots is seen at low
frequencies.   The pulsator is in a time-delay binary system with a 460\,d
orbital period.  In support of its binary nature, the observed RV range,
$\Delta$RV=5.1 km/s, agrees well with the predicted $K_1 = 5$ km/s.
\vskip0.3truecm

{\noindent \bf  KIC\,4662336} (*14) - The frequencies of largest amplitude
(above 500 $\mu$mag) are all sub-Nyquist and should be readily detectable from
the ground.  No obvious frequency modulation (FM) sidelobes are seen.  A system
of low-frequency peaks at around 1.7\,d$^{-1}$ (possibly $g$ modes) is striking
because of the almost equal spacing;  the frequencies, periods and spacings are
identified in {\bf Table 7}.  Although neither the frequencies nor the periods
are equally spaced, the change in spacing forms a nice linear trend; similar
such trends are seen in $\gamma$\,Dor stars (e.g. Bedding {\it et al.} 2014).
Most of the peaks are concentrated in the  range $10 < \nu < 22$\,d$^{-1}$, but
there is no obvious pattern or common separation in this range.
\vskip0.3truecm

\begin{table*} \centering \label{Table6} \caption{Pulsation frequencies,
$\nu_n$\,($d^{-1}$), and amplitudes, $A_n$\,($\mu$mag = ppm), for the
periodogram peaks of largest amplitude and greatest significance.  The numbers
are based on analysis of the SC data if available, otherwise of the Q0-Q17 LC
data.   When the amplitude of the super-Nyquist peak (at $\nu >
24.4615\,d^{-1}$) was greater than that of the corresponding sub-Nyquist peak
the former was assumed to be the real frequency and the latter an alias
frequency. Binary systems are identified with {\bf boldface} KIC numbers, and
frequencies used in the time-delay analyses have been \underline{underlined},
and rotation (and/or orbital) frequencies are in {\color{red}{\it red italics}}.  }

\begin{tabular}{lllllll}
\hline
\multicolumn{1}{c}{ Star } & \multicolumn{1}{c}{$\nu_1$($A_1$)}   & \multicolumn{1}{c}{$\nu_2$($A_2$)}  & \multicolumn{1}{c}{$\nu_3$($A_3$)}  & 
\multicolumn{1}{c}{$\nu_4$($A_4$)}  & \multicolumn{1}{c}{$\nu_5$($A_5$)}  & \multicolumn{1}{c}{ $\nu_6$($A_6$) }     \\
\multicolumn{1}{c}{      }  & \multicolumn{1}{c}{$\nu_7$($A_7$)}   & \multicolumn{1}{c}{$\nu_8$($A_8$)}  & \multicolumn{1}{c}{$\nu_9$($A_9$)}  & 
\multicolumn{1}{c}{$\nu_{10}$($A_{10}$)}  & \multicolumn{1}{c}{$\nu_{11}$($A_{11}$)}  & \multicolumn{1}{c}{  $\dots$ etc.  }     \\
\hline
1162150 (*15)       & 16.40793\,(1601) &  17.23122\,(816)  & 18.49044\,(648)  & 16.40005\,(563) &  16.18382\,(523) & 15.22177\,(285)\\
3456605 (*24)       & 13.87706\,(4640) &  14.22773\,(3559) & 10.69431\,(1650) & 13.87217\,(1167) & 12.94050\,(789) & 13.25948\,(721)  \\  
4168579 (*23)       &  5.10953\,(234)  &   4.74623\,(199)  &  {\color{red}{\it 2.49131}}\,(165)  & 17.21841\,(75)   &  2.38202\,(79)  &  7.60083\,(73)   \\  
{\bf 4243461} (*4)  & 21.45819\,(1300) &  \underline{13.36151}\,(817)  & \underline{20.01540}\,(650)  & \underline{18.04597}\,(461)  & 
                    \underline{19.13558}\,(306) & \underline{18.11525}\,(291)  \\ 
                    & \underline{16.67878}\,(209)  &  \underline{17.35007}\,(208)  &       &    &   &  \\  
4662336 (*14)       & 12.51132\,(4037) &  13.55432\,(3428) & 12.19013\,(1854) & 13.90794\,(1672) & 14.26849\,(1277)& 20.84469\,(1249)\\ 
4756040 (*20)       & 19.80761\,(2566) &  26.51536\,(1455) & 20.41056\,(1183) & 21.96026\,(1092) &  2.23450\,(447) & 20.67829\,(363)  \\  
5036493 (*26)       & 28.80941\,(1439) &  23.89670\,(1164) & 23.73328\,(568)  & 30.06947\,(531)  & 26.39105\,(535) & 29.88074\,(502)  \\  
5390069             &  7.94464\,(183)  &   8.65358\,(181)  &  8.03156\,(143)  &  7.83105\,(130)  &  8.08036\,(112) &  1.60096\,(186)  \\  
                    &  2.27398\,(116)  &   1.66293\,(93)   &  0.1593\,(40)    &  0.3187\,(45)   &  0.4710\,(48)  & 0.6220\,(55)   \\   
{\bf 5705575} (*22) & \underline{20.53586}\,(1387) &  26.66737\,(872)  & \underline{20.27854}\,(751)  & 25.27748\,(706)  & 25.36201\,(469) & 
                      22.39538\,(317)  \\ 
                    & 22.31121\,(317)  &  \underline{21.35656}\,(276)  & \underline{23.99475}\,(211)  & \underline{24.78404}\,(211)  &                                    \\ 
6130500 (*9)        & 16.96920\,(4988) &  15.65390\,(3785) & 17.19123\,(2164) & 17.58345\,(1700) & 17.19123\,(1396)& 17.58005\,(1371) \\  
                    & 20.02412\,(1023) &  18.04089\,(998)  & 21.69512\,(654)  & 16.65137\,(537)  & 11.39845\,(537) &  2.07024\,(2005) \\  
                    & {\color{red}{\it 1.15158}}\,(1767) &   1.02188\,(1413) &  1.44498\,(1325) &  1.11645\,(901)  &  2.29761\,(821) &  1.27516\,(605)  \\   
6227118 (*27)       & 33.71604\,(304)  &  33.71445\,(159)  & 35.39488\,(134)  & 36.09299\,(99)   & 37.05546\,(88)  & 30.14356\,(85)   \\ 
6445601 (*2)        & 13.06792\,(1696) &  14.32617\,(838)  & 13.41806\,(487)  & 13.73040\,(108)  & 13.70703\,(93)  & 16.73954\,(49) \\  
                    & 10.50716\,(40)   &   8.12804\,(36)   &  1.62431\,(136)  &  1.08296\,(107)  &  {\color{red}{\it 0.5416}}\,(76)  &                \\  
6520969 (*21)       & 39.19536\,(265)  &  54.74894\,(230)  & 39.15203\,(198)  & 39.06451\,(69)   &                 &                 \\  
{\bf 6780873} (*5)  & \underline{14.18757}\,(1635) &  \underline{13.43627}\,(912)  &  0.74535\,(1507) &  0.70228\,(1360) &  0.86415\,(1358) & 0.67020\,(1154) \\  
                    &  1.04091\,(1077) &   1.08573\,(945)  &  0.99881\,(947)  &  1.06323\,(919)  &                  &                 \\ 
7020707 (*16)       & 15.53514\,(1428) &  18.20383\,(1332) & 16.70110\,(986)  & 17.20074\,(496)  & 16.05242\,(454)  & 17.83731\,(244)  \\   
                    & 17.96163\,(196)  &  16.78914\,(169)  & 22.02607\,(182)  & 22.31278\,(170)  & 21.15035\,(151)  & 17.03594\,(145)  \\  
                    & 14.78336\,(141)  &  10.70760\,(105)  & 15.04419\,(100)  &  0.33502\,(153)  &  0.84609\,(97)   &  2.61767\,(87)  \\  
{\bf 7174372} (*8)  & 16.02162\,(3692) &  11.52111\,(2560) & 11.68348\,(1601) & 15.62332\,(1419) & 12.15787\,(957)  & 15.84335\,(907) \\   
                    & 10.70399\,(823)  &    {\color{red}{\it 0.2500}}\,(2700)  &  0.5000\,(1028) &   0.7501\,(642)     \\   
{\bf 7300184}       & \underline{11.65981}\,(28895)&  20.76654\,(4324) & 20.72239\,(2982) & 13.71237\,(2524) & 20.11222\,(1941) & 23.31961\,(1637) \\ 
                    & 22.48131\,(960)  &   0.25325\,(2081) &  0.29738\,(1348) &                  &                  &                  \\ 
7301640 (*10)       & 21.93607\,(2483) &  26.50609\,(977)  & 23.86695\,(1007) & 23.57156\,(702)  & 19.63272\,(652)  & 26.72637\,(498) \\   
                    & 20.36786\,(373)  &  20.49722\,(339)  & 26.98485\,(286)  & 27.18351\,(250)  &  3.36753\,(317)  &  2.73281\,(311)   \\
7621759 (*6)        &  4.66961\,(1487) &   5.41867\,(741)  &  3.96393\,(693)  &  5.05030\,(685)  &  4.71298\,(616)  &  5.25911\,(572)  \\   
                    & 18.51980\,(564)  &   6.01497\,(532)  &  2.62024\,(531)  &  0.83635\,(515)  &  0.74907\,(486)  & 18.03528\,(427)  \\ 
7765585 (*28)       & 19.07862\,(873)  &  24.16114\,(765)  & 23.98715\,(622)  & 22.27004\,(609)  & 21.02492\,(552)  & 21.61366\,(353)  \\  
                    & 14.70550\,(317)  &  10.58127\,(263)  & 23.53583\,(197)  & 20.15150\,(191)  &  2.64222\,(188)  &  0.84329\,(112)  \\ 
{\bf 7819024} (*19) & \underline{16.29036}\,(332)  &  \underline{21.06214}\,(315)  & \underline{19.42795}\,(297)  & 
                      \underline{16.49367}\,(158)  &  \underline{21.76567}\,(89)   &  {\color{red}{\it  1.4066}}\,(339)              \\ 
                    &  2.83717\,(322)  &  5.40761\,(252)   &  3.76815\,(228)  &  2.59583\,(214)  &  0.8252\,(206)  &                   \\ 
{\bf 8004558} (*1)  & \underline{25.52139}\,(11886)& \underline{24.91583}\,(5381)  & \underline{26.32943}\,(2689) & 
                      \underline{25.68562}\,(916)  & \underline{27.30128}\,(864)  &  0.60555\,(604)   \\   
8110941 (*29)       & 16.24003\,(941)  & 24.91933\,(788)   & 16.29438\,(547)  & 12.41964\,(320)  & 13.06036\,(299)  & 23.95900\,(289)  \\    
                    & 13.34605\,(285)  & 13.50247\,(283)   &  1.36114\,(265)  &  {\color{red}{\it 0.68793}}\,(204)  &  1.37657\,(173)  &  0.64287\,(127)  \\  
8196006 (*30)       & 15.51033\,(701)  &  {\color{red}{\it 1.35986}}\,(49)    &  2.72124\,(40)   &  4.08092\,(14)   &  5.44320\,(8)    &   \\   
8330910 (*3)        & 24.14147\,(547)  & 21.99215\,(445)   & 19.63965\,(464)  & 16.94367\,(175)  & 15.75556\,(115)  & 18.84284\,(108)  \\  
                    &  {\color{red}{\it 2.2724}}\,(69)      &  4.5513\,(64)  &  6.8246\,(24)     \\  
9244992 (*7)        & 12.33937\,(6515) & 12.92006\,(4642)  & 15.99298\,(502)  & 12.31246\,(495)  & 14.96647\,(359)  & 12.35230\,(313)  \\  
                    &  1.09100\,(2987) &  1.07591\,(2580)  &  1.15851\,(2265) &  1.17432\,(2256) &  1.21147\,(2046) &  1.29332\,(2019)  \\ 
{\bf 9267042} (*12) & \underline{24.6637}\,(9099) & \underline{22.4487}\,(3617)  & \underline{20.4008}\,(2872) & 23.2384\,(1864) & 
                      \underline{21.4236}\,(1800) & 19.5155\,(1650)  \\  
{\bf 9535881} [*25] &  2.08152\,(409)  &  4.16301\,(972)   &  6.24452\,(219)  &  8.32602\,(347)  & 10.40754\,(79)   & 12.48903\,(106) \\ 
                    &  {\color{red}{\it 0.48885}}\,(172)  &   0.42386\,(140)  &  0.94685\,(133)  &  1.44971\,(157)  &  1.38111\,(125)  &  1.48714\,(102)    \\  
{\bf 9966976} (*31) & \underline{17.8582}\,(376)  &  \underline{20.4512}\,(302)  & \underline{33.7040}\,(218)  & \underline{30.4500}\,(165)  & 
                      20.2785\,(168)   & 16.0114\,(165)  \\   
                    &  {\it 0.1145}\,(197)   & 0.228\,(50)       & 0.346\,(40)      &   1.8407\,(82)   &  2.4505\,(71)   &                  \\ 
{\bf 10989032} (*32)& {\color{red}{\bf {\it 0.43382}}}\,(1415) &   0.86764\,(1790) &   29.31930\,(768)  &  35.83161\,(709) & 32.42100\,(494)  & 33.45078\,(451) \\  
                    & 34.21988\,(449)  &  31.79224\,(466)  & 35.45244\,(404)  &  36.28355\,(334) & 31.01511\,(258)  & 36.20560\,(249)  \\ 
11649497 (*11)      & 22.48921\,(1161) &  19.40347\,(495)  & 27.94219\,(384)  & 24.85830\,(340)  & 26.99322\,(263)  & 27.93027\,(207)   \\   
{\bf 11754974} (*13)& \underline{16.34477}\,(67903)& \underline{21.39904}\,(14121) & \underline{20.90745}\,(10566)& \underline{20.94360}\,(7289) & 
                      \underline{22.66036}\,(2263) &                      \\ 
{\bf 12643489} [*17]&  0.07080\,(61)   &  0.12817\,(75)    &  0.33354\,(61)   &  0.33868\,(54)   &  0.70823\,(30)   &  0.88463\,(35)  \\   
                    &  1.41678\,(48)   &  2.12519\,(10)    &  2.83359\,(46)   &  3.54204\,(13)   &  4.25041\,(46)   &  4.95881\,(12)  \\  
                    &  5.66722\,(37)   &  6.37565\,(12)    &  7.08401\,(38)   &  7.79239\,(10)   &  8.50086\,(32)   &  9.20922\,(12)  \\  
12688835 (*18)      &  {\color{red}{\it 1.67430}}\,(264)  &  1.78784\,(239)   &  1.78263\,(178)  &  3.5704\,(100)   &  3.3798\,(90)    &  3.3389\,(85)   \\ 
                    &  5.0452\,(40)    &  23.98174\,(178)  & 27.66717\,(111)   & 19.01177\,(79)   & 29.83500\,(78) & 34.56846\,(72)    \\ 
\hline
\end{tabular}
\end{table*}

\begin{table}
\centering
\label{Table7}
\caption{KIC\,4662336 (*14) -  sequence of periodogram peaks around
1.7\,d$^{-1}$, at frequencies $\nu_n$\,(d$^{-1}$) with amplitudes 
$A_n$\,($\mu$mag), consecutive frequency spacings $\Delta\nu = \nu_{n+1} - \nu_n$
(d$^{-1}$),  periods $P_n$\,(day) and consecutive period spacings $\Delta P = P_n
- P_{n+1}$ (d$^{-1}$). Frequencies 1,2,4 appear to be aliases of 3, while
5,7,8 appear to be aliases of 6; and in general the frequency spacings are increasing. }
\begin{tabular}{rrrrrr}
\hline
 $n$ & $\nu_n$ & $A_n$ & $\Delta\nu$ & $P_n$ & $\Delta P$  \\
\hline
 1 &   1.5486 & 197 & 0.0236 &    0.6457 &   0.0096 \\
 2 &   1.5722 & 275 & 0.0251 &    0.6361 &   0.0100 \\
 3 &   1.5973 & 957 & 0.0269 &    0.6261 &   0.0104 \\
 4 &   1.6242 & 192 & \multicolumn{1}{c}{--}  &  0.6157 & \multicolumn{1}{c}{--}    \\[5pt]
 5 &   1.6843 & 702 & 0.0339 &    0.5937 &   0.0117 \\
 6 &   1.7181 & 754 & 0.0368 &    0.5820 &   0.0122 \\
 7 &   1.7550 & 643 & 0.0400 &    0.5698 &   0.0127 \\
 8 &   1.7950 & 401 & 0.0435 &    0.5571 &   0.0132 \\
 9 &   1.8385 & 494 & 0.0477 &    0.5439 &   0.0137 \\
10 &   1.8861 & 425 &        &    0.5302 &          \\
\hline
\end{tabular}
\end{table}

{\noindent  \bf  KIC\,4756040} (*20) - Shortward of the LC Nyquist frequency
the four highest amplitude peaks occur at 19.808, 22.424, 20.410 and
21.960\,d$^{-1}$; the second of these is an alias of the higher-amplitude
super-Nyquist frequency 26.515\,d$^{-1}$.  One also sees many low-amplitude
peaks  in the range 19--24\,d$^{-1}$, and rich but low-amplitude peak systems
in the frequency ranges 0.5--2.7 ($g$-modes?) and 7.3--10.0\,d$^{-1}$.  There
are no apparent equidistant spacings.  About half of the peaks between 15 and
24.4 d$^{-1}$ are alias peaks, and no frequencies of significant amplitude are
seen above 35\,d$^{-1}$.  \vskip0.3truecm

{\noindent  \bf  KIC\,5036493} (*26) -  The 0-50\,d$^{-1}$ LC periodogram is
dominated by many peaks with frequencies between 18 and 31\,d$^{-1}$ and
amplitudes in excess of 300 ppm; several of the super-Nyquist frequencies are
real.  The dominant real frequency is at 28.809\,d$^{-1}$, with an alias at
20.130\,d$^{-1}$.  Low-amplitude peaks are seen in the range 12-15\,d$^{-1}$,
many of which have a spacing of 0.148\,d$^{-1}$.   There are very few peaks
with $\nu < 5$\,d$^{-1}$, though there might be a long-term variation of $\nu =
0.135$\,d$^{-1}$ (similar to the frequency spacing just
mentioned).\vskip0.3truecm

{\noindent \bf   KIC\,5390069} - The periodogram shows a clump of peaks between
7 and 9\,d$^{-1}$, and four prominent frequencies between 1.5 and 2.5 d$^{-1}$
(at 1.601, 1.663, 2.081 and 2.274\,d$^{-1}$), all of which  have amplitudes
smaller than 200 $\mu$mag.  A series of equally-spaced peaks seen at 0.1593,
0.3187, 0.4710 and 0.6220\,d$^{-1}$ may correspond to the rotation frequency and its
harmonics, but the amplitudes increase with frequency.  \vskip0.3truecm

{\noindent \bf KIC\,5705575} (*22) -  In the 0-50\,d$^{-1}$ LC periodogram
nearly all the real peaks are located between 16 and 32 \,d$^{-1}$, some of
which are super-Nyquist.  Several low-amplitude peaks can be seen at low
frequencies (5.3314, 1.1367, 1.3427, 2.5202\,d$^{-1}$), but there is no
evidence for the rotation frequency ({\it i.e.}, no harmonic peaks).
Time-delays suggest an orbital period close to 538 days (see $\S3.3$).  The
nine spectra taken at seven epochs clearly show RV variability
($\Delta$RV$\sim$17 km/s) consistent with the photometry and with the
photometrically-derived RV curve.  \vskip0.3truecm

{\noindent  \bf  KIC\,6130500} (*9) -  The pulsation is dominated by several
$p$-mode frequencies between 15 and 20\,d$^{-1}$, in addition to a rich
spectrum of $g$-mode frequencies between 1 and 4\,d$^{-1}$.  There are no
obvious super-Nyquist frequencies, no frequency modulation sidelobes, and no
obvious spacings.   Ground-based observations probably would classify this star
as a hybrid as there are several peaks with amplitudes greater than 1\,mmag.
\vskip0.3truecm

{\noindent  \bf  KIC\,6227118} (*27) -  The dominant peaks, none of which have
amplitudes exceeding  $\sim$300\,$\mu$mag, are all above the LC Nyquist
frequency, and amplitude modulation is visible in the light curve.  All the
major peaks exhibit a similar multiplet structure, caused mainly by aliasing
with the 372.5\,d orbital period of the {\it Kepler} telescope.  The dominant
peak, at $\nu_1 = 33.7160$\,d$^{-1}$, is flanked by the second highest peak at
$\nu_2 = 33.7145$\,d$^{-1}$.  This is the only system in which almost all the
power is beyond the LC Nyquist frequency.  \vskip0.3truecm

{\noindent  \bf KIC\,6445601} (*2) - The periodogram is relatively simple, with
high $p$-mode peaks at 13.07, 14.33 and 13.42\,d$^{-1}$, and several
low-frequency equally-spaced peaks corresponding to the rotation frequency
0.5415 d$^{-1}$ (Neilsen {\it et al.} 2013) and its harmonics at 1.083, 1.624,
2.166\,d$^{-1}$.   No FM sidelobes are seen, but there exists the possibility of
amplitude variability.  \vskip0.3truecm

{\noindent   \bf  KIC\,6520969} (*21) - Two {\it real} super-Nyquist
frequencies, at $\nu_1$=39.195\,d$^{-1}$ and $\nu_2$=54.749\,d$^{-1}$, both of
which are single and well-defined, dominate the 0-60\,d$^{-1}$ periodogram.
The latter is greater than twice the LC Nyquist frequency ({\it i.e.},
$>$48.92\,d$^{-1}$).   The high peaks at 5.812 and 9.74\,d$^{-1}$ (and at
43.126 and 58.68\,d$^{-1}$) are aliases of $\nu_1$ and $\nu_2$, respectively.
Near the $\nu_1$ peak are  less significant single peaks at 39.1520 and
39.0645\,d$^{-1}$.  Other significant super-Nyquist frequencies include 41.44,
and 51.01\,d$^{-1}$.  In the vicinity of 43.13, 54.72, 58.68, 58.72, and
58.81\,d$^{-1}$ one sees frequency quintuplets with separations of
$\sim$0.0027\,d$^{-1}$; the splittings most likely are due to aliasing with the
372.5~d orbital period of the {\it Kepler} telescope.  Below 4\,d$^{-1}$ the
dominant frequencies are at 0.99, 1.53, 1.58, 2.07 and 3.73\,d$^{-1}$ and show
no evidence of rotational harmonics.  \vskip0.3truecm

{\noindent \bf KIC\,6780873} (*5) - The periodogram shows only two prominent
peaks in the $\delta$\,Sct range, at 14.188 and 13.436\,d$^{-1}$, and a very
rich low-frequency spectrum with many dozens of peaks concentrated around
0.20--1.25, 2.25--3.33 and 3.33--4.45\,d$^{-1}$, with amplitudes in the range
300--1625\,ppm.  No obvious frequency spacings can be detected.  Except for the
two $p$-mode peaks this star would be classified as a $\gamma$~Dor star.  In
$\S$2.1.2 the RV variations revealed that this star is a double-lined
spectroscopic binary, consistent with a time-delay analysis of the Q0-Q17 LC
photometry ($\S$3.3) that reveals it to be a close binary system with an
orbital period $P_{\rm orb} = 9.16 \pm0.03$~d.   \vskip0.3truecm

{\noindent  \bf  KIC\,7020707} (*16) - The dominant frequencies are between 15
and 20\,d$^{-1}$.  The three highest peaks, at 15.54, 18.20 and
16.70\,d$^{-1}$, are real ({\it i.e.}, not aliases) and show no evidence of
multiplet structure.  At low frequencies a few peaks are seen (all with
amplitudes less than 150\,$\mu$mag), two of which appear to be  $\nu_{\rm rot}
= 0.43$\,d$^{-1}$ and its harmonic at 0.85\,d$^{-1}$.
\vskip0.3truecm


{\noindent  \bf  KIC\,7174372} (*8) -  The LC periodogram exhibits two sets of
complex $p$-mode pulsations (near 12\,d$^{-1}$ and 16\,d$^{-1}$), a distinct
triplet of high-SNR peaks at 0.25, 0.50 and 0.75\,d$^{-1}$, and many
low-amplitude peaks distributed over a wide range of frequencies.   In the
light curve the 0.25\,d$^{-1}$ peak shows up as broad dips every four days,
which, if not fictitious, may correspond to the rotation period claimed by
Nielsen {\it et al.} (2013), or may be due to 4-d orbital motion in a close
`ellipsoidal' binary (see $\S3$).  The multi-periodic pulsations all occur at
amplitudes below 0.01 mag.  \vskip0.3truecm

{\noindent \bf  KIC\,7300184} -  The periodogram shows an extreme peak at
11.66\,d$^{-1}$, the amplitude of which, $28900$\,$\mu$mag, is exceeded only by
the $57690$\,$\mu$mag  dominant peak for KIC\,11754974.  Features with smaller
amplitudes ($\sim$4300-2500\,$\mu$mag) are  seen at 20.76, 20.72 and
13.71\,d$^{-1}$.   Four other peaks are also seen at 20.11, 20.19, 20.27 and
20.35\,d$^{-1}$, all of which are separated by $\sim$0.08\,d$^{-1}$.  A
binarogram analysis shows the SX~Phe star to be in a wide binary system with an
orbital period of 640$\pm$60 days.  \vskip0.3truecm

{\noindent \bf  KIC\,7301640} (*10) - Most of the power is at frequencies
between 19 and 28\,d$^{-1}$, including  super-Nyquist frequencies 26.51, 26.73,
27.18 and 27.89\,d$^{-1}$.   At low frequencies there are  many scattered
low-amplitude peaks ($<$400\,ppm),  the highest amplitude being at 3.37 and
2.73\,d$^{-1}$.  There are no obvious FM sidelobes, and there is no evidence
for $\nu_{\rm rot}$ harmonics.  \vskip0.3truecm

{\noindent  \bf KIC\,7621759} (*6) -   The highest peak is at 4.67\,d$^{-1}$,
and most of the peaks in the periodogram are of low frequency, with hundreds of
peaks having $\nu < 8$\,d$^{-1}$ (g-modes?).   The high frequencies at 18.52 and
18.04\,d$^{-1}$ are of intermediate amplitude.  There are no obvious equal
spacings and no obvious FM sidelobes.  \vskip0.3truecm

{\noindent \bf  KIC\,7765585} (*28) - Many frequencies are present, with most
of the power above 18\,d$^{-1}$ and no evidence for super-Nyquist frequencies.
A few low-frequency peaks with moderate amplitudes are present (0.5992,
2.6422\,d$^{-1}$) along with many dozens of low-amplitude peaks.  Although $v
\sin i$ is large there is no evidence at short frequencies for rotational
harmonics.   \vskip0.3truecm 

{\noindent \bf KIC\,7819024}  (*19) - The 0-25\,d$^{-1}$ LC periodogram shows
several high frequency peaks, almost all having amplitudes higher than their
super-Nyquist counterparts.  Much of the power is at frequencies below
6\,d$^{-1}$, including significant peaks at 1.41 and 2.84\,d$^{-1}$.
Frequency modulations due to binary motion suggest $P_{\rm orb}\sim$216 days.
\vskip0.3truecm

{\noindent \bf KIC\,8004558} (*1) -  All the highest peaks in the
0-50\,d$^{-1}$ LC periodogram occur at super-Nyquist frequencies.  A variable
beat pattern is seen in the light curve, due mainly to the three strongest
frequencies at 25.521, 24.916 and 26.329\,d$^{-1}$.  At frequencies below
4\,d$^{-1}$ the greatest power occurs at 0.6056\,d$^{-1}$.  Several other
$g$-mode peaks between 0.6 and 3.0\,d$^{-1}$ are seen, with no evidence for
$\nu_{\rm rot}$ harmonics.  This high-velocity star is a time-delay binary
system with an orbital period of $\sim$262 days.  \vskip0.3truecm

{\noindent \bf KIC\,8110941} (*29) - The 0-50\,d$^{-1}$ LC periodogram reveals
two well-resolved triplets, at 16.2400, 16.2943 and 16.1856\,d$^{-1}$, and at
24.9193, 24.9800 and 24.8578\,d$^{-1}$.   The latter triplet is located beyond
the LC Nyquist frequency and so alias peaks appear at 24.0197, 23.9590 and
24.0812\,d$^{-1}$.  One also sees a distinct set of many closely-spaced peaks
between 12.1--13.9\,d$^{-1}$ and some scattered peaks at high frequencies. 
Below 3\,d$^{-1}$ one sees two sets of peaks, at $\sim$0.66\,d$^{-1}$ and near 
1.36\,d$^{-1}$, where each set contains many closely-spaced
unresolved peaks within a band of about 0.06\,d$^{-1}$.
These correspond to the $\nu_{\rm rot} = 0.6456$\,d$^{-1}$ (and its harmonic) given
in the Nielsen {\it et al.} (2013) catalogue.   A possible third harmonic at
2.07\,d$^{-1}$ is also weakly visible. 
\vskip0.3truecm

{\noindent \bf KIC\,8196006} (*30) - Almost all the pulsation power is at
15.51\,d$^{-1}$ with no evidence for $g$-mode frequencies.  At low frequencies
four low-amplitude peaks dominate, three of which (at 2.72, 4.08 and
5.44\,d$^{-1}$) are harmonics of the rotation frequency 1.360\,d$^{-1}$ (Balona
2013).    \vskip0.3truecm

{\noindent \bf KIC\,8330910} (*3) - The six highest peaks in the 0-50\,d$^{-1}$
LC periodogram all occur at frequencies below  the LC Nyquist frequency.  No
obvious FM sidelobes are seen, and possibly the amplitudes are variable.  The
spacing and amplitudes of the peaks at 2.2725, 4.551 and 6.825\,d$^{-1}$
suggest that the first of these is the rotation frequency and the other two are
its harmonics.  Such a high $\nu_{\rm rot}$ is consistent with the high
$<$$v\sin i$$>$\,=224$\pm$3 km/s.  \vskip0.3truecm

{\bf \noindent KIC\,9244992} (*7) -  The 0-50\,d$^{-1}$ frequency spectrum is
dominated by two $p$-mode peaks at 12.339 and 12.920\,d$^{-1}$, which, after
pre-whitening, reveal a ``plethora'' of $p$-mode peaks between 11 and
17\,d$^{-1}$,  and a complex high-amplitude group of $g$-mode peaks between 0.9
and 1.3\,d$^{-1}$.  A detailed asteroseismic analysis of the pulsations, and an
asteroseismic application of their use for determining physical
characteristics, including the rotation profile, was made by Saio {\it et al.}
(2015). Their best-fit model was:  $\mathscr{M}$=1.45\,$\mathscr{M}_{\odot}$;
$T_{\rm eff}=6622$\,K; $L$=7.14\,$L_{\odot}$; $R =2.03 R_{\odot}$; $\log
g$=3.982; age=1.9 Gyr;  and (X,Y,Z)=(0.724, 0.266, 0.010), corresponding to
[Fe/H] = --0.15 dex.  No peak in the low-frequency periodogram at the
well-established asteroseismic rotation frequency of 0.015\,d$^{-1}$ is seen,
perhaps suggesting an absence of spots or other surface features.
\vskip0.3truecm

{\bf \noindent KIC\,9267042} (*12)  - The SC periodogram (middle panel of
Fig.\,15) reveals that all of the significant real $p$-mode frequencies are
higher than 19\,d$^{-1}$ and that the highest peak is super-Nyquist at
$\nu_1=24.664$\,d$^{-1}$.  Thus, in the LC periodogram the sub-Nyquist
high-amplitude peak at 24.26\,d$^{-1}$ is an alias.   The next highest real
peaks occur at 22.449 and 20.401\,d$^{-1}$ and are mainly responsible for the
complex beat pattern seen in the {\it Kepler} light curve.   The peak
amplitudes may be variable.   Time-delay analyses suggest that the pulsator
{\it may} be in a binary (triple?) system with $P_{\rm orb}$ longer than the
four year duration of the {\it Kepler} photometry (see $\S3.3$).
\vskip0.3truecm

\begin{figure} 
\centering 
\begin{overpic}[width=7.5cm]{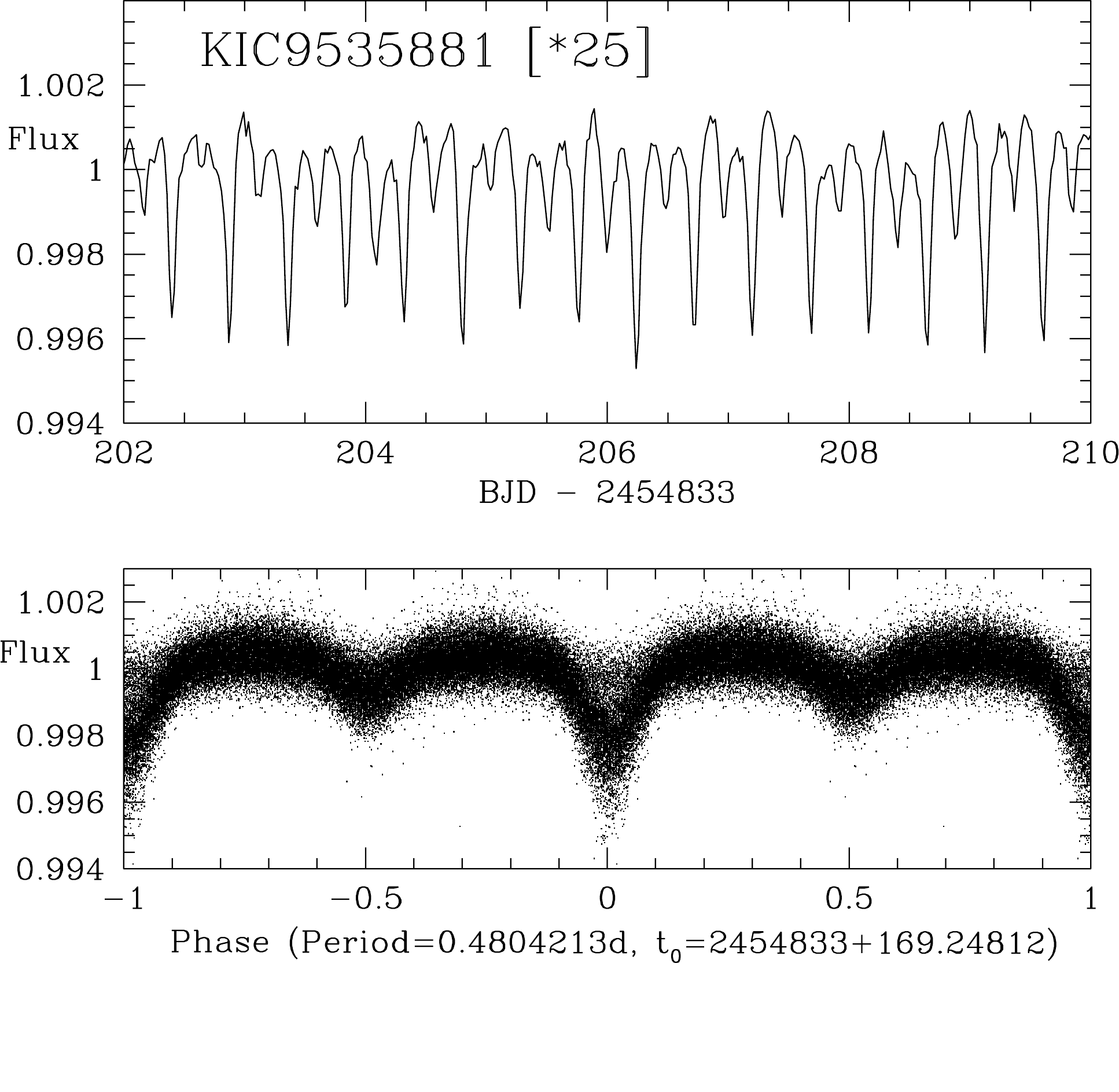} \end{overpic}
\vskip-0.6truecm 
\caption{Light curves  for  the W\,UMa-type non-SX\,Phe binary KIC\,9535881
[*25].  The top panel shows an eight-day portion of the light curve, and the
bottom panel shows all 63416 LC data points phased with the 0.48\,d orbital
period  given in the Villanova EB catalog. } \label{BN25} \end{figure}

{\bf \noindent KIC\,9535881} [*25, KOI\,4228] - This W~UMa-type interacting
binary system is one of the two mistakenly classified SX~Phe candidates.
Initially it was flagged as a {\it Kepler} Object of Interest (KOI) but it is now
considered to be a `false positive'.  The Villanova EB catalog lists it as
having an orbital period of 0.4804213\,d and a mid-time for the primary eclipse
of BJD$_0$=2455002.248121.   In the periodogram this period corresponds to the
peaks at 2.0815\,d$^{-1}$ and 4.1630\,d$^{-1}$, with additional harmonics of
these at 6.244, 8.326, 10.407, 12.489, 14.571 and 16.652\,d$^{-1}$.  The
primary and secondary minima  both exhibit variable depths, shown in {\bf
Figure\,16}.   At frequencies below 2\,d$^{-1}$ several low-amplitude peaks are
seen, possibly indicating that at least one component is a $\gamma$~Dor
variable, or that $\nu_{\rm rot}\sim0.49$\,d$^{-1}$ ({\it i.e.}, $P_{\rm
rot}$$\sim$2.1\,d).  \vskip0.3truecm

{\bf \noindent KIC\,9966976} (*31) - The pulsations are of very low amplitude
($<$400\,$\mu$mag), with  most peaks occurring at frequencies higher than
12\,d$^{-1}$, including real super-Nyquist frequencies at 33.704 and
30.450\,d$^{-1}$, and a fairly sparse low-frequency spectrum.  The strongest
low-$\nu$ peak occurs at 0.1146\,d$^{-1}$ (with harmonics at 0.228 and
0.346\,d$^{-1}$), identified as $\nu_{\rm rot}$ by Balona (2013);  all three
peaks exhibit an underlying broad Gaussian feature somewhat similar to those
seen in Figure\,6 of Balona (2013).  Additional low-frequency peaks are seen at
1.84 and 2.45\,d$^{-1}$ ($g$-modes?).  R\o mer time-delay analyses (see $\S3.3$)
suggest that the pulsator is in a binary system with an orbital period longer
than four years.  \vskip0.3truecm

{\bf \noindent  KIC\,10989032} (*32, KOI\,7397) - The {\it Kepler} light curve
shows both eclipses and pulsations.  These appear as two distinct sets of
frequencies in the 0-25\,d$^{-1}$ long cadence periodogram:  (1) two high peaks
associated with the 2.3\,day orbital motion, one peak at $\nu_{\rm
orb}$=0.43375\,d$^{-1}$ and another at twice this, accompanied by a series of
equally-spaced harmonic peaks between 1.73 and $\sim$12\,d$^{-1}$ -- the
harmonics that are even multiples of $\nu_{\rm orb}$ have much larger
amplitudes than those at odd multiples of $\nu_{\rm orb}$; and (2) a group of
intermediate-amplitude pulsation peaks at frequencies between 12 and
20\,d$^{-1}$, almost all of which are aliases of real $p$-mode pulsations at
frequencies between 29 and 37 d$^{-1}$.  As a  KOI  it is
characterised as a ``false positive with a significant secondary event and a
centroid offset''.  The spectral type is based on an excellent match to the
A5\,V standard star HD\,23194.  \vskip0.3truecm

\vskip0.3truecm {\noindent \bf KIC\,11649497} (*11) -  Almost all of the
significant real frequencies are above 15 d$^{-1}$, with  many in the super-Nyquist
region.   All the major peaks show substructure due to aliasing caused by
the orbital periods of {\it Kepler} and the Earth.  `Low-amplitude eclipses'
were noted by BN12.  The analysis is complicated by location on Module 3 and by
amplitude modulation.  \vskip0.3truecm

{\noindent \bf KIC\,11754974} (*13) - Murphy {\it et al.} (2013b) discovered
(from a careful asteroseismic investigation using the LC:Q0-Q13 and SC:Q6-Q7
data) that the  SX\,Phe pulsator is in a 343-day non-eclipsing binary system.
An independent pulsation and time-delay analysis using all the LC:Q0-17 data
was later conducted by Balona (2014b).  These studies identified five
independent $p$-mode frequencies between 16 and 23\,d$^{-1}$, the most powerful
of which is at 16.34475\,d$^{-1}$.  Also seen were many combination
frequencies, including some very distinctive low-amplitude quintuplets.  The
strong peak at 0.09524\,d$^{-1}$ seen in the binarogram of Balona (2014) is
confirmed as fictitious; neither it nor the weaker peak at  16.24951\,d$^{-1}$
(=16.34475--0.09524)  is seen in the SC periodogram (see Fig.\,15).   No peaks
are seen in the low-frequency LC or SC periodograms at the asteroseismic
rotation frequency of 0.383\,d$^{-1}$, perhaps suggesting the absence of spots
or other surface features.  \vskip0.3truecm

\begin{figure} \begin{center}
\begin{overpic}[width=8.0cm]{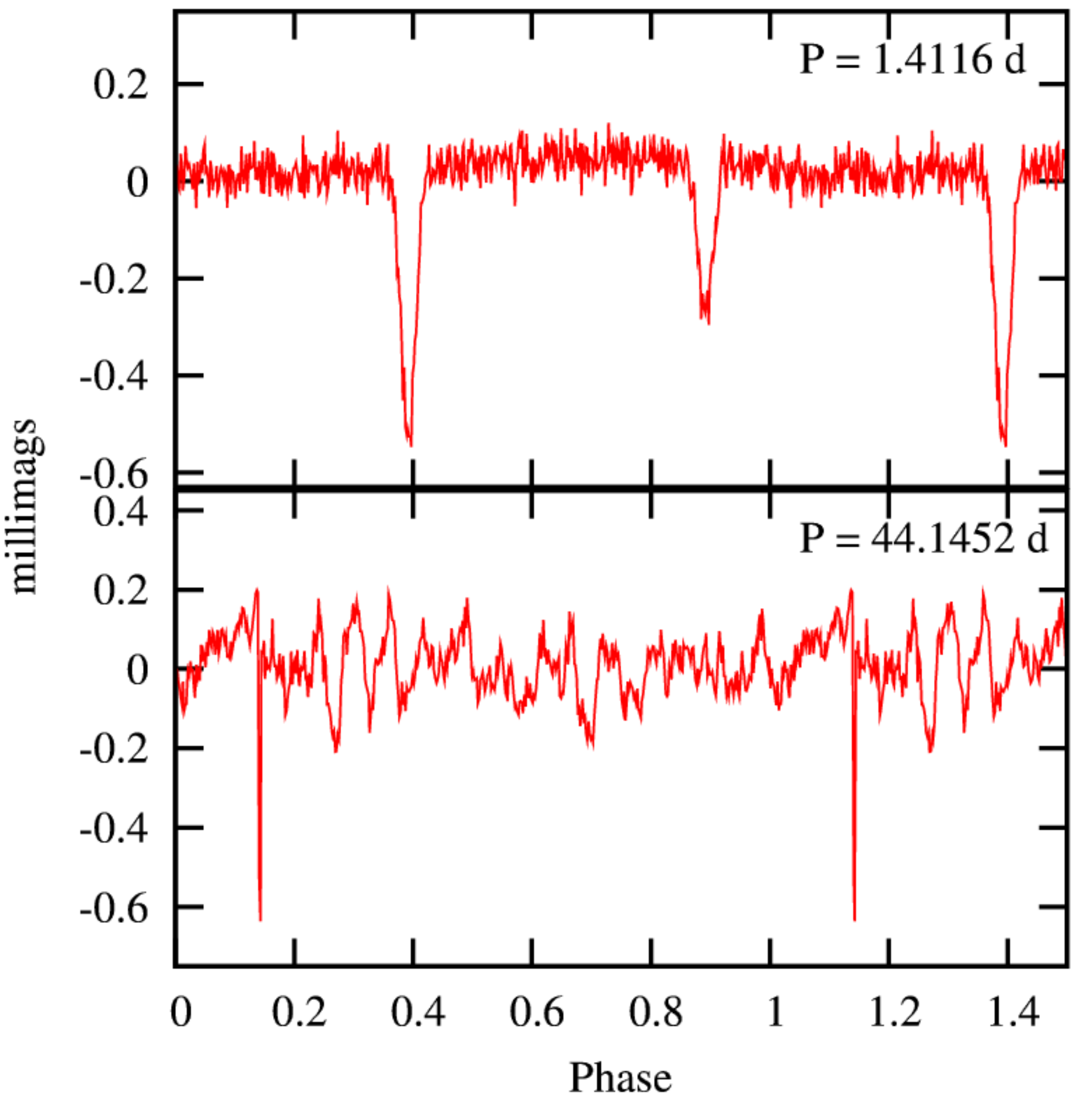}  \put(45,102){KIC\,12643589} \end{overpic}
\end{center}
\caption{Light curves  for the eclipsing binary KIC\,12643589 [*17], 
based on the {\it Kepler} Q1-17 LC photometry, and phased
with the orbital period 1.4116~d (upper panel) and with the 44.1452\,d transit
period (lower panel).   } \label{K12643589_LC_transit} \end{figure}

{\noindent \bf KIC\,12643589} [*17, KOI\,376] - This close binary (along with
KIC\,9535881) was mistakenly classified as an SX~Phe star.   The {\it Kepler}
light curve plotted in {\bf Figure\,17} shows eclipses and transits that agree
with the Villanova EB catalog orbital periods of 1.4116278~d and 44.1471769~d.
As a KOI it was identified as having two `false positive' planet
identifications.  The well-defined harmonic series seen in Fig.\,14 is due to
the binarity; the harmonics with higher peaks are seen at even multiples of the
orbital frequency, $\nu_{\rm orb}$=0.70840\,d$^{-1}$, and those with lower
peaks are seen at odd multiples of $\nu_{\rm orb}$.  A secondary series of
low-amplitude harmonics, equally spaced by an amount 0.02265\,d$^{-1}$ and
caused by the transits, is seen in the periodogram out to $\sim$6\,d$^{-1}$.
Two complex multi-peaked gaussians of unknown origin are seen centered on 0.13
and 0.33\,d$^{-1}$.  No evidence for pulsations is seen.  \vskip0.3truecm


{\noindent \bf KIC\,12688835} (*18) - The periodogram is dominated by two sets
of complex, relatively low-amplitude ($\sim$150-250 $\mu$mag) peaks, at 1.67
and 1.78\,d$^{-1}$, with possible harmonics (also complex) at 3.35 and
3.56\,d$^{-1}$, and at 5.05 and 5.35\,d$^{-1}$.   If the harmonics are real and
due to rotation then $P_{\rm rot} \sim 0.60$\,d (Balona 2013).  Differential
rotation cannot be excluded.  Concerning the pulsations, the low-frequency
spectrum does not resemble that of a typical $\gamma$~Dor star.  Several of the
strongest $p$-mode frequencies are in the super-Nyquist region.

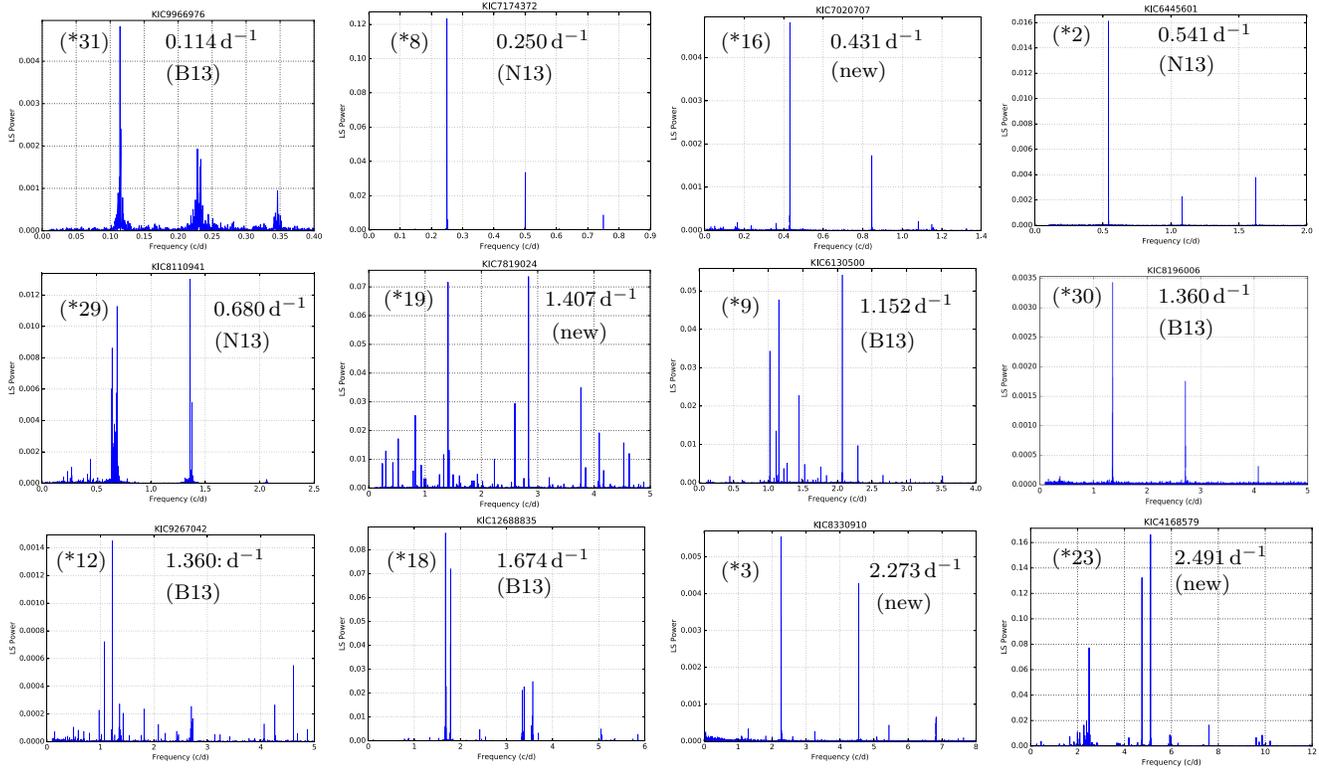
\begin{figure*} \centering 
\begin{overpic}[width=4.3cm]{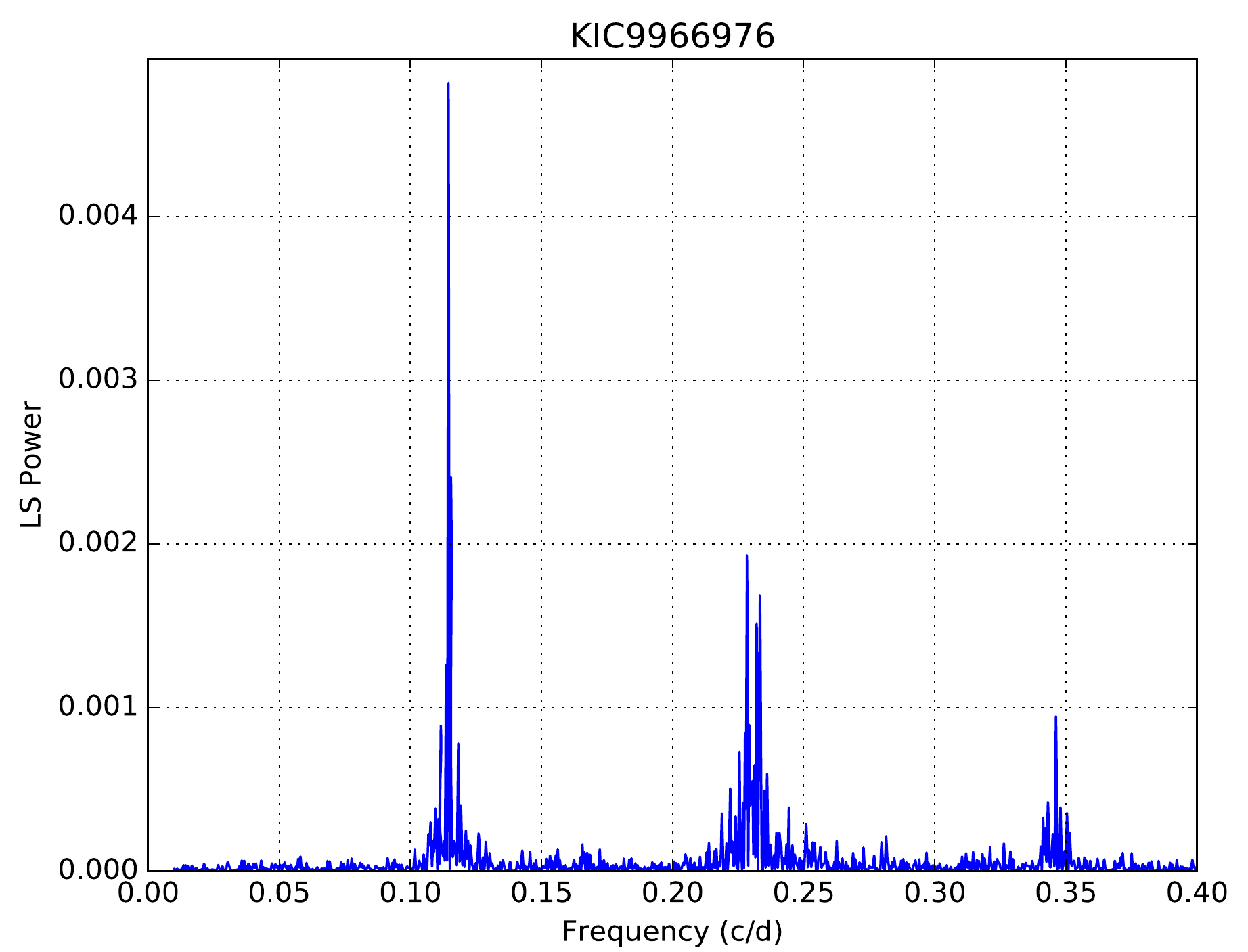}   \put(17,63){(*31)}   \put(50,63){0.114\,d$^{-1}$ }       \put(50,53){(B13)}  \end{overpic}
\begin{overpic}[width=4.3cm]{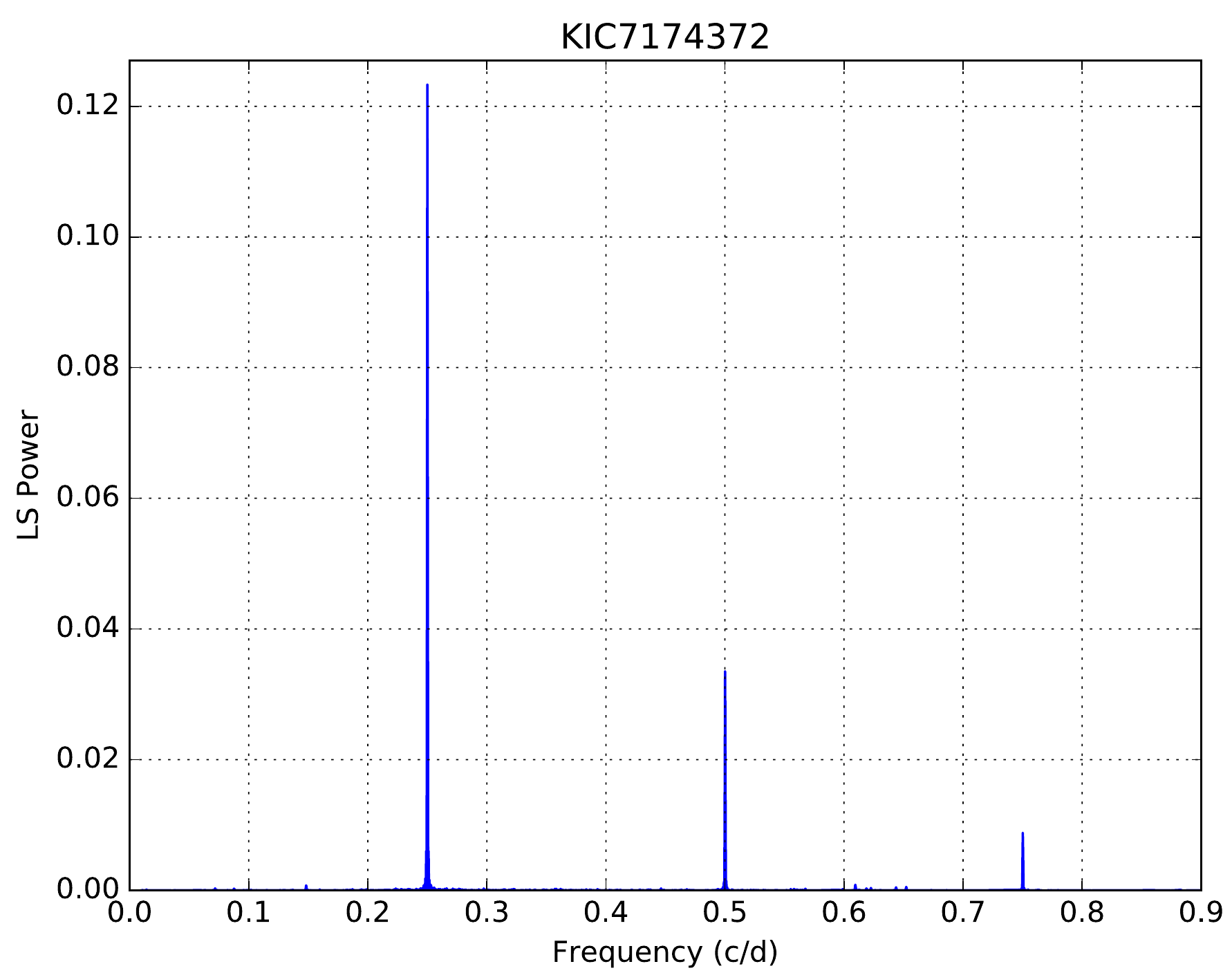}   \put(17,63){(*8)}    \put(50,63){0.250\,d$^{-1}$}        \put(50,53){(N13)}    \end{overpic}
\begin{overpic}[width=4.3cm]{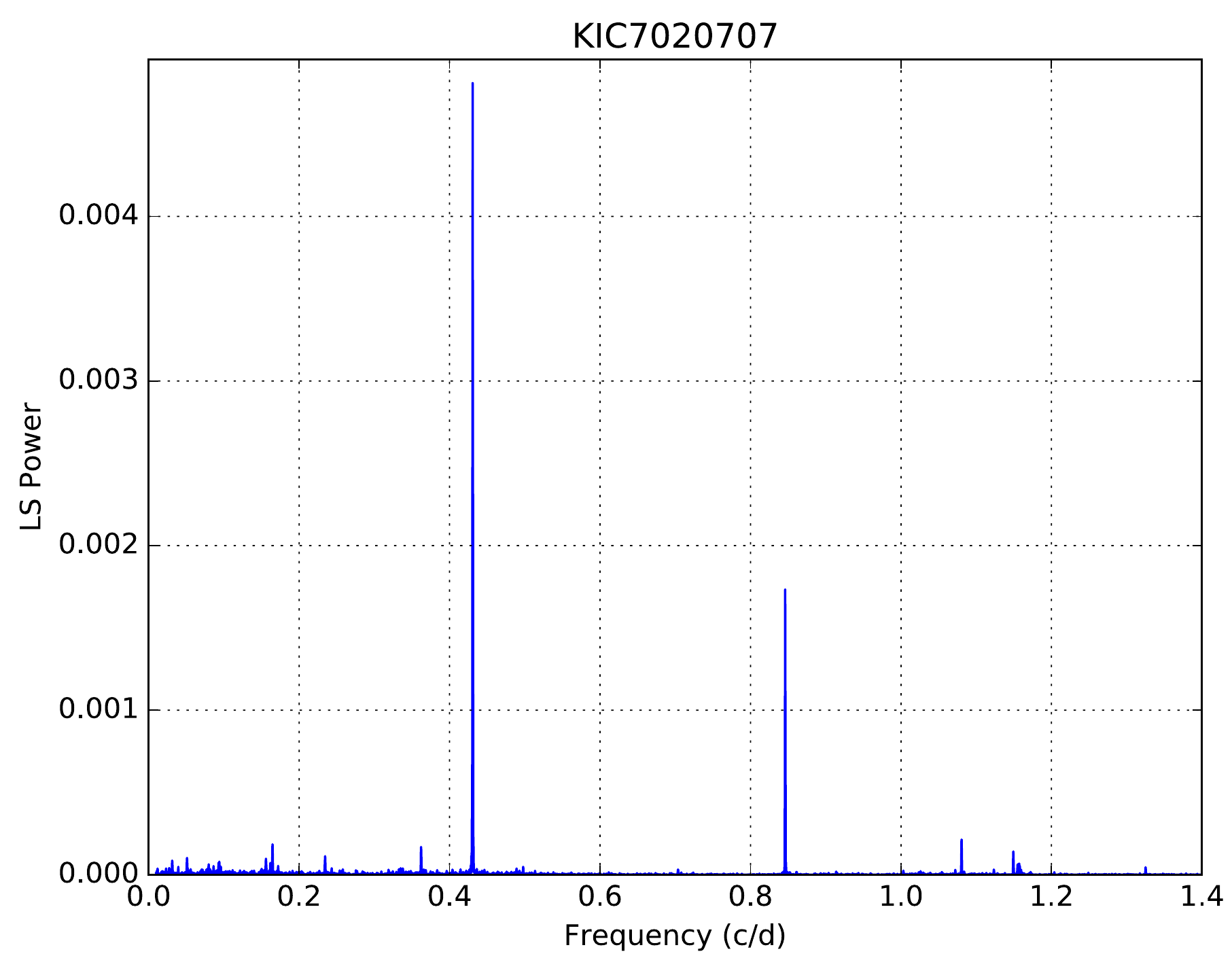}   \put(17,63){(*16)}   \put(51,63){0.431\,d$^{-1}$}        \put(51,54){(new)}   \end{overpic}
\begin{overpic}[width=4.3cm]{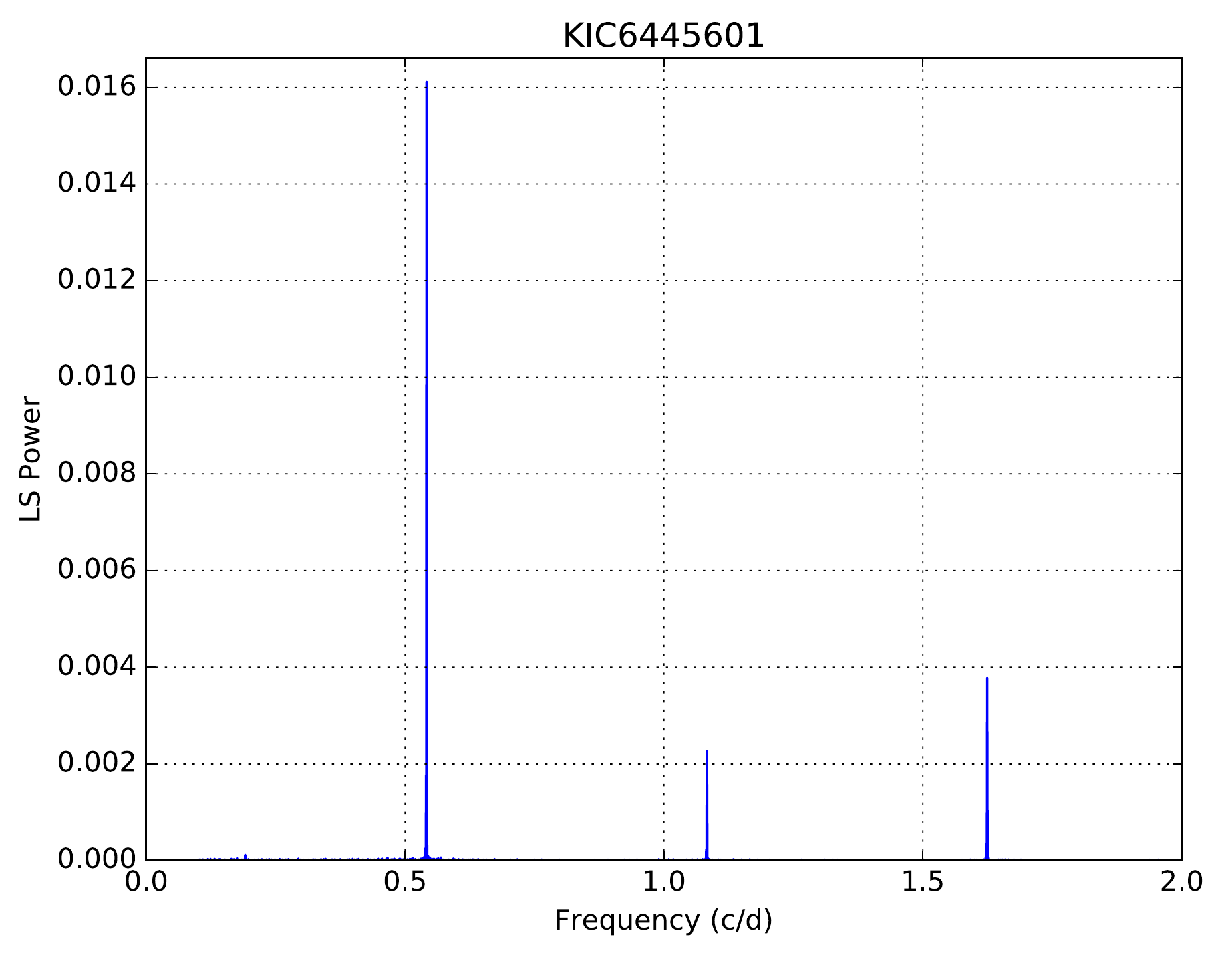}   \put(17,65){(*2)}    \put(50,65){0.541\,d$^{-1}$}        \put(50,56){(N13)}  \end{overpic}
\begin{overpic}[width=4.3cm]{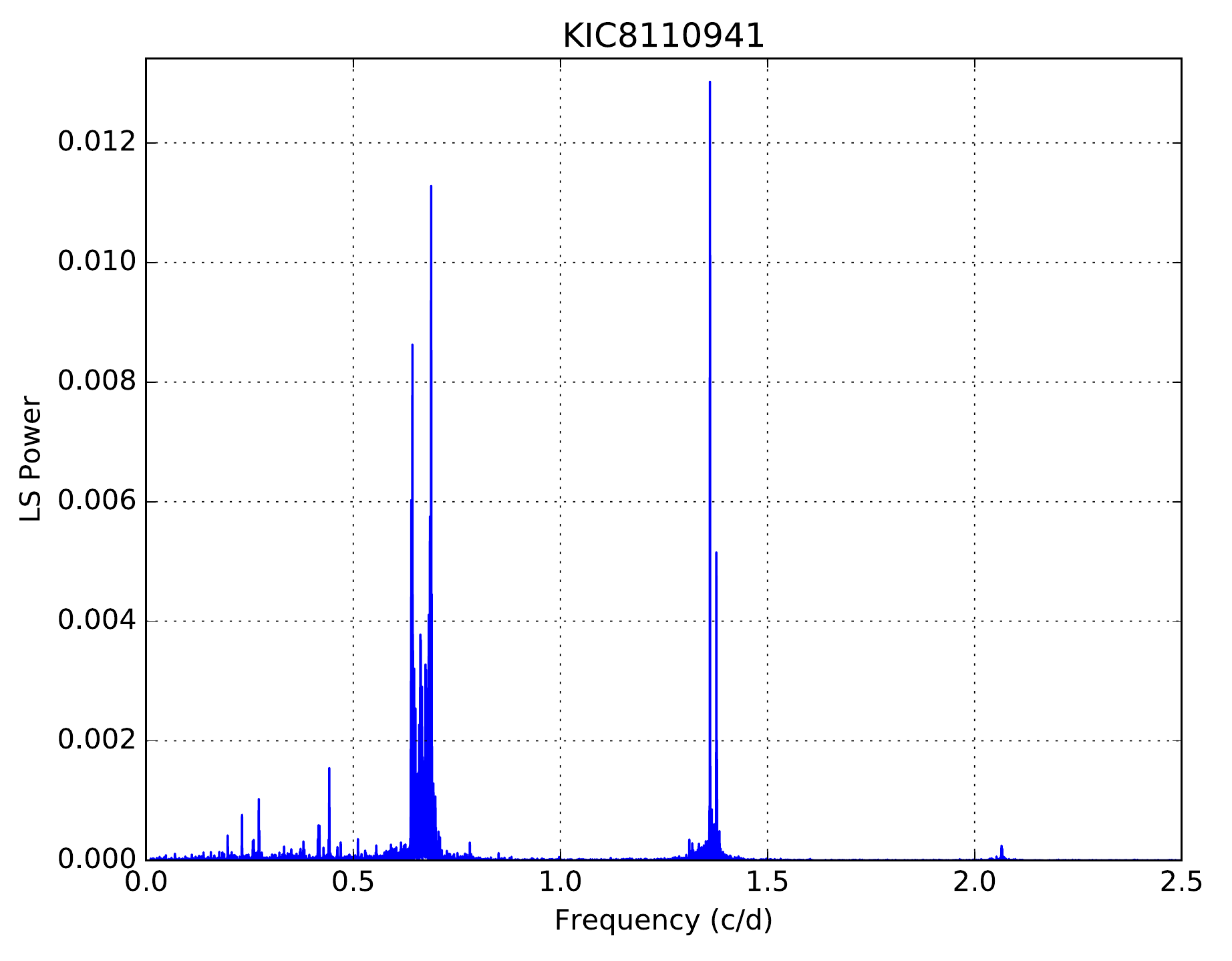}   \put(17,60){(*29)}   \put(65,60){0.680\,d$^{-1}$}        \put(65,50){(N13)}   \end{overpic}
\begin{overpic}[width=4.3cm]{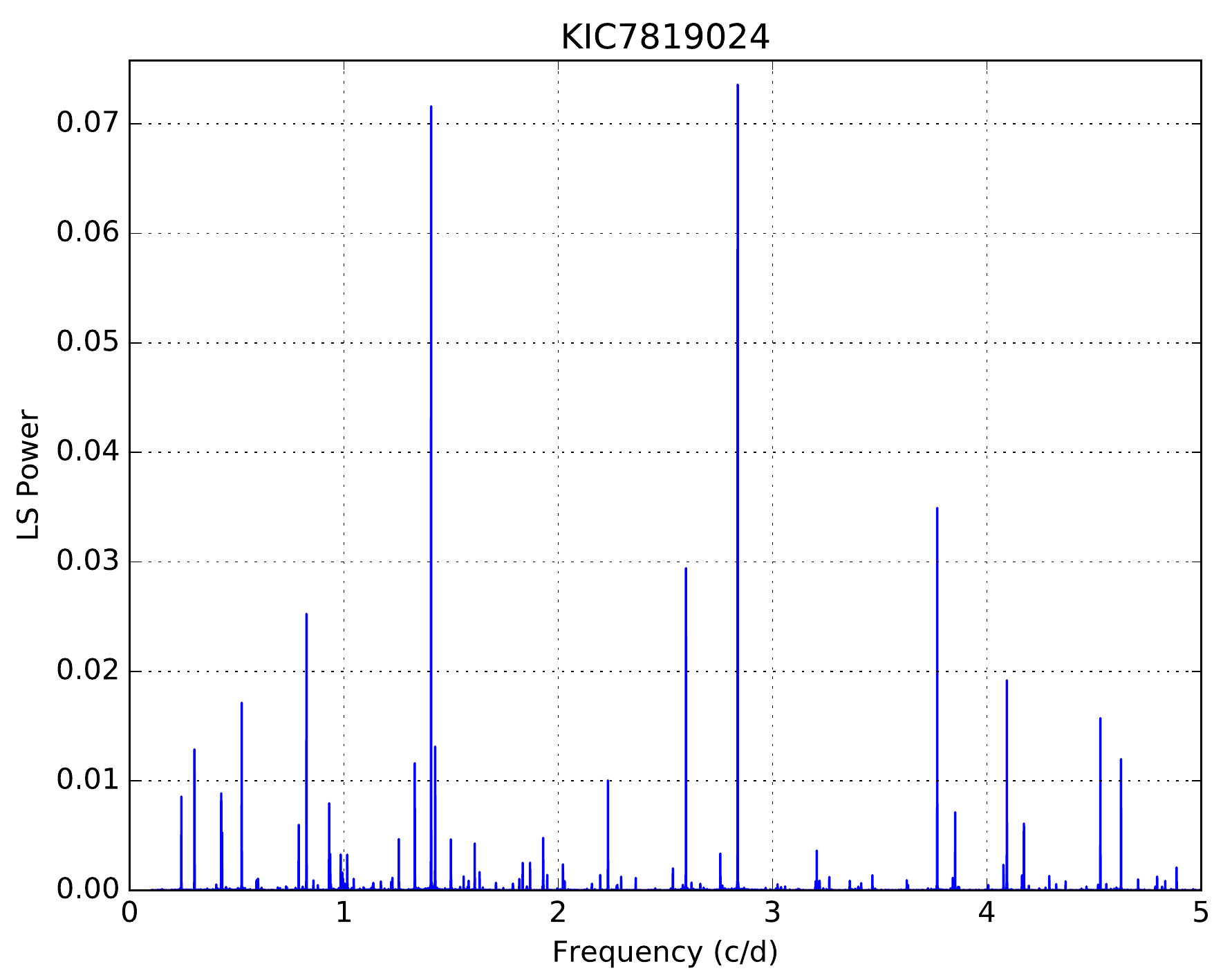}   \put(15,63){(*19)}   \put(65,63){1.407\,d$^{-1}$}  \put(67,53){(new)}  \end{overpic}
\begin{overpic}[width=4.3cm]{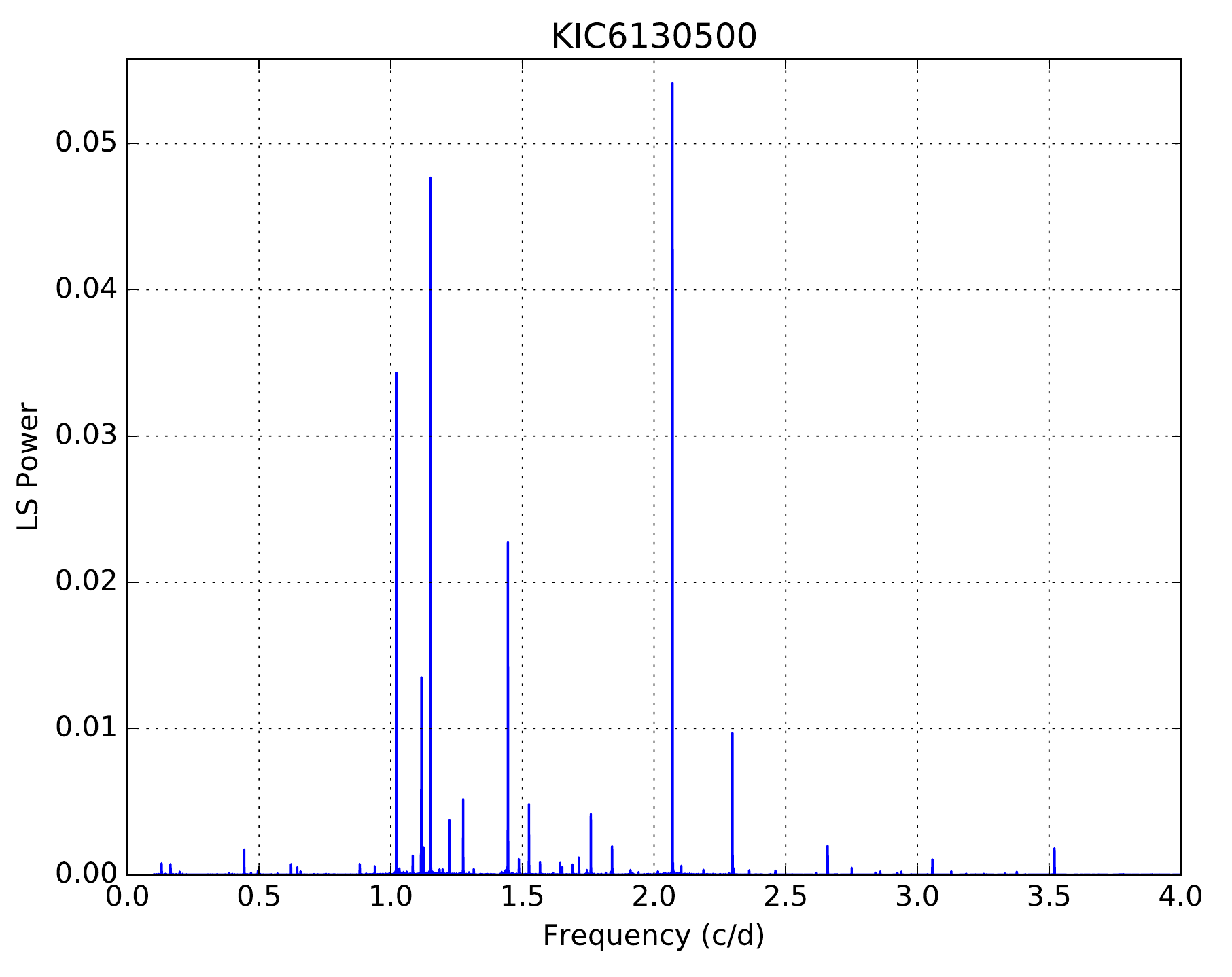}   \put(17,61){(*9)}    \put(60,61){1.152\,d$^{-1}$ }       \put(60,50){(B13)}  \end{overpic}
\begin{overpic}[width=4.3cm]{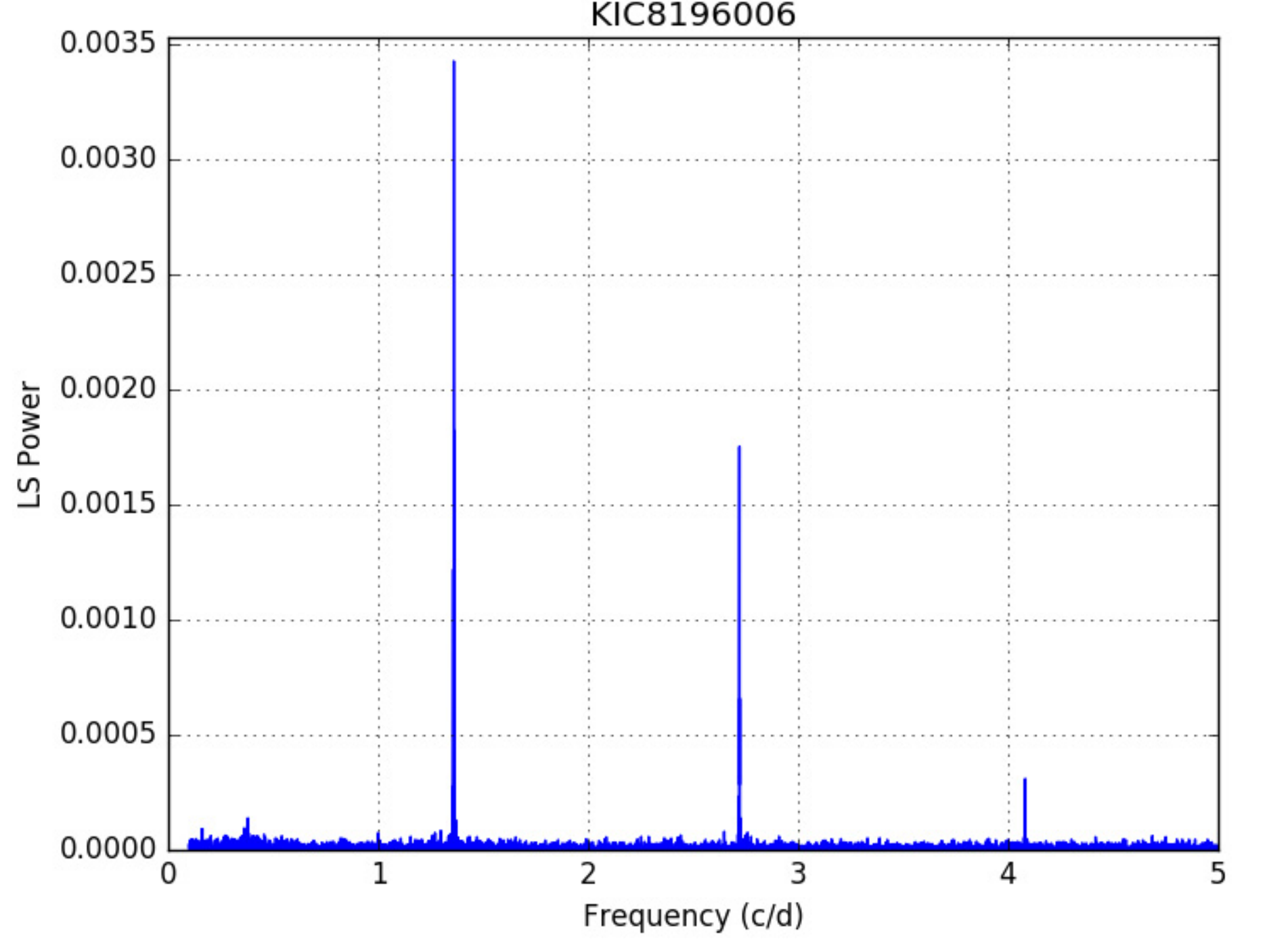}   \put(17,64){(*30)}   \put(50,64){1.360\,d$^{-1}$}        \put(50,54){(B13)} \end{overpic}
\begin{overpic}[width=4.3cm]{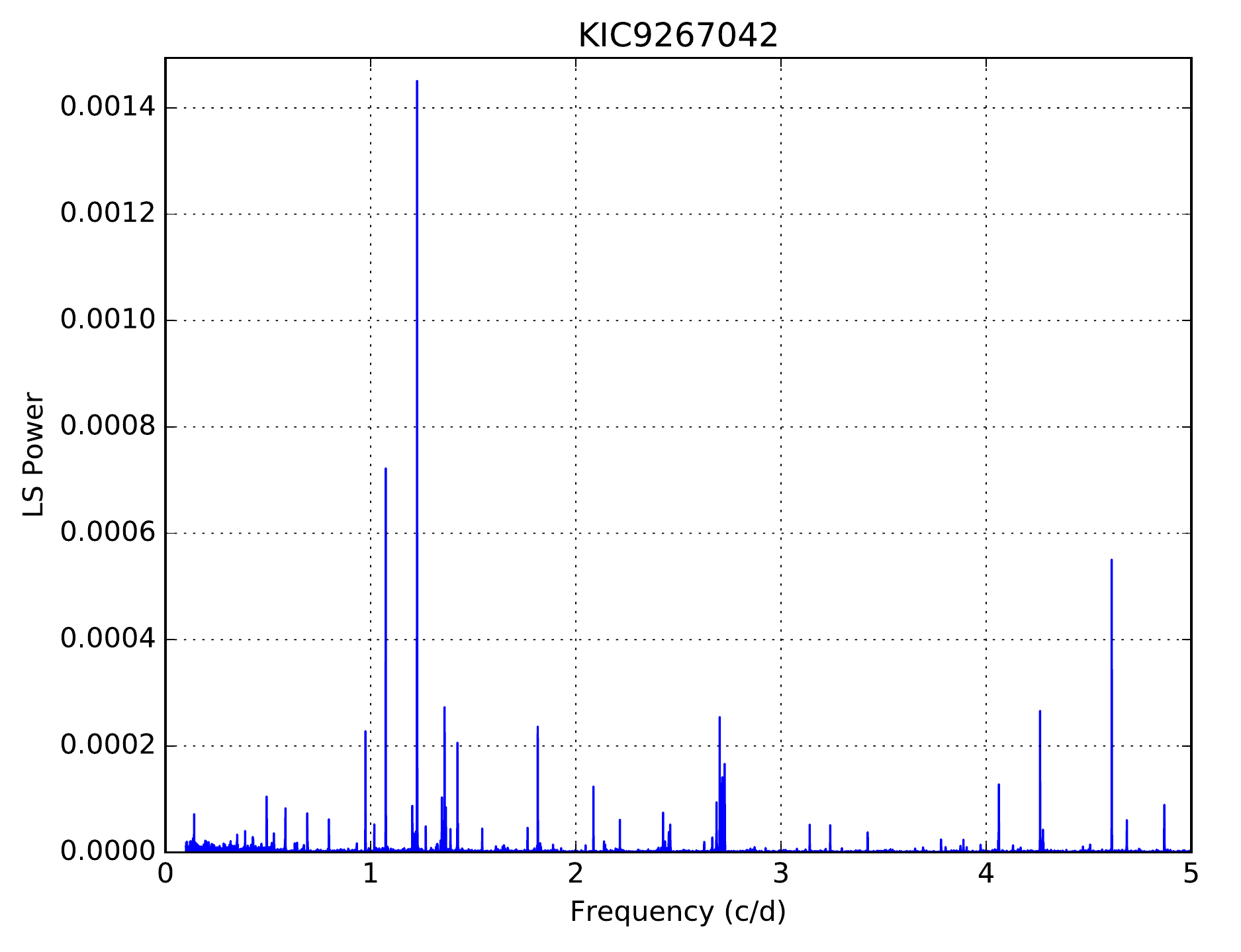}\put(16,62){(*12)}   \put(50,62){1.360:\,d$^{-1}$}        \put(50,52){(B13)}  \end{overpic}
\begin{overpic}[width=4.3cm]{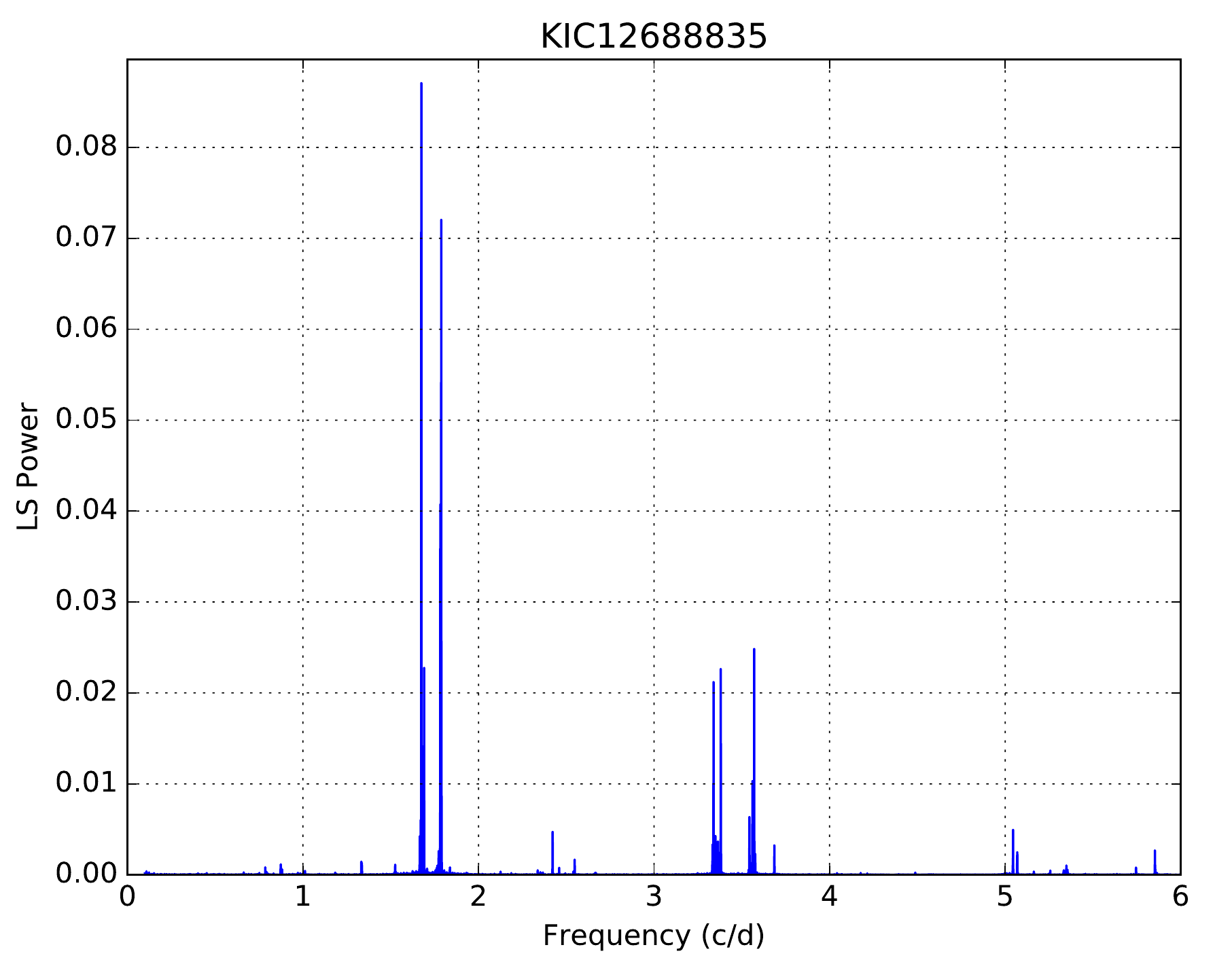}  \put(16,62){(*18)}   \put(50,62){1.674\,d$^{-1}$}        \put(50,54){(B13)}  \end{overpic}
\begin{overpic}[width=4.3cm]{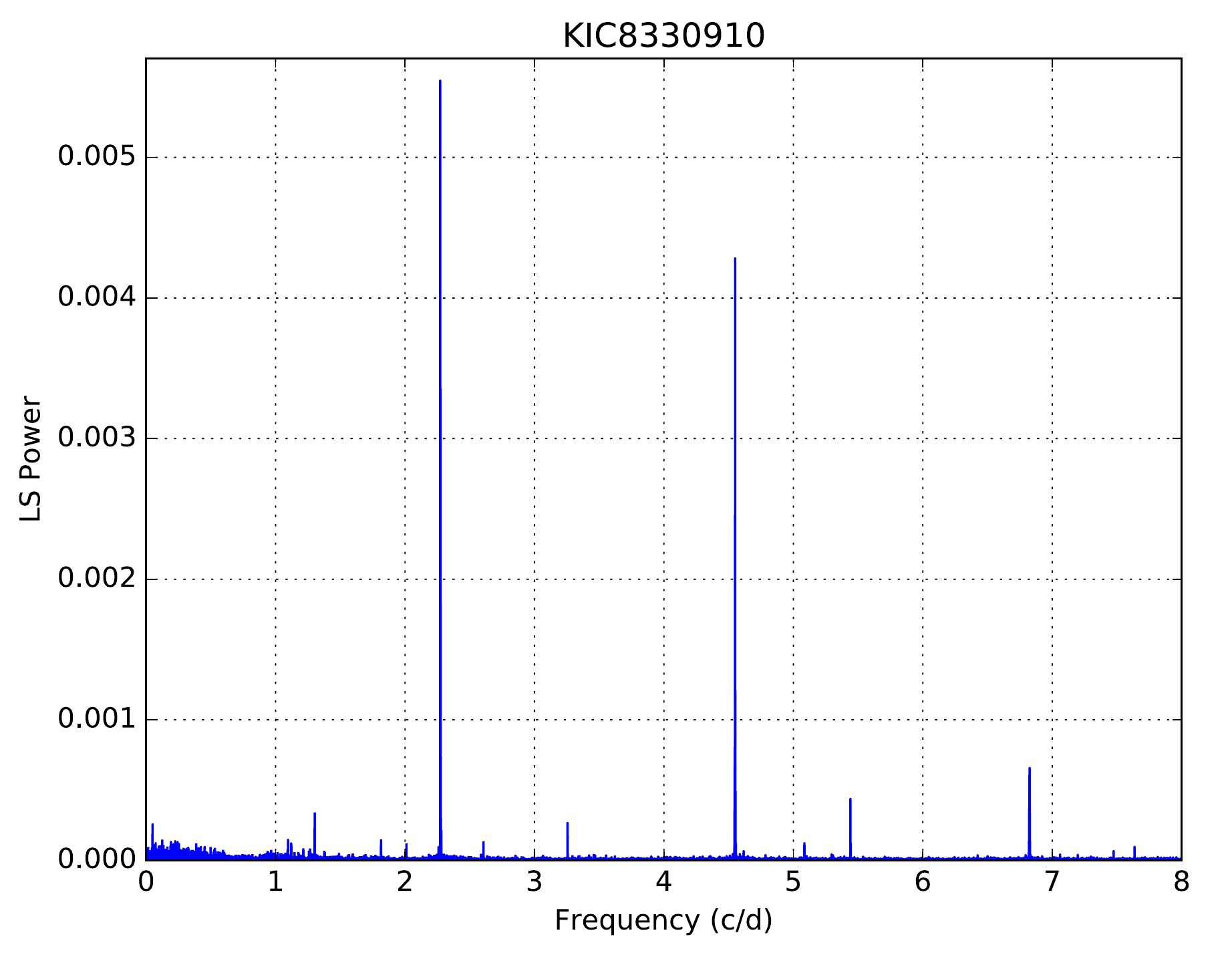}  \put(17,59){(*3)}    \put(63,59){2.273\,d$^{-1}$}        \put(65,49){(new)}  \end{overpic}
\begin{overpic}[width=4.3cm]{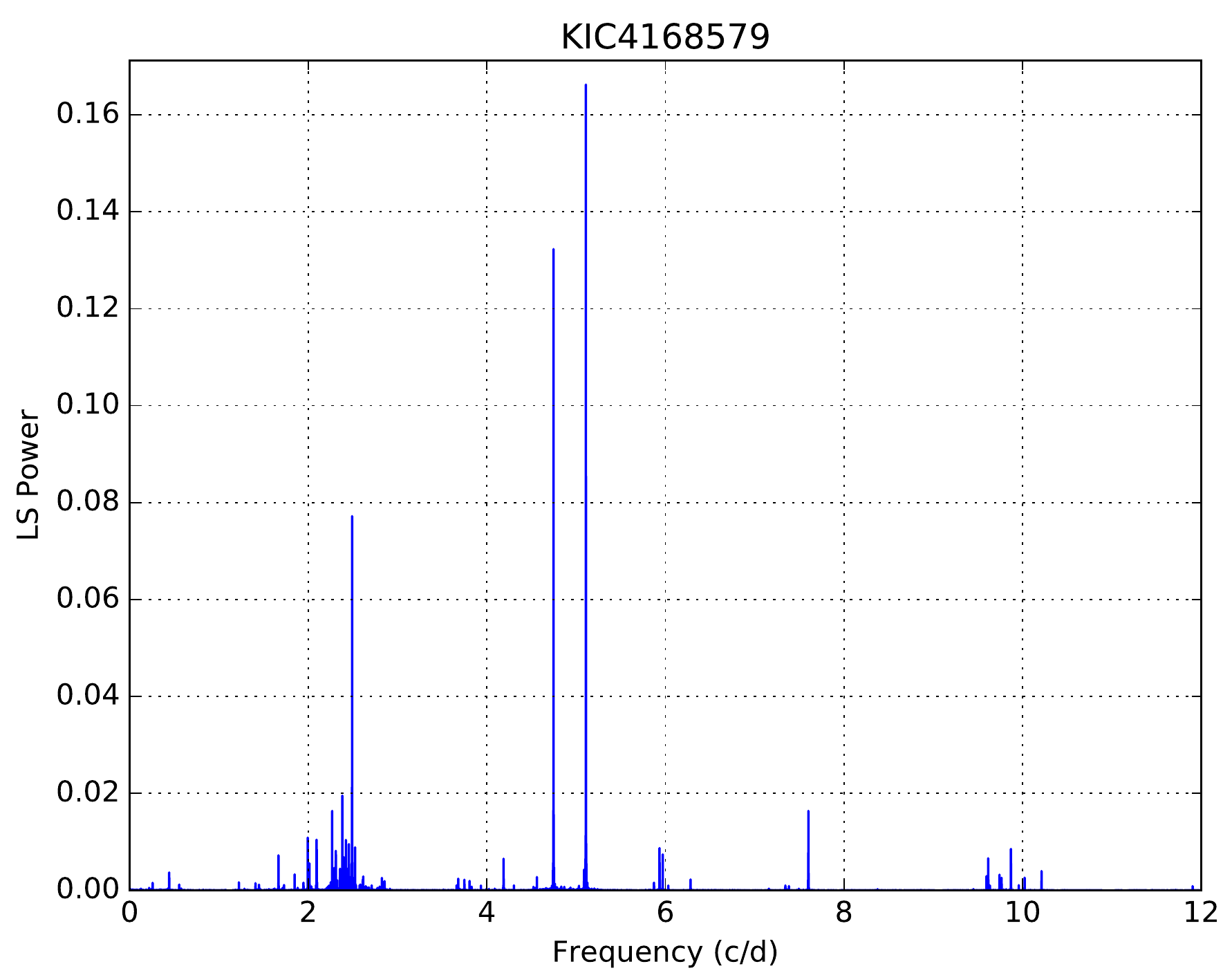}   \put(17,63){(*23)}   \put(55,63){2.491\,d$^{-1}$}        \put(55,55){(new)} \end{overpic}
\caption{Peaks and possible harmonics seen at low frequencies in {\it Kepler} Q0-Q17 LC periodograms. 
KIC number (on top),  CFHT star number (in parentheses),  rotation frequency, $\nu_{\rm rot}$ (rotations
per day), and source of the rotation frequency (see Table\,8), are noted in
each panel.  } \label{fig16} \end{figure*}

\subsection{Rotation}

The {\it Kepler} data have proven to be useful for establishing surface and
interior rotation periods for many stars across the HR diagram (see review by
Aerts 2015), including the two stars in our sample that were subjected to
detailed asteroseismic investigations.  For KIC\,11754974\,(*13), the
inclination angle of the rotation axis (= $47^{+7}_{-15}$\,deg), the equatorial
rotation velocity (= 34.18\,km/s), and the radius (= 1.764\,$R_{\odot}$)  were
determined by Murphy {\it et al.} (2013b).  These values  imply  $\nu_{\rm rot}
= 0.383$\,d$^{-1}$ and $v \sin i = 25.0^{+2.7}_{-6.9}$\,km/s,  the latter being
in close agreement with the mean projected rotation velocity measured from the
spectra, 28.8$\pm$1.7 km/s.  For KIC\,9244992\,(*7),   the core and envelope
rotation rates (both assumed to be constant) are available from  Saio {\it et
al.} (2015), who deduced from rotational splitting of its pulsation
frequencies, a rotation period of 66.2$\pm$0.6\,d  ({\it i.e.}, $\nu_{\rm
rot}=0.015$\,d$^{-1}$) at its surface and 63.9$\pm$0.2\,d at its core;  such
slow rotation corresponds to an equatorial rotation velocity of only 1.5 km/s,
which is  consistent with the measured  $v\sin i < 7$\,km/s.  Owing to the
importance of rotation further such detailed investigations ought to be carried
out.  


Surface rotation frequencies have also been derived by examining low frequency
periodograms.  If a star exhibits light variation due to the presence of a
surface spot or other co-rotating surface feature, estimation  of the rotation
period may be possible.   Light variations due to surface features are unlikely
to be sinusoidal and therefore harmonics might be expected in the low-frequency
periodogram.  Thus, detection of a low-frequency peak and its harmonic(s) might
reasonably   be attributed to a  revolving surface feature, in which case the
rotation period of the star can be deduced (see Balona 2011, and Balona \&
Dziembowski 2011).  By identifying low-frequency peaks and their associated
harmonics in {\it Kepler} periodograms, Balona (2013) derived rotation periods
for 875 {\it Kepler} A-type stars, six of which are among the stars being
considered here.  Rotation periods for four other BN12 stars were similarly
derived by Nielsen {\it et al.} (2013) as part of their program to measure
rotation periods for 12000 {\it Kepler}-field main-sequence
stars\footnote{Nielsen {\it et al.} (2013) assume that their sample stars are
Sun-like with convective envelopes and spots due to dynamo-driven magnetic
fields, a situation which may not apply to the radiative envelopes of SX~Phe
stars.}.  Because the $\nu_{\rm rot}$ range investigated by Nielsen {\it et
al.} was limited to frequencies between 0.03 and 1.0\,d$^{-1}$ none of the
stars in their catalog has a rotation period longer than 33 days or shorter
than one day.  Of course other non-rotational interpretations of low-frequency
peaks are possible, including that some or all of the low-frequency peaks have
a pulsation origin and are combination frequencies (e.g., Kurtz {\it et al.}
2015), an interpretation favoured by SJM.

\begin{table} \label{rotf} \caption{Rotation rates for the program SX~Phe
stars.  Columns (2)-(4) contain the rotational frequency, $\nu_{\rm rot}$,  the
rotational period, $P_{\rm rot}$, and a reference label, where B13, N13, M13 and S15
represent Balona (2013), Nielsen {\it et al.} (2013), Murphy {\it et al.}
(2013) and Saio {\it et al.} (2015), respectively.  The last column contains the
spectroscopically-measured mean projected rotational velocity from Table\,3.
{\bf Boldface} numbers identify binaries (see $\S3.3$), and values followed by
colons are considered uncertain. }

\centering
\begin{tabular}{lllcc}
\hline
\multicolumn{1}{c}{ KIC  }   & \multicolumn{1}{c}{$\nu_{\rm rot}$} & \multicolumn{1}{c}{ $P_{\rm rot}$}  &  Ref.  &  $<$$v \sin i$$>$ \\
\multicolumn{1}{c}{ (CFHT) } & \multicolumn{1}{c}{[$d^{-1}$]}      & \multicolumn{1}{c}{ [$d$] }   &    &   [km\,s$^{-1}$] \\
\multicolumn{1}{c}{(1)} & \multicolumn{1}{c}{(2)} & \multicolumn{1}{c}{(3)} & (4) & (5)    \\
\hline
9244992\,(*7)         &  0.015  & 66.2  & S15   &  $<$6.7$\pm$0.3 \\
{\bf 9966976}\,(*31)  &  0.114  & 8.77  & B13   &   123$\pm$1 \\
{\bf 7174372}\,(*8)   &  0.2501 & 3.998 & N13   &    41.6$\pm$1.0 \\
{\bf 11754974}\,(*13) &  0.383  & 2.61  & M13   &    28.8$\pm$1.7 \\
7020707\,(*16)        &  0.431  & 2.32  & new   &   105$\pm$2 \\ 
{\bf 10989032}\,(*32) &  0.434  & 2.30  & B13   &    45.0$\pm$0.9 \\
6445601\,(*2)         &  0.5415 & 1.847 & N13   &    71.3$\pm$0.4 \\
8110941\,(*29)        &  0.6456 & 1.549 & N13   &  $<$\,7.5$\pm$0.2 \\
{\bf 7819024}\,(*19)  &  0.825: & 1.21: & N13   &    95.1$\pm$1.3 \\  
                      &  1.407  & 0.711 & new   &    95.1$\pm$1.3 \\ 
6130500\,(*9)         &  1.152  & 0.87  & B13   &    49.3$\pm$1.1 \\
8196006\,(*30)        &  1.360  & 0.74  & B13   &    92.6$\pm$1.3 \\
{\bf 9267042}\,(*12)  &  1.360: & 0.74: & B13   &   106$\pm$3 \\  
12688835\,(*18)       &  1.674  & 0.60  & B13   &   230$\pm$5 \\
8330910\,(*3)         &  2.273  & 0.44  & new &   224$\pm$3 \\  
4168579\,(*23)        &  2.491  & 0.40  & new &   197$\pm$4 \\  
\hline

\end{tabular}
\end{table}

In {\bf Table\,8}  rotation rates from the papers by Balona, by Nielsen {\it et
al.}, and from the two abovementioned asteroseismic investigations, have been
summarized.  Note that all of the rotation periods, except that for
KIC\,9244992, are shorter than 10 days, as might be expected for stars with
spectral types earlier than $\sim$F5 (see Fig.2 of  Nielsen {\it et al.}).   

A systematic search of the low-frequency periodograms of all the BN12 stars was
conducted to identify low-frequency  `peaks and harmonics'.   For most of the
stars, including KIC\,11754974 and KIC\,9244992,  no obvious $\nu_{\rm rot}$
harmonics' were detected.  However,  possible rotation frequencies  were
identified for three previously unmeasured stars, KIC\,7020707, KIC\,8330910
and KIC\,4168579, and for KIC\,7819024, all fast rotators with $v \sin i$
values greater than 90\,km\,s$^{-1}$;   these too are given in Table\,8.

Low-frequency periodograms for 12 of the surveyed stars are plotted in {\bf
Figure\,18}.   The locations of the highest peaks, all with high S/N ratios,
usually correspond to the rotation frequency (or its harmonic) given by Balona
or Nielsen {\it et al.}   An exception is KIC\,7819024\,(*19).  For it the
moderate-amplitude peak at the Nielsen {\it et al.} rotation frequency,
0.825\,d$^{-1}$, shows no harmonic at 1.65\,d$^{-1}$ and thus 0.825\,d$^{-1}$
is probably  spurious;  on the other hand, large peaks at 1.41\,d$^{-1}$ and
2.84\,d$^{-1}$ suggest  $\nu_{\rm rot}$ is 1.41\,d$^{-1}$.

Another discrepancy was found for KIC\,9267042 (*12).  Below 3\,d$^{-1}$ several
significant, but low amplitude,  peaks are seen, the strongest of which occur
at 1.226, 1.0737, 1.3604 and 2.701\,d$^{-1}$, where the last two  correspond to
the value $\nu_{\rm rot}$=1.360\,d$^{-1}$ suggested by Balona (2013), and its
harmonic.  Since the first two frequencies have significantly higher amplitudes
$\nu_{\rm rot}$ = 1.360 d$^{-1}$  must be considered questionable.



\begin{figure} \centering 
\begin{overpic}[width=7.9cm]{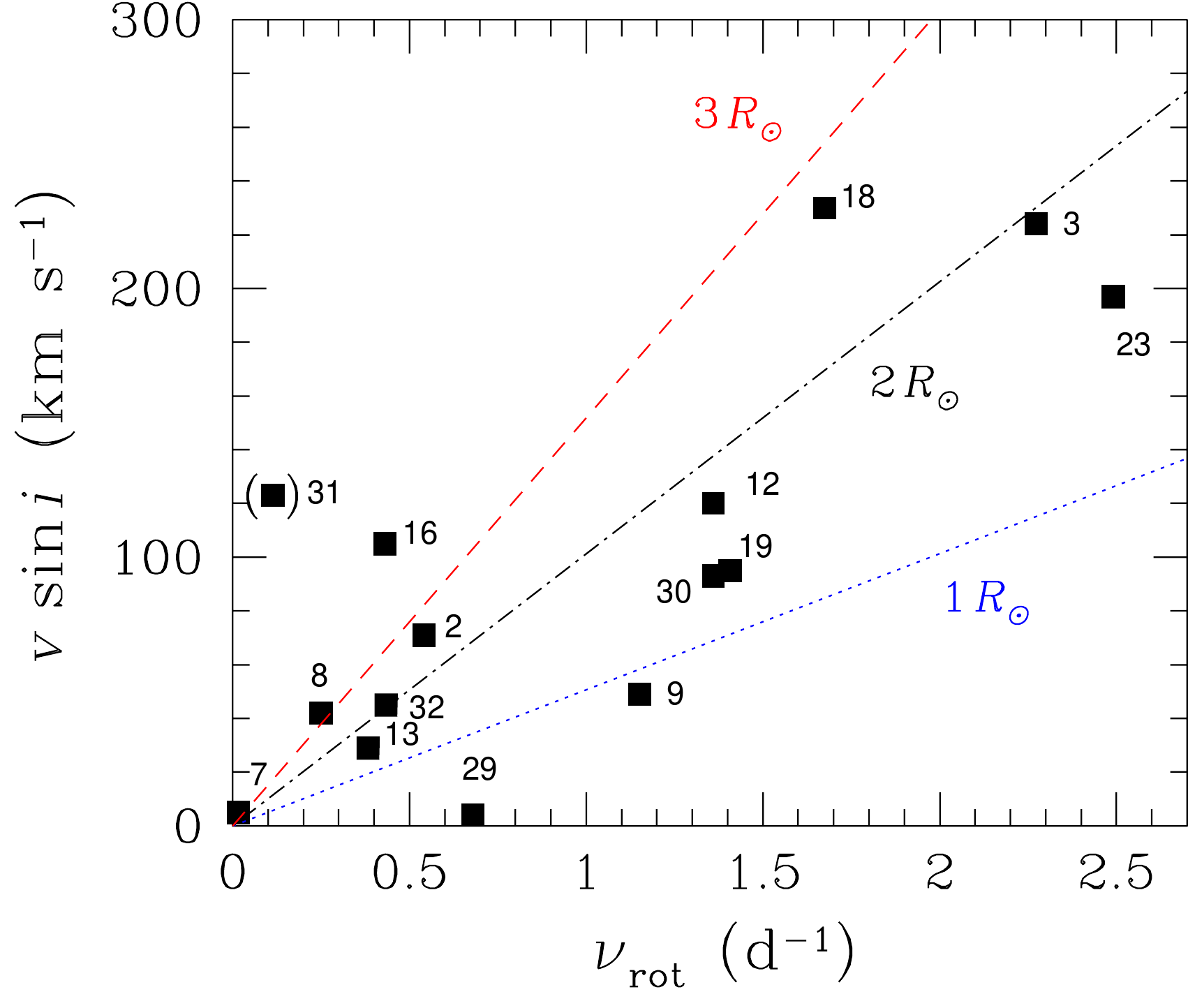} \end{overpic} 
\caption{Projected rotational velocity
as a function of  rotational frequency for the SX~Phe stars in the {\it Kepler}
field (see Table\,8).    The three lines show the equatorial velocities ({\it
i.e.}, $v \sin i$ when  $i = 90^{\circ}$)  as a function of $\nu_{\rm rot}$ for
typical A-type stars having radii of $1\,R_\odot$ (blue dotted line),
$2\,R_\odot$ (black dash-dot line)  and $3\,R_\odot$ (red dashed line). The
labels are the CFHT star numbers.  } \label{Fig19}
\end{figure}

Further inspection of the periodograms shows that only five stars (CFHT nos.
8, 16, 2, 30 and 3) exhibit single, well-defined peaks at the presumptive
$\nu_{\rm rot}$, with lower-amplitude harmonics at two (and sometimes three)
times $\nu_{\rm rot}$.  KIC\,9966976 (*31) has a dominant peak at
0.114\,d$^{-1}$,  the $\nu_{\rm rot}$ value given by Balona (2013),  with
apparent harmonics of lower amplitude at two and three times $\nu_{\rm rot}$,
but the peaks are  broad and made up of several distinct frequencies, possibly
due to differential rotation in a spotted star, or due to combination
frequencies.   Such groupings of two, three or more frequencies at the peaks
and harmonics, rather than single frequencies, also are seen for CFHT nos. 29,
9, 12, 18 and 23 (Fig.\,18).

Since $v \sin i = 2 \pi R \sin i  \,\, \nu_{\rm rot}$ one might expect to
observe a positive correlation  between  $\nu_{\rm rot}$ and  $v \sin i$, where
the degree of  correlation depends on the distributions of the unknown radii
and unknown inclination angles of the stars.   {\bf Figure\,19} is a plot of $v
\sin i$ versus $\nu_{\rm rot}$ for the stars in Table\,8,  where the location
of any given point below, say, the $R$=2\,$R_{\odot}$ line, can be understood
as arising from either a smaller radius or an inclination angle less than
90$^{\circ}$,  or some combination of the two possibilities.  For instance, the
location  of KIC\,8110941 (*29), a narrow-lined star with  $v \sin i <
7\pm1$\,km/s,  is readily explained if its inclination angle is near 0$^\circ$
({\it i.e.}, it is  observed pole-on).   On the other hand, the estimated
rotation frequency of KIC\,9966976 (*31), $\nu_{\rm rot} = 0.114$\,d$^{-1}$, is
too low to account for a $v \sin i$ as large as the measured value,  which
suggests that the features seen in its low frequency periodogram may not be
attributable to a spot or other co-rotating feature, but could be an
artifact attributable to the long-term trends and jumps in {\it Kepler}
photometry, or due to combination frequencies (see Kurtz {\it et al.} 2015).

\subsection{Binary Systems}

SX~Phe stars are Pop.II blue stragglers (BSs) in the lower instability strip, and  BSs are
believed to be binary (or triple) systems that have experienced mass transfer
or possibly coalescence (Hoyle 1964; McCrea 1964; Iben \& Tutukov 1984; Iben
1986; Nemec \& Mateo 1990b; Mateo {\it et al.} 1990; Bailyn 1995; Ferraro {\it
et al.} 2006, 2014; Perets \& Fabrycky 2009), from which it follows that SX~Phe
stars likewise have a binary (or triple star) nature.  Of course, if
coalescence has occurred then the original binary nature may no longer be
visible.   In any case, since only a few SX~Phe stars in binary systems are
known (see  $\S$3.3.3) determination of the fraction of binaries among the BN12
stars and their properties is of considerable interest. 

During the course of this investigation 11 SX~Phe binaries have been
identified.    Three of these stars are close binaries with orbital periods
less than 10 days, and the other eight stars were discovered, from R\o mer
time-delay analyses, to be in wider binaries with orbital periods
$\sim$200-1800 days.  Orbital periods and other significant characteristics for
the eleven SX~Phe binaries are summarized in {\bf Table\,9}.  The quantities
recorded are:  the orbital period, $P_{\rm orb}$; the half-range of the primary
star's RV variation, $K_1$; the orbital eccentricity, $e$; the argument of
periastron, $\omega$; the time of periastron passage, $T_{\rm per}$; the
projected semi-major axis of the primary, $a_1 \sin i $; and the mass function
for the secondary star, $f_2(\mathscr{M}_1, \mathscr{M}_2, \sin i)$.  Table\,9
also gives the orbital characteristics for the two misclassified non-pulsating
binaries:   KIC\,9535881\,[*25] is a W\,UMa system with $P_{\rm orb}=0.4804$\,d
(and  possible  $\gamma$~Dor pulsations);  and  KIC~12643589\, [*17] is a
triple system consisting of a 1.41\,d eclipsing binary with additional transits
every $P_{\rm orb}$=44\,d (see bottom panel of Fig.17).   Both systems are of
interest in their own right because they are binaries belonging to the galactic
halo stellar population;  however, we are concerned in this  paper primarily
with the SX~Phe stars so little more will be said about these two stars.

\begin{table*} \label{tabrv} \caption{Orbital elements for the    {\it
Kepler}-field  SX~Phe stars in binary systems, ordered by orbital period.
The columns contain: (1) KIC and CFHT star numbers; (2) orbital period; (3)
half-range of the RV variation of the primary star ({\it i.e.}, the
more-massive SX~Phe star);  (4) projected semi-major axis of the orbit of the
primary star; (5) orbital eccentricity; (6) argument of periastron; (7) time of
periastron passage;  and (8) mass function of the (less-massive) secondary
star, $f_2$.  The last column indicates the source of the information in that
row, where `TD' stands for the time-delay (binarogram) method of Balona (2014b),
`FM' and `PM' refer to the frequency and phase modulation methods described by
Shibahashi \& Kurtz (2012) and Murphy {\it et al.} (2014), `EB' refers to the
Villanova EB catalog,   and `RV' refers to a `Spectroscopic Binary Solver'
(Johnson 2004) solution using the RVs in Table\,1. Also given in the table are
the previously published orbital elements for KIC\,11754974 from Murphy {\it et
al.} (2013b, `M13') and from Balona (2014b, `B14'), and the results of
simultaneously solving for the phase modulations given the observed radial
velocities (`PM+RV').    }

\begin{tabular}{lcrcclcll}
\hline
\multicolumn{1}{c}{KIC} &\multicolumn{1}{c}{$P_{\rm orb}$}& \multicolumn{1}{c}{ $K_1$} &    $a_1 \sin i$ & \multicolumn{1}{c}{$e$} &\multicolumn{1}{c}{$\omega$} 
                        & \multicolumn{1}{c}{ $T_{\rm per}$}  &\multicolumn{1}{c}{ $f_2(\mathscr{M}_1, \mathscr{M}_2,  i)$ }  & Method  \\
\multicolumn{1}{c}{(CFHT)}  & \multicolumn{1}{c}{ [days] }  & \multicolumn{1}{c}{ [km\,s$^{-1}$]} & [AU]  &  & \multicolumn{1}{c}{ [deg] }   & \multicolumn{1}{c}{2450000+} &  \multicolumn{1}{c}{ [$\mathscr{M}_\odot$] } &   \\
\multicolumn{1}{c}{(1)}    &\multicolumn{1}{c}{(2)} & \multicolumn{1}{c}{(3)}  & \multicolumn{1}{c}{(4)} & \multicolumn{1}{c}{(5)}  & \multicolumn{1}{c}{(6)} &
\multicolumn{1}{c}{(7)}    & \multicolumn{1}{c}{(8)} &  \multicolumn{1}{c}{ (9)}   \\
\hline

\multicolumn{9}{c}{\bf (a) Close binaries (orbital periods $<$ 10 days)}  \\[2pt]

{\bf 10989032}&  (2.3050976)      & $18.3 \pm 0.5$  & $3.86 \times 10^{-3}$  & $0.08 \pm 0.03$   &$ 211\pm15$        &$6547.6\pm0.1$ &   $(1.45 \pm 0.18) \times 10^{-3}$ &  RV  \\  
(*32)         &  2.3050976        &                 &                        &                   &                     &                             &                          & EB  \\  [3pt]

{\bf 7174372} & $4.0   \pm  0.1$  & $1.3 \pm 0.5$   &  $4.9 \times 10^{-4}$ &  0.0       &  undef.            &   undef.      & $(9.5 \pm 0.1) \times 10^{-7}$ & FM,RV  \\  
(*8)          &                   &                 &                       &            &                    &               &                                &        \\  [3pt]

{\bf 6780873} & 9.1547$\pm$.0003 & 38.73$\pm$.12    & 0.0325$\pm$.0001 & (5$\pm$5)$\times10^{-4}$ &   undef.     &    undef.            &  (5.51$\pm$0.05)$\times10^{-2}$   & PM+RV  \\ 
(*5)          & $9.156 \pm 0.046$ & $44.4\pm6.0$    & $0.037 \pm 0.005$ &                   &                    &               &  $(8.0 \pm 3.0) \times 10^{-2}$    & TD  \\  
              & $9.16  \pm 0.03 $ &                 &                  &                   &                    & $6531 \pm 1$  &   $(7.2 \pm 12.8) \times 10^{-2}$   &  FM \\  
              & $9.161 \pm 0.001$ & $38.7\pm0.9$    & $0.033 \pm 0.001$ & $ 0.04 \pm 0.02$ & $ 311 \pm 15$      & $6518.7\pm0.3$&   $(5.5 \pm 0.2) \times 10^{-2}$    & RV \\ [6pt]  

\multicolumn{9}{c}{\bf (b) R{\o}mer time-delay (wide) binaries} \\[2pt]
{\bf 7819024} & $216.6  \pm 0.7$  & $ 7.27 \pm0.43$ & $0.144 \pm 0.007$  & $ 0.09 \pm 0.08$ & $299\pm82$        & $5050\pm50$ &   $(8.48\pm 1.24) \times 10^{-3}$  & PM+RV \\  
(*19)         & $216.2  \pm 3.9$  & $ 7.37 \pm0.95$ & $0.147 \pm 0.019$  &                  &                   &            &  $(8.96 \pm 3.42) \times 10^{-3}$    & TD \\   
              & $216.2  \pm 3.1$  &                  &                   &                  &                   &            &  $(9.2 \pm 3.2) \times 10^{-3}$   & FM  \\  [3pt] 
      
{\bf 8004558} & $259.8  \pm 0.2$  & $ \,\,9.70\pm0.08$ &  $0.231 \pm 0.002$  & $0.020 \pm 0.012$ & $287\pm11$         & $4993\pm8$   &  $(2.45 \pm 0.06) \times 10^{-2}$  & PM+RV \\  
(*1)          & $262.1  \pm 0.2$  & $10.57\pm0.08$   & $0.255 \pm 0.002$  &                  &                    &             &  $(3.21 \pm 0.07) \times 10^{-2}$   & TD \\   
              & $261.2  \pm 0.9$  &                &                 &                  &                    &                &   $(2.83 \pm 0.04) \times 10^{-2}$   & FM  \\ [3pt]  

{\bf 11754974}& $342.6 \pm 0.5$   & $ 8.39 \pm0.08$ & $0.264 \pm 0.003$ & $ 0.013 \pm 0.012$ & $37 \pm 8$        & $5110 \pm 6$  &  $(2.09 \pm 0.06) \times 10^{-2}$  & PM+RV  \\  
(*13)         & $344.8  \pm 0.2$  & $ 8.19 \pm0.06$ & $0.259 \pm 0.002$ &                   &                    &               &  $(1.96 \pm 0.04) \times 10^{-2}$   & TD,B14   \\   
              & $343.3  \pm 0.3$  & $ 8.35 \pm0.04$ & $0.263 \pm 0.002$ & $ 0.01 \pm 0.01$  & $ 102 \pm 38$      & $4999 \pm 37$ &  $(2.07 \pm0.03) \times 10^{-2}$    & M13 \\  
              & $343.1  \pm 0.8$  &                 &                  &                    &              &                   &    $(2.06 \pm 0.07) \times 10^{-2}$  & FM \\ [4pt]  

{\bf 4243461} & $481.6  \pm 1.8$  & $ 5.29 \pm0.12$  & $0.233 \pm 0.005$ &  $0.045\pm 0.031$ & $ 163 \pm 36$      & $5296 \pm 48$  & $(7.35 \pm 0.41) \times 10^{-3}$   & PM+RV  \\  
(*4)          & $488.8  \pm 2.9$  & $ 5.34 \pm0.28$  & $0.240 \pm 0.013$ &                  &                    &              &   $(7.69 \pm 0.12) \times 10^{-3}$   & TD      \\  
              & $475.5  \pm 6.2$  &                  &                   &                  &                    &             &   $(8.17 \pm 1.09) \times 10^{-3}$     & FM \\  [3pt]  

{\bf 5705575} & $537.7  \pm 0.9$  & $ 6.70 \pm0.04$ & $0.331 \pm 0.003$  & $0.018 \pm 0.014$& $52 \pm 10$    & $5367 \pm 16$ & $(1.68 \pm 0.04) \times 10^{-2}$    & PM+RV  \\  
(*22)         & $537.2 \pm 1.5$   & $ 6.60 \pm0.18$  & $0.326\pm0.009$  &                  &                    &              & $(1.60\pm0.13) \times 10^{-2}$  & TD  \\  
              & $539.5  \pm 4.1$  &                &                   &                  &                    &             &  $(1.57 \pm 0.09) \times 10^{-2}$     & FM  \\   
             &   (537.5)         & $ 6.55 \pm0.83$  & $0.324\pm0.040$  &\multicolumn{1}{c}{(0.0)}&    undef.         &  undef.      & $(1.57\pm0.59) \times 10^{-2}$   & RV   \\ [3pt]  

{\bf 7300184} & $635 \pm 20$      & $ 0.02 \pm0.02$  & $0.001 \pm 0.001$   &                  &            &             & $(3.2 \pm 3.2) \times 10^{-11}$  & PM  \\  
              & $640 \pm 60$      & $ 0.07 \pm0.04$  & $0.004 \pm 0.002$   &                  &            &             & $(1.8 \pm 1.8) \times 10^{-8}$   & TD  \\ [3pt]  

{\bf 9966976} &      $>$1500      & $0.62 \pm 0.05$ & $0.088 \pm 0.006$ &                 &                    &              & $ (3.8  \pm 1.0) \times 10^{-5}$ & PM  \\ 
(*31)         & $1750 \pm 100$    & $0.66 \pm 0.08$ & $0.106 \pm 0.011$ &                 &                    &              & $ (5.2  \pm 1.7) \times 10^{-5}$ & TD  \\ [3pt]  

{\bf 9267042} &  ($>$1500?)       & $0.72 \pm 0.02$ &  $0.091 \pm 0.003$ &                &                    &               & $ (5.4  \pm 0.5) \times 10^{-5}$ & PM  \\ 
(*12)         & $1760 \pm  70$?   & $0.70 \pm 0.03$ &  $0.113 \pm 0.004$ &               &                     &               & $ (6.3  \pm 0.6) \times 10^{-5}$ & TD \\ [6pt]  

\multicolumn{9}{c}{\bf (c) Close eclipsing binaries (Pop.II) that do not contain an SX\,Phe star}  \\[2pt]

{\bf 9535881} & 0.4804213         & $ 2.1 \pm 0.2$  &  $9.2 \times 10^{-5}$   &   0.0            &   undef.           &  undef.       &   $(4.4  \pm 0.9) \times 10^{-7}$  & EB,RV \\  
$[*25]$ \\
{\bf 12643589}& 1.4116278         & $ 1.7 \pm 0.1$  &  $2.2 \times 10^{-4}$   &  0.0            &   undef.           &  undef.       &    $(7.0 \pm 1.0) \times 10^{-7}$  & EB,RV \\  
$[*17]$       & 44.147177         &                 &                  &                    &               &                          &                                  & EB  \\
\hline
\end{tabular}
\end{table*}

\subsubsection{\bf  Close Binaries }

Based on the appearance of their light curves, their observed RV variations,
and their broad CCFs, it is clear that KIC\,10989032\,(*32), KIC\,7174372\,(*8)
and KIC\,6780873\,(*5) have orbital periods less than 10 days and small
separations.  Their orbital and other properties are discussed next.

\begin{figure} \begin{center}
\begin{overpic}[width=7.7cm]{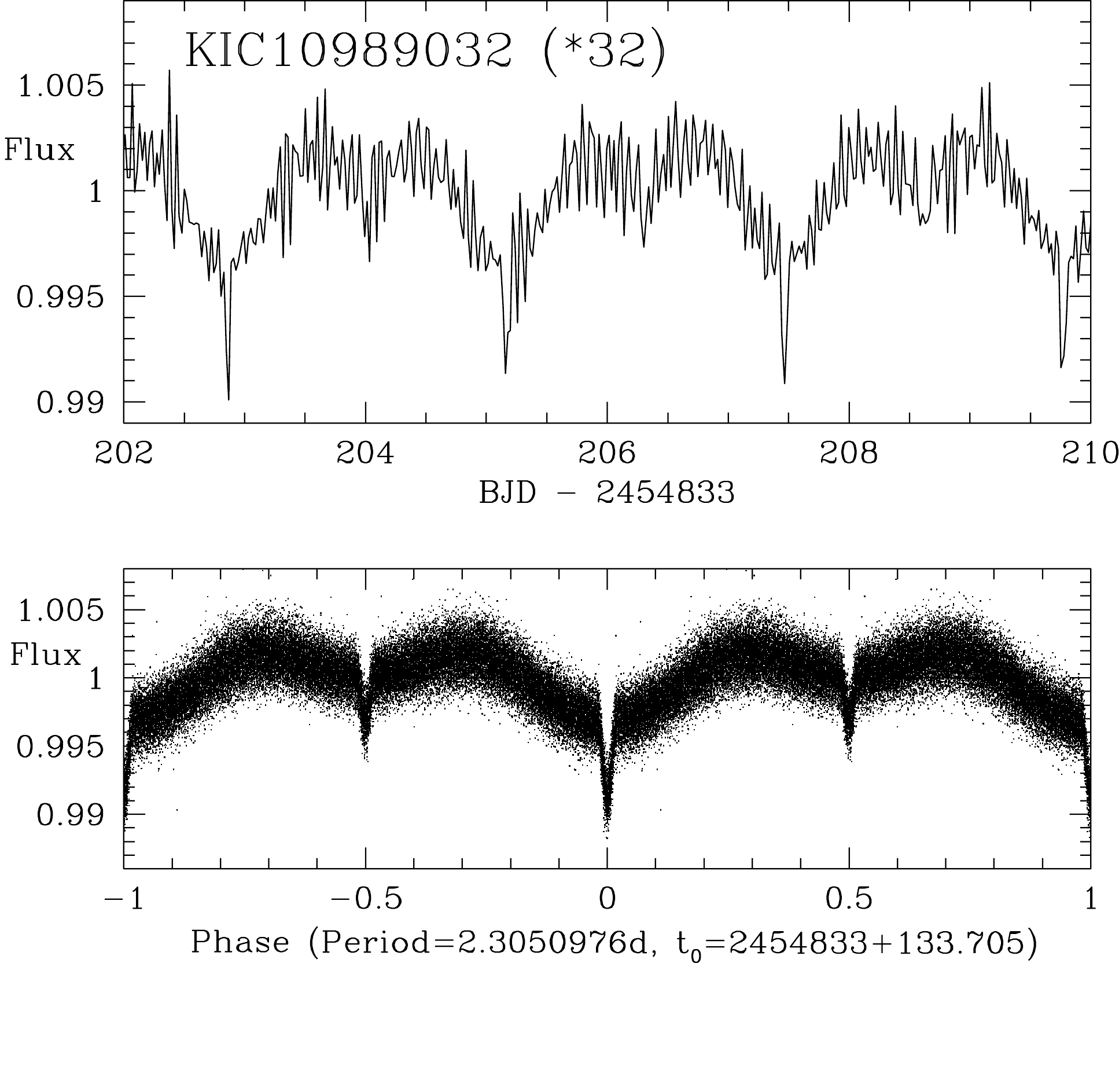} \end{overpic}
\vskip-0.5truecm
\begin{overpic}[width=7.0cm]{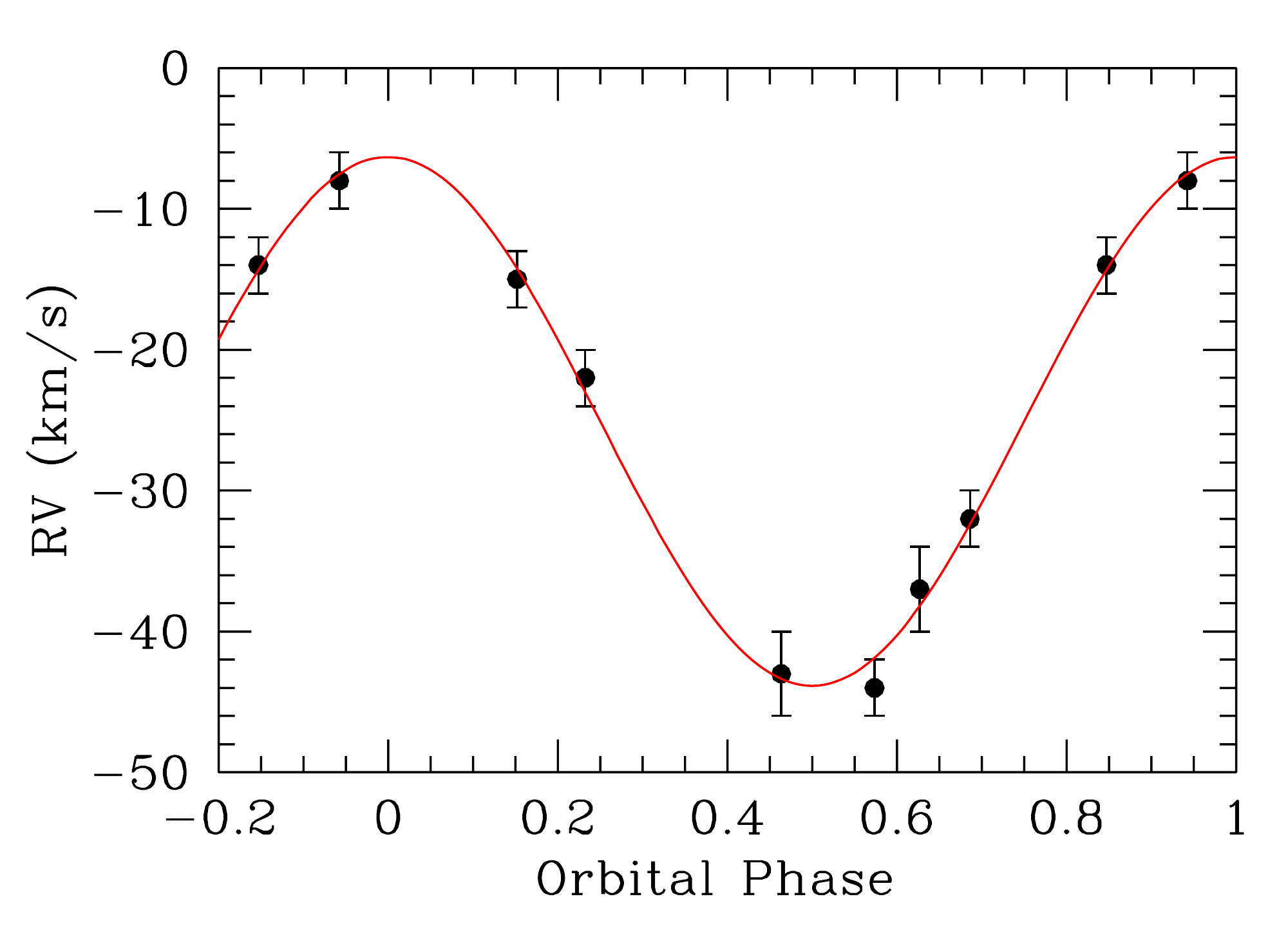} \end{overpic}
\end{center} 
\caption{{\it Kepler} light curves (top, normalized LC fluxes) and
radial velocity curve (bottom) for KIC~10989032 (*32), phased with the 2.3-day
orbital period and mid-eclipse time of the primary eclipse given in the
Villanova EB catalog.  The fitted cosine curve is shown in red.   }
\label{VillanovaLCs} \end{figure}

\vskip0.5truecm \noindent  {\bf KIC\,10989032} (*32) -- This eclipsing system
is listed in the Villanova EB catalog as having  $P_{\rm orb} =
2.3050976(35)$\,d, and a `morphology classification parameter', 0.53,
indicating that it is probably  semi-detached (morph$\sim$\,0.5-0.7) but is
close to being detached (morph $\lesssim$\,0.5).  Other basic characteristics
include: mid-eclipse time for the primary eclipse,
BJD$_0$=2454966.705$\pm$0.013; primary and secondary eclipse depths of 0.0095
and 0.0043 ({\it i.e.}, very low amplitude grazing eclipses);  primary and
secondary phase widths of 0.025 and 0.034; and a separation in phase between
the primary and secondary eclipses of 0.499, suggesting a near-circular orbit
(unless the major axis of the orbital ellipse is orientated towards us, {\it
i.e.} $\omega = \pm 90^{\circ}$).  Additional parameters of the system include
(see Table\,1 in Pr\u sa {\it et al.} 2011): temperature ratio,
$T_2/T_1=0.454$;  sum of the fractional radii, $(R_1 + R_2) / a$ = 0.208 (where
$a$ is the semi-major axis of the orbit,  equal to the separation of the
two stars,  $a_1 + a_2$); orbital inclination angle, $\sin i = 0.9947$  ({\it
i.e.}, nearly edge-on, $i$ = 84.10$^{\circ}$); and  radial and tangential
components of the orbital eccentricity, $e \sin \omega$ = --0.241 and $e \cos
\omega$ = 0.459. 

{\bf Figure\,20} shows an eight-day portion of the {\it Kepler} light curve
(top panel),  the long cadence fluxes  phased with the Villanova $P_{\rm orb}$
and BJD$_0$ (middle panel), and the RV curve (bottom panel).     The radial
velocities are well fit with a cosine function.   From the optimum RV fit the
eccentricity  $e$=0.08$\pm$0.03, with systemic velocity $\gamma =
-24.96\pm0.31$ km/s, RV amplitude $K_1$ = $18.28 \pm 0.45$\,km/s, argument of
periastron $\omega = 211\pm16^{\circ}$\, and time of periastron passage
2456547.61$\pm$0.09.  With these values the mass function for the secondary,
$f_2$ = $( \mathscr{M}_{2} \sin i)^3 / ( \mathscr{M}_1 + \mathscr{M}_2 )^2 $,
which can also be expressed as $P_{\rm orb} K_1^3  (1 - e^2)^{3/2} / 2 \pi G$,
has the value $0.00145\pm0.00018$\,$\mathscr{M}_\odot$;  also, the projected
semi-major axes, $a \sin i =8.60\,R_\odot$, and $a_1 \sin i = 0.83\,R_\odot$.
Since $\sin i$ is so close to 1.0 this mass function is practically identical
to the de-projected mass function, $f_2^{\prime} = 0.00149$, and the observed
RV curve is practically identical to the curve that would be seen if the system
were viewed edge-on.  Combining the separation, $a = 8.65\,R_\odot$, with the
sum of the fractional radii given above, implies $R_1 + R_2 = 1.8\,R_\odot$.
From the cosine fit to the observed RVs the time of maximum RV occurs at BJD =
2456548.563$\pm$0.011, from which one can infer that eclipses occurred
$\frac{1}{4} P_{\rm orb}$ earlier and later, {\it i.e.,} at BJD 2456547.986 and
2456549.139, respectively.  These `observed' eclipse times are to be compared
with the primary eclipse time predicted from the Villanova EB catalog: BJD$_0$
+ 686$P_{\rm orb}$ = 2456548.002, the time difference amounting to only
0.016\,day or 23\,min.  This difference is much smaller than the error expected
from the propagated uncertainties and is consistent with the observed lack of
eclipse timing variations (see Conroy {\it et al.} 2014, and the online EB
catalog).  Finally, with no RV information for the secondary, the mass ratio,
$q$, and the individual masses are unknown.  However, if the primary of
KIC\,10989032 were a typical A5\,V star (see Table 3) it might be expected to
have a mass $\sim$2.0\,$\mathscr{M}_\odot$ and a radius near 1.7\,$R_\odot$
(Gray 2005).  Models of  Pop.\,II blue stragglers that have undergone mass
transfer suggest lower total masses, possibly in the range 1.1 to 1.5
$\mathscr{M}_\odot$ (see Fiorentino {\it et al.} 2014, 2015).  Assuming
1.5\,$\mathscr{M}_\odot$ for the mass of the primary, the observed $K_1 =
18.28$ km/s suggests $q=0.11$ and a mass for the secondary of
0.16\,$\mathscr{M}_\odot$.  The mass-radius relation of Demory {\it et al.}
(2009) suggests that such a low-mass star might be expected to have a radius
$R_2$\,$\sim$ 0.15-0.20\,$R_{\odot}$, consistent with the above discussion of
sizes.

\vskip0.5truecm  \noindent {\bf KIC\,7174372} (*8) -- As noted earlier, the
{\it Kepler} light curve shows low-amplitude, high-frequency oscillations
superimposed on periodic  low-amplitude dips.  It is  possible that  the four
day period of the dips is due to rotation (Nielsen {\it et al.} 2013); however,
the dips may also be due to orbital motion in an ellipsoidal binary system,
{\it i.e.},  a system comprising ``non-eclipsing close binaries whose
components are distorted by their mutual gravitation''  where ``the light
variations are due to the changing cross-sectional areas and surface
luminosities that the distorted stars present to the observer at different
phases'' (Morris 1985).  The star is not listed in the Villanova EB catalog.
Owing to large uncertainties in the measured RVs (caused by the broad spectral
lines), it is difficult to tell if the RVs are varying ($\Delta RV = 4.9 \pm
4.8$\,km/s); however, considerable broadening is seen in the CCFs (Fig.\,5).
If both $P_{\rm orb}$ and $P_{\rm rot}$ are near 4.0\,d then the system is
tidally synchronized; such synchronization is common in binary systems with
short orbital periods (Duquennoy \& Mayor 1991).

\begin{table} \centering \caption{Example orbital properties for the non-eclipsing SB2 system KIC\,6780873 (*5), for four possible
primary-star masses.  The quantities were calculated assuming the following
observational constraints:  $P_{\rm orb}$=9.161$\pm$0.001\,d;
$e$=0.04$\pm$0.02; $a \sin i = 16.48$\,$R_{\odot}$; $\gamma$=10.07$\pm$0.26
km/s;  $K_1$=38.8$\pm$0.9 km/s; $K_2$=52.4$\pm$1.0 km/s;    $q$ =
0.74$\pm$0.03;  time of periastron passage, $T_{\rm per}$=2455518.73$\pm$0.33; argument
of periastron for the primary and the secondary, $\omega_1$=311$\pm$14${^\circ}$ and
$\omega_2$=131$\pm$14$^{\circ}$. The velocities in the last two rows were calculated
assuming circular orbits.  }
 
\label{tab:BN5}
\begin{tabular}{lcccc}
\hline
\multicolumn{1}{l}{ $\mathscr{M}_1$ = } & 0.9\,$\mathscr{M}_\odot$  & 1.2\,$\mathscr{M}_\odot$   & 1.5\,$\mathscr{M}_\odot$   & 1.8\,$\mathscr{M}_\odot$  \\
\midrule
$\mathscr{M}_2$ [$\mathscr{M}_\odot$]  & 0.67    & 0.89   & 1.11   & 1.33   \\
$i$ [deg]              &  50.40  & 44.43  & 40.53  & 37.70  \\
$a$ [$R_\odot$]        & 21.39  & 23.56  & 25.38  & 26.97  \\
$a_1$ [$R_\odot$]      & 9.10  & 10.02  & 10.79  & 11.47  \\
$a_2$ [$R_\odot$]      & 12.29 & 13.54  & 14.59  & 15.50  \\
$v_1$ [km/s]           & 50.3 & 55.4   &  59.6  & 63.4   \\   
$v_2$ [km/s]           & 67.9  & 74.8   &  80.6  & 85.7   \\   
\hline
\end{tabular}
\end{table}

\vskip0.5truecm

\vskip0.4truecm   \noindent {\bf KIC\,6780873} (*5) - This double-lined
spectroscopic binary (see $\S2.1$) exhibits the largest RV range of all the
program stars, $\Delta RV = 69.9\pm0.3$\,km/s.    Six of the nine CCFs  resolve
into two components, with  heights that  clearly identify the primary and
secondary components (see Fig.\,3).   From an FM analysis the orbital period
was initially estimated  to be 9.16$\pm$0.03\,d.  Similar periods were derived
from a time-delay binarogram analysis, from a PM analysis (where a sampling
window of 3\,d instead of the usual 10\,d was needed), and from an analysis
based only on the 18 RVs (using the `Spectroscopic Binary Solver' [SBS] program
of Johnson 2004).  The current best estimate of the orbital period, $P_{\rm
orb} =  9.1547 \pm 0.0003$\,d,  combines the results of the PM method with RV
information (see Murphy {\it et al.} 2016).  With this period the symmetric RV
curves were well fit with eccentricity, $e$=0.0, and gamma velocity ({\it
i.e.}, median RV), $\gamma =10.07\pm0.26$ km/s (see Figure\,4).  The observed
amplitudes of the two RV curves,  $K_1$ = 38.8$\pm$0.9 km/s and $K_2$ =
52.4$\pm$1.0 km/s, give a mass ratio, $q$ = 0.74$\pm$0.03, which is independent
of the orbital inclination angle.  The  projected semi-major axis, $a \sin i$ =
16.48 $R_\odot$ (=0.077\,AU), follows  from  $K_1 (P_{\rm orb}/2 \pi) ((1 +
q)/q) $.  Assuming the parameters given in the caption for Fig.\,4 the
predicted RV curves fit well the measured individual RVs, with standard
deviations of the O-C residuals $\sim$0.62 km/s for both components.  Since the
photometry shows  no evidence of  eclipses, the individual masses and the
orbital inclination angle, $i$, could not be determined;   only the mass
functions for the individual components, $f_1 = 0.136 \mathscr{M}_\odot$  and
$f_2 = 0.055 \mathscr{M}_\odot$ could be estimated.

To resolve  the degeneracy between  mass and inclination angle four different
masses were assumed for the SX Phe (primary) star, $\mathscr{M}_1$ = 0.9, 1.2,
1.5 and 1.8 $\mathscr{M}_\odot$, and inclination angles were calculated by
constraining the  predicted RV curves to have the observed $K_1$ and $K_2$
values.  Some properties of the binary system for the assumed masses are
summarized in  {\bf Table\,10}, where:  $\mathscr{M}_2$ follows from the
observed mass ratio $q$ (=$K_1/K_2$); the semi-major axis, $a$, follows from
Kepler's third law, $a^3 = G (\mathscr{M}_1 + \mathscr{M}_2) P^2 / (4 \pi^2)$;
and the individual distances from the centre of mass, $a_1$ and $a_2$, were
computed from the $a$ and $K$ values.  RV curves were  computed for each mass,
from which the best-fit inclination angles, $i$, were found.   Since $e$ is
near zero,  the individual velocities, $v_1$ and $v_2$, which follow from the
$a_1$, $a_2$ and $P_{\rm orb}$ values, are very nearly constant.  Note that for
a doubling of $\mathscr{M}_1$, from 0.9 to 1.8\,$\mathscr{M}_\odot$, the
inclination angle changed little, from 50.4 to 37.7 degrees.

\subsubsection{\bf R\o mer Time-Delay (Wide) Binaries}

Shibahashi {\it et al.} (2012, 2015),  Telting {\it et al.} (2012),  Murphy
{\it et al.} (2013b, 2014, 2015a,b), Balona (2014b), Balona {\it et al.}
(2015), Kurtz {\it et al.} (2015), Koen (2014) and others have shown that it is
possible, using only {\it Kepler} time-series photometry, to derive RV curves
and orbital elements for binary systems in which one or more of the stars is
pulsating.    As the pulsating star orbits the barycentre of the binary system
its distance changes, resulting in R{\o}mer time delays of the pulsation
frequencies (also seen as phase modulations), which are equivalent to frequency
modulations.  When working in the time domain these  methods have been
variously referred to as the `time-delay' (TD) method (Telting {\it et al.}
2012, Balona 2014b), the `phase modulation' (PM) method (Murphy {\it et al.}
2014, 2015b, 2016), and the `O-C (and light travel time)' method (Koen 2014);
and in the frequency domain the `frequency modulation' (FM) method (Shibahashi
{\it et al.} 2012, 2015).    Successful application of these methods requires:
(1) high precision photometry (in order to derive accurate pulsation
frequencies and to detect contamination by nearby frequencies); (2) photometry
with a time baseline sufficiently long to resolve long orbital periods, and of
sufficiently short cadence to resolve close binaries with short orbital
periods;  and (3) one or more members of the binary system pulsating with one
or more high-amplitude frequencies that are uncontaminated by close frequencies
({\it i.e.}, no significant neighbouring peaks in the periodogram) and which
exhibit {\it steady} pulsation ({\it i.e.}, no amplitude or intrinsic
variations).  The 32 SX~Phe candidates and associated {\it Kepler} photometry
meet all these requirements  in the case of  binary systems that have  orbital
periods in the range $\sim$10 to $\sim$1700 days. 

One prior  application of the time-delay method was the detailed asteroseismic
analysis of KIC\,11754974 (*13) by Murphy {\it et al.} (2013b).  Time delays of
up to $\sim$180\,s  revealed the SX~Phe star to be the primary in a
non-eclipsing 343\,d binary system that has  an almost circular orbit;
assuming 1.5\,$\mathscr{M}_\odot$ for the mass of the primary and orbital
inclination angle $i=90^{\circ}$,  gives a  secondary  mass of
0.6$\pm$0.2\,$\mathscr{M}_{\odot}$, which is characteristic of a K- or early-M
main-sequence star.   More recent analyses of KIC\,11754974 using the complete
{\it Kepler} Q0-17 data set have been made by Balona (2014b) and  Murphy {\it et
al.} (2014).

For the present paper both time- and frequency-domain methods were used to
search  the BN12 stars for SX~Phe pulsators residing in binary systems.  The
analyses were carried out using the TD (binarogram) method described by Balona
(2014b), and the FM and PM methods described by Shibahashi {\it et al.} (2012,
2015) and Murphy {\it et al.} (2014, 2015b, 2016).   All the available Q0-Q17
{\it Kepler} photometry was utilized, and for the time-domain analyses,  a
sampling interval of 10\,d was adopted, which  corresponds to a Nyquist
frequency of 0.05\,d$^{-1}$ and precludes  measurement of orbital periods
shorter than $\sim$20\,d.  Care was taken to select only real frequencies with
amplitudes high enough to allow   detection of binary motion (use of an
incorrect alias frequency often results in an apparent ``orbital period'' close
to that of the {\it Kepler} satellite, {\it i.e.}, 372\,d).  For the FM
analyses the short-period limit was reduced  to $\sim$4\,d, and led to the
initial discovery  of the 9.16\,d orbital period for KIC\,6780873 and the
4.0\,d orbital period for KIC\,7174372.  Sensitivity to short periods is the
biggest advantage that the frequency domain has over the time domain.  Orbital
elements derived from the current analysis are summarized in Table\,9.

Observed phase shifts from the time-delay analyses are shown in {\bf
Figure\,21} for six of the stars.   The graphs for each star give, for the
frequencies noted on the right side of each graph, the phase variation as a
function of orbital phase.  The points in each graph represent the measured
pulsation phase (10\,d segments) of the photometry.  The six stars have orbital
periods between 208 and 670 days and therefore  can be classified as `wide'
binaries.  The high signal-to-noise ratios of the phase variations leave little
doubt about the binary nature for five of the six stars.  In the case of
KIC\,7300184 only one pulsation frequency was employed in the analysis, therefore
the orbital parameters $e$, $\omega$ and $T_{\rm per}$ could not be calculated.

\begin{figure*} \centering 
\begin{overpic}[width=5.5cm]{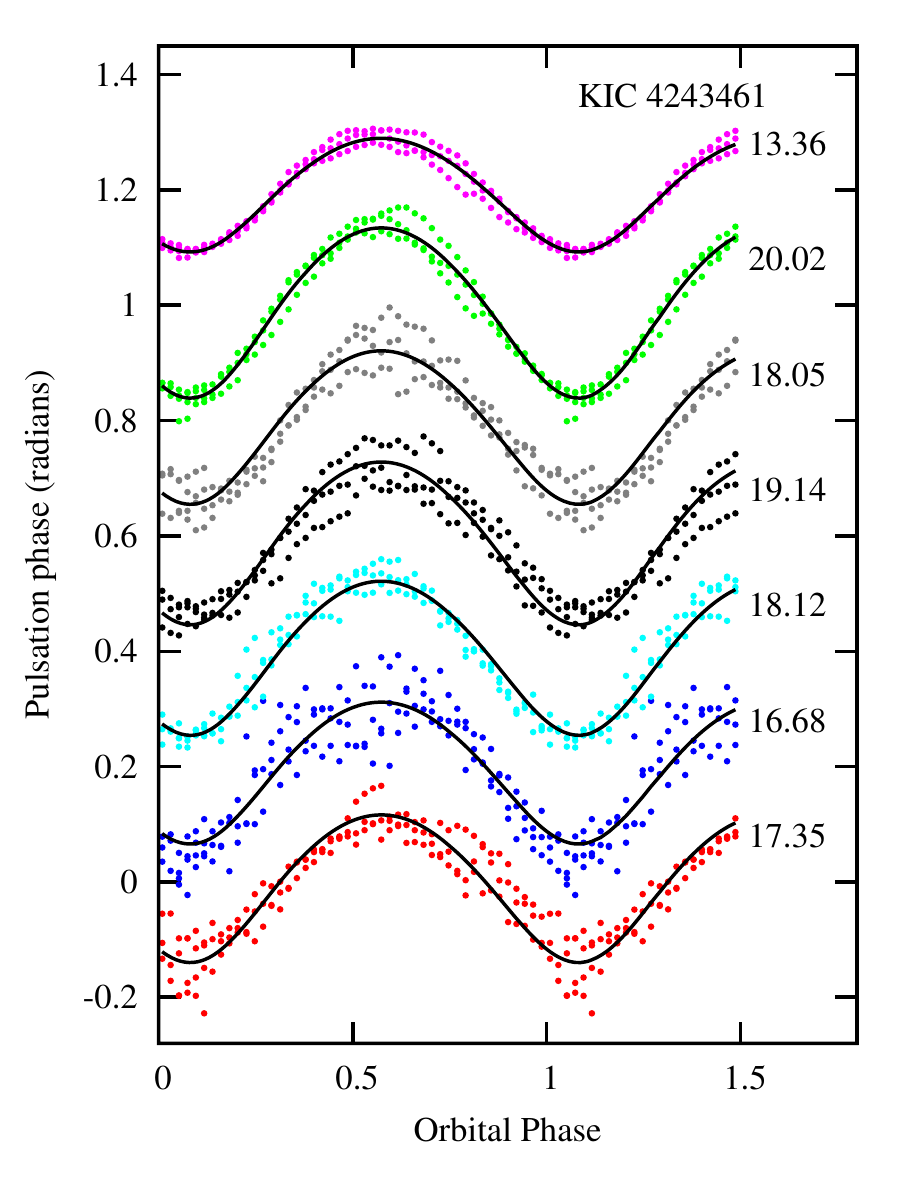} \end{overpic}
\begin{overpic}[width=5.5cm]{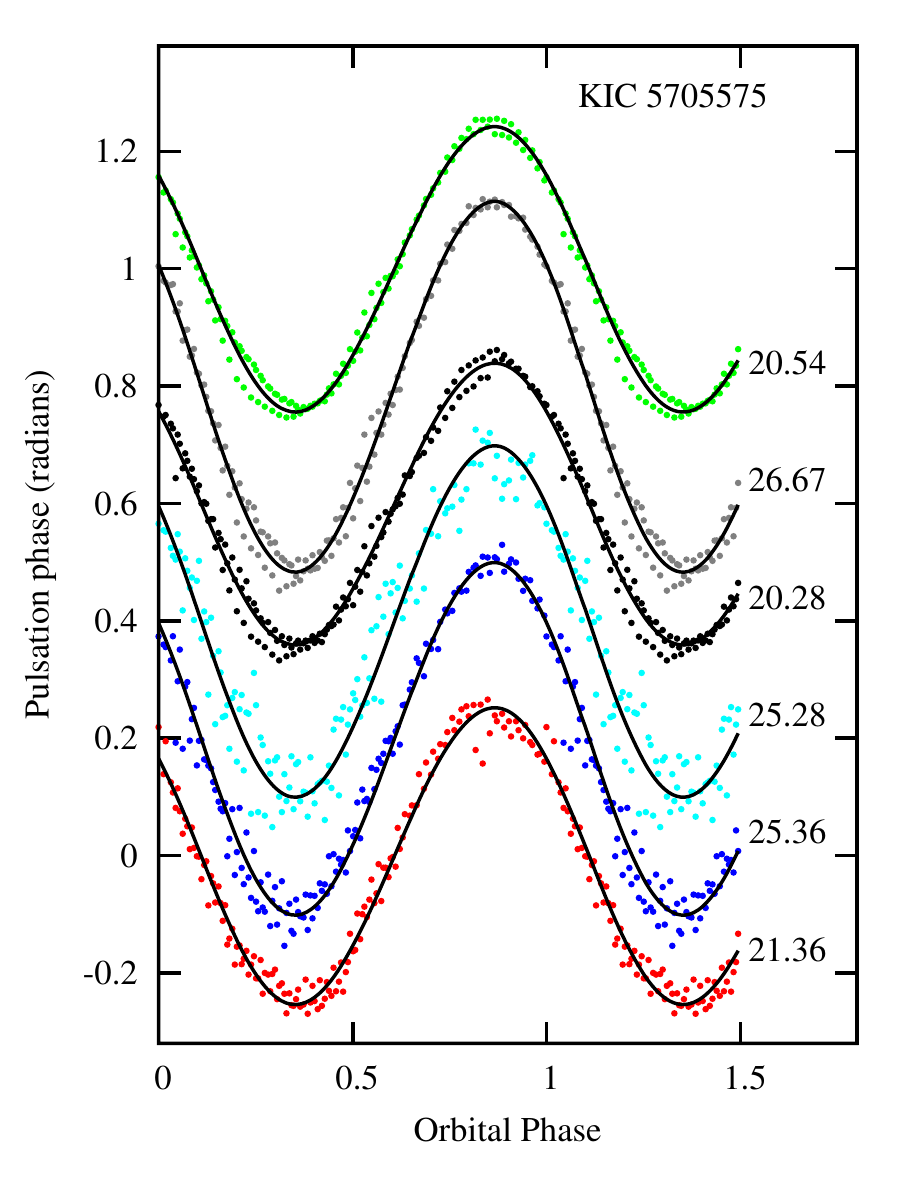} \end{overpic}
\begin{overpic}[width=5.5cm]{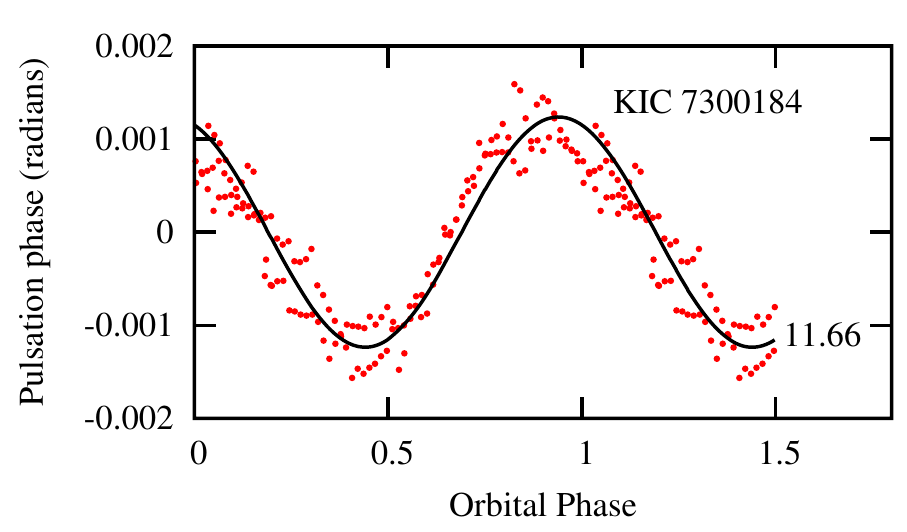} \end{overpic}
\begin{overpic}[width=5.5cm]{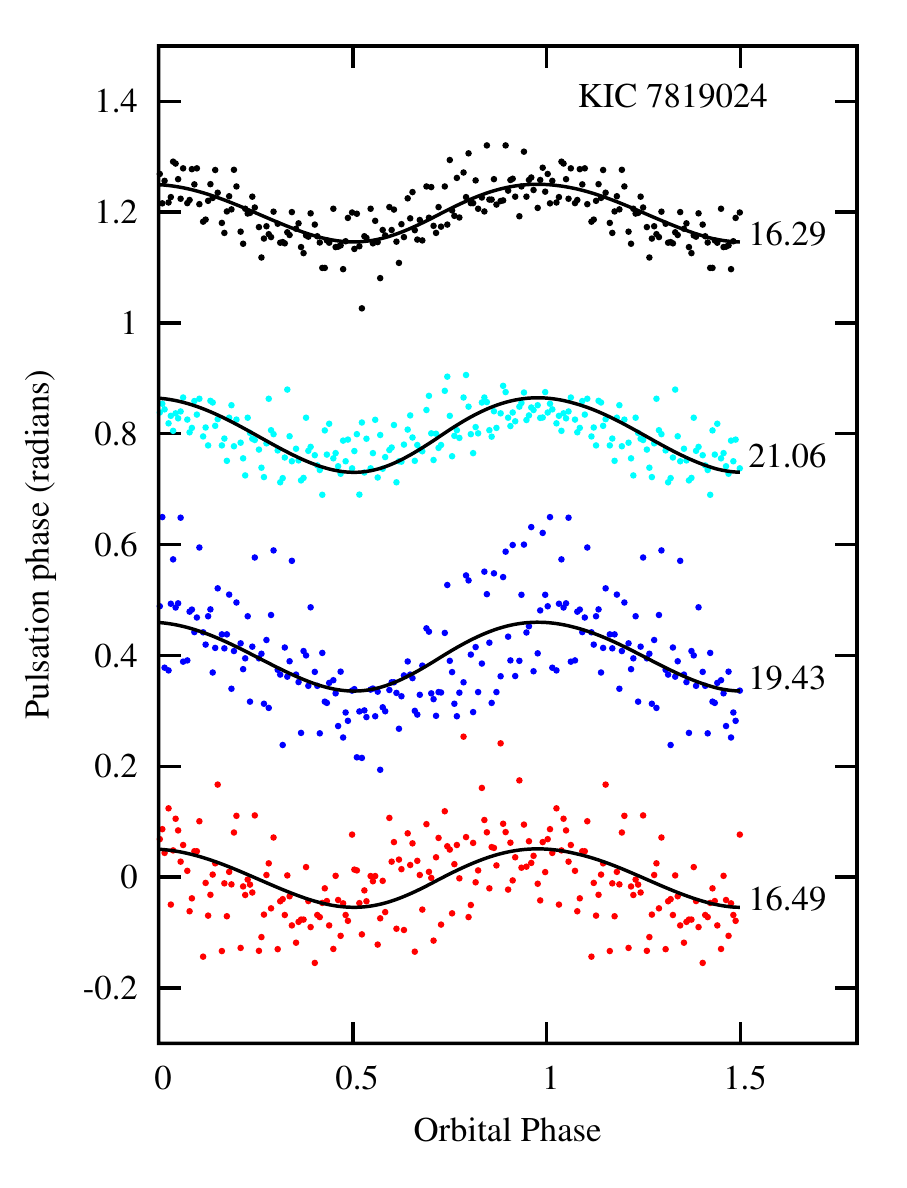} \end{overpic}
\begin{overpic}[width=5.5cm]{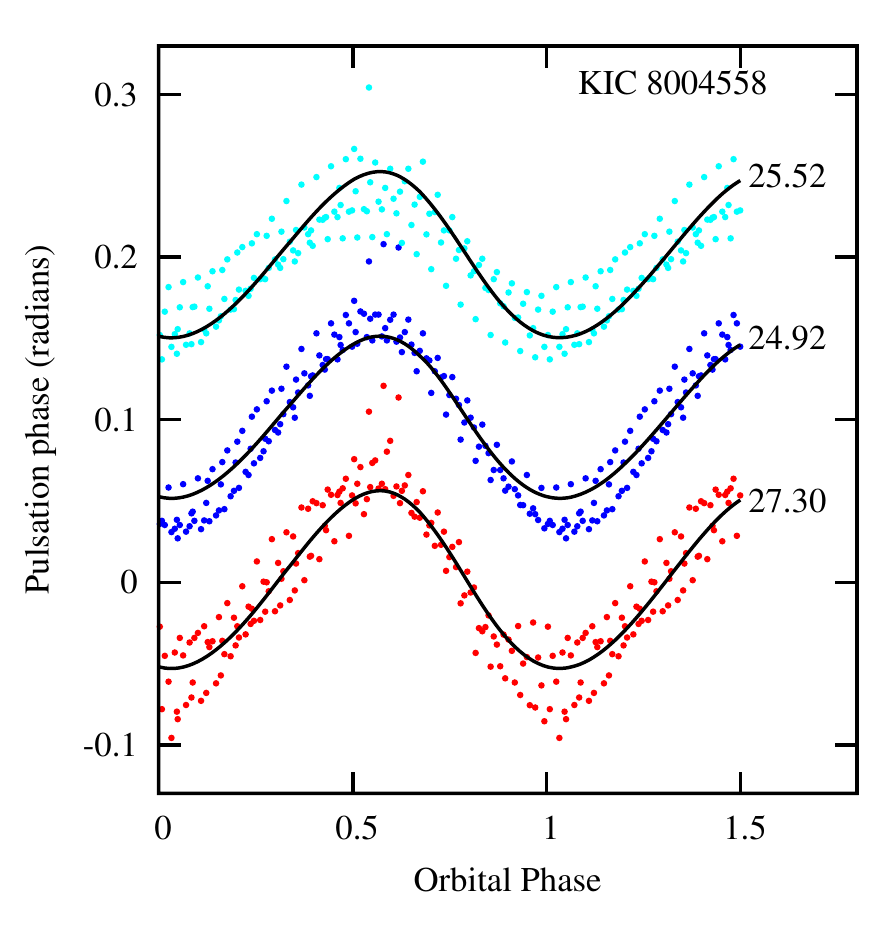} \end{overpic}
\begin{overpic}[width=5.5cm]{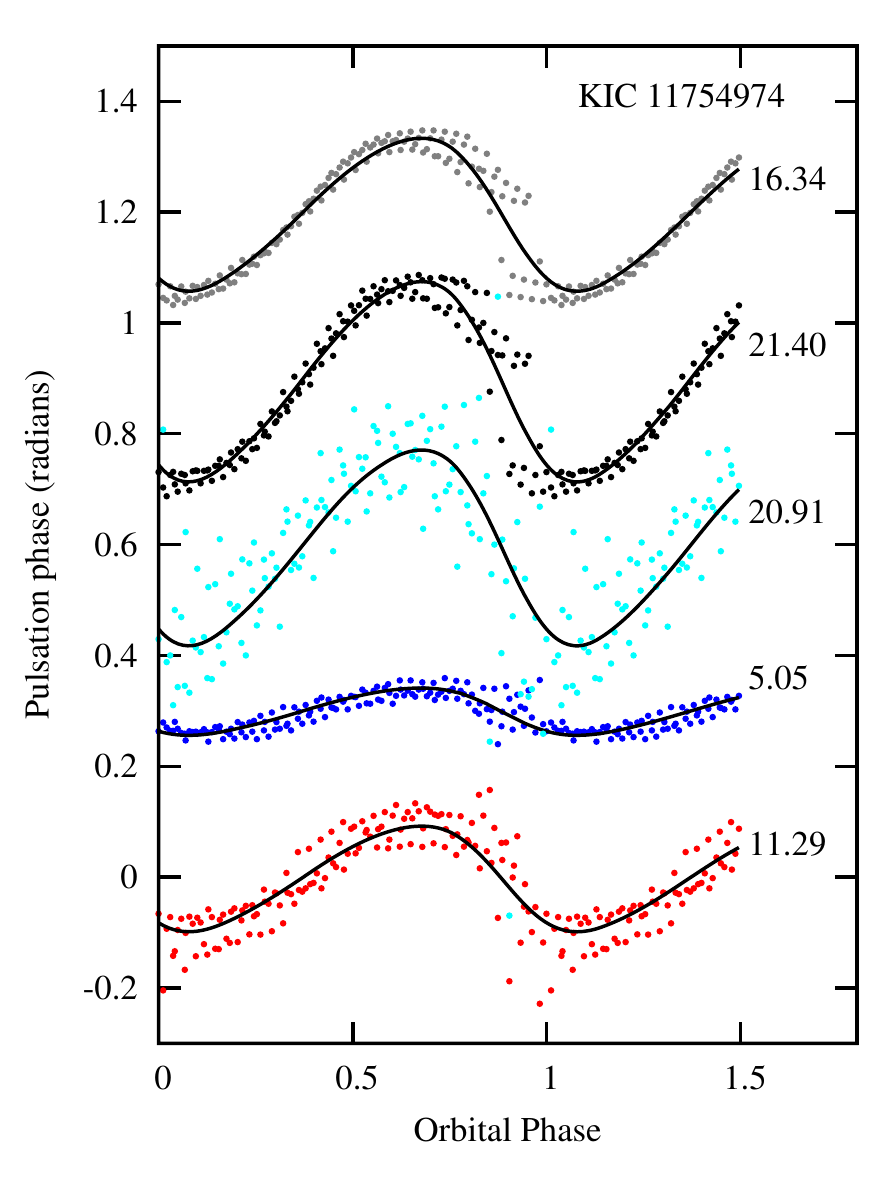} \end{overpic}
\caption{Time delay (TD) results for six binary stars: phase variation of the
pulsation modes as a function of the orbital phase.  For clarity an arbitrary
vertical displacement has been made (and different colour used) for each
pulsation frequency, and the particular frequency (units of d$^{-1}$) is given
on the right.  For KIC\,7300184 the periodogram shows only one pulsation
frequency of sufficiently high amplitude to reveal possible pulsation phase variations.
The solid (black) curves  are the fitted phase variations assuming the orbital
parameters given in Table~10. } \label{p4243461} 
\end{figure*}

\begin{figure*}
\begin{center}

\begin{overpic}[width=8.0cm]{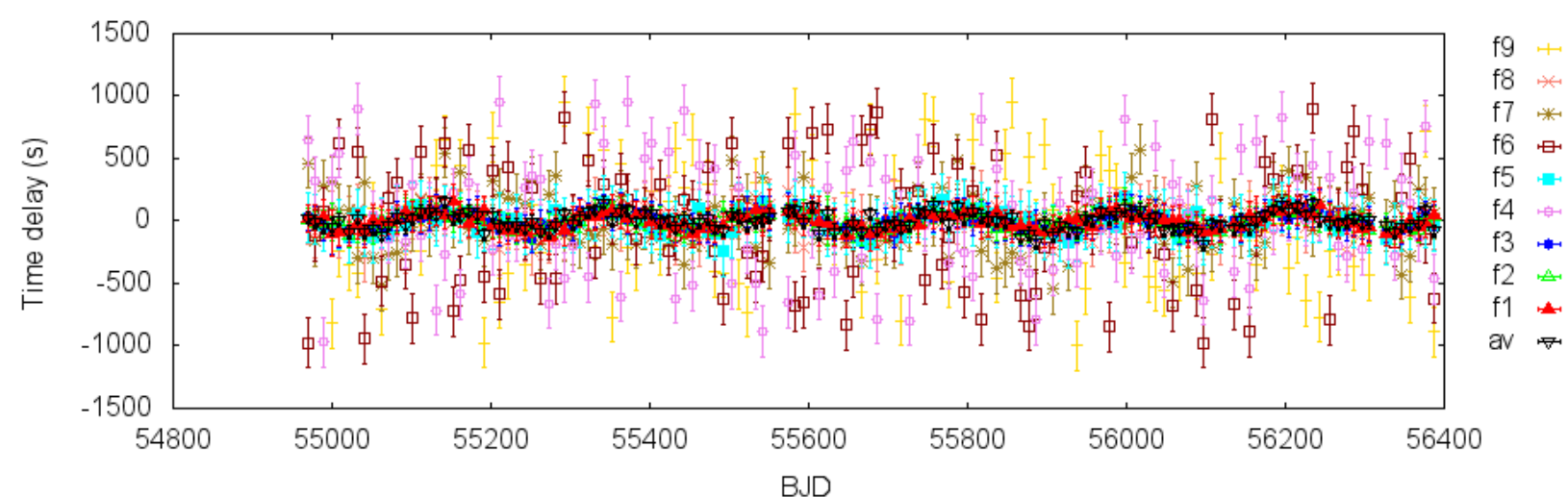} \end{overpic}
\begin{overpic}[width=8.0cm]{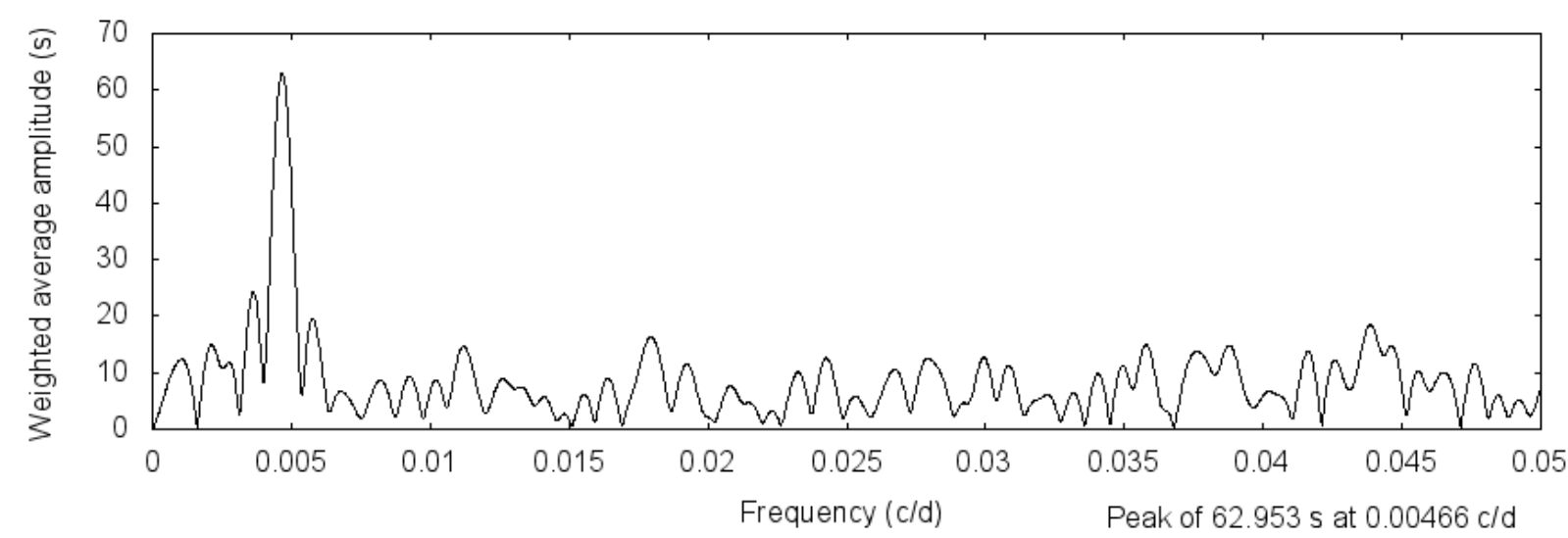} \put(35,20){KIC\,7819024 (*19) - $P_{\rm orb}$=215\,d} \end{overpic} 

\begin{overpic}[width=8.0cm]{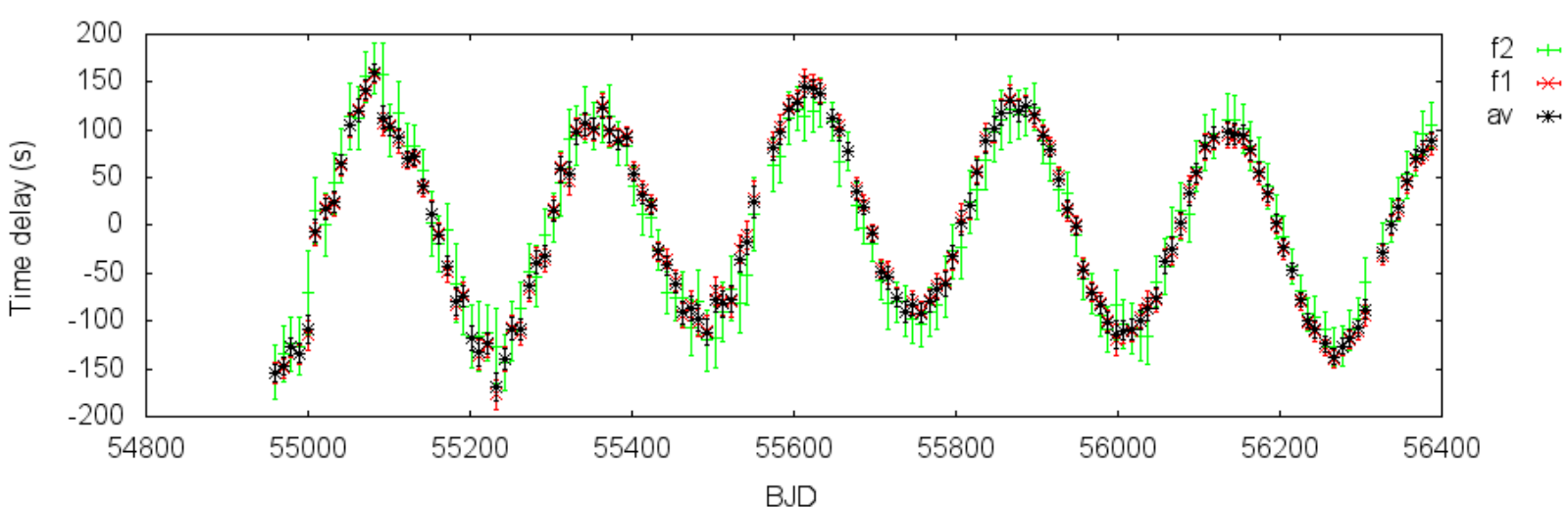} \end{overpic}
\begin{overpic}[width=8.0cm]{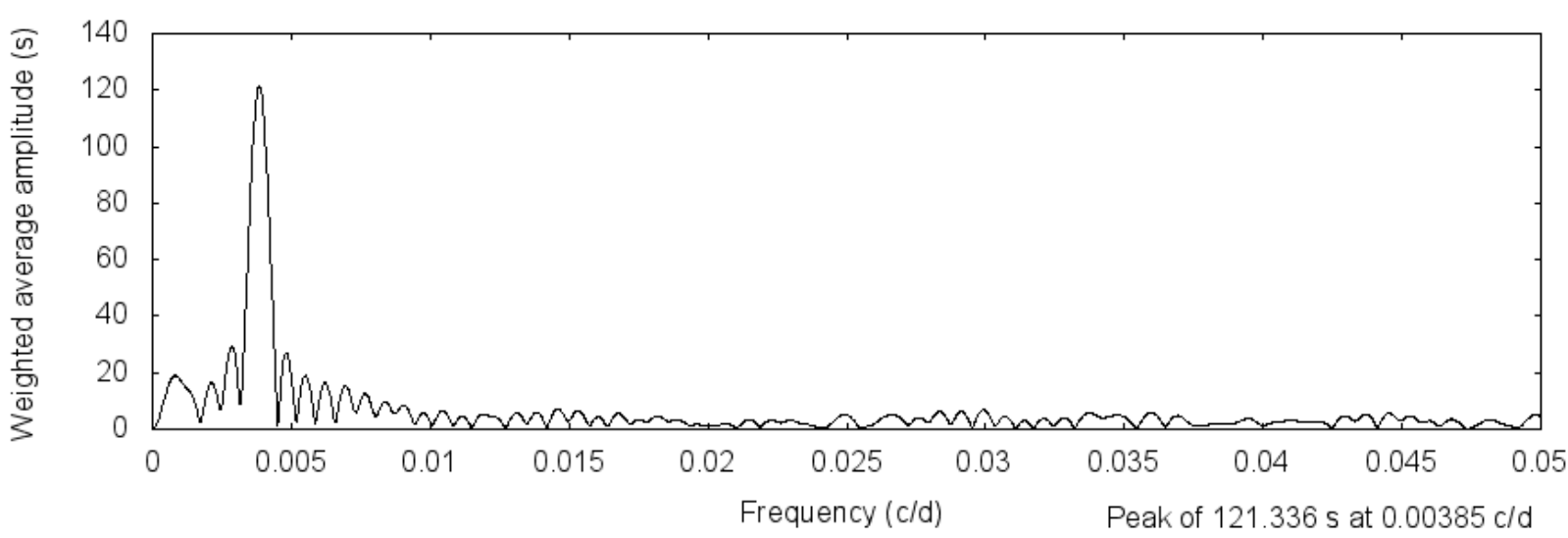} \put(35,20){KIC\,8004558 (*1) - 260\,d} \end{overpic} 

\begin{overpic}[width=8.0cm]{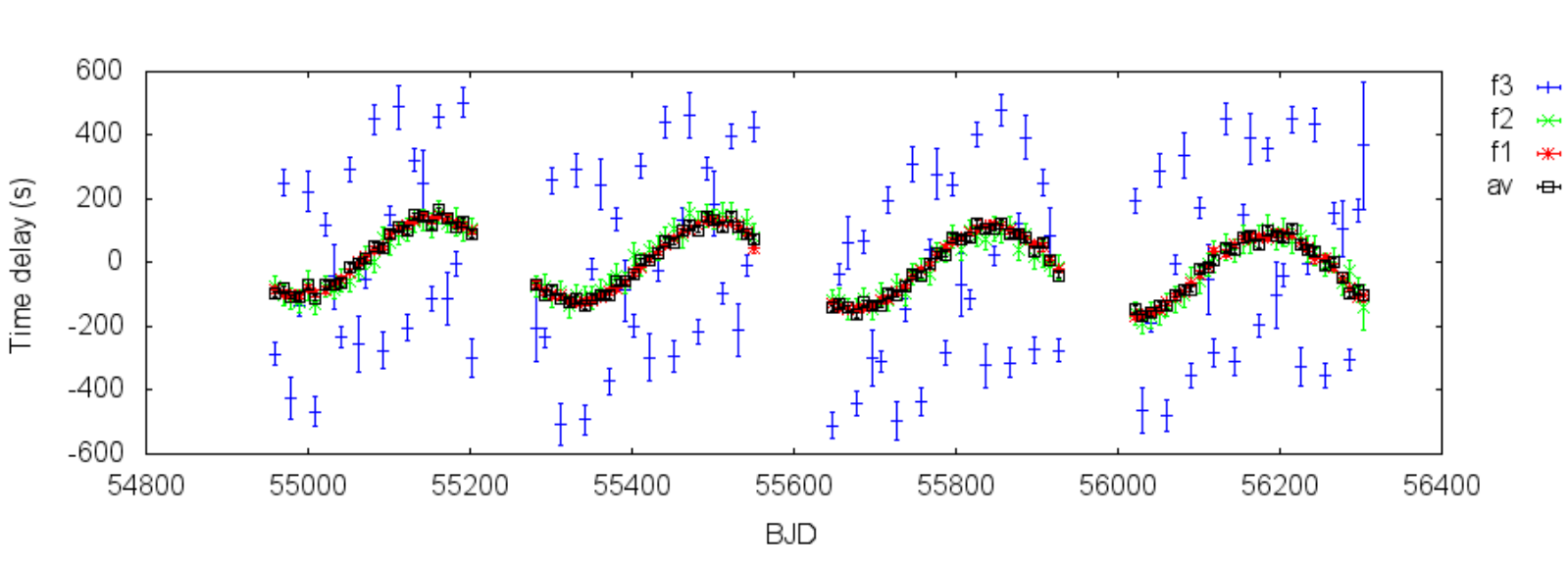} \end{overpic}
\begin{overpic}[width=8.0cm]{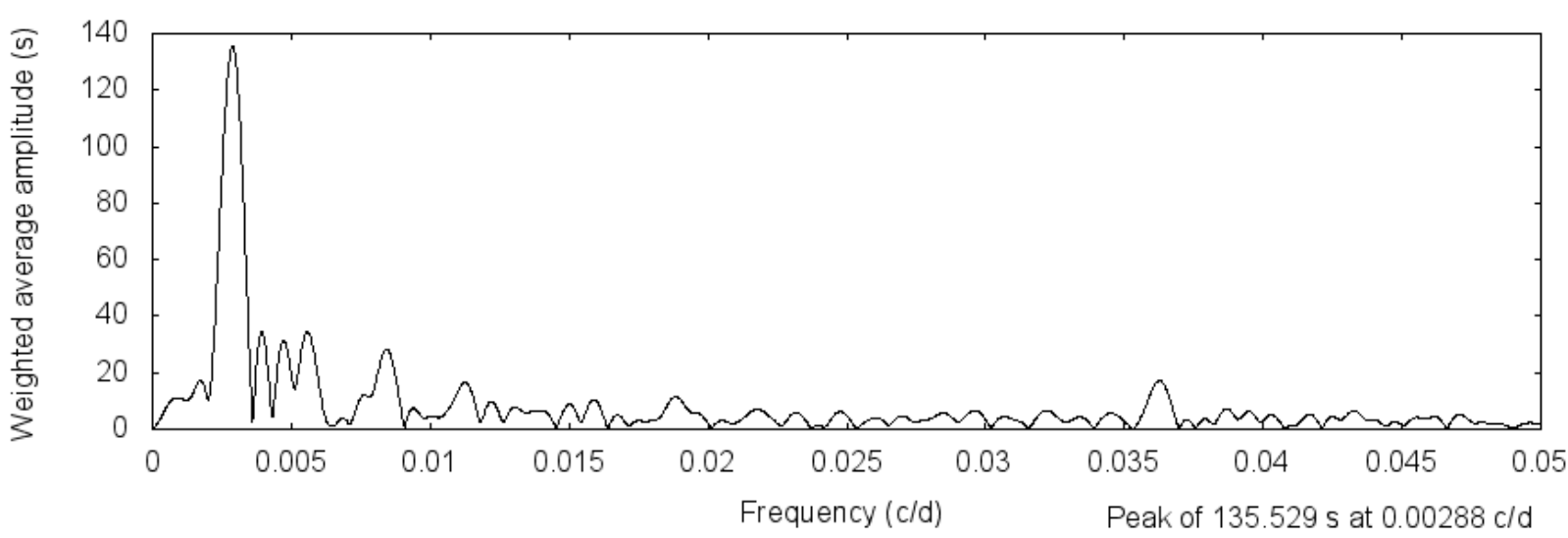} \put(35,20){KIC\,11754974 (*13) - 347\,d} \end{overpic} 

\begin{overpic}[width=8.0cm]{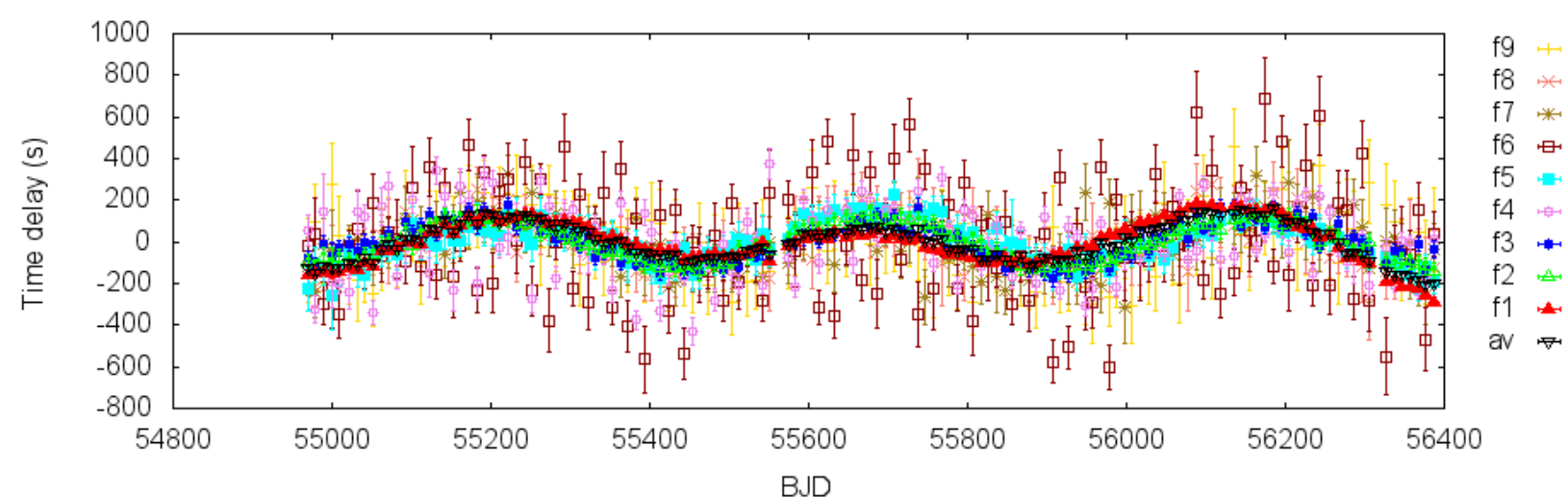} \end{overpic}
\begin{overpic}[width=8.0cm]  {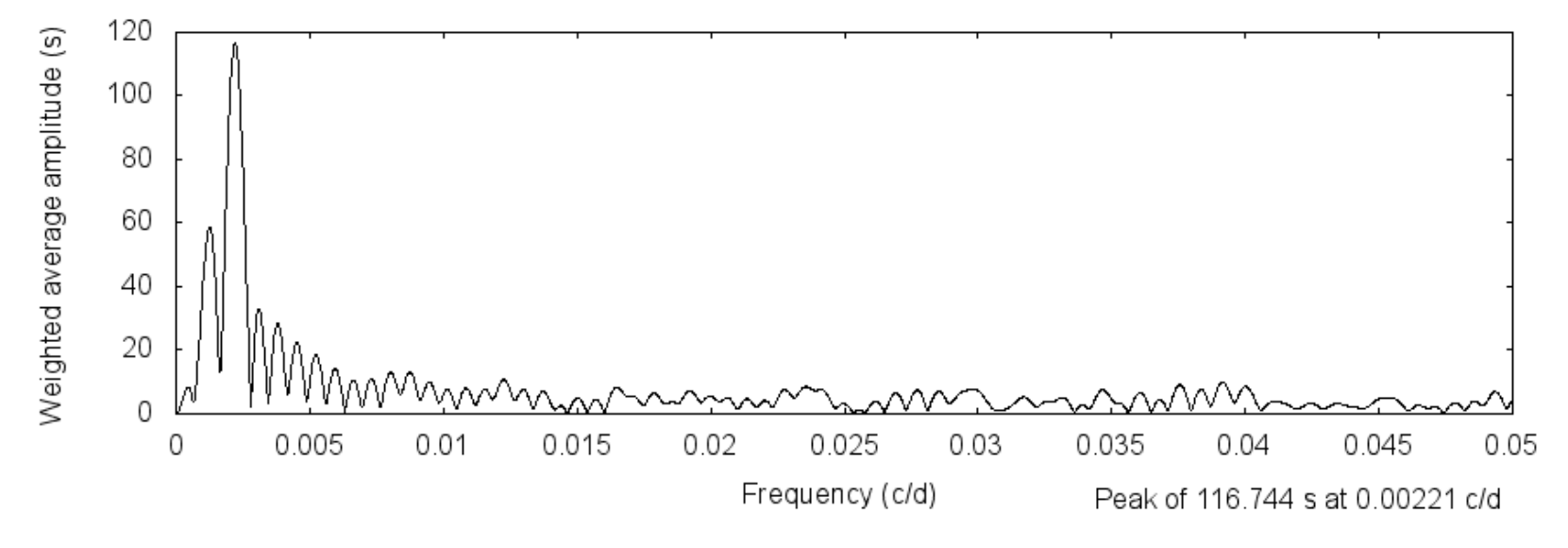} \put(35,20){KIC\,4243461 (*4) - 452\,d} \end{overpic} 

\begin{overpic}[width=8.0cm]{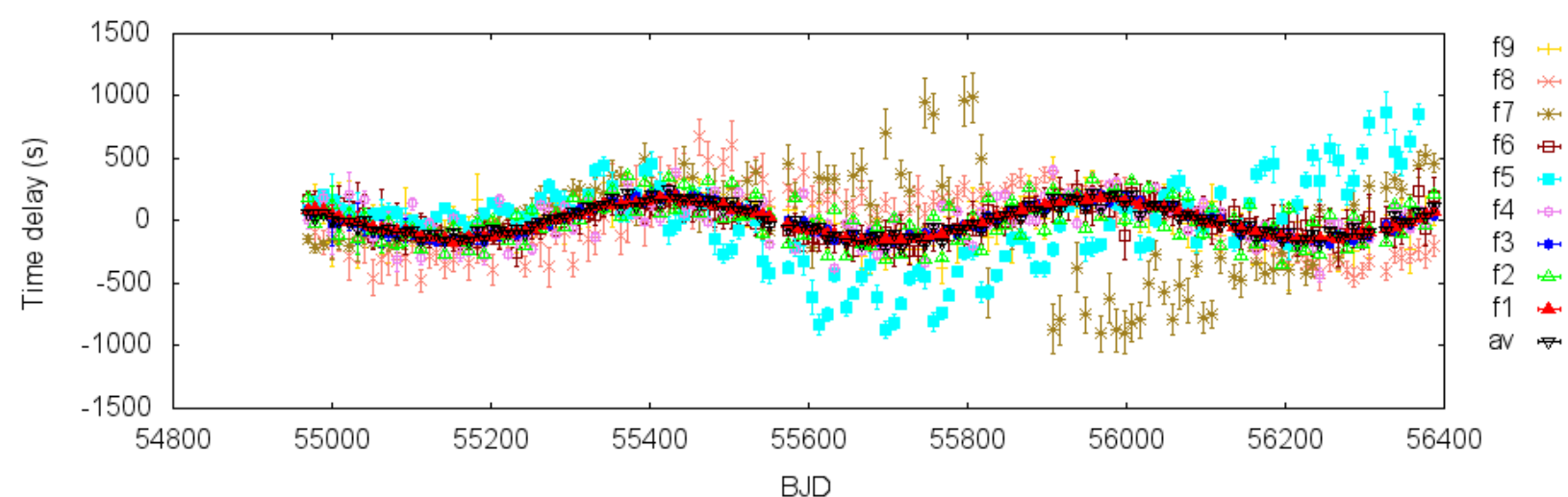} \end{overpic}
\begin{overpic}[width=8.0cm]{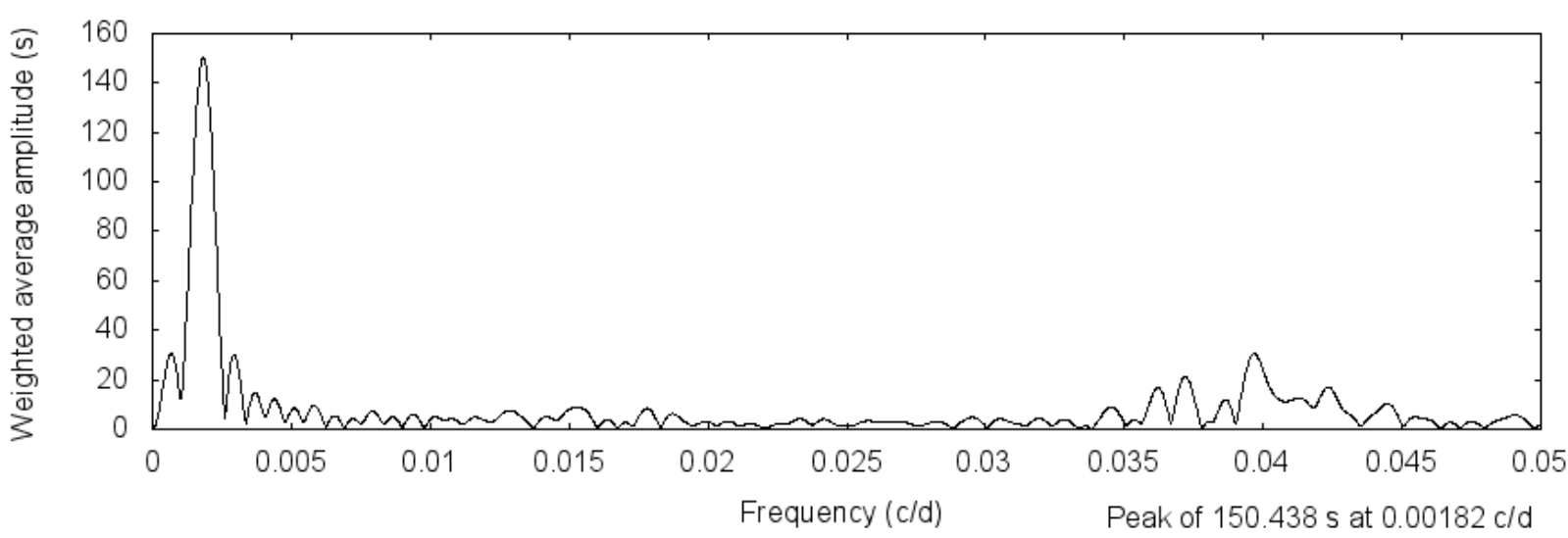} \put(35,20){KIC\,5705575 (*22) - 549\,d} \end{overpic} 

\begin{overpic}[width=8.0cm]{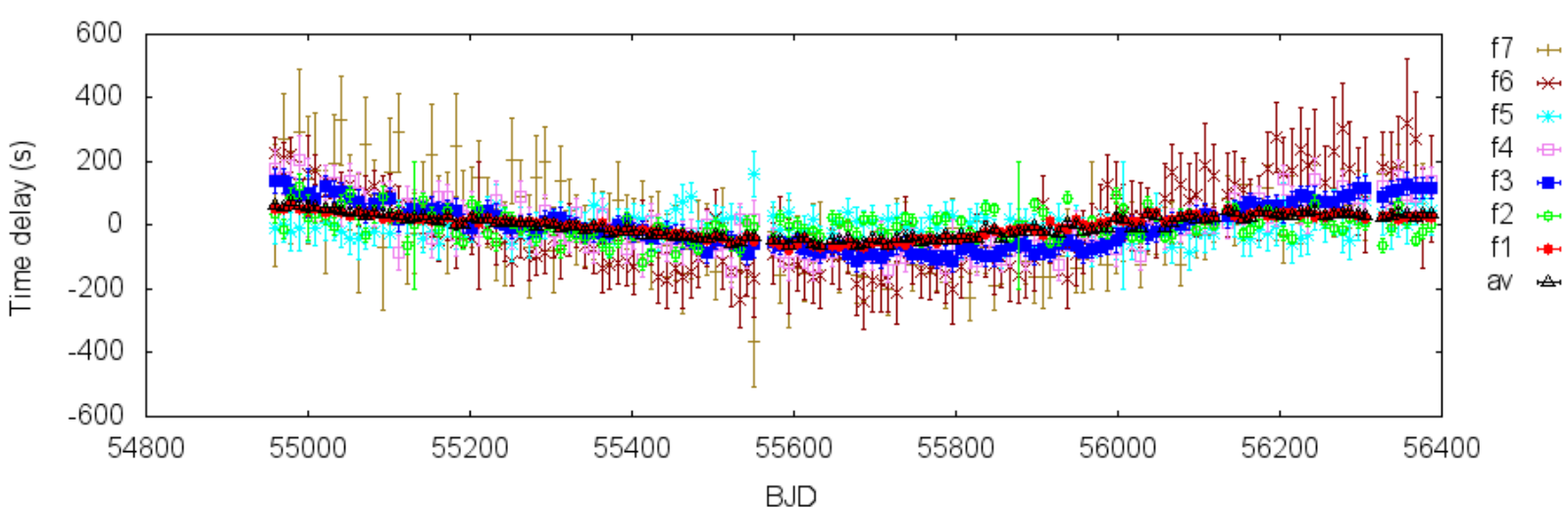} \end{overpic}
\begin{overpic}[width=8.0cm]{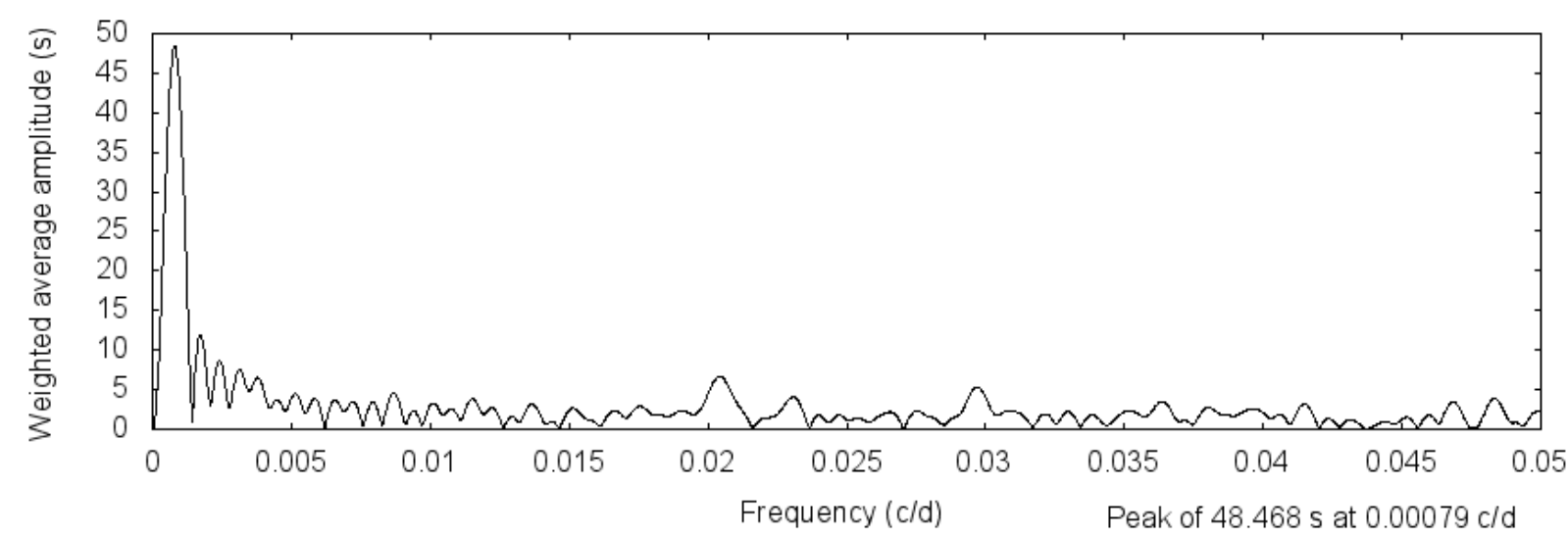} \put(35,20){KIC\,9267042 (*12) - $P_{\rm orb}$$>$1500\,d?} \end{overpic} 

\begin{overpic}[width=8.0cm]{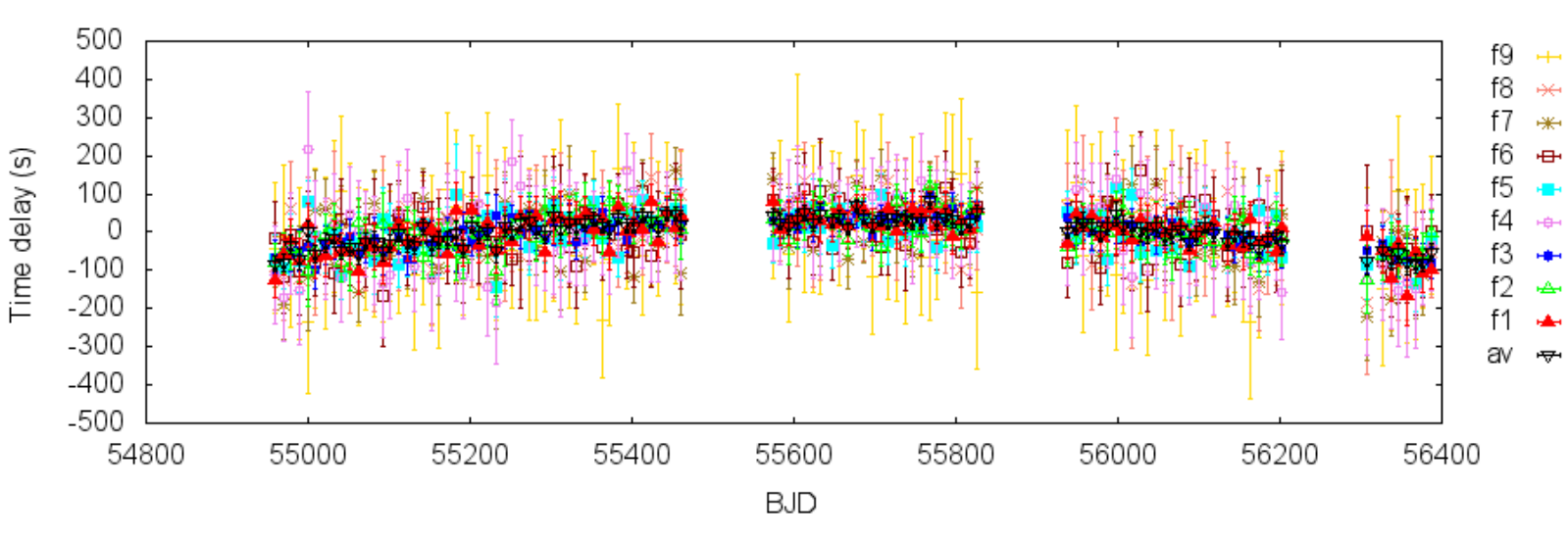} \end{overpic}
\begin{overpic}[width=8.0cm]{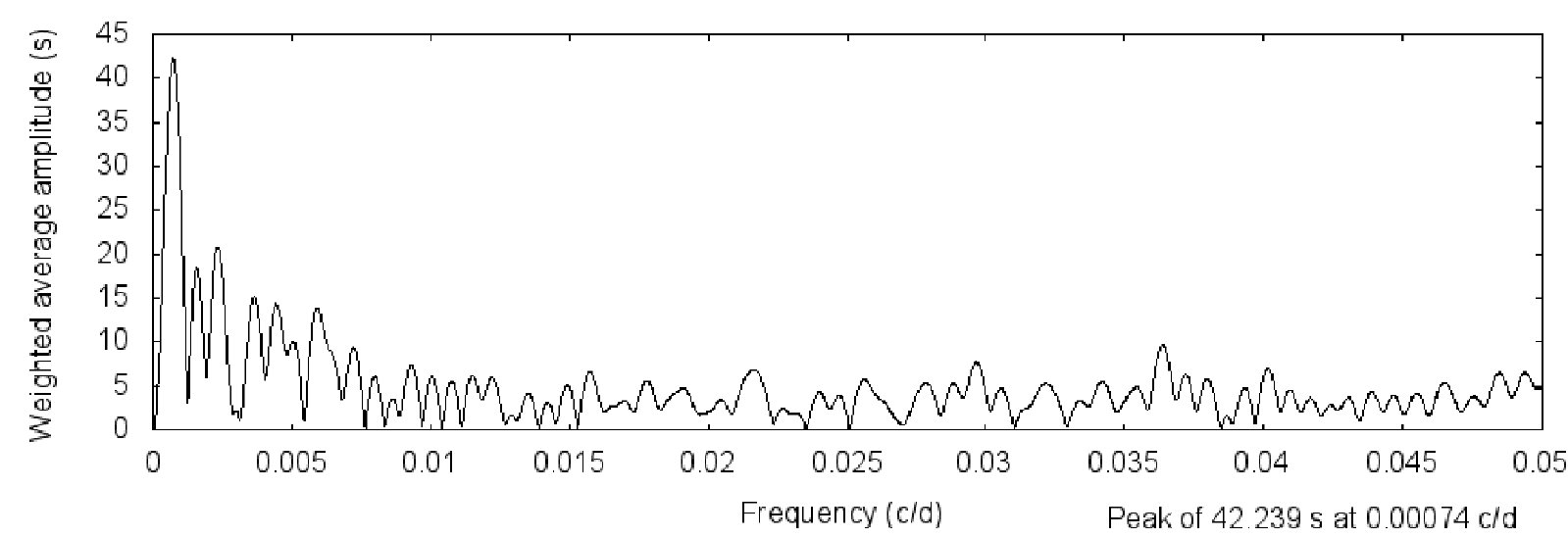} \put(35,20){KIC\,9966976 (*31) - $P_{\rm orb}$$>$1500\,d} \end{overpic} 

\caption{Phase modulation (PM) analysis results:  time-delay diagrams (left) and Fourier transforms of the weighted average time delays (right) 
for six of the photometric binary systems, including the two long-period stars.  The ordering (top to bottom) is by increasing orbital period.
For each star  the weighted-average time delays are shown with black symbols (inverted open triangles).  
} \end{center} 
\end{figure*}

In {\bf Figure\,22} time-delay diagrams and Fourier transforms of the weighted
average time delays from PM analyses are plotted for seven of the wide
binaries.    The panels are ordered by orbital period, shortest at the top and
longest at the bottom.   For each time-delay graph (left side) individual
extracted frequencies that were used for the analysis are plotted with
different symbols (see legend on right).   For all but the two longest period
binaries the null hypothesis that there is no time delay can be ruled out with
considerable confidence.  The panel for KIC\,11754974 (third row from the top)
shows  that  dominant pulsation frequencies do not always lead to the same
time-delay (or phase modulation) curve.  In  some cases  the chosen pulsation
frequencies may have been  contaminated by other close frequency peaks, which
would have had the effect of perturbing the phases at the sampling times and
obscuring the frequency variation due to binarity;  by analyzing multiple
frequencies this problem was largely overcome.

The derived  orbital periods for KIC\,9267042 (*12) and KIC\,9966976 (*31), if
real, are longer than the $\sim$4-year duration of the {\it Kepler} data set
and are  too long to be resolved by the FM method.  However,  phase
variations indicative of binarity were found using both the PM and TD methods
(see Table\,9).  The strongest argument that can be made for the reality of the
measured time delays is that in Fig.\,22 similar curvatures can be seen for all
of the  pulsation frequencies employed for the analyses, a finding that is
consistent with variation  caused by binary motion.  When KIC\,9966976 was
analyzed using the binarogram method the same $P_{\rm orb}$ was found for the
four frequencies that were considered, and similar values of $a \sin i$ were
derived for all four frequencies, which lends  considerable support to the
conclusion of its binary nature.  The evidence was less convincing for
KIC\,92670423;    while the same dominant peak ({\it i.e.}, $P_{\rm orb}$) was
found for the four test  frequencies, a range was seen in the amplitudes of the
associated peaks ({\it i.e.}, $a \sin i$) thereby reducing  the likelihood of
the binary conclusion.

{\bf Figure\,23} shows the results of PM+RV analyses for six  time-delay
binaries where both the phase modulations {\it and} the RVs were included in
the orbit solutions.  The graphs were created using the method outlined in
Murphy {\it et al.} (2016).  Inclusion of the RVs extended the time baselines
by almost two years, and added weight to several of the orbital solutions.  The
biggest improvement compared to the single-method results was for
KIC\,6780873\,(*5), the 9.1\,day double-lined spectroscopic binary;  in this
case,  the precision of the orbital period improved to a few tens of seconds
for a light travel time (across the orbit) of only a few tens of milli-seconds,
and the uncertainty of its orbital eccentricity improved to within 0.0005 of
zero (signifying circular orbits).  For  the binaries with large $v \sin i$
values the addition of the RVs to the solutions resulted in  minimal
improvement over the PM-only solutions;  and for KIC\,11754974 (*13) the
maximum time delay is now seen to be $\sim$140 sec, down from the $\sim$180\,s
noted earlier.

Masses and radii for the program stars, taken from Huber {\it et al.} (2014),
are given in {\bf Table\,11}.   The average mass, 1.70\,$\mathscr{M}_\odot$,
and  average radius, 2.25${R}_\odot$ (excluding  the extreme outlier
KIC\,6227118),  were adopted for the two stars that are missing from the Huber
{\it et al.} sample,  KIC\,5390069 and KIC\,7300184.  For the binary systems,
lower limits for the masses of the secondary stars are also in the table.
These were derived by assuming an orbital inclination angle, $i=90^{\circ}$
({\it i.e.}, edge-on), adopting the Huber {\it et al.} mass for the primary
star, and using the mass function $f_2(\mathscr{M}_1, \mathscr{M}_2, \sin i)$
given in Table\,9 to solve for the minimum mass of the secondary.  Assuming
1.5\,$\mathscr{M}_\odot$ for the KIC\,11754974 primary star mass,  instead of
the H14 value of 1.73\,$\mathscr{M}_\odot$, gives a slightly lower minimum mass
for the secondary, 0.43$\pm$0.15\,$\mathscr{M}_{\odot}$, still characteristic
of a K- or early-M main-sequence star.   The secondaries for
KIC\,7174372\,(*8), KIC\,9267042\,(*12) and KIC\,9966976\,(*31)  are seen to
have $\mathscr{M}_2$ lower limits below 0.08\,$\mathscr{M}_\odot$, which offers
the possibility that the companions are brown dwarfs.  

Based on our findings,  eight of the SX~Phe stars appear to have companions
with orbital periods between 200 and 1800 days, seven of which are new
discoveries (the binarity of KIC\,11754974 having already been discovered by
Murphy {\it et al.} 2013b).  The orbits for these stars are  probably  too
large for there to have been the mass transfer necessary to create the SX\,Phe
primary ({\it i.e.}, a pulsating blue straggler).  Of course, if one star
already entered the red giant phase it could now exist as an extreme horizontal
branch star with an exposed core, {\it i.e.}, an sdB or WD star.  It seems most
likely that the primary of these `wide' binaries is a coalesced star while the
secondary is the third star  in a triple system.   Indirect evidence for the
coalescence hypothesis comes from the presence in HR diagrams of various types
of binary systems seen among the BSs found in globular clusters and dwarf
galaxies (Mateo {\it et al.} 1990; Kaluzny {\it et al.}  2007, 2013;  Thompson {\it et al.}
2010; Rozyczka {\it et al.} 2010).   It would be interesting to know if any of
the program stars contain white dwarf companions of the sort recently observed
in three BSs in the 7\,Gyr old open cluster NGC~188 (Gosnell {\it et al.} 2014).

\begin{figure} 
\begin{center}
\begin{overpic}[width=8.0cm]{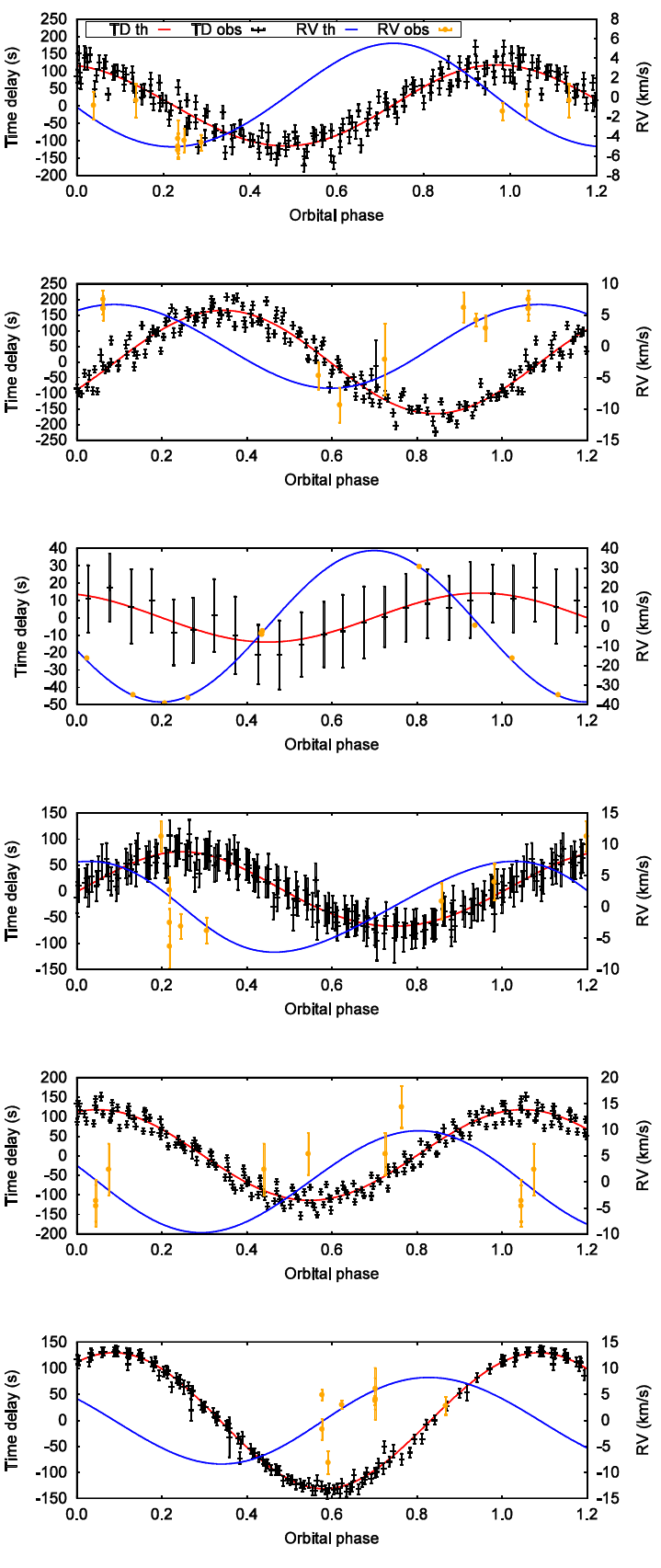} \end{overpic}
\end{center}

\caption{Phase modulation plus radial velocity (PM+RV) analysis results for six
wide binaries.  From top to bottom the curves shown are for the stars
KIC\,4243461 (*4), KIC\,5705575 (*22), KIC\,6780873 (*5), KIC\,7819024 (*19),
KIC\,8004558 (*1) and KIC\,11754974 (*13).  For each star time-delay and
radial-velocity curves are plotted, phased with the orbital periods given in
Table\,9.  The RVs have been offset by the systemic RVs.  Both  measured and
fitted time-delays (black points and red curves), and measured and fitted RVs
(ochre points and blue curves), are plotted. } \label{PM+RV} \end{figure}

\begin{table} \label{rotf} \caption{Radii and masses  for the 34 program stars.
The radius and mass of the primary, $R_1$ and $\mathscr{M}_1$, are from
Table\,5 of Huber {\it et al.} (2014).   The mass of the secondary,
$\mathscr{M}_2$, was calculated by combining the mass function,
$f_2$($\mathscr{M}_1$, $\mathscr{M}_2$, $\sin i$), from Table\,9, with the
H14 primary mass from column (4), and assuming $i=90^{\circ}$; thus the $\mathscr{M}_2$
are lower limits. The KIC numbers of the binary systems are in {\bf boldface}.  }
\centering
\begin{tabular}{lcccc}
\hline
\multicolumn{1}{c}{ KIC  }   & CFHT  & $R_1/R_{\odot}$ &  $\mathscr{M}_1/\mathscr{M}_\odot$  &   $\mathscr{M}_2/\mathscr{M}_\odot$  \\
\multicolumn{1}{c}{ no. }    &  no.  &   (H14)         &          (H14)                      & (this paper)  \\
\multicolumn{1}{c}{(1)} & (2) & (3) & (4) & (5)    \\
\hline
    1162150     & 15 & 4.31    & 2.09 & \\ 
    3456605     & 24 & 2.27    & 1.64 & \\
    4168579     & 23 & 2.78    & 1.94 & \\
  {\bf 4243461} & 4  & 1.55    & 1.39 & $>$\,0.27 \\ 
    4662336     & 14 & 2.54    & 1.77 & \\
    4756040     & 20 & 1.97    & 1.74 & \\
    5036493     & 26 & 2.42    & 2.00 & \\
    5390069     & -- & [2.25]  &[1.70]& \\
  {\bf 5705575} & 22 & 2.08    & 1.68 & $>$\,0.42 \\
    6130500     & 9  & 1.90    & 1.71 & \\
    6227118     & 27 & 11.32:  & 3.24:& \\
    6445601     & 2  & 1.80    & 1.52 & \\
    6520969     & 21 & 3.03    & 2.01 & \\
  {\bf 6780873} & 5  & 1.11    & 1.01 & $>$\,0.50  \\
    7020707     & 16 & 2.32    & 1.73 & \\
  {\bf 7174372} & 8  & 1.77    & 1.46 & $>$\,0.01 \\
  {\bf 7300184} & -- &[2.25]   &[1.70]&           \\
    7301640     & 10 & 1.78    & 1.56 & \\
    7621759     & 6  & 1.99    & 1.52 & \\
    7765585     & 28 & 1.37    & 1.40 & \\
  {\bf 7819024} & 19 & 1.83    & 1.53 & $>$\,0.31 \\
  {\bf 8004558} & 1  & 2.45    & 1.71 & $>$\,0.49 \\
    8110941     & 29 & 1.74    & 1.50 & \\
    8196006     & 30 & 1.35    & 1.27 & \\
    8330910     & 3  & 1.90    & 1.55 & \\
    9244992     & 7  & 3.93    & 1.87 & \\
  {\bf 9267042} & 12 & 3.12    & 2.03 & $>$\,0.06\\
 {\bf 9535881}  &[25]& 1.65    & 1.46 & $>$\,0.01 \\
 {\bf 9966976}  & 31 & 2.54    & 1.97 & $>$\,0.05\\
 {\bf 10989032} & 32 & 2.18    & 2.11 & $>$\,0.20 \\
   11649497     & 11 & 2.16    & 1.81 & \\
 {\bf 11754974} & 13 & 2.23    & 1.73 & $>$\,0.46  \\
 {\bf 12643589} &[17]& 1.12    & 1.15 & $>$\,0.01 \\
   12688835     & 18 & 2.88    & 1.97 & \\
\hline
\end{tabular}
\end{table}

\subsubsection{\bf Other SX\,Phe Binaries}

Approximately one third of the BN12 sample of {\it Kepler}-field SX\,Phe stars
(eleven of 30 stars) have been found to be in binary systems with orbital
periods ranging from a few days to several years.  While these are not the
first SX~Phe stars in binary systems to be identified, the long baseline (four
years) and high-precision of the photometry has allowed us to define with great
precision the orbits of several of the stars and their oscillations.  

Other presently known SX~Phe binaries (which exhibit a similarly large range of
separations and types) include:  (1) four ``blue metal poor''
(BMP) halo stars found by George Preston and his collaborators --
CS\,22966-043  (see Preston \& Landolt 1998, 1999), which has an orbital period
of $\sim$319 days and pulsations closely resembling those of the SX~Phe stars
in the very metal-poor globular cluster NGC~5053 (Nemec {\it et al.} 1995), and
CS\,22871-040, CS\,22896-103 and CS\,29499-057 (Preston \& Sneden 2000).
Curiously CS\,22896-103 is not metal-poor but has a metallicity similar to the
Sun, [Fe/H]=$-0.10$\,dex, and is one of $\sim$17 metal-rich BSs found among the
sample of BMP stars studied by Preston \& Sneden (see their Tables\,1 and 5).

(2) BL~Cam, which from an analysis of its O-C diagram, is  a multi-periodic
low-metallicity high-amplitude SX~Phe star, and was  found to be in a 144.2\,d close
binary system, possibly orbited by a brown dwarf having an orbital period $\sim$9.3
years  (Fauvaud {\it et al.} 2006, 2010); 

(3) QU\,Sge in M71 is the first SX~Phe binary found in a globular cluster
(Jeon {\it et al.} 2006; see McCormac {\it et al.} 2014).  It is a
semi-detached Algol-type eclipsing binary (orbital period 3.8 days) with ``the
secondary component fully filling its Roche lobe and the primary filling its
Roche lobe by about 33\%''.  After subtracting the eclipses the primary star
was seen to be an SX~Phe star  with two close frequencies, 35.883 d$^{-1}$ and
39.867 d$^{-1}$ (corresponding to periods of 40.2 and 36.1~min, respectively)
indicative of  non-radial pulsations.  Given that the mean [Fe/H] of M71 is
near $-0.80$ dex, with a small (0.03 dex) standard deviation that suggests
considerable homogeneity (Cordero {\it et al.} 2015, and references therein),
it is reasonable to assume that QU\,Sge too has a similar metal abundance.

Relaxing the distinction between SX~Phe and $\delta$~Sct stars, $\delta$~Sct
binary systems were first identified over 40 years ago.  The first such
identification was AB~Cas, a $\delta$~Sct star in a 1.37\,d Algol-type system
(Tempesti 1971; Rodr\'iguez {\it et al.} 1994, 1998), where  the $\delta$~Sct
star has a small amplitude range, $\Delta$$V$$\sim$0.05 mag, and appears to be
monoperiodic with $f=17.16$\,d$^{-1}$.  Another early identification was Y~Cam
(Broglia 1973; Broglia \& Conconi 1984; Rodriguez {\it et al.} 2010), which
shows several significant frequencies between 15 and 18 d$^{-1}$ (Kim {\it et
al.} 2002).      More recently, numerous  low-amplitude $\delta$~Sct stars in
eclipsing binary systems have been identified by Rodriguez \& Breger (2001),
Mkrtichian {\it et al.} (2005) and Soydugan {\it et al.} (2006).  The last of
these papers identified 25 EBs containing low-amplitude (0.007 to 0.02 mag)
$\delta$~Sct stars.  Derekas {\it et al.} (2009) have investigated the question
of binarity and multiperiodity in 10 $\delta$~Sct (HADS) stars.  Other recent
studies  include those by Southworth {\it et al.} (2011),  Lampens {\it et al.}
(2011),  Hambleton {\it et al.} (2013), Debosscher {\it et al.} (2013), and
Lee {\it et al.} (2016).

\section{SUMMARY}

Our goal was to characterize better the sample of 34 {\it Kepler}-field SX~Phe
stars identified by BN12.  High-resolution spectra for 32 of the 34 stars were
acquired and analyzed, and all the available Q0-Q17 long- and short-cadence
{\it Kepler} photometry for the 34 stars were re-analyzed. 

Radial velocities were derived from 184 program star spectra (calibrated with
24 spectra of IAU standard stars).   Approximately half of the stars show some
evidence for RV variability.  By combining the measured mean RVs with the
tangential motions,  $U,V,W$ space motions were derived.  Five  of the stars
were found to have large negative $V$-velocities, and 29 of the 32 stars have a
total space motion $T > 300$\,km/s.  All  of the stars lie in the `galactic
halo' region of the Toomre diagram (except possibly KIC\,6227118 which appears
to lie in the `thick disk' region).   

Also derived from the spectra were projected rotation velocities and
macroturbulent velocities.  Two thirds of the stars are fast rotators with
$v$\,sin\,$i$ $>$ 50~km/s, including four stars with $v$\,sin\,$i$ $\geq$ 200
km/s.  Several  stars were found to have  macroturbulent velocities in the range
10-30 km/s;  such turbulence may be an even larger contributor to  observed
broadening of  spectral lines than  rotation.  

Other atmospheric parameters that were measured  include $T_{\rm eff}$,
$\log\,g$, $v_{\rm mic}$ and [Fe/H];  these spectroscopic estimates improve
upon previous values, such as those that rely only on calibrations involving
photometry (KIC, H14).  The spectral types range from A2-F2, corresponding to
surface temperatures in the range 8600-6900\,K.  The mean metallicity of the
sample is near solar, with only a few of the stars having a marked metal
weakness;  in this sense the stars resemble the metal-rich A-stars (Perry 1969,
Preston 2015).  By analogy with metal-rich RR~Lyrae stars, such as those found
in metal-rich globular clusters the BN12 sample can be characterized as
`metal-rich SX Phe stars'.  In fact, the existence of metal-rich SX~Phe
stars complements the recent finding by Torrealba {\it et al.} 2015) that
suggests a higher concentration of more metal-rich stars near the galactic disk
(see their Figs.12-13).

SX~Phe stars have, until now, been thought of as ``metal-poor Population\,II
variables that have high space velocities and fall in the blue straggler domain
of the color-magnitude diagrams of globular clusters'' (McNamara 1995).  The
identification in this paper   of {\it bona fide} metal-rich SX~Phe stars suggests that
the definition of such stars should not be limited to ``metal-poor'' pulsators.

\section*{Acknowledgments} 

This paper is based on spectra acquired with the Canada-France-Hawaii 3.6-m
telescope (CFHT) and  the 3.5-m telescope at the Apache Point Observatory
(APO), and on photometry from NASA's {\it Kepler} mission and  the Korean 1.8-m
Bohyunsan Observatory.   CFHT is operated by the National Research Council of
Canada, the Institut National des Sciences de l'Univers of the Centre National
de la Recherche Scientifique of France, and the University of Hawaii.  The CFHT
pipeline, Upena, uses J.-F. Donati's software Libre-ESpRIT (Donati et al.
1997).  We wish to thank the CFHT time allocation committee for its generous
awards of observing time, and Nadine Manset and her team of service observers
for making the ESPaDOnS observations.  We also wish to thank Dr. Suzanne
Hawley, the director of the APO 3.5-m telescope which is  owned and operated by
the Astrophysical Research Consortium, for observing time, and the telescope
operators for their assistance.  Funding for the {\it Kepler} mission is
provided by the NASA Science Mission directorate.  The authors thank the {\it
Kepler} team for their generosity in allowing the data to be released and for
their outstanding efforts which have made these results possible.  JMN thanks
Federico Gonz\'alez for his `rcros' software (see D\'iaz {\it et al.} 2011)
that was used for deriving the $v$sin$i$ values,  Dr.\,Karen Kinemuchi for her
hospitality at the Apache Point Observatory and at New Mexico State University,
Dr. Sim\'on-Diaz for making public his IACOB software, and the referee for
useful comments.  A special thanks is owed to Dr.\,Amanda Linnell Nemec for
helpful suggestions and for making critical readings of the paper.  We also
thank Drs.\,Don Kurtz, Andrzej Pugulski, Conny Aerts and Patricia Lampens for
useful disussions.  JMN is grateful to the Camosun College Faculty Association
for financial assistance.  LAB wishes to thank the National Research Foundation
and the South African Astronomical Observatory for financial support.  SJM is
supported by the Australian Research Council, and by the Danish National
Research Foundation and ASTERISK project (grant agreement numbers: DNRF106 and
267864, respectively).  Some of the data presented in this paper were obtained
from the Multimission Archive at the Space Telescope Science Institute (MAST).
STScI is operated by the Association of Universities for Research in Astronomy,
Inc., under NASA contract NAS5-26555. Support for MAST for non-HST data is
provided by the NASA Office of Space Science via grant NNX09AF08G and by other
grants and contracts.  This research has also made use of the SIMBAD database,
operated at CDS, Strasbourg, France, of the Aladin Sky Atlas, and of the VALD3
database (operated at Uppsala University, the Institute of Astronomy RAS in
Moscow, and the University of Vienna).

\end{document}